\documentclass[
headinclude=true, 
headsepline=true,
twoside=true,
BCOR=10mm, 
    fontsize=10 pt,
paper=17cm:24cm, 
pagesize, 
DIV=14, 
headings=normal, 
appendixprefix=true,
draft=false,
numbers=noendperiod, 
toc=bibliography, 
parskip=false, 
captions=bottombeside,
version=last 
]{scrbook}

\setkomafont{disposition}{\normalcolor\bfseries} 

\usepackage{ifdraft}
\ifdraft{
	\usepackage[notref,notcite]{showkeys}
	\usepackage[obeyspaces,hyphens,spaces]{url}
	\renewcommand*{\showkeyslabelformat}[1]{%
		\fbox{\parbox{1.15cm}{\normalfont\small\ttfamily\url{#1}}}}
}{}

\addtokomafont{caption}{\small}

\renewenvironment{thebibliography}[1]{
  \begin{oldthebibliography}{#1}
    \setlength{\itemsep}{.3em}
    \setlength{\parskip}{0em}
}
{
  \end{oldthebibliography}
}

\makeatletter
\DeclareRobustCommand*{\bfseries}{%
   \not@math@alphabet\bfseries\mathbf
   \fontseries\bfdefault\selectfont
   \boldmath
}
\makeatother

\usepackage{tabularx}

\makeatletter
\renewcommand{\@chapapp}{}

\makeatother

\usepackage{subfiles}
\usepackage[dutch,american]{babel}
\usepackage[T1]{fontenc}
\usepackage{lmodern, microtype}
\usepackage{wrapfig}
\setcapindent{0pt}
\usepackage[labelfont=bf]{caption}

\usepackage{mathrsfs}
\usepackage[hang, flushmargin]{footmisc}
\usepackage[pdftex]{hyperref}
\usepackage{graphicx}
\usepackage{amssymb}
\usepackage{amsmath}
\usepackage{verbatim}
\usepackage{bm}
\usepackage[separate-uncertainty=true,multi-part-units=single]{siunitx}
\usepackage{cancel}
\usepackage{cite}
\usepackage{makecell}
\usepackage{booktabs}
\usepackage{arydshln}
\usepackage{braket}
\usepackage{multirow}
\usepackage{pdfpages}

\mathchardef\mhyphen="2D
\newcommand{\onlyinsubfile}[1]{#1}
\newcommand{\notinsubfile}[1]{}

\usepackage{footmisc} 
\usepackage{appendix}
\usepackage{chngcntr}
\usepackage{etoolbox}
\AtBeginEnvironment{subappendices}{
\section*{Appendices}
\counterwithin{figure}{chapter}
\counterwithin{table}{chapter}
}

\def\blfootnote{\xdef\@thefnmark{}\@footnotetext}
\long\def\symbolfootnote[#1]#2{\begingroup%
\def\thefootnote{\fnsymbol{footnote}}\footnote[#1]{#2}\endgroup}

\usepackage[dvipsnames,xcdraw]{xcolor}

\DeclareOldFontCommand{\rm}{\normalfont\rmfamily}{\mathrm}
\DeclareOldFontCommand{\sf}{\normalfont\sffamily}{\mathsf}
\DeclareOldFontCommand{\tt}{\normalfont\ttfamily}{\mathtt}
\DeclareOldFontCommand{\bf}{\normalfont\bfseries}{\mathbf}
\DeclareOldFontCommand{\it}{\normalfont\itshape}{\mathit}
\DeclareOldFontCommand{\sl}{\normalfont\slshape}{\@nomath\sl}
\DeclareOldFontCommand{\sc}{\normalfont\scshape}{\@nomath\sc}

\newcommand{\beq}{\begin{equation}}  \newcommand{\eeq}{\end{equation}}
\newcommand{\bal}{\begin{aligned}}   \newcommand{\eal}{\end{aligned}}
\newcommand{\bea}{\begin{eqnarray}}  \newcommand{\eea}{\end{eqnarray}}

\newcommand{\bmat}{\left(\begin{array}}
\newcommand{\emat}{\end{array}\right)}

\newcommand{\bbC}{\mathbb{C}}
\newcommand{\bbR}{\mathbb{R}}
\newcommand{\bbH}{\mathbb{H}}
\newcommand{\bbQ}{\mathbb{Q}}
\newcommand{\bbN}{\mathbb{N}}

\newcommand{\cO}{\mathcal{O}}
\newcommand{\cU}{\mathcal{U}}
\newcommand{\cT}{\mathcal{T}}
\newcommand{\cE}{\mathcal{E}}

\newcommand{\cC}{\mathcal{C}}
\newcommand{\cD}{\mathcal{D}}
\newcommand{\cL}{\mathcal{L}}
\newcommand{\cS}{\mathcal{S}}
\newcommand{\cK}{\mathcal{K}}
\newcommand{\cN}{\mathcal{N}}
\newcommand{\cX}{\mathcal{X}}
\newcommand{\cW}{\mathcal{W}}
\newcommand{\cG}{\mathcal{G}}
\newcommand{\cA}{\mathcal{A}}

\newcommand{\cB}{\mathcal{B}}
\newcommand{\cF}{\mathcal{F}}

\newcommand{\cR}{\mathcal{R}}

\newcommand{\cV}{\mathcal{V}}

\newcommand{\cM}{\mathcal M}

\newcommand{\eps}{\varepsilon}

\newcommand{\fg}{\mathfrak{g}}

\newcommand{\fk}{\mathfrak{k}}
\newcommand{\fp}{\mathfrak{p}}

\newcommand{\pd}{\partial}

\renewcommand{\Im}{\mathrm{Im}\,}
\renewcommand{\Re}{\mathrm{Re}\,}

\usepackage{cancel}

\newcommand{\be}{\begin{equation}}
\newcommand{\ee}{\end{equation}}

\newcommand{\half}{\frac{1}{2}}

\newcommand{\bbZ}{\mathbb{Z}}

\usepackage{graphbox}
\usepackage{bbm}

\newcommand{\dd}{\mathrm{d}}

\usepackage{ytableau}

\usepackage{dsfont}

\usepackage{tensor}
\usepackage{amsthm}
\usepackage{quiver}
\usepackage{tikz-cd}
\usepackage{physics}
\usepackage{slashed}
\usetikzlibrary{babel}
\usepackage{mathtools}
\usepackage{adjustbox}
\usepackage{epigraph}

\usepackage[margin=0pt,font=small,labelfont=normalfont,skip=22pt]{subcaption}

\def\Im{\mathop{\mathrm{Im}}\nolimits}
\def\Re{\mathop{\mathrm{Re}}\nolimits}

\usepackage{thmtools}
\usepackage{mdframed}

\usepackage{enumerate}
\usepackage{tkz-euclide}
\usepackage{pifont}
\usepackage[many]{tcolorbox}

\tcbset {
	base/.style={
		arc=0mm, 
		bottomtitle=0.5mm,
		boxrule=0mm,
		colbacktitle=black!10!white, 
		coltitle=black, 
		fonttitle=\bfseries, 
		left=2.5mm,
		leftrule=1mm,
		right=3.5mm,
		title={#1},
		toptitle=0.75mm,
		breakable
	}
}

\definecolor{brandblue}{rgb}{0.34, 0.7, 1}
\definecolor{carmine}{rgb}{0.59, 0.0, 0.09}

\newtcolorbox{redbox}[1]{
	colframe=carmine, 
	base={#1}
}

\newtcolorbox{mainbox}[1]{
	colframe=brandred, 
	base={#1}
}

\newtcolorbox{subbox}[1]{
	colframe=black!30!white,
	base={#1}
}

\setkomafont{chapter}{\normalfont\huge\bfseries}
\setkomafont{chapterprefix}{\normalfont\Large}

\renewcommand{\chapterformat}{%
  Chapter~\thechapter%
}

\renewcommand*{\chapterheadendvskip}{\vspace{1cm}}

\RedeclareSectionCommand[
  beforeskip=0cm,
  afterskip=1cm,
  font=\LARGE\bfseries,
]{chapter}

\makeatletter
\renewcommand*{\@@makechapterhead}[1]{%
  {\parindent \z@ \raggedright
   \normalfont
   \centering \usekomafont{chapterprefix}\chapterformat\par
   \vskip .5\baselineskip
   \hrule
   \vskip .5\baselineskip
   \centering \usekomafont{chapter}#1\par
   \nobreak
   \vskip .5\baselineskip
   \hrule
   \chapterheadendvskip
  }}
\makeatother

\begin{document}

\includepdf[pages=1]{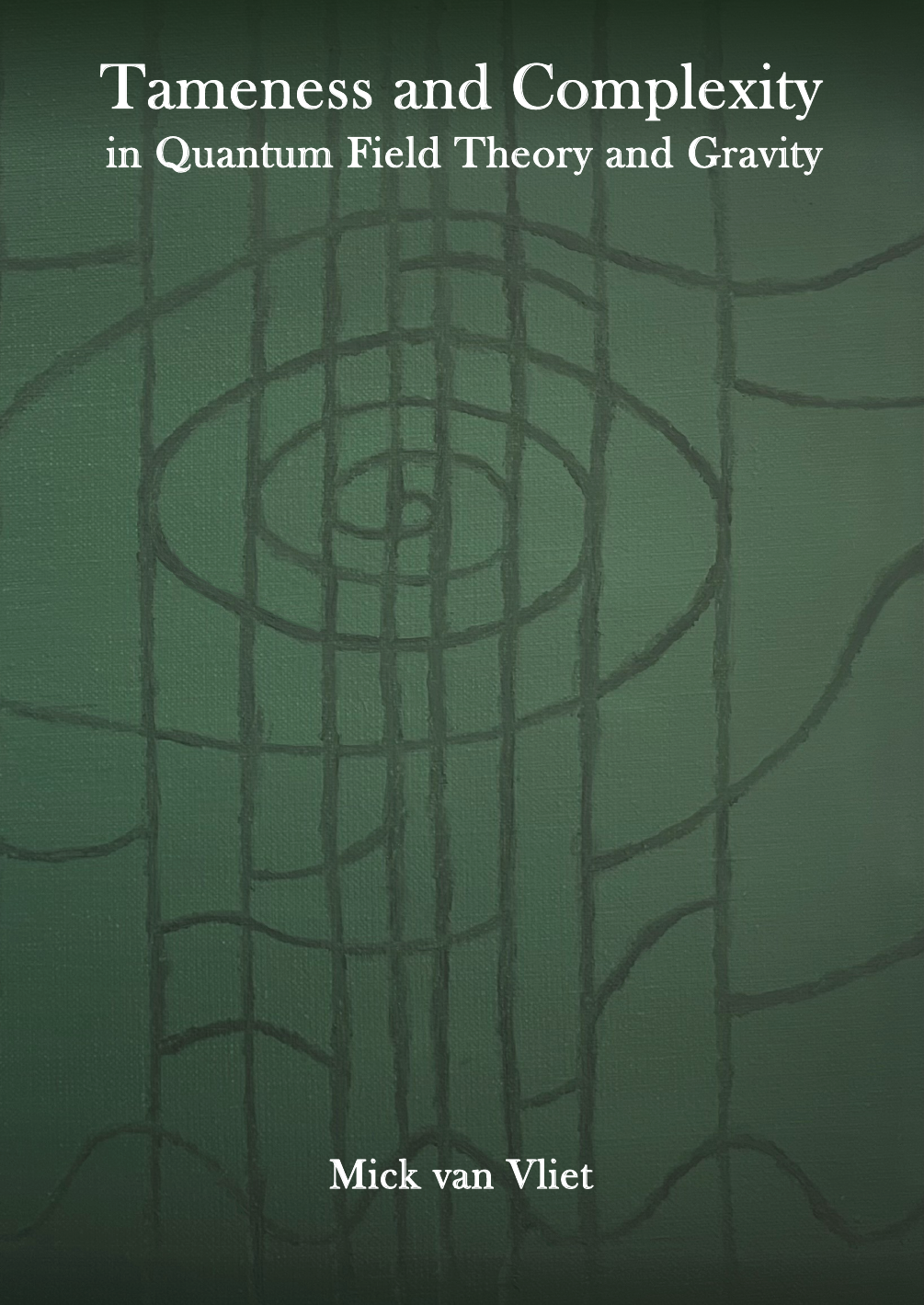}

\renewcommand{\onlyinsubfile}[1]{}
\renewcommand{\notinsubfile}[1]{#1}

\thispagestyle{empty}
\vspace*{4cm}

{\centering
{\huge \textbf{Tameness and Complexity}}\vspace{.2cm} \\
{\Large \textbf{in Quantum Field Theory and Gravity}}

\vspace{1.45cm}
{\LARGE Mick van Vliet}\\
}
\newpage~\thispagestyle{empty}
\vfill
\noindent 
PhD thesis, Utrecht University, September 2026
\vspace{1cm}

\noindent ISBN: 978-94-6536-208-3
\vspace{1cm}

\noindent DOI: 10.33540/3736
\vspace{1cm}

\vspace{0.5cm}

\noindent\textbf{About the cover:} In the theory of tame geometry, mathematical structures can be divided into simple pieces called \textit{cells}. 
The cover illustrates a cell decomposition, partitioning various geometric shapes such as spirals and oscillating curves into simple segments. Mathematically, these curves could extend endlessly, but they are limited by the finite size and resolution of the cover.
The composition thereby symbolizes the contrast between infinity in mathematics and finiteness of complexity in physics, which is a central theme of this thesis. Painted by the author.

\newpage

\thispagestyle{empty}
\vspace*{0.2cm}

\begin{center}
{\huge \textbf{Tameness and Complexity \\
\vspace{2mm}
\Large in Quantum Field Theory and Gravity} }\vspace{.2cm} \\
\vspace{1.45cm}

{\huge Tamheid en Complexiteit \\ 
\vspace{2mm}
\Large in Kwantumveldentheorie en Zwaartekracht}
\vspace{.2cm}  \\

(met een samenvatting in het Nederlands)\vspace{2.2cm}\\
 
{\LARGE Proefschrift}\vspace{0.8cm}\\
\end{center}

{\large{{ \noindent ter verkrijging van de graad van doctor aan de Universiteit Utrecht op gezag van de rector magnificus, prof.~dr.~ir.~W.~Hazeleger, ingevolge het besluit van het College voor Promoties in het openbaar te verdedigen op vrijdag 11 september des middags te 12:15 uur}}}\vspace{0.4cm}\\
{\centering
{\large door}\vspace{0.6cm}\\
{\LARGE Mick 
van Vliet}\vspace{.6cm}\\
\par}

{
 \newpage \thispagestyle{empty}
 {\raggedright
{\Large \textbf{Promotoren:}}\\
{\large Prof.~dr.~T.W.~Grimm}\\
{\large Prof.~dr.~S.J.G.~Vandoren}}


\vspace{1cm}
\noindent
{\Large \textbf{Beoordelingscommissie:}}\\
{\large Prof.~dr.~V.~Balasubramanian}\\
{\large Prof.~dr.~G.R.~Cavalcanti}\\ 
{\large Dr.~N.E.~Chisari}\\ 
{\large Prof.~dr.~R.~Cluckers}\\ 
{\large Dr.~M.~Montero}\\ 
}
\vfill



\newpage

\section*{Abstract}
\noindent 
Finiteness appears to play a fundamental role in the mathematical structures underlying the laws of physics.
Tame geometry, defined by the theory of o-minimality, provides a framework for making this idea precise. In this thesis we study the concept of tameness and finiteness of complexity across applications in quantum field theory and quantum gravity, and demonstrate that many classes of physical functions and theories are tame. To quantify these ideas, we use the novel theory of sharp o-minimality to measure the complexity of physical objects. The analysis covers wavefunctions in quantum mechanics, classical field configurations, non-perturbative observables in zero-dimensional quantum field theories, cosmological correlation functions, and Lagrangians of effective field theories.
Finally, we argue that the emergence of tameness in physics aligns with manifestations of finiteness in quantum gravity, converging towards the idea that effective theories of quantum gravity admit a description of finite complexity. 
\thispagestyle{empty}

\frontmatter

\chapter*{Publications}
\noindent The ideas presented in this thesis arose in collaboration with Thomas Grimm, Arno Hoefnagels, David Prieto, Giovanni Ravazzini, and Lorenz Schlechter.\\

\noindent\textbf{Part I} of this thesis provides an introduction to tameness and complexity, made precise by the theory of o-minimality. It is based on existing mathematical literature, but also contains new perspectives drawing from the works \cite{Grimm:2023xqy,Grimm:2024hdx,Grimm:2024mbw,Grimm:2024elq,Grimm:2025lip,Grimm:2026haa,Grimm:2025zhv} listed below.
\\

\noindent\textbf{Part II} of this thesis concerns the application of tameness and complexity to quantum field theory. It is based on the following publications:
\begin{itemize}
	\item[\cite{Grimm:2023xqy}] Thomas W. Grimm, Lorenz Schlechter, Mick van Vliet: \emph{Complexity in tame quantum theories}, \textbf{JHEP 05 (2024) 001},  \href{https://arxiv.org/abs/2310.01484}{\textbf{[arXiv: 2310.01484]}}
    \item[\cite{Grimm:2024hdx}] Thomas W. Grimm, Giovanni Ravazzini, Mick van Vliet: \emph{Taming non-analyticities in QFT observables}, \textbf{JHEP 02 (2025) 009},  \href{https://arxiv.org/abs/2407.08815}{\textbf{[arXiv: 2407.08815]}}
    \item[\cite{Grimm:2024mbw}] Thomas W. Grimm, Arno Hoefnagels, Mick van Vliet: \emph{Structure and complexity of cosmological correlators}, \textbf{Phys. Rev. D 110 (2024) 123531}, \\ \href{https://arxiv.org/abs/2404.03716}{\textbf{[arXiv: 2404.03716]}}
    \item[\cite{Grimm:2024elq}] Thomas W. Grimm, Mick van Vliet: \emph{On the complexity of quantum field theory}, \textbf{JHEP 06 (2025) 215},  \href{https://arxiv.org/abs/2410.23338}{\textbf{[arXiv: 2410.23338]}}
\end{itemize}

\noindent\textbf{Part III} of this thesis is devoted to the connection between tameness, complexity and quantum gravity. It is based on the following publications:
\begin{itemize}
	\item[\cite{Grimm:2025lip}] Thomas W. Grimm, David Prieto, Mick van Vliet: \emph{Tame embeddings, volume growth, and complexity of moduli spaces}, \textbf{Phys. Rev. D 112 (2025) 106015}, \href{https://arxiv.org/abs/2503.15601}{\textbf{[arXiv: 2503.156015]}}
    \item[\cite{Grimm:2026haa}] Thomas W. Grimm, David Prieto, Mick van Vliet: \emph{Tame complexity of effective field theories
    in the quantum gravity landscape}, \textbf{JHEP 06 (2026) 192}, \href{https://arxiv.org/abs/2601.18863}{\textbf{[arXiv: 2601.18863]}}
\end{itemize}

\newpage

\noindent Another publication on tameness and complexity to which the author contributed, which is not covered in detail in this thesis, is the following:
\begin{itemize}
\item[\cite{Grimm:2025zhv}] Thomas W. Grimm, Arno Hoefnagels, Mick van Vliet: \emph{A reduction algorithm for cosmological correlators: cuts, contractions, and complexity}, \textbf{JHEP 03 (2026) 208},  \href{https://arxiv.org/abs/2503.05866}{\textbf{[arXiv: 2503.05866]}}
\end{itemize}

\tableofcontents

\newpage 
\vspace{1cm}

\mainmatter


{
\renewcommand{\chapterformat}{\relax} 
\addchap{Introduction}
}
\setlength{\parindent}{0pt}
The ultimate goal of physics is to understand and describe the laws of nature which govern all phenomena in our universe. Our understanding advances
through empirical observation and theoretical reasoning,
and to record our understanding, we require a language in which these physical laws can be formulated. 
A natural philosophical question which then emerges is:
\[
\textit{what is the language of physics?}
\]

This is an ambitious question with a long history. 
An answer has already been given four centuries ago by Galileo, who stated that the universe is written in the language of mathematics. 
The striking capability of mathematics to describe the universe has been emphasized by Wigner as the \textit{unreasonable effectiveness of mathematics in the natural sciences}, and the symbiosis between these two fields has arguably never been as successful as it is today.
Despite this incredibly strong connection, physics and mathematics remain fundamentally different.  
Ultimately, physics is anchored in reality, while mathematics in principle 
has no such limitations. With this in mind, we may expect that only a subset of mathematics is needed to formulate the laws of nature, and that the answer to our question 
can be refined.\\

The limitations which distinguish physics from the generality of mathematics can be characterized by \textit{finiteness}. At an operational level, we only have access to finitely many experimental measurements, providing us with only a finite amount of information on the physical laws. Already from this angle we might therefore hope that these laws can be mathematically described with finite information. An example of this limitation is the distinction between the renormalizable quantum field theories and the non-renormalizable ones, whose uncontrollable quantum \mbox{corrections} cannot be mastered with a finite number of parameters. In addition to these pragmatic considerations, the finiteness of information in physics appears in fact to be fundamental. This is demonstrated by the Bekenstein bound,
which states that the entropy of a physical configuration with a bounded radius and energy is finite; it cannot exceed the Bekenstein-Hawking entropy of a black hole. Ideas of this type indicate that the combination of quantum mechanics and gravity enforce a finite information density in physical systems.\\

\begin{figure}[h]
    \centering
    \includegraphics[width=1\linewidth]{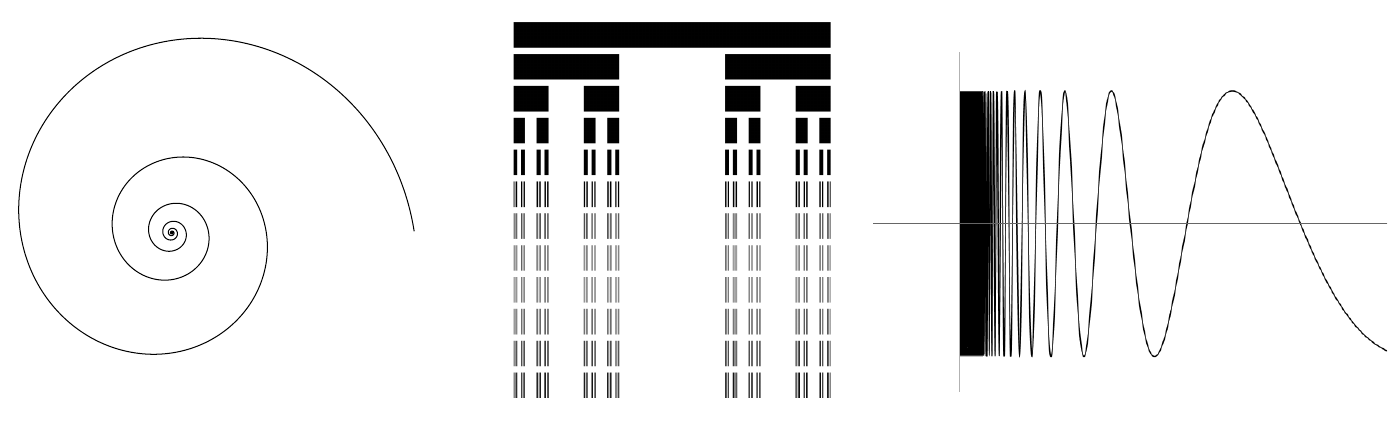}
    \caption{Examples of wildly behaving mathematical objects: an infinite spiral, the Cantor set, and the topologist's sine curve.}
    \label{fig:Wildsets}
\end{figure}

In contrast, mathematics admits wild objects with pathological infinities contained within a bounded region.
Infinitely complex objects such as those in the figure above do not seem to appear in physics. Instead, it seems that geometric objects in physics have a certain regularity which avoids pathologies of infinite complexity. On these grounds it appears that the mathematical language of physics should be constrained by a notion of \textit{tameness}, which makes the finiteness of information in physics precise. \\

A natural framework in which to formalize our question is model theory,
a branch of mathematical logic. In this setting the notion of language has a precise \mbox{meaning}, and is essentially defined as a collection of symbols and relations from which logical sentences may be formed.
Model theory then takes a bird's-eye view on \mbox{mathematics}, and seeks to organize mathematical languages and theories according to logical principles. One of the most striking logical principles is G\"odel's concept of incompleteness, which essentially states that whenever a consistent mathematical theory is strong enough to describe the arithmetic of integers, it inevitably contains statements whose truth cannot be decided within the theory.
Developments of this type motivated model theorists in the 1980s to formalize a concept of tameness for mathematical theories, by which the bizarre logical consequences of \mbox{incompleteness} and undecidability can be avoided.
These aspirations align with the ideas of Grothendieck, 
who envisioned around the same time that wild mathematical objects such as those of figure \ref{fig:Wildsets} are pathological and should be excluded from a proper framework of geometry \cite{Grothendieck_1997}. \\




\newpage
Out of these considerations emerged a theory which had a revolutionary impact on model theory and mathematics beyond: the theory of \textit{o-minimal structures},
introduced by Van den Dries \cite{VdDriesFirst} and formalized by Pillay and Steinhorn \cite{PillaySteinhorn}. 
From a geometric perspective, these structures are collections of subsets of Euclidean space, closed under basic set-theoretic operations, and constrained by a finiteness axiom called o-minimality, which demands that all sets should consist of finitely many connected pieces. This modest axiom yields an incredibly rich framework of tame geometry, as reviewed by Van den Dries \cite{VdDries}. 
The geometric objects living in these structures are constrained by strong theorems, which together lead to the picture that they have a finite geometric complexity. These theorems have led to remarkable new insights in various areas of mathematics, including algebraic geometry, analytic geometry, and number theory. The principle of o-minimality thus provides a powerful criterion to distinguish finiteness and infinity, and thereby yields a promising notion of tameness for mathematical languages. \\












The finiteness which lies at the core of o-minimality is qualitative, and the theory can be made even more powerful by promoting tameness to a \textit{quantitative} notion. This refinement was developed recently
by Binyamini and Novikov, who introduced the theory of \textit{sharp o-minimality} in which an o-minimal structure is supplemented by a measure of complexity \cite{binyamini_tameness_2023}. In this theory, the complexity of a geometric object is essentially defined through the information contained in its description, and controls the finiteness of its geometric and computational features through explicit complexity bounds. This theory thus has the potential to reinforce the applications of tame geometry, by replacing abstract finiteness by precise numbers. 
This resonates with the way in which finiteness of information manifests itself in physics, through sharp bounds like the Bekenstein bound. \\









The aim of this thesis is to explore tame geometry as the language of physics. The idea that tameness could be a physical principle was first proposed by Grimm \cite{Grimm:2021vpn}, and extended in \cite{Douglas:2022ynw,Douglas:2023fcg}. The research presented in this thesis builds further upon this idea, and focuses on quantum field theory, one of the pillars of modern physics. We demonstrate that many objects arising in quantum field theory are tame and admit a description of finite complexity in the language of o-minimal structures. While the framework of quantum field theory is in principle not fundamentally constrained by finiteness, we are ultimately only interested in those theories which can consistently describe the laws of physics of our universe. \\

This is where gravity enters. Gravity, described by Einstein's theory of general relativity, forms another pillar of modern physics, complementary to quantum field theory. One of the outstanding challenges in physics is to unite these two pillars into a single theory, \textit{quantum gravity}, which provides a unified description of all forces of nature. 
Attempts at quantizing the gravitational field using perturbative quantum field theory fundamentally fail, as the theory is non-renormalizable and has no control over the quantum corrections at high energies.
This signals that a complete theory of quantum gravity requires new physics. The precise formulation of this physics, despite the existence of strong candidates, remains elusive at present.
Nevertheless, quantum gravity can be described as an \textit{effective} field theory valid up to a finite energy cutoff scale. At this effective level, the issue of non-renormalizability is avoided by restricting the theory to low energies. 
In this thesis, we will not be concerned with the language of quantum gravity itself, but with the language of the low-energy effective theories that arise from it. \\

The task at hand is then to understand the effective field theories of quantum gravity. These theories are described by an action containing the Einstein-Hilbert term as well as an infinite series of higher-order interactions consistent with the specified matter content and symmetries. 
From a low-energy perspective, the generality of this construction leads to an enormous space of seemingly consistent effective theories of quantum gravity. 
However, in the past years it has become increasingly clear that only a relatively tiny subset of these theories actually arise as a low-energy limit of quantum gravity. The insight that consistency with quantum gravity imposes strong constraints on effective theories led to the birth of the swampland program, initiated by Vafa \cite{Vafa:2005ui}. This program seeks to formulate criteria which distinguish the landscape of effective theories which admit an ultraviolet completion to quantum gravity from the swampland of theories which ultimately fail to be consistent with gravity. Many of these quantum gravity criteria center around finiteness, constraining the size of the landscape, the matter spectrum, geometric features of field spaces, and amplitudes. \\





It is compelling to draw a parallel between the swampland program's search for the finiteness principles underlying the quantum gravity landscape, and model theory's search for the finiteness principles characterizing tame mathematical theories. The connection between tameness and the swampland was made by Grimm's proposal of the \textit{tameness conjecture}, which asserts that all field spaces, coupling functions, and parameter spaces appearing in effective theories of quantum gravity must be definable within an o-minimal structure \cite{Grimm:2021vpn}. 
This thesis develops this perspective further, by introducing a measure of complexity for quantum field theories based on sharp o-minimality and analyzing the tameness and complexity of their observables. Despite the presence of infinitely many higher-order corrections, we find that certain classes of effective theories can be represented with finite information. The research of this thesis culminates in the idea that quantum gravity enforces a description of finite complexity on effective theories, and that sharp o-minimality provides a natural candidate for the language of the low-energy physics of our universe.


\subsection*{Outline of the thesis}
The outline of this thesis is as follows. 
\begin{enumerate}
    \item[(\textbf{I})] \textbf{Tame geometry.} The purpose of the first part of the thesis is to provide an introduction to tameness and complexity defined by the mathematical theory of o-minimality. In chapter~\ref{ch:tameness} we motivate and introduce o-minimal structures and explain how these structures yield a framework of tame geometry. In chapter~\ref{ch:complexity} we introduce the complexity theory of sharply o-minimal structures, which provides a quantitative version of tame geometry. 
    \item[(\textbf{II})] \textbf{Quantum field theory.} In the second part of the thesis we study applications of tameness and complexity to quantum field theory. Chapter~\ref{ch:classical} begins with a general discussion of tameness in physics and studies examples in quantum mechanics and classical field theory. In chapter~\ref{ch:observables} we analyze the tameness and complexity of non-perturbative observables in quantum field theory, focusing on zero-dimensional theories and using techniques based on differential equations and resummation. Chapter~\ref{ch:CCC} concerns the complexity of tree-level observables in cosmological quantum field theories. Finally, in chapter~\ref{ch:complexityQFT} we explore how sharp o-minimality can be used to define an intrinsic measure of complexity for quantum field theories. 
    \item[(\textbf{III})] \textbf{Quantum gravity.} In the third part of the thesis we study the connection between tame geometry and quantum gravity. Chapter~\ref{ch:string} briefly reviews aspects of quantum gravity, string theory, and the swampland program. In chapter~\ref{ch:volumes} we investigate the tameness of the geometry of moduli spaces, showing that highly supersymmetric theories are tame and demonstrating a general link between quantum gravity, duality, and tame isometric embeddings. Finally, in chapter~\ref{ch:EFTcomplexity} we argue that consistency with quantum gravity imposes that effective field theories admit a description of finite complexity.
\end{enumerate}

\newpage

\RedeclareSectionCommand[ 
beforeskip=12ex,
            ]{part}
\setpartpreamble[u][\textwidth]{
\vspace*{1cm}
\hrulefill 
\vspace*{0.5cm}

This part of the thesis provides a detailed introduction to tame geometry, and thereby lays the mathematical foundations of this thesis. The central concept of part I, and ultimately of the entire thesis, is that of an \textit{o-minimal structure}. These structures are defined in chapter \ref{ch:tameness}, and it is explained how they yield a beautiful framework of tame geometry in which geometric objects are regulated by finiteness. Chapter \ref{ch:complexity} discusses the theory of sharply o-minimal structures, in which tame geometric objects admit a measure of complexity. The implementation of this measure of complexity in physical theories underlies a significant part of the thesis.

\vspace*{0.5cm}
\hrulefill }

\part{Tame geometry}\label{part1}

\chapter{Tameness and o-minimality}\label{ch:tameness}
\setlength{\parindent}{0pt}

In this chapter we begin our venture into the theory of o-minimal structures. 
We start with an introductory section which motivates the theory from various angles. With this motivation in mind, we develop the main definitions of o-minimality, and review the essential theorems which govern the resulting geometric framework. These theorems together demonstrate what makes the sets inside an o-minimal structure tame. The scope of this chapter is to review tame geometry with a balance between  covering the fundamental aspects of the theory and remaining focused towards the implementation in the rest of the thesis.

\section{Introduction and motivation}
\label{sec:TamenessMotivation}
\subsection{Foundational motivation}
\begin{figure}[h]
    \centering
    \includegraphics[width=0.8\linewidth]{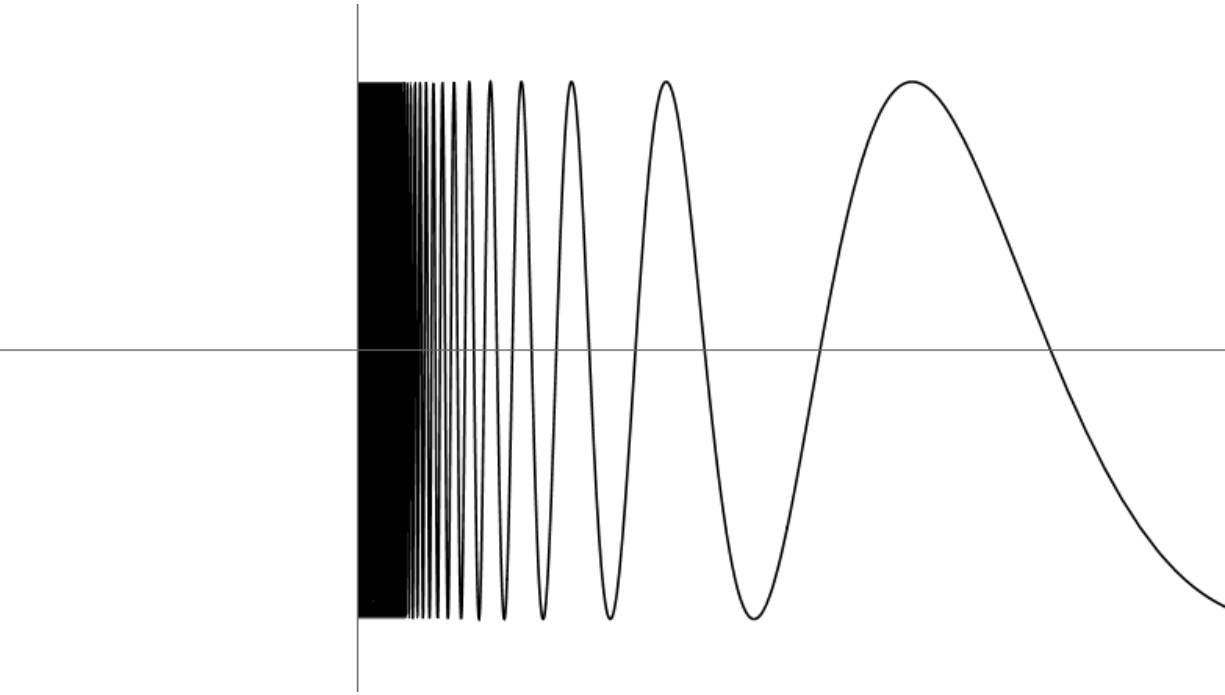}
    \caption{The topologist's sine curve, defined as the graph of the function $x\mapsto \sin(1/x)$ for $x>0$.}
    \label{fig:TopologistSine}
\end{figure}

The origins of tameness in mathematics can be traced back to Grothendieck's visionary research proposal \textit{Esquisse d'un programme} written in 1984, in which he identified various future research directions for mathematics \cite{Grothendieck_1997}. Among these directions is a proposal for a new framework of topology. He argued that topology is not the most natural foundation for geometry, and had instead been created for the purposes of mathematical analysis.
This concern comes from the fact that topology contains certain pathological mathematical objects which defy geometric intuition.
The classic example is the topologist's sine curve depicted in figure~\ref{fig:TopologistSine}. Towards $x\to 0$, the oscillations of the sine function seem to get infinitely dense, and the geometry of this curve behaves wildly. One manifestation of this is that near $x=0$ the  dimension of this set breaks down.
Formally speaking, the boundary of this set is one-dimensional, while the set itself is also one-dimensional. In contrast, in geometry we expect that spaces have a well-defined notion of dimension, and that the boundaries have a strictly lower dimension. Grothendieck envisioned that in the foundations of geometry, topology should be replaced by a \textit{tame topology} which excludes these pathological objects.\\

\subsection{Motivation from model theory}
Another front where mathematicians searched for a notion of tameness is logic and model theory. G\"odel's incompleteness theorems showed that mathematical theories which include the arithmetic of the integers have drastic logical consequences: these theories are undecidable, in the sense that they contain true statements whose truth cannot be determined.
This fact motivated the search for logical structures with tame mathematical behavior. 
A fundamental discovery which shows that the incompleteness theorems can be evaded was made by Tarski and Seidenberg, who found that the theory which contains only the algebraic properties of real numbers \textit{is} decidable \cite{Tarski,Seidenberg}. The key difference is that this theory cannot define the set of integers, or in fact any infinite discrete object, avoiding the setting of G\"odel's theorems. It was subsequently conjectured by Tarski that decidability persists when exponentiation of real numbers is added to the theory. While this conjecture remains unsolved, it was noted by Van den Dries that all geometric objects in this theory satisfy remarkable finiteness properties \cite{VdDriesFirst}. These finiteness properties were formalized in \cite{PillaySteinhorn,PillaySteinhornKnight}, leading to the definition of o-minimality.  
Eventually, it was realized that this concept not only formalizes a notion of tameness in logic, but also provide a realization of Grothendieck's geometric idea of tameness,\footnote{We remark that there exist several other notions of tameness in model theory. Variations on the o-minimality axiom, such as d-minimality, yield alternative frameworks of tame geometry  \cite{HieronymiNotes}.} marking the birth of tame geometry \cite{VdDries}.

\subsection{Contemporary motivation}
Aside from this intrinsic interest, techniques from o-minimality have recently led to significant progress in several areas of mathematics, most notably in algebraic and arithmetic geometry. One of the most remarkable examples is the recent proof of the long-standing André-Oort conjecture \cite{AndreOortProof,AndreOortRev},\footnote{In fact, in July 2026 Jacob Tsimerman received the Fields Medal for his contribution in the recasting of o-minimality as a fundamental method of arithmetic and complex algebraic geometry.} which strongly relies on one of the theorems in o-minimality reviewed in section~\ref{sec:tamenesstheorems}. 
Another example is Hodge theory, where tameness has made a natural appearance as a geometric framework lying at the interface of algebraic and analytic geometry \cite{BKT}. 
While we will not review these developments in detail here, it serves to illustrate that o-minimality has become valuable tool in contemporary mathematics. 
\vspace{-1em}

\subsection{Towards o-minimality}
The previous subsections motivate the relevance of o-minimal structures. 
Before presenting the technicalities, it is instructive to first briefly envision what form the definition of an o-minimal structure should take, guided by the following ideas:
\begin{subbox}{Motivating ideas for o-minimality}
    \begin{itemize}
    \item We require a mathematical principle which can discern whether a geometric object is tame or not. 
    \item In order to achieve this, we start by specifying a collection of subsets of Euclidean space, to be thought of as \textit{tame sets}, forming the building blocks of tame geometry.
    \item To provide a meaningful notion of geometry, we must be able to construct new tame sets from existing ones in a controlled way, by demanding that this collection is closed under basic geometric operations. 
    \item To have any power as a geometric framework, it must at least include simple geometric objects which can be constructed algebraically.
    \item In order to ensure that this collection of sets defines a tame geometry, it must be constrained by a finiteness axiom. 
\end{itemize}
\end{subbox}

With these basic ideas in mind, we now continue with the technical details of o-minimality. We refer to the foundational book \cite{VdDries} for an extensive review. 

\section{Defining o-minimality}\label{sec:defomin}
\subsection{O-minimal structures, tame sets, and tame functions}

The central definition of this chapter is the following. 
\begin{subbox}{Definition: o-minimal structure}
An \textit{o-minimal structure} on the real numbers $\bbR$ is a collection $\mathcal{S}=(\mathcal{S}_n)_{n\geq 1}$, with $\mathcal{S}_n$ consisting of subsets of $\bbR^n$, satisfying the following conditions.
\begin{enumerate}
 \item[(i)] \textbf{(Unions and intersections)} 
  If $A,B\in \cS_n$, then 
  \begin{equation}
      A\cup B  ,\, A\cap B \in \cS_n \,.
  \end{equation}
    \item[(ii)] \textbf{(Complements, projections, and products)} 
    If $A\in \cS_n$, then 
   \begin{equation}
       \bbR^n \setminus A \in \cS^n\,,\,\,\pi(A)\in \cS_{n-1} \,,
   \end{equation}
    for any linear projection $\pi:\bbR^{n}\to\bbR^{n-1}$. Moreover, if $B\in \cS_m$, then
    \begin{equation}
        A\times B \in \cS_{n+m} \,.
    \end{equation}
    \item[(iii)] \textbf{(Algebraic sets)} If $P$ is a real polynomial in $n$ variables, then 
    \begin{equation}
        \{P=0\}\in \cS_n \,.
    \end{equation}
    \item[(iv)] \textbf{(O-minimality)}
    all sets in $\cS_1$ have finitely many connected components.
\end{enumerate}
\end{subbox}
Let us unravel the components of this definition step by step. The first part of the definition indicates that an o-minimal structure consists of sets in Euclidean space, closed under elementary operations that can be performed on sets as specified by axiom (i) and (ii). Axiom (iii) demands that this collection of sets at least contains all zero sets of polynomials, so that the resulting geometric framework will at least include algebraic sets, which form the building blocks of algebraic geometry. Collections of sets satisfying axioms (i), (ii), and (iii) are known as \textit{structures} on the real numbers. With a choice of structure one specifies which sets may be defined
within the theory, and therefore sets that belong to a structure $\cS$ are called $\cS$-\textit{definable}, or \textit{definable} if the underlying structure is left implicit. \\

Finally, axiom (iv) is the finiteness condition that, as we will see in the rest of this chapter, leads to the tame geometry of definable sets in o-minimal structures. It demands that the definable subsets in $\bbR$ consist of finite unions of points and intervals. The assumption that o-minimal structures are closed under linear projections then ensures that this finiteness condition extends to definable sets in higher dimensions: by projecting down to the real line $\bbR$, any definable set in $\bbR^n$ must ultimately reduce to a finite union of points and intervals. This principle has enormous consequences, which are demonstrated by the various tameness theorems that definable sets satisfy, as we discuss in section~\ref{sec:tamenesstheorems}.\\

We refer to definable sets in an o-minimal structure as \textit{tame sets}. A function $f:A\to B$ between two tame sets $A$ and $B$ is tame if the graph 
\begin{equation}
\Gamma(f) = \{ (x,y)\,|\,f(x)=y\}
\end{equation}
is a tame subset of the product $A\times B$. For example, the graph of a polynomial
\begin{equation}
    f(x)= a_nx^n + a_{n-1}x^{n-1} \ldots a_1 x +a_0 
\end{equation}
is precisely the zero set $\{P=0\}\subseteq\bbR^2$ of the two-variable polynomial 
\begin{equation}
P(x,y) =  y-a_nx^n + a_{n-1}x^{n-1} \cdots a_1 x +a_0 \,,
\end{equation}
which is an algebraic set. This shows that polynomials are tame functions, definable in any o-minimal structure, as a consequence of axiom (iii). Tame functions satisfy various basic properties, such as being closed under sums, products, and compositions. To illustrate this in an example, it is instructive to show that images and preimages of tame sets are tame sets using the axioms of structures.
If $f:A\to B$ is a tame function and $U\subseteq A$ and $V\subseteq B$ are tame sets, then the image of $U$ and the preimage of $V$ can be expressed as 
\begin{equation}
f(U) = \pi_B\big((U\times B) \cap\Gamma(f)\big)\,, \quad f^{-1}(V)=\pi_A\big( (A\times V)\cap \Gamma(f) \big) \,.
\end{equation}
Here $\pi_A$ and $\pi_B$ are the linear projections from $A\times B$ to $A$ and $B$, respectively. Since structures are closed under projections, intersections, and products, it follows that $f(U)$ and $f^{-1}(V)$ are tame sets.

\subsection{First examples of o-minimal structures}
To get familiar with o-minimal structures, let us discuss a few first examples. 

\subsubsection*{Semi-algebraic sets}
Since any structure must at least contain the algebraic sets, it is a natural question whether the collection of all algebraic sets forms a structure. This turns out to be false: for example, consider the linear projection $(x,y)\mapsto y$ of the parabola $\{(x,y)\in \bbR^2 \,|\, x^2=y\}\subseteq \bbR^2$. The resulting image is the interval $[0,\infty)$, which cannot be described using equalities of polynomials. Instead, it is defined by an inequality $y\geq 0$. This shows that structures must also contain sets defined by inequalities of polynomials in order for the axioms to hold. These sets are known as \textit{semi-algebraic sets}, and in general they are obtained by taking finite unions, intersections, and complements of sets of the form 
\begin{equation}
    \big\{(x_1,\ldots,x_n) \in \bbR^n \,\big|\, P(x_1,\ldots,x_n) =0 , \, Q(x_1,\ldots,x_n)>0 \big\}
\end{equation}
for real polynomials $P$ and $Q$. The collection of all semi-algebraic subsets of $\bbR^n$ for $n\geq 1$ forms a structure denoted by $\bbR_{\rm alg}$. This relies on the fact that semi-algebraic sets are closed under linear projections, which is a deep result in model theory due to Tarski and Seidenberg \cite{Tarski,Seidenberg}. The semi-algebraic subsets of $\bbR$ are precisely the finite unions of points and intervals, which makes $\bbR_{\rm alg}$ our first example of an o-minimal structure. Since any o-minimal structure must contain the semi-algebraic sets, this also shows that $\bbR_{\rm alg}$ is the smallest possible collection of sets forming an o-minimal structure.

\subsubsection*{Generating structures from classes of functions}
The natural follow-up question is whether there exist o-minimal structures beyond $\bbR_{\rm alg}$.
The standard method of constructing new structures is by starting with a selection of sets which we would like to be definable in our geometry, and by iteratively applying axioms (i) and (ii) to this selection to produce a larger collection of definable sets.
In this sense, a selection of subsets of Euclidean space \textit{generates} a structure. Typically, the original selection consist of graphs of functions that we would like to be definable.
By choosing a sufficiently well-behaved set of functions we generate a candidate o-minimal structure, for which it remains to verify axiom (iv). 
Essentially all o-minimal structures used in  tame geometry are obtained in this manner. An important example of this principle is the structure generated by the real exponential function $\text{exp}:\bbR\to\bbR$, denoted by $\bbR_{\rm exp}$.
It was shown by Wilkie that the structure $\bbR_{\rm \exp}$ is o-minimal, marking another deep result in model theory \cite{WilkieRExp}. A list of examples of o-minimal structures is discussed in section~\ref{sec:catalog}.

\subsubsection*{Non-tame sets and functions}
The o-minimality axiom is a strong condition, and many objects are immediately ruled out of tame geometry. The signature example of a set which is not definable in any o-minimal structures is the set of integers $\bbZ\subseteq \bbR$, since it consists of an infinite union of disconnected points.\footnote{From the point of view of G\"odel's incompleteness theorems, it is a natural expectation that the set of integers is incompatible with a mathematical notion of tameness.} Likewise, any set consisting of infinitely many disconnected components cannot be definable in any o-minimal structure.
Consequently, functions with an infinite amount of geometric features are not tame. For example, the graph of $\sin:\bbR\to\bbR$ intersected with the horizontal axis defines the set of integers, and is therefore incompatible with o-minimality.

\subsection{Duality of logic and geometry}\label{sec:dualitylogic}
The formulation of o-minimality is completely geometric in nature, but as a brief aside, it is insightful to discuss how the axioms in the definition of a structure reflect logical operations. There is a simple duality between logic and geometry coming from the fact that a set $X\subseteq \bbR^n$ can be specified as the set of points $x=(x_1,\ldots,x_n)$ for which a certain logical formula $F(x)$ is true, which we denote by $X =  \{ x\in  \bbR^n \,|\,F(x)  \}$. In many cases, these logical formulas are simply equations or inequalities, but they can also take more complicated forms. Generally, the relevant logical formulas in this setting are the so-called \textit{first-order formulas}, which are generated by the logical symbols $=$, $>$, $\Rightarrow$ (implies), $\wedge$ (and), $\vee$ (or), $\neg$ (not), $\exists$ (there exists), and $\forall$ (for all), as well as addition and multiplication of real numbers. Logical operations performed on these first-order formulas can then be translated to geometric operations on the resulting sets, providing a useful correspondence:

\begin{subbox}{Correspondence between logical and geometric operations}
\begin{itemize}
    \item $\{x\,|\,F(x) \wedge G(x) \} = \{x\,|\,F(x)\} \cap \{x\,|\,G(x)\}$
    \item $\{x\,|\,F(x) \vee G(x) \} = \{x\,|\,F(x)\} \cup \{x\,|\,G(x)\}$
    \item $\{x\,|\,\neg\, F(x)\} = \bbR^n\setminus \{x\,|\, F(x)\}  $
    \item $\{x\,|\, F(x)\Rightarrow G(x) \} =  \bbR^n\setminus \{x\,|\, F(x) \wedge\neg G(x)\}  $
    \item $\{x\,|\,\exists y\in Y \,:\, F(x,y)\} = \pi\big( (\bbR^n\times Y) \cap \{(x,y) \,|\,F(x,y) \}\big)  $
    \item $\{x\,|\,\forall y\in Y,\, F(x,y)\} = \bbR^n\setminus \pi\big(    (\bbR^n\times Y )\cap (\bbR^{n+m}\setminus  \{(x,y)\,|\, F(x,y)\}   ) \big)  $
\end{itemize}
\end{subbox}

For example, the negation operation ($\neg$) in logic corresponds to the complement operation ($\setminus$) in geometry.
The geometric interpretation of the quantifier symbols $\exists$ and $\forall$ takes a somewhat complicated form, and relies on products and linear projections $\pi:\bbR^{n+m}\to\bbR^n$. Comparing the operations in this correspondence with the axioms of an o-minimal structure, we observe that structures are closed under precisely these geometric operations. The point of this correspondence is therefore not to explicitly translate logic to geometry, but to show that simple logical operations preserve definability in an o-minimal structure. \\

Let us discuss an example to illustrate this idea. Consider the topological closure $\overline X$ of a set $X\in \bbR^n$, which is defined as 
\begin{equation}
    \overline X = \big\{ y \in \bbR^n \,\big|\,\forall \eps>0 \,\exists x\in X \,:\, \norm{x-y}^2 <\eps \big\} \,.
\end{equation}
In particular, the closure can be described by means of a first-order formula. By the duality of logic and geometry it then follows that if $X$ is definable, then so is $\overline X$, since it can be obtained by the geometric operations under which an o-minimal structure is closed. 
In a similar vein, it follows that many constructions in real analysis which can be expressed using first-order formulas, such as limits and differentiation, can be performed within tame geometry.\\

From this point onward, the development of the theory splits into two parallel directions. In section~\ref{sec:tamenesstheorems} we proceed by reviewing the geometric properties of tame sets, and in section~\ref{sec:catalog} we discuss a wealth of examples of o-minimal structures. These sections can be read independently.

\section{Essential theorems in o-minimality}\label{sec:tamenesstheorems}
From the o-minimality axiom it is evident that finiteness is at the core of the theory, but the tameness of sets in an o-minimal structure truly comes to light by considering some of the important theorems which they satisfy. These theorems, which we will collectively refer to as \textit{tameness theorems} in this thesis, determine the universal tame features of tame sets and lie at the heart of applications of o-minimality.
We will review these essential theorems in this section. 
Many of these theorems and their proofs can be found in the foundational book by Van den Dries \cite{VdDries}, and where it is needed we refer to further literature.

\subsection{Monotonicity}\label{sec:monotonicity}
The consequences of o-minimality for tame functions of one variable are captured by the following theorem, which is a cornerstone of tame geometry \cite{VdDries}.

\begin{subbox}{Monotonicity theorem}
Let $f:(a,b)\to\bbR$ be a tame function. Then there is a finite number of points
\begin{equation}
    a_0 <a_1 < \cdots < a_n < a_{n+1} 
\end{equation}
with $a=a_0$ and $b=a_{n+1}$ such that the restrictions $f|_{(a_i,a_{i+1})}$ are continuous and either constant or strictly monotonic.
\end{subbox}

\begin{figure}[h]
    \centering
    \includegraphics[width=0.9\linewidth]{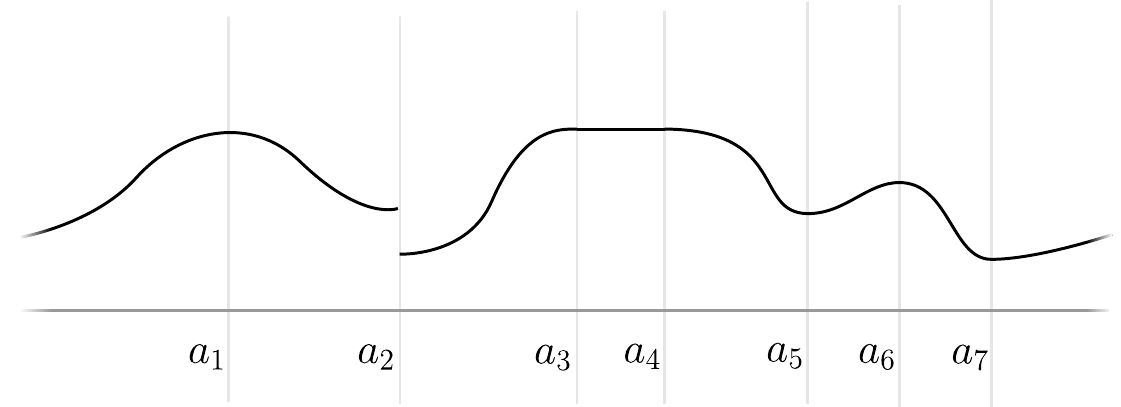}
    \caption{An example of a tame function on the real line. There are only finitely many discontinuities and isolated critical points. The function is either constant or strictly monotonic and continuous on each of the intervals $(a_i,a_{i+1})$.}
    \label{fig:Mono1}
\end{figure}

Essentially, the theorem states that tame functions have finitely many discontinuities and that they change from increasing to decreasing and vice versa finitely many times. It is remarkable that such a strong structural statement follows only from the assumption that $f$ is definable in an o-minimal structure. Figure \ref{fig:Mono1} shows the graph of a generic tame function, controlled by the monotonicity theorem.

As a consequence of the monotonicity theorem, tame functions have a very controlled asymptotic behavior. In particular, if $f:\bbR\to\bbR$ is a tame function, then there is a point $a_n \in \bbR$ such that $f$ is monotonic or constant on $(a_n,\infty)$. This means that the limit of $f$ towards infinity is always well-defined, in the sense that it either converges to a finite value or diverges to $\pm \infty$; it cannot oscillate at infinity. This fact implies that differentiation interacts surprisingly well with o-minimality. \\

Based on the monotonicity theorem it can be shown that any tame function $f:\bbR\to\bbR$ is differentiable at all but finitely many points. If a tame function is differentiable to begin with, then it follows that its derivative is tame as well. \\

While we will not study the proof of the monotonicity theorem, it is insightful to briefly think about how such a result can be obtained from the condition that tame subsets of $\bbR$ are finite unions of intervals and points. The essential idea is that the set of critical points of $f$ can itself be expressed as a definable subset of the real line, since it can be described from first-order formulas depending on $f$.  The critical points of $f$ therefore form a finite union of intervals and points: the intervals correspond to $f$ being constant, and the points correspond to $f$ changing between being increasing and decreasing. 

\subsection{Cell decomposition}
\label{sec:celldec}
The monotonicity theorem is the stepping stone to a far more powerful theorem, the \textit{cell decomposition theorem}. Intuitively, this theorem states that any tame set can be decomposed into finitely many simple components called cells, and the theorem has been referred to as the fundamental theorem of o-minimality. \\

\begin{figure}[h]
    \centering
    \includegraphics[width=0.7\linewidth]{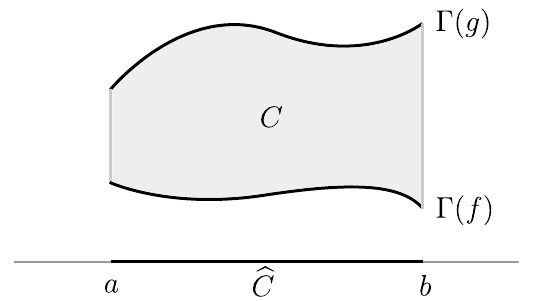}
    \caption{An example of a two-dimensional band cell $C\subseteq\bbR^2$. The cell projects down to a one-dimensional cell $\widehat C = (a,b) \subseteq \bbR$, and is defined as the area between the graphs of two tame functions $f,g:(a,b)\to\bbR$.}
    \label{fig:Cell1}
\end{figure}

To formally state the theorem, we first introduce the definition of cells, which proceeds by induction on the dimension of the ambient space. In the one-dimensional case, a cell $C\subseteq \bbR$ is either a single point or a (finite or infinite) open interval. In higher dimensions, a cell $C \subseteq \bbR^{n+1}$ is a set of one of the following two forms:
\begin{align*}
      \{ (x,y)\in \widehat C\times \bbR \,&|\,  f(x)=y \}   &&\text{(graph)}\,, \\
     \{ (x,y)\in \widehat C\times \bbR \,&|\,  f(x)<y<g(x) \}   &&\text{(band)}\,,
\end{align*}
where $\widehat C$ is a cell in $\bbR^n$ and $f$ and $g$ are tame functions on $\widehat C$.
For the band-shaped cells, it is assumed that $f<g$, and the functions may be set to $-\infty$ and $+\infty$ respectively. An important aspect of the definition is that the cells are cylindrical, in the sense that the image of a cell under the linear projection $\bbR^{n+1}\to\bbR^n$ onto the first $n$ coordinates is again a cell. Cells therefore have a well-defined notion of dimension: graph cells preserve the dimension of the underlying cell, and band cells increase the dimension by one. The topology of cells is simple: an $n$-dimensional cell is homeomorphic to an $n$-cube. 
Some examples of cells are shown in  figure~\ref{fig:Cell1}. \\

\begin{figure}[h]
    \centering
    \includegraphics[width=0.65\linewidth]{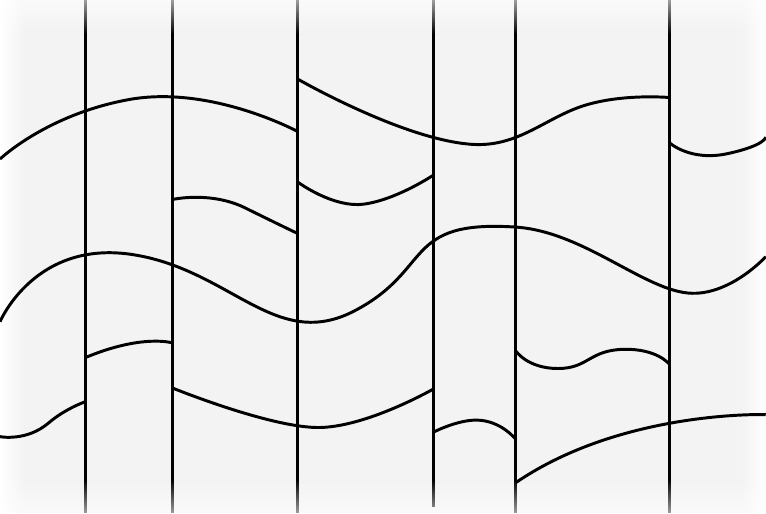}
    \caption{An example of a cell decomposition of $\bbR^2$. The cells at the edge of the figure extend to infinity.}
    \label{fig:CellDec1}
\end{figure}

A \textit{cell decomposition} of $\bbR^n$ is a partition of $\bbR^n$ into cells, that is, a finite collection of cells $C_1,\ldots,C_N$ such that 
\begin{equation}
    \bigcup_{i=1}^N C_i = \bbR^n \quad \text{and} \quad C_i \cap C_j =\emptyset \quad \text{for} \quad i\neq j\,.
\end{equation}
Figure \ref{fig:CellDec1} shows an example of a cell decomposition of the Euclidean plane.\\

With this definition of cell decomposition, we have the following universal structural theorem for tame sets \cite{VdDries}. 

\begin{subbox}{Cell decomposition theorem}
Let $X_1,\ldots,X_m\subseteq \bbR^n$ be tame sets. Then there exists a  cell decomposition of $\bbR^n$ such that each $X_k$ is a finite union of cells. Additionally, if $f:X\to \bbR$ is tame, there exists a cell decomposition of $\bbR^n$ for which $X$ is a union of cells and the restriction of $f$ to each cell is continuous. 
\end{subbox}

This theorem shows that cells may be regarded as the building blocks of o-minimal geometry. Since the geometry of a cell is simple to describe and each definable set consists of finitely many cells, it demonstrates how strongly the definable sets in an o-minimal structure are constrained by tameness. Furthermore, in technical applications of tame geometry, the theorem implies that geometric problems concerning a tame set can be reduced to finitely many cells. Applications of this form rely on the fact that there exist explicit geometric algorithms which find the cell decomposition of a given tame set. Finally, the cell decomposition theorem also highlights the fact that the tameness of sets and the tameness of functions are fundamentally equivalent: we can think of the graph of a tame function as a tame set, but conversely we can now think of a tame set as being delimited by tame functions. An illustration of the cell decomposition theorem is given in figure~\ref{fig:Celldec2}.

\begin{figure}[h]
    \centering
    \includegraphics[width=1\linewidth]{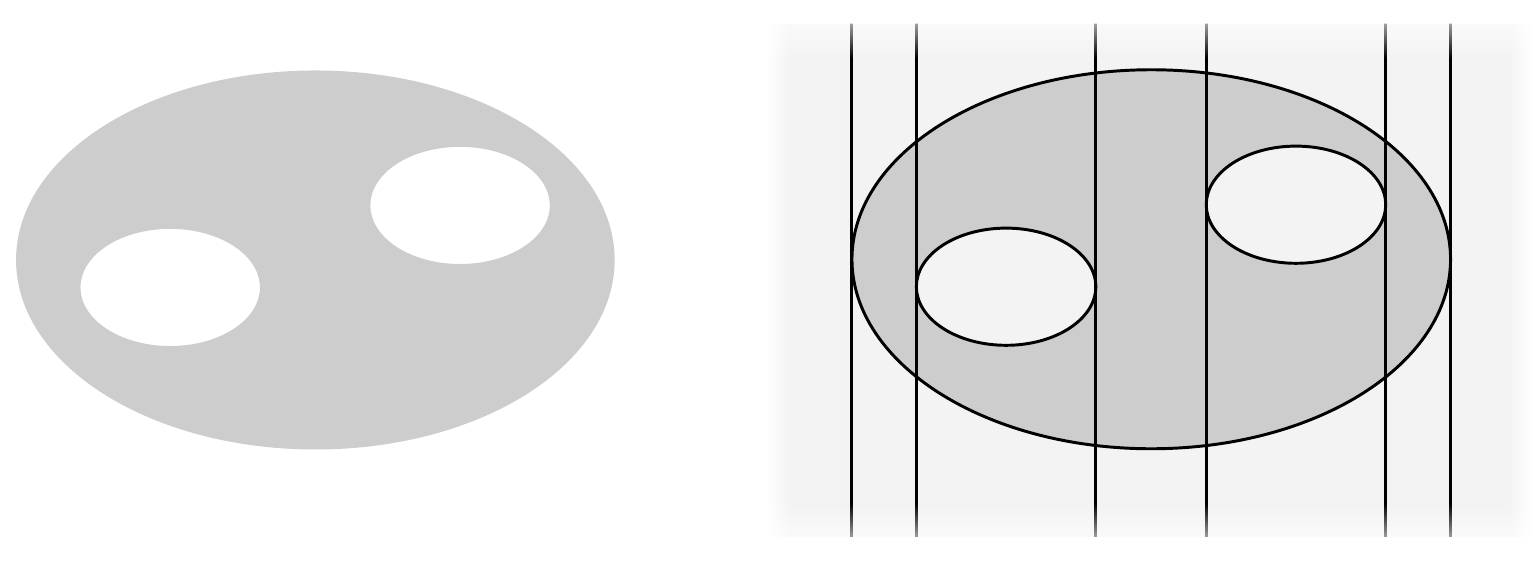}
    \caption{An example of a tame set $X\subseteq \bbR^2$ (left) and a cell decomposition of $\bbR^2$ adapted to $X$ (right). Here $X$ is obtained as the union of seven two-dimensional cells.}
    \label{fig:Celldec2}
\end{figure} 

\subsection{Triangulation}
A similar theorem which controls the topological structure of tame sets is the \textit{triangulation theorem}. It essentially states that the topology of tame sets can always be obtained by gluing together finitely many points, lines, triangles, and higher-dimensional simplices. To state the theorem formally, recall that a $k$-simplex in $\bbR^n$ is the convex hull of $k$ affinely independent points, i.e.~a set of the form
\begin{equation}
    \langle a_0,\ldots,a_k \rangle = \bigg\{ \sum_j t_ja_j \,\bigg|\,t_0,\ldots,t_k>0 \text{ and } \sum_jt_j=1 \bigg\},
\end{equation}
where $a_1-a_0,a_2-a_0,\ldots,a_k-a_0\in\bbR^n$ are linearly independent vectors. A face of a simplex $\langle a_0,\ldots,a_k\rangle$ is a simplex spanned by a non-empty proper subset of $\{a_0,\ldots,a_k\}$. Finally, a \textit{simplicial complex} in $\bbR^n$ is a finite collection $\cK$ of simplices in $\bbR^n$ satisfying the following technical property: for any two simplices $\sigma,\sigma'\in\cK$, the intersection $\overline\sigma\cap \overline\sigma'$ is either empty or equal to the closure $\overline\tau$ of a common face $\tau$ of $\sigma $ and $\sigma'$. This assumption ensures that the simplices in a simplicial complex align geometrically. The statement of the triangulation theorem is then as follows.  
\begin{subbox}{Triangulation theorem}
Let $X\subseteq \bbR^n$ be a tame set. Then there exists a definable homeomorphism between $X$ and a finite simplicial complex in $\bbR^n$.
\end{subbox}

\begin{figure}[h]
    \centering
    \includegraphics[width=0.9\linewidth]{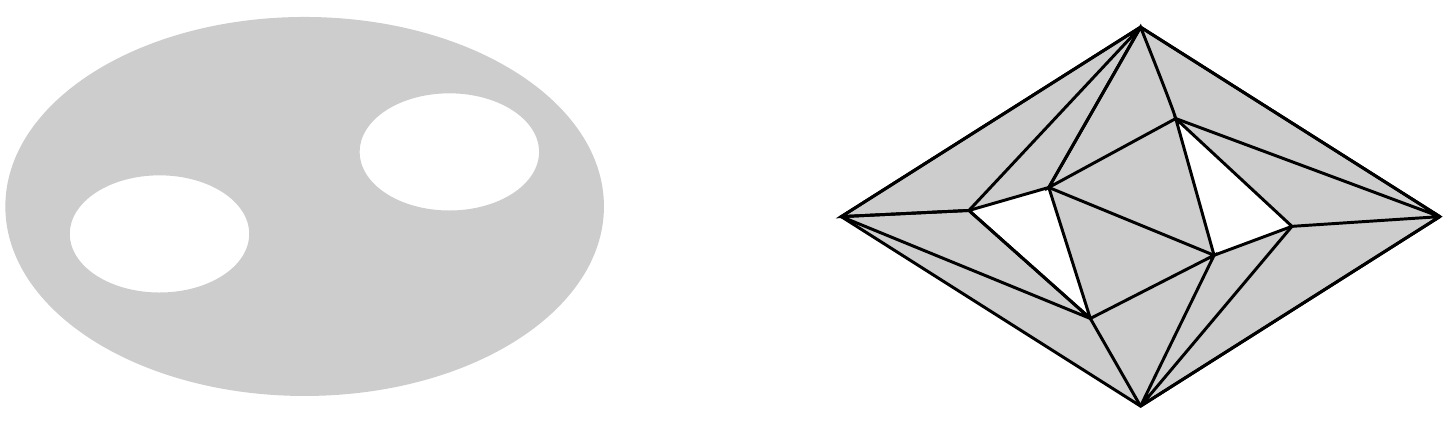}
    \caption{An example of a tame set $X\subseteq \bbR^2$ (left) and a simplicial complex representing a triangulation of $X$ (right).}
    \label{fig:Celldec2}
\end{figure} 

This theorem captures the idea that definable objects have a finite topological complexity, as
a consequence of the o-minimality assumption. It allows one to describe the topology of a definable set
or function with a finite amount of combinatorial data. As a result, many topological invariants associated to tame spaces, such as ranks of homology, cohomology, and homotopy groups, are finite.

\subsection{Algebraization}
The three tameness theorems discussed so far are intrinsic to o-minimality, but there are also theorems which indicate that tame geometry has deep ties to other geometric settings. This holds in particular on the interface of complex\footnote{For the complex numbers, o-minimality is defined by identifying the complex plane $\bbC$ with $\bbR^2$.} algebraic and analytic geometry, as illustrated by the following theorem \cite{ominChow}.

\begin{subbox}{Peterzil-Starchenko theorem}
Let $X\subseteq \bbC^n$ be a closed complex analytic subvariety. If $X$ is a tame set, then $X$ is algebraic.
\end{subbox}

The theorem essentially states that a tame set defined by holomorphic equations can actually be defined by algebraic equations, and thus it is known as an \textit{algebraization} theorem. It is a variation of the Chow theorem, a foundational theorem in complex algebraic geometry, which states that a closed analytic subvariety of the complex projective space $\mathbb{CP}^n$ is algebraic. The key difference is that the Chow theorem  relies on the compactness $\mathbb{CP}^n$, whereas the Peterzil-Starchenko theorem holds in the non-compact space $\bbC^n$. The assumption that the set $X$ is definable in an o-minimal structure turns out to be precisely strong enough to accommodate for the generalization to a non-compact ambient space. 
There exist significant extensions of this result, realizing an o-minimal version of Serre's algebraic/analytic correspondence \cite{GriffithsConj}. While we do not discuss these results further here, it highlights that the interaction with tameness and complex geometry is remarkably powerful.

\subsection{Counting rational points}\label{sec:pilawilkie}
Another geometric setting to which tameness has strong connections is arithmetic geometry. 
One of the classic problems in this setting is to characterize solutions to equations in the rational numbers. Phrased geometrically, the problem is to find the rational points $X(\bbQ)=X \cap \bbQ^n$ of a set $X\subseteq \bbR^n$. If $X$ is an algebraic set given by solutions to polynomial equations, then one expects to structurally encounter points in $X(\bbQ)$ for arithmetic and algebraic reasons. On the contrary, if $X$ is non-algebraic, or \textit{transcendental}, rational points are more mysterious and structureless, and only seem to appear incidentally. A remarkable theorem of Pila and Wilkie \cite{PilaWilkie} shows that rational points in a transcendental set $X$ are exceptionally rare if $X$ has a tame geometry, providing a deep link between tameness and arithmetic. \\

The statement of the theorem requires us to introduce some terminology for rational points. We define the \textit{height} of a rational number $a/b$, with $a$ and $b$ coprime integers, as $H(a/b)=\max(|a|,|b|)$. The height of a point  $x\in\bbQ^n$ is defined as the maximum of the heights of the coordinates of $x$. For a set $X\subseteq \bbR^n$, the algebraic part $X^{\rm alg}$ is defined as the union of all connected positive-dimensional semi-algebraic subsets of $X$, and can be thought of as the part of $X$ which can be described using polynomials. In contrast, the transcendental part $X^{\rm tr}=X\backslash X^{\rm alg}$ is the part of $X$ which is non-algebraic. The Pila-Wilkie theorem is a statement on the rarity of rational points of bounded height in $X^{\rm tr}$.

\begin{subbox}{Pila-Wilkie theorem}
Let $X\subseteq \bbR^n$ be a tame set. Then for every $\eps>0$ there exists a constant $C>0$ such that for every $H\geq 1$, we have 
\begin{equation}
  \#\{ X^{\rm tr}( \bbQ) \,|\,H(x)\leq H \} \leq C H^\eps \,.
\end{equation}
\end{subbox}

In other words, if the geometry of $X$ is tame, then the number of rational points in the transcendental part of $X$ grows subpolynomially with the height $H$. These rational points are therefore extremely rare, only appearing sporadically as the height is increased, as measured by the constant $C$. \\

To illlustrate why tameness is necessary in this theorem, consider the graph of the sine function $f(x) = \sin(\pi x)$, which is a transcendental set. Every period of the sine function structurally hits rational points, for example the points $(k,0)\in\bbQ^2$ for every $k\in\bbZ$.
The number of rational points thus grows linearly with the height $H$. For non-tame functions, even faster growth in $H$ is possible, by considering wild functions like the topologist's sine function. In contrast, the number of rational points on the graph of a tame function grows very slowly. For example, for an exponential function such as $f(x)=2^x$ the growth is logarithmic in $H$.\\

The Pila-Wilkie theorem has played a pivotal role in applications of o-minimality to other areas of mathematics, and has been used to prove or make progress on several long-standing conjectures, such as the André-Oort conjecture, the Manin-Mumford conjecture, and the Zilber-Pink conjecture; we refer to \cite{AndreOortRev} for a review. In these applications, the theorem is often used in reverse: if a tame set contains many rational points, then it must be algebraic.

\section{Examples of o-minimal structures}\label{sec:catalog}
In this section we give a brief overview of some of the most prominent o-minimal structures. 
Each o-minimal structure discussed below is generated by a class of functions, by the procedure explained in section~\ref{sec:defomin}. In this sense, the discussion in this chapter provides an answer to the following question:
\[
\textit{what classes of functions generate a tame geometry}?
\]
Any set described in terms of functions of this type will have a geometry constrained by the tameness theorems of the previous section.\\

Instead of considering individual examples of o-minimal structures, it may be tempting to search for the largest class of functions generating an o-minimal structure, since this structure would in principle host every tame function. A surprising fact, however, is that there exists no maximal o-minimal structure: there exist mutually incompatible o-minimal structures, in the sense that the o-minimality axiom would fail if these structures were combined \cite{DenjoyCarleman}. This remarkable conclusion indicates that the individual o-minimal structures discussed in this section each have their own value within the framework.

\subsection{Polynomials -- \texorpdfstring{$\bbR_{\rm alg}$}{}}
As introduced previously, the polynomials generate the smallest o-minimal structure known as $\bbR_{\rm alg}$. The sets in this structure are the semi-algebraic sets, defined by means of equations and inequalities of polynomials. More explicitly, set definable in $\bbR_{\rm alg}$ can be represented in the form
\begin{equation}\label{eq:semialgset}
    X = \bigcup_{i=1}^p \bigcap_{j=1}^{q_i} \{x \,|\, P_{ij}(x)=0\,,\,\,Q_{ij}(x)>0 \}  \,
\end{equation}
where $P_{ij}$ and $Q_{ij}$ are polynomials. \\

The structure $\bbR_{\rm alg}$ has a special significance in logic, as highlighted in section~\ref{sec:TamenessMotivation}: it is the only o-minimal structure which is known to give rise to a \textit{decidable} mathematical theory \cite{Tarski,Seidenberg}. This means that for every mathematical statement that can be expressed in terms of semi-algebraic sets and first-order logic, there exists an effective algorithm which decides whether this statement is true or false.

\subsection{The real exponential -- \texorpdfstring{$\bbR_{\rm exp}$}{}}
When the graph of the real exponential function $\exp:\bbR\to\bbR$ is added to $\bbR_{\rm alg}$, a structure known as $\bbR_{\rm exp}$ is generated. The o-minimality of this structure was established by Wilkie \cite{WilkieRExp}. The sets which are definable in this structure have a similar representation to equation~\eqref{eq:semialgset}, if we replace polynomials by \textit{exponential polynomials}, i.e.~functions of the form
\begin{equation}
f(x_1,\ldots,x_n) = P(x_1,\ldots,x_n,e^{x_1},\ldots,e^{x_n})
\end{equation}
for a polynomial $P$ in $2n$ variables. The crux of Wilkie's result is to show that these sets are closed under linear projections. It is conjectured by Tarski that the theory of the real exponential field is decidable as well. It is important to note that this structure only defines the \textit{real} exponential, and not the complex exponential $\exp:\bbC\to\bbC$, since the complex exponential has infinitely many oscillations on the non-real directions in the complex plane.

\subsection{Restricted analytic functions -- $\bbR_{\rm an}$}
A \textit{restricted analytic function} is defined as the restriction to a closed ball $B$ of a real analytic function defined on an open set containing $B$. The collection of all such functions generate a structure denoted by $\bbR_{\rm an}$, which was established to be o-minimal by results of Gabrielov \cite{gabrielov1968projections}. In contrast to the case of the real exponential function, this o-minimal structure is generated by an extremely large class of functions. The restriction to a compact domain is essential. On unbounded domains, analytic functions can easily fail to be tame, as illustrated by the sine and cosine functions. Even on bounded but non-compact domains, such as open intervals, analyticity does not guarantee tameness, as exemplified by the topologist's sine curve shown in figure \ref{fig:TopologistSine}. In the literature, the geometry generated by the sets in $\bbR_{\rm an}$ is often referred to as globally subanalytic geometry.

\subsection{Restricted analytic functions and exponential -- \texorpdfstring{$\bbR_{\rm an,exp}$}{}}
The previous two examples can be combined into one: the real exponential function and the restricted analytic functions together generate the structure $\bbR_{\rm an,exp}$. The o-minimality of this structure was shown by Van den Dries and Miller in \cite{VdDriesMiller}. This structure has the strongest presence in geometric applications of tame geometry, since it turns out that the combination of analytic functions on compact domains and the unbounded exponential are sufficient to describe many phenomena in algebraic and analytic geometry. This is marked by the fact that the \textit{period map}, the central object in Hodge theory, was shown to be definable in $\bbR_{\rm an,exp}$ \cite{BKT}. We revisit this point later in this chapter.

\subsection{Pfaffian functions -- \texorpdfstring{$\bbR_{\rm Pf}$}{} and \texorpdfstring{$\bbR_{\rm rPf}$}{}}
For reasons to be uncovered in chapter~\ref{ch:complexity}, one of the most important structures in this thesis is the one generated by \textit{Pfaffian functions}. To define these, we need the concept of a Pfaffian chain. 

\begin{subbox}{Pfaffian chain} 
A \textit{Pfaffian chain} is a finite sequence of functions $\zeta_1,\ldots,\zeta_r:U\to\bbR^n$ defined on a product of open intervals $U \subseteq\bbR^n$ satisfying a system of first-order differential equations of the form
\begin{equation} \label{eq:Pfaffianchain}
    \frac{\pd \zeta_i}{\pd x_j} = P_{ij}(x_1,\ldots,x_n,\zeta_1,\ldots,\zeta_i) \quad\text{for} \quad i=1,\ldots,r, \quad j=1,\ldots, n
\end{equation}
where each $P_{ij}$ is a polynomial in $n+i$ variables.  
\end{subbox}

Crucially, this system has a certain triangularity requirement, namely that the differential equations for $\zeta_i$ only depend on the previous functions in the chain $\zeta_1,\ldots,\zeta_i$, and not on $\zeta_{i+1},\ldots,\zeta_r$. Given such a chain formed by functions $\zeta_1,\ldots,\zeta_r$, a \textit{Pfaffian function} is a function $f:U\to\bbR$ of the form
\begin{equation}\label{eq:Pfaffianfunction}
f(x_1,\ldots,x_n) = P(x_1,\ldots,x_n,\zeta_1,\ldots,\zeta_r)\,,
\end{equation}
where $P$ is a polynomial in $n+r$ variables. If the domain $U$ is bounded, i.e.~none of the intervals extend to infinity, then the Pfaffian chain is \textit{restricted}. \\ 

Pfaffian functions satisfy remarkable global finiteness properties \cite{Khovanskii}, using which it was shown that the structure $\bbR_{\rm Pf}$ which they generate is o-minimal \cite{WilkieRPfaff}. The smaller structure generated by restricted Pfaffian functions, denoted by $\bbR_{\rm rPf}$, is therefore o-minimal as well. As we will see in the next chapter, Pfaffian functions admit a notion of complexity, and therefore play a central role in this thesis. 
We postpone a further development of the theory of Pfaffian functions to  section~\ref{sec:Pfaff}.

\subsection{Restricted Noetherian functions -- \texorpdfstring{$\bbR_{\rm rN}$}{}}
A class of functions lying between the restricted Pfaffian functions and the restricted analytic functions is the class of \textit{Noetherian functions}. The definition is almost identical to that of Pfaffian functions, the only difference being that the triangularity requirement of a Pfaffian chain is omitted. Explicitly, given an open box $U\subseteq \bbR^n$, a Noetherian chain consists of functions $\zeta_1,\ldots,\zeta_r:U\to\bbR$ satisfying a system of differential equations of the form 
\begin{equation}
\frac{\partial \zeta_i}{\partial x_j} =  P_{ij} (x_1,\ldots,x_n,\zeta_1,\ldots,\zeta_r)  \quad\text{for} \quad i=1,\ldots,r, \quad j=1,\ldots, n \,,
\end{equation}
where each $P_{ij}$ is a polynomial in $n+r$ variables. Analogous to Pfaffian functions, a Noetherian function is then defined by a polynomial in the variables $x_1,\ldots,x_n$ and the functions $\zeta_1,\ldots,\zeta_r$, and a Noetherian function is called restricted if the domain $U$ is bounded. The collection of restricted Noetherian functions generates a structure $\bbR_{\rm rN}$ which is contained in $\bbR_{\rm an}$ and therefore o-minimal. The reason for considering this structure comes from the fact that they are under much more control than the general class of restricted analytic functions, while still being fairly flexible and applicable in many settings.

\subsection{Gevrey functions -- \texorpdfstring{$\bbR_{\rm G}$}{}}
While the structures considered so far have all been generated by classes of analytic functions, there also exist o-minimal structures defined by means of non-analytic functions. An important example is the structure $\bbR_{\rm G}$ generated by the class of \textit{Gevrey functions} introduced in \cite{DRIES_SPEISSEGGER_2000}. These functions play an important role in the theory of resummation, which appears in chapter~\ref{ch:observables}. Since the definition of Gevrey functions is quite technical, and since they are most naturally introduced in parallel with resummation, we defer a detailed exposition to chapter~\ref{ch:observables}.

\subsection{The Pfaffian closure}\label{sec:Pfaffclosure}
The class of Pfaffian functions satisfies strong global finiteness properties, which are essentially inherited from the intrinsic finiteness of the polynomials which appear in the defining system of differential equations. This principle can be used to construct new o-minimal structures from existing ones, in the following way.\\

Given an o-minimal structure $\cS$, we consider functions $\zeta_1,\ldots,\zeta_r:U\to \bbR$ satisfying a differential chain 
\begin{equation}
    \frac{\pd \zeta_i}{\pd x_j} = F_{ij}(x_1,\ldots,x_n,\zeta_1,\ldots,\zeta_i) \, ,
\end{equation}
where the $F_{ij}$ are differentiable tame functions definable in $\cS$. The functions obtained in this way turn out to inherit the tameness of the function in $\cS$, and generate a new o-minimal structure called the \textit{Pfaffian closure} of $\cS$, denoted by $\cS_{\rm Pf}$ \cite{Speisegger99}. This remarkable fact implies that any o-minimal structure can be enhanced by including these generalized Pfaffian functions. \\

To conclude this section, we note that it is remarkable that the variety of function classes considered here all give rise to a tame geometry, constrained by the powerful tameness theorems of section~\ref{sec:tamenesstheorems}.


\section{Tame geometry beyond Euclidean space}
Up until this point we have considered tameness of sets inside Euclidean space. By taking tame sets as a local model, o-minimality can be extended to spaces which are not naturally embedded in Euclidean space. This greatly extends the scope of tame geometry, and will be essential for our applications to physics.

\subsection{Tame spaces}

The first step in generalizing tame geometry is the following definition.

\begin{subbox}{Definition: $\cS$-definable space}
Given an o-minimal structure $\cS$, an $\mathcal{S}$-\textit{definable space} is a topological space $X$ with a finite open cover $\{U_i\}$ together with homeomorphisms \mbox{$\phi_i:U_i\to A_i \subseteq \bbR^{n_i}$} such that the sets $A_i$, the intersections $A_{ij}=\phi_i(U_i\cap U_j)$, and the transition functions \mbox{$\phi_{ij}=\phi_j\circ \phi_i^{-1}:A_{ij}\to A_{ji} $} are $\mathcal{S}$-definable. 
Such a cover is called a \textit{definable atlas}, and the maps $\phi_i$ are the charts of the definable atlas. 
\end{subbox}

The definition is analogous to that of a manifold or variety, but a crucial distinction is that a tame space is required to have a \textit{finite} atlas. This is necessary to preserve the essential finiteness of o-minimal structures. We will often not specify the underlying o-minimal structure, 
and refer to such a space as a \textit{tame space}. \\

Tame spaces arise naturally in geometry, as illustrated by the following two examples:
\begin{itemize}
\item \textbf{Algebraic varieties.} A real algebraic variety $X$ is an $\bbR_{\rm alg}$-definable space, since $X$ can be covered by finitely many affine varieties which serve as an $\bbR_\text{alg}$-definable atlas. Due to the compatibility of different affine covers, the definable structure on $X$ is uniquely determined by the algebraic variety structure. The construction also works if $X$ is a complex algebraic variety, since we may then view $X$ as a real algebraic variety of twice the dimension.
\item \textbf{Compact real analytic manifolds.} A real analytic manifold $X$ can be given the structure of an $\bbR_{\rm an}$-definable space if it is compact. The manifold comes by definition with an atlas of real analytic coordinate charts, and by compactness this atlas can be chosen to be finite. In addition, compactness guarantees that the transition functions can be chosen to be restricted analytic, so that they are definable in $\bbR_{\rm an}$.
\end{itemize}
We will encounter more examples of tame spaces throughout this thesis. \\

This generalization extends to subsets of tame spaces and functions defined on tame spaces, in the following way. A subset $A$ of a tame space $X$ is tame if the image $\phi_i(A\cap U_i)$ is a definable subset of $\bbR^{n_i}$ for each chart. Subsequently, a function between tame spaces $f:X\to Y$ is tame if its graph $\Gamma(f)$ is a tame subset of the product $X\times Y$. In this manner, many of the essential properties of tame sets and tame functions on Euclidean space, such as the cell decomposition and the triangulation theorem, extend to the global setting of tame spaces \cite{VdDries}.

\subsection{Tame manifolds and tame differential geometry}\label{sec:tameDG}
By strengthening the assumptions in the definition of a tame space, we come to the interface of tame geometry and differential geometry, which is of great relevance for the applications of tameness to physics. While o-minimality has fundamental links to algebraic geometry and arithmetic geometry, the connection to differential geometry is not as thoroughly developed. For the purposes of this thesis, it provides the natural language in which to frame the geometric applications of tameness. A more detailed exposition is given in \cite{ominDR}.

\begin{subbox}{Definition: tame manifold}
A \textit{tame manifold} $X$ is a tame space for which the transition functions in the finite definable atlas of charts are smooth.
\end{subbox}
These charts may be defined to be $\cC^k$-differentiable, but we restrict ourselves to the $\cC^\infty$ case. In the spirit of the triangulation theorem, tame manifolds can be thought of as smooth manifolds with a finite topological complexity. This includes compact analytic manifolds, but also non-compact manifolds which have a tame geometry at infinity.\\

Recall from the discussion in subsection~\ref{sec:monotonicity} that the derivative of a tame function is tame. This means that tame geometry and differential geometry are remarkably compatible. For example, by unwinding the definitions and checking that definability is preserved in each step, it can be shown that the tangent bundle $TX$ of a tame manifold $X$ is a tame manifold. In a similar vein, bundles of differential forms and tensor fields of a tame manifold are tame as well. The tameness of sections of these bundles can be characterized in the following way \cite{ominDR}.

\begin{subbox}{Tameness of tensor fields}
Given a set of tame coordinates $x^\mu$ on a local patch $U$ of a tame manifold $X$, a tensor field 
\begin{equation}
    \cT = \cT_{\mu_1\cdots \mu_p}^{\nu_1\cdots \nu_q} \,\dd x^{\mu_1}\otimes \cdots \otimes \dd x^{\mu_p} \otimes \pd_{\nu_1}\otimes \cdots \otimes \pd_{\nu_q} 
\end{equation}
is tame (as a local section of $T^{p,q}X$) if and only if each coefficient $\cT_{\mu_1\cdots \mu_p}^{\nu_1\cdots \nu_q}$ is a tame function on $U$.
\end{subbox}
This characterization is particularly useful for vector fields $v=v^\mu \pd_\mu$ and differential $1$-forms $\omega = \omega_\mu \dd x^\mu$. Any tame tensor field has a graph whose geometry has finitely many geometric features, controlled by the theorems of section~\ref{sec:tamenesstheorems}. By specializing to $(0,2)$-tensor fields, one can define a notion of tame Riemannian manifold as a tame manifold equipped with a tame metric $g= g_{\mu\nu }\dd x^\mu \otimes \dd x^\nu$. The resulting theory at the interface of o-minimality and Riemannian geometry is not developed in the literature, and as we will show in chapter~\ref{ch:volumes}, there are challenging open mathematical problems in this direction. 

\subsection{Tameness and Hodge theory}\label{sec:hodge}
We now briefly highlight a specialized mathematical application of o-minimality in the field of Hodge theory. The reason for doing so is twofold. Firstly, it emphasizes the remarkable strength of the framework in other geometric settings. Secondly, it underlies several of the more advanced applications of tameness to physics \cite{Grimm:2021vpn,Douglas:2022ynw,Bakker:2023xkt}. 

\newpage 
The complex cohomology groups of a compact K\"ahler manifold $X$ decompose as
\begin{equation}
 H^k(X;\bbC) = \bigoplus_{p+q=k}H^{p,q}(X)\,,
\end{equation}
with $H^{p,q}(X)$ consisting of cohomology classes represented by forms of degree $(p,q)$. This decomposition is called a \textit{Hodge structure}. When the geometry of $X$ is deformed, parametrized by a moduli space $\cM$, then the associated Hodge structure varies as well. This \textit{variation of Hodge structure} is captured by a map 
\begin{equation}
    \Phi: \cM \to \cD \,,
\end{equation}
from the moduli space $\cM$ into a space $\cD$ which classifies different Hodge structures. This map is called the \textit{period map}, and it is the central object of study in Hodge theory and complex algebraic geometry. A geometric interpretation of the period map is that it measures integrals of holomorphic forms over different topological cycles in the manifold $X$. The period map appears in several physical settings: it is of great relevance in the study of compactifications of string theory, discussed in chapter~\ref{ch:string}, where geometric deformations of a higher-dimensional spacetime manifold correspond to dynamical fields in the lower-dimensional theory. 
Another physical setting where Hodge theory appears is the evaluation of Feynman integrals in quantum field theory, mentioned in chapter~\ref{ch:observables}. We refer to \cite{Li:2022mhy,vandeHeisteeg:2022gsp,Monnee:2024gsq} for reviews of physical applications of Hodge theory. \\

A remarkable result of Bakker, Klingler, and Tsimerman is that the space $\cD$ is an $\bbR_{\rm alg}$-definable manifold, and that the period map $\Phi$ is definable in $\bbR_{\rm an,exp}$ \cite{BKT}. This means that the period map, despite being a complicated transcendental function, has a tame geometry. This is especially non-trivial in the asymptotic regions of the moduli space $\cM$, where the geometry of $X$ may degenerate.  
We refer to \cite{FresanNotes} for a review of these developments. \\

The theory of o-minimality provides a rich framework of tame geometry with many intriguing facets, and we hope that this chapter gives the reader an idea of what the field is about.
There are many aspects of the theory which can be developed further, part of which can be found in the cited literature.
One of these aspects, regarding the quantitative nature of o-minimality, is the topic of the next chapter.

\chapter{Complexity and sharp o-minimality}\label{ch:complexity}
\setlength{\parindent}{0pt}

The aim of this chapter is to introduce a notion of \textit{complexity} in tame geometry, made precise by the theory of sharp o-minimality. The measure of complexity developed in this chapter will play a crucial role in the applications to quantum field theory and quantum gravity in part II and part III of the thesis. As in the previous chapter, we start with an introductory section intended to provide context and motivation for the technical details that will follow in the rest of the chapter. 
We then proceed with the definition of sharply o-minimal structures, discuss the main examples and features of these structures, and describe how the complexity of tame sets can be characterized.

\section{Introduction and motivation}\label{sec:introcomplexity}
\subsection{Quantifying tameness}
In the previous chapter we have explained how the concept of o-minimality yields a geometric framework centered around the idea of finiteness, and how the tame geometry of sets and functions inside this framework is made manifest by numerous tameness theorems.
While o-minimality thus provides a powerful and elegant distinction between finite and infinite, it is a purely qualitative principle, and leaves the matter of \textit{how finite} unaddressed. It is therefore natural to ask whether o-minimality can be elevated to a quantitative principle, replacing the abstract finiteness appearing in the tameness theorems by explicitly computable numbers. \\

Answering this question turns out to be ambitious,  
and the first steps towards establishing a quantitative refinement of tame geometry were taken by Binyamini and Novikov, who introduced and developed the notion of \textit{sharp o-minimality} \cite{binyamini_tameness_2023}. The idea behind sharp o-minimality is to assign an axiomatic measure of complexity to every tame set, which controls their number of connected components and thereby lies at the root of the finiteness in tame geometry. There are several reasons why defining such a measure of complexity is a profound mathematical challenge. First, it has to apply to the vast collection of tame sets contained in an o-minimal structure, and must therefore be extremely general. Secondly, in order for the complexity to be meaningful and consistent, it has to behave in a controlled manner under all the geometric operations which can be performed in tame geometry.  
Finally, it has to incorporate the idea that a tame set or function may admit many different descriptions. In this chapter, we explain how these challenges can be addressed.

\subsection{On notions of information and complexity}
To set the stage for a definition of complexity in tame geometry, let us first zoom out and discuss some general aspects of information and complexity. The question of what precisely constitutes the information conveyed by an object $X$ is subtle and admits several interpretations. 
Inspired by a seminal paper of Kolmogorov \cite{Kolmogorov}, we distinguish three fundamentally different perspectives on quantifying information: 
\begin{itemize}
    \item \textbf{Combinatorial.} The combinatorial approach is the most simplistic, and counts the information needed to specify one object $X$ among a selected finite set of objects $\cX$, regarding each object as equally complex. The number of binary units required to do this, which scales as $\log |\cX|$, then defines the combinatorial information of $X$. This measure of information is not intrinsic to $X$ but depends on the set $\cX$, and it is combinatorial in the sense that it is concerned with counting the number of permissible objects $|\cX|$. 
    \item \textbf{Probabilistic.} The probabilistic approach incorporates randomness, and applies when there is a probability distribution $p(X)$ on the set of objects $\cX$. It measures the information by the entropy of the distribution,
    \begin{equation}\label{eq:entropy}
    S = - \sum_{X\in\cX} p(X) \log p(X)   \,,
    \end{equation}
    which is the expected amount of combinatorial information conveyed through selecting an object $X$ from $\cX$ at random. When the probability distribution is uniform, i.e.~$p(X) = 1/|\cX|$, it reduces to the combinatorial case. The probabilistic nature of this approach makes it of great relevance to physics, manifesting itself as thermodynamic entropy in statistical mechanics and as von Neumann entropy in quantum mechanics. 
    \item \textbf{Descriptive.} The descriptive approach characterizes the information of an object $X$ as its minimal description within a fixed language. 
    It is therefore intrinsic to $X$, and captures the internal structure of the object rather than its relation to an ensemble of external objects.
    Nonetheless, this information is sensitive to the chosen language in which $X$ is described. 
    The foundational example of this approach is Kolmogorov's algorithmic complexity, which defines the complexity of a string of symbols as the length of the shortest algorithm which produces that string. An example in physics is quantum computational complexity, which characterizes the number of elementary quantum gates required to construct a desired unitary operator.
\end{itemize}

Many measures of information exist, but most of them are a variation on one of these three themes. For the purposes of this thesis, the objects $X$ are tame sets, and the collection $\cX$ is an o-minimal structure. Since the collection $\cX$ is far too large to characterize combinatorially, and since there is no randomness involved, it appears that the descriptive notion of information is the natural point of view to take in tame geometry. Therefore, the descriptive perspective on information lies at the heart of what we will think of as `complexity' in this thesis. \\

In this conception, a measure of complexity within a class of objects can generally be defined roughly through the following steps:
    \begin{itemize}
    \item identify a set of elementary constituents from which the objects are comprised;
    \item define a set of compositional rules by which the objects are constructed from these constituents;
    \item specify how the complexity can be calculated from the compositional rules. 
\end{itemize}
As an example, in quantum computational complexity the elementary constituents are the chosen quantum gates, the compositional rule is operator multiplication, and complexity can be calculated by counting the number of gates, by Nielsen's geometric approach \cite{nielsen2005geometricapproachquantumcircuit}, or by other methods. Any complexity measure defined in this manner is inherently sensitive to the chosen elementary constituents and compositional rules, and hence suffers from a defect of canonicity. Instead of \mbox{regarding} complexity as a precise quantity, it is therefore sometimes better viewed through scaling laws characterizing the growth of complexity with the number of constituents. These scaling laws are especially useful for comparing the complexity of different objects, and for capturing their universal features.\\ 

\newpage
In this chapter, we will see how to implement this prescription for tame geometry.
Arguably, at the most atomic level the elementary constituents in tame geometry are points on the real line, or equivalently individual real numbers.
The compositional rules in tame geometry should then be the geometric operations under which an o-minimal structure is closed, reflected in axioms (i), (ii), and (iii). Equivalently, invoking the duality of logic and geometry, the compositional rules may be framed as first-order logical formulas over the real numbers. \\

The assignment of a measure of complexity to tame sets in an o-minimal structure is a subtle issue. 
As a basis, we might regard a single real number as having one unit of information.\footnote{Here we view the complexity of a real number from a geometric perspective, as defining a featureless point on the real line $\bbR$. In contrast, from a fundamental perspective a real number may in fact require an infinite amount of information to be specified. While this is a fascinating topic in its own right, we do not explore it further here. For an interesting discussion on the complexity of numbers in the context of periods we refer to \cite{Kontsevich2001}.}
Subsequently, viewing a tame set as defined in terms of logical formulas, a meaningful notion of complexity should reflect how complicated this logical formula is, which in turn should be deducible from the amount of logical symbols and real numbers appearing in the formula. Here, the duality of logic and geometry reveals a powerful aspect: since a logical operation can be translated to a geometric operation, the descriptive logical complexity of a tame set 
propagates to its topological and geometric complexity, through the geometric operations underlying the tameness theorems. 
This fundamental fact, which we will explore in detail in this chapter, is one of the main virtues of complexity in tame geometry.

\subsection{Complexity in real algebraic geometry}\label{sec:geomcompAG}

Having acquired a basic understanding of how we should characterize complexity in a geometric setting, let us see how these ideas are realized in algebraic geometry in Euclidean space. Since semi-algebraic sets constitute the simplest o-minimal structure, this discussion will serve as a prototype for a quantitative tame geometry. \\

The basic building blocks of algebraic geometry are polynomials, so we begin by attempting to characterize the complexity of a real polynomial
\begin{equation}
    P(x_1,\ldots,x_N) = \sum_{I,\,|I|\leq D} a_I x^I\, 
\end{equation}
of degree $D$ in $N$ variables. In this notation, $I=(I_1,\ldots,I_N)$ is a multi-index, and
\begin{equation}
    |I|=I_1+\cdots+I_N, \quad a_I = a_{I_1\cdots I_N}, \quad \text{and}\quad x^I = x_1^{I_1} \cdots x_N^{I_N}\, .
\end{equation}
The maximal number of terms of such a polynomial depends on $N$ and $D$ as
\begin{equation}\label{eq:Polyterms}
    \cC =  \frac{(N+D)!}{N!D!} \,.
\end{equation}
Regarding a real number as a fundamental unit of information, the descriptive point of view suggests that this number is a good measure of the complexity of $P$, since it is the number of independent real numbers $a_I$ required to specify $P$. However, as noted in the previous section, in tame geometry we wish to link the descriptive measure of complexity to geometric complexity. While $N$ and $D$ appear on the same footing in equation \eqref{eq:Polyterms}, they play a radically different role in geometry. 
This is formalized by the following fundamental theorem. 

\begin{subbox}{Bézout's theorem}
Let $P_1,\ldots,P_N$ be polynomials in $N$ real variables of degrees $D_1,\ldots,D_N$ respectively. Then the number of non-degenerate solutions of the system of equations $P_1=\ldots=P_N=0$ is bounded by $\prod_{k=1}^N D_k$.
\end{subbox}

A consequence\footnote{This can be shown by applying Bézout's theorem to $P$ and its partial derivatives, and using ideas from Morse theory.} of the Bézout bound is that the number of connected components of the zero set $\{P=0\}$ is bounded by $D^N$. Therefore, the geometric complexity of algebraic sets depends \textit{polynomially} on the degree $D$, while it depends \textit{exponentially} on the number of variables $N$.\footnote{While the focus on this chapter lies on geometric complexity, we emphasize that this is inextricably linked to computational complexity, since the Bézout bound controls the number of solutions to algebraic equations.} A measure of descriptive complexity of $P$ which accurately controls its geometric features should reflect this striking difference.\\ 

A resolution is as follows: instead of combining $N$ and $D$ into a single number, we disentangle them and measure the complexity of a polynomial by \textit{two numbers}. While by doing so we lose the ability to straightforwardly compare the complexities of two polynomials, viewing $(N,D)$ as the descriptive complexity of a polynomial ensures the geometric complexity is under much more control. In addition, it allows us to better keep track of the interaction between complexity and geometric operations.\\

To illustrate this, let $P_1,\ldots,P_n$ be real polynomials of degree $D_1,\ldots,D_n$ in $N_1,\ldots,N_n$ variables respectively, and consider the zero sets $X_i=\{P_i=0 \}\subseteq \bbR^{N_i}$ for $i=1,\ldots,n$. We then have $b_0(X_i)\leq D_i^{N_i}$, and from Bézout's theorem we can infer that the number of components of the union of the $X_i$ are bounded by 
\begin{equation}
    b_0\Big(\bigcup_{i=1}^n X_i \Big) = b_0\Big( \Big\{\prod_{i=1}^nP_i =0 \Big\} \Big)  \leq \Big( \sum_{i=1}^n D_i \Big)^{\max_i (N_i)} \,.
\end{equation}
This bound indicates that the number of variables $N_i$ and the degree $D_i$ behave rather differently under a geometric operation: we must take the maximum of the number of variables $\max_i (N_i)$ in order to make the union well-defined in $\bbR^{\max_i(N_i)}$, whereas the degree appears to be an additive quantity under unions. A similar conclusion can be drawn from considering the intersection of algebraic sets. 
These observations serve as a prelude to sharp o-minimality.

\subsection{Towards sharp o-minimality}
With the discussion of this section, we are almost in a position to define sharp o-minimality in a natural way. Let us reiterate the core ideas:

\begin{subbox}{Motivating ideas for sharp o-minimality}
    \begin{itemize}
    \item We wish to refine the axioms of o-minimal structures so that the tameness theorems become quantitative.
    \item To do so, we define a notion of complexity for tame sets. It should measure the descriptive information of a tame set, while effectively controlling the number of connected components.
    \item This measure should behave in a controlled manner under the geometric operations of o-minimal structures, reflecting the corresponding operations on logical formulas.  
    \item In order to disentangle the polynomial and super-polynomial dependence, the complexity should be measured by two numbers. 
\end{itemize}
\end{subbox}

With these ideas in mind, we are prepared for the technical definition of sharp o-minimality, which provides an axiomatic realization of this discussion.

\section{Defining sharp o-minimality}\label{sec:defsharp}

\subsection{Format-degree filtrations and sharply o-minimal structures}
Let us now turn to the technical definition of sharp o-minimality. The measure of complexity is implemented by organizing the sets in an o-minimal structure $\cS$ into collections $\Omega_{\cF,\cD}$, indexed by two integers $\cF,\cD\in\bbN$ called \textit{format} and \textit{degree}. These collections are assumed to form a filtration, meaning that
\begin{equation}\label{eq:FDfiltration}
    \Omega_{\cF,\cD} \subseteq \Omega_{\cF,\cD+1}, \quad \Omega_{\cF,\cD} \subseteq \Omega_{\cF+1,\cD} 
    \quad \text{and}\quad  
    \bigcup_{\cF,\cD}\Omega_{\cF,\cD}=\cS  \, . 
\end{equation}
The collection $\Omega = \{\Omega_{\cF,\cD}\}$ is called a \textit{format-degree filtration}, and the subsets in $\Omega_{\cF,\cD}$ are thought of as having complexity $(\cF,\cD)$. In order for this definition of complexity to be meaningful, we have to impose a number of consistency requirements with the axioms of o-minimality, leading us to the central definition of this chapter introduced by Binyamini and Novikov \cite{binyamini_tameness_2023}.
\begin{subbox}{Definition: sharply o-minimal structure}
A \textit{sharply o-minimal structure} consists of an o-minimal structure $\cS$ equipped with a format-degree filtration $\Omega$ satisfying the following conditions.
\begin{enumerate}
    \item[(i)] \textbf{(Unions and intersections)} If $A_i \in \Omega_{\cF_i,\cD_i}$ for $i=1,\ldots,k$, then
     \begin{equation}
     \bigcup_{i=1}^ k A_i \,,\,\,  \bigcap_{i=1}^ k A_i \in \Omega_{\cF,\cD} \,,
     \end{equation}
     where $\cF=\max_i\{ \cF_i\}$ and $\cD=\sum_i \cD_i$.
    \item[(ii)] \textbf{(Complements, projections, and products)} If $A \in \Omega_{\cF,\cD}$, $A\subseteq \bbR^n$, then 
     \begin{equation}
       \bbR^n\backslash A\,, \,\,  \pi(A) \in \Omega_{\cF,\cD} \,,
     \end{equation}
     for any linear projection $\pi:\bbR^n\to\bbR^{n-1}$, and
     \begin{equation}
     A\times \bbR \,, \,\,\bbR\times A \in\Omega_{\cF+1,\cD} \,.
     \end{equation}
    \item[(iii)] \textbf{(Algebraic sets and dimension)} For any real $n$-variable polynomial $P$, 
    \begin{equation}
        \{P=0\} \in \Omega_{n,\deg P} \,.
    \end{equation}
    Moreover, if $A \in \Omega_{\cF,\cD}$ and $\bbR^n$, then $\cF \geq n$.
    \item[(iv)] \textbf{(Sharp o-minimality)} For every $\cF$ there is a polynomial 
    $P_\cF$ such that if $A \in \Omega_{\cF,\cD}$ and $A\subseteq \bbR$, then $A$ has at most $P_\cF(\cD)$ connected components.
\end{enumerate}
\end{subbox}
Note how these axioms mirror those of o-minimal structures. The axioms (i) and (ii) control how the complexity of tame sets behaves under elementary set-theoretic operations. In particular, it shows that for tame sets constructed by taking unions and intersections, the format $\cF$ serves as a bound on the formats of the constituent sets, whereas the degree $\cD$ is an additive quantity that grows when more sets are added. Meanwhile, the complexity $(\cF,\cD)$ remains constant under linear projections and complements.  Axiom (iii) implements the idea that format and degree generalize the dimension and degree from algebraic geometry. \\

Finally, axiom (iv) is the crucial axiom of sharp o-minimality which makes tameness quantitative. It assumes that the o-minimal structure $\cS$ with its format-degree filtration $\Omega$ comes equipped with a collection of polynomials $\{P_\cF\}_{\cF\in\bbN}$ which control the number of connected components of one-dimensional tame sets in terms of their format and degree $(\cF,\cD)$. 
The closure under geometric operations, and in particular under linear projections, ensures that every tame set inherits the quantitative finiteness from the sharp o-minimality assumption.

\subsection{Complexity in sharply o-minimal structures}
Let us now discuss the notion of complexity defined through the format-degree filtration in more detail. We begin by noting that a sharply o-minimal structure is defined abstractly, as an axiomatization of a complexity measure in tame geometry. In particular, it only provides a measure of complexity after selecting the format-degree filtration. We will discuss examples of how format-degree filtrations are constructed later in this chapter.

\subsubsection*{The complexity spectrum of a tame set}
\begin{figure}[h]
    \centering
    \includegraphics[width=0.67\linewidth]{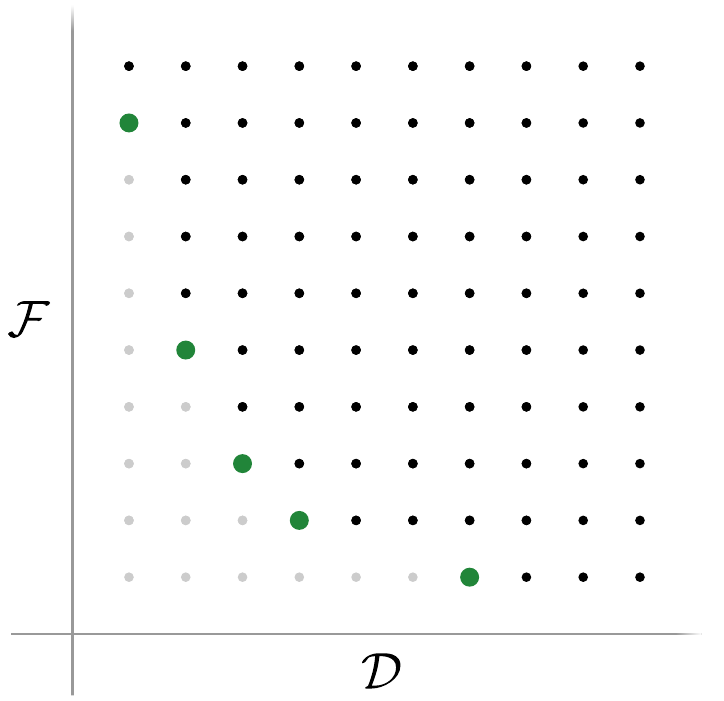}
    \caption{An example of the complexity spectrum of a tame set $X$. This spectrum consists of the points $(\cF, \cD)$ for which $X \in\Omega_{\cF,\cD}$, and is represented by the black and green dots. The extremal points, illustrated by the green dots, form a finite set $\cC_\Omega(X)$ which characterizes the complexity of $X$.}
    \label{fig:FDfiltration}
\end{figure}

Assuming that we have a sharply o-minimal structure $\cS$ with a format-degree filtration $\Omega$, consider a tame set $X$ definable in $\cS$. The complexity of $X$ is then characterized by the format-degree pairs $(\cF,\cD)$ for which $X\in\Omega_{\cF,\cD}$. There is no unique such pair, and instead the filtration structure of $\{\Omega_{\cF,\cD}\}$ implies that there are infinitely many pairs $(\cF,\cD)$ with $X\in\Omega_{\cF,\cD}$. We call the collection of these pairs $\{(\cF,\cD)\,|\,X\in\Omega_{\cF,\cD}\}$ the \textit{complexity spectrum} of $X$. This concept is visualized in figure \ref{fig:FDfiltration}.
Instead of viewing the apparent non-uniqueness of format and degree as an obstacle, we recall that this is a fundamental feature of descriptive complexity: the complexity of an object is not purely inherent to the object itself, but also determined by the representation of the object. \\

The complexity spectrum is uniquely determined by only finitely many points, namely the extremal points $(\cF,\cD)$ with the property that
\begin{equation}
    X \notin \Omega_{\cF,\cD-1} \quad\text{and}\quad X \notin \Omega_{\cF-1,\cD} \,.
\end{equation}
The relevance of these points is that there is no trivial way to reduce $(\cF,\cD)$, and hence these correspond to the optimal representations of $X$. With this in mind, the finite set 
\begin{equation}
    \cC_{\Omega}(X) = \big\{(\cF,\cD) \,|\, X\in \Omega_{\cF,\cD}, \,X\notin \Omega_{\cF,\cD-1},\, X\notin \Omega_{F-1,D} \big\} \,.
\end{equation}
characterizes the complexity of a tame set $X$ in a sharply o-minimal structure.  \\

\newpage 

The interpretation of format and degree as the complexity of a tame set comes from the polynomials $\{P_\cF\}$ that are part of the definition of a sharply o-minimal structure. The fact that the number of connected components of a tame set $X\subseteq \bbR$ with $X\in\Omega_{\cF,\cD}$ is bounded by $P_\cF(\cD)$ implies that the finiteness in o-minimality, captured by the tameness theorems, can be bounded by an $\cF$-dependent polynomial in $\cD$ as well. This will be discussed in more detail in section~\ref{sec:tamenesstheorems2}. 
Given a geometric bound in terms of $\cF$ and $\cD$, whose form may depend on the geometric property under consideration, the optimal representation of a tame set minimizes this bound. In this way, these geometric bounds can be used to select a minimal complexity in the complexity spectrum of a tame set. In the spirit of dualities in physics, the $(\cF,\cD)$-pairs in $\cC_\Omega(X)$ representing a tame set can be interpreted as duality frames, with the preferred optimal duality frame depending on the situation.

\subsubsection*{Example: complexity of points}
To illustrate the complexity in a sharply o-minimal structure, let us compute it for a number of elementary examples. We focus on the sharply o-minimal structure $\bbR_{\rm alg}$, where the degrees of sets are determined by the degrees of underlying polynomials. First, consider a set $\{a\}$ consisting of a single point in $\bbR$. This is an algebraic set, namely the zero set of the polynomial $P(x)=x-a$. It follows from axiom (iv) that $\{a\}\in \Omega_{1,1}$, and therefore $\cC_\Omega(\{a\})=\{(1,1)\}$. More generally, consider a set of $n$ points $\{a_1,\ldots,a_n\}$ in $\bbR$. This set can be obtained as the zero set of a degree $n$ polynomial,
\begin{equation}
    \{a_1,\ldots,a_n\} = \bigg\{ \prod_{j=1}^n(x-a_j)=0 \bigg\}\,.
\end{equation}
It thus follows that $\{a_1,\ldots,a_n\}\in\Omega_{1,n}$. In general, whether the pair $(1,n)$ is part of the extremal points in the complexity spectrum $\cC_\Omega(\{a_1,\ldots,a_n\})$ depends on the precise form of the format-degree filtration. Since we are currently focusing on $\bbR_\text{alg}$, the definable sets can only be constructed by polynomials, so in this structure we indeed have $(1,n) \in \cC_\Omega(\{a_1,\ldots,a_n\})$. Now let us consider alternative representations of $\{a_1,\ldots,a_n\}$ with a higher format $\cF$. Depending on the number of points $n$, we can obtain this finite set by projecting a higher-dimensional set of lower degree. B\'ezout's theorem implies that the intersection of $N$ degree $D$ hypersurfaces in $\bbR^N$ consists of at most $D^N$ isolated points. Projecting this intersection down to the real line preserves the complexity by the axioms of sharp o-minimality, so if $n\leq D^N$ we have $\{a_1,\ldots,a_n\}\in \Omega_{N,ND}$. This example shows that even for the simplest sets, the complexity spectrum contains multiple extremal pairs $(\cF,\cD)$.

\subsubsection*{Example: topological closure of a tame set}
To further illustrate how the format and degree of a tame set behaves under geometric operations, let $X\subseteq \bbR^n$ be a tame set with $X \in \Omega_{\cF,\cD}$ and consider its topological closure 
\begin{equation}
    \overline X = \big\{ y \in \bbR^n \,\big|\,\forall \eps>0 \,\exists x\in X \,:\, \norm{x-y}^2 <\eps \big\} \,,
\end{equation}
which was shown to be definable in subsection~\ref{sec:dualitylogic}. Invoking the correspondence between logical formulas and geometric operations, we can express this set as
\begin{equation}\label{eq:closureproj}
\overline X  = \bbR^n\setminus \pi_{\bbR^n}\left( \bbR^{n+1} \setminus \pi_{\bbR^n\times \bbR}\Big( \big\{(y,\eps,x) \in \bbR^n\times \bbR\times A \,\big|\, \norm{x-y}^2 <\eps \big\} \Big)\right)\,,
\end{equation}
using the linear projections $\pi_{\bbR^n}:\bbR^n\times\bbR\to \bbR^n$ and $\pi_{\bbR^n\times\bbR}:\bbR^n\times\bbR \times X\to \bbR^n\times\bbR$. The axioms of a sharply o-minimal structure now instruct us how to calculate the format and degree of $\overline X$, and from equation~\eqref{eq:closureproj} we infer that 
\begin{equation}
    \overline{X} \in \Omega_{\cF+n+2,\cD+2}\,.
\end{equation}
We leave the details of this calculation as an instructive exercise for the reader who wishes to get familiar with geometric manipulations in sharp o-minimality. Note that if $X$ is a closed set, then $\overline X =X$ so that $\overline X\in \Omega_{\cF,\cD}$ as well. This shows that the format and degree obtained by tracing through geometric operations, though they hold universally, may not be the most efficient representation of a tame set. 

\subsubsection*{Reductions of format-degree filtrations}
While the format-degree filtration in a sharply o-minimal structure is a choice of additional data on top of an o-minimal structure, there exists a notion of \textit{reduction} of format-degree filtrations which makes it possible to compare different filtrations \cite{binyamini_tameness_2023}. Given two format-degree filtrations $\Omega$ and $\widehat\Omega$ on a structure $\cS$, $\Omega$ is said to be reducible to $\widehat\Omega$ if there exists a function $q:\bbN\to\bbN$ and a collection of polynomials $\{Q_\cF\}_{\cF\in\bbN}$ with integer coefficients such that 
\begin{equation}
    \Omega_{\cF,\cD} \subseteq \Omega'_{q(\cF),Q_\cF(\cD)} \quad \text{for all } \cF,\cD\in\bbN \,.
\end{equation}
This condition entails that the format and degree with respect to the two different filtrations have comparable scaling properties, encoded by a function $q$ shifting the format and an $\cF$-polynomial $Q_\cF(\cD)$ universally modifying the polynomial growth in the degree. In technical applications of sharp o-minimality this construction is essential to compare different format-degree filtrations  \cite{binyamini_tameness_2023,binyamini2022sharply}.

\subsection{Examples, non-examples, and conjectural examples}
While the theory of o-minimality has been established for decades with many known examples, sharp o-minimality is still a novel concept which is under active development, and consequently only few examples are known at present.
In this section we discuss the key examples and non-examples of function classes which generate a sharply o-minimal structure, as well as structures which are conjectured to admit a format-degree filtration satisfying the axioms of sharp o-minimality.

\subsubsection*{Algebraic functions}
The simplest o-minimal structure is the structure $\bbR_{\rm alg}$ consisting of semi-algebraic sets, and in view of the discussion of subsection~\ref{sec:geomcompAG} on complexity in real algebraic geometry, it constitutes the first example of a sharply o-minimal structure \cite{binyamini_tameness_2023}. The format-degree filtration is generated by assigning a complexity $(\cF,\cD)=(N,D)$ to the zero set $\{P=0\}$ of any $N$-variable polynomial of degree $D$. Even though $\bbR_\text{alg}$ is a relatively simple o-minimal structure, this result is non-trivial and relies on effective cell decomposition algorithms. The computational complexity of these algorithms determine the polynomials $\{P_\cF\}$ needed in the definition of sharp o-minimality.  The form of these polynomials generalizes the Bézout bound $\cD^\cF$.

\subsubsection*{Pfaffian functions}
The definition of sharp o-minimality is rooted in the complexity of algebraic sets, and therefore its novelty lies mostly in examples beyond $\bbR_{\rm alg}$. The main example of a sharply o-minimal structure beyond $\bbR_{\rm alg}$ is the structure generated by Pfaffian functions, encountered in section~\ref{sec:catalog} \cite{binyamini2022sharply}. In essence, this is because Pfaffian functions have a strong similarity to polynomials, despite being a fairly general class of analytic functions. Specifically, the finiteness properties of Pfaffian functions are quantitative in nature, in the sense that there are sharp bounds on geometric features of graphs of Pfaffian functions \cite{GabVor04}. This structure is of great relevance for this thesis, and therefore has an extended discussion in section~\ref{sec:Pfaff}. 

\subsubsection*{Log-Noetherian functions}
There also exist o-minimal structures which are known to satisfy controllable quantitative finiteness properties, of which it is nonetheless not yet known whether they admit a format-degree filtration making them into a sharply o-minimal structure. One such structure which is conjectured to be sharply o-minimal, is generated by the class of log-Noetherian functions introduced by Binyamini \cite{binyamini2024lognoetherianfunctions}. These functions are defined on certain complex cellular domains by a differential chains similar to Pfaffian and Noetherian chains, generalized by allowing for logarithmic singularities. The definition of this structure is rather technical, and postponed to subsection~\ref{sec:LNfunctions}.

\subsubsection*{Restricted analytic functions}
The main example of an o-minimal structure which is \textit{not} sharply o-minimal is the structure $\bbR_{\rm an}$ generated by restricted analytic functions \cite{binyamini_tameness_2023}. Intuitively, the information required to specify a generic analytic function is too much to admit polynomially controlled effective bounds. An example which illustrates this is the $\bbR_{\rm an}$-definable holomorphic function 
    \begin{equation}
        f(z) = \sum_{k=1}^\infty z^{\omega_k}
    \end{equation}
on the disk of radius $1/2$, where the sequence $\omega_k$ is inductively defined by $\omega_k=2^{\omega_{k-1}}$ and $\omega_1=1$. For $N\geq 1$ and $\eps>0$, consider the subsets of $\bbC$ given by 
\begin{equation}
    X_N(\eps) = \Big\{ z\in \bbC \,\Big|\, f(z) = \eps + \sum_{k=1}^N z^{\omega_k}  \Big\} \,.
\end{equation}
For $0<\eps\ll 1$, the number of points in this set is $|X_N(\eps)|=\omega_{N+1}$, since for $\eps\to 0 $ the equation defining this set becomes $z^{\omega_{N+1}} \approx \eps$. Suppose that $\bbR_{\rm an}$  admits a format-degree filtration satisfying the axioms of sharp o-minimality. The number of points in $X_N(\eps)$ would then have to grow polynomially in $\omega_{N}$, with the coefficients of the polynomial depending on the complexity of $f$. Since $|X_N(\eps)|\sim \omega_{N+1} =2^{\omega_N}$ for large $N$ and small $\eps$, this yields a contradiction, proving that $\bbR_{\rm an}$ cannot be sharply o-minimal \cite{binyamini_tameness_2023}. This conclusion extends to any o-minimal structure containing the restricted analytic functions, such as $\bbR_{\rm an,exp}$, which shows that sharp o-minimality is a strong condition, incompatible with some of the important o-minimal structures.

\section{Pfaffian functions}\label{sec:Pfaff}
The prime example of sharp o-minimality beyond algebraic functions comes from the class of Pfaffian functions, introduced in section \ref{sec:catalog}. In this section we explain how a complexity can be associated to these functions, which provides a way of defining a format and degree for tame sets in Pfaffian structures.

\subsection{Pfaffian chains and Khovanskii's theorem}
For the convenience of the reader we recall that a Pfaffian chain on an open box $U\subseteq \bbR^n$ is a finite sequence of analytic functions $\zeta_1,\ldots,\zeta_r:U\to\bbR$ which satisfies a system of first-order differential equations of the form 
\begin{equation}\label{eq:Pchain2}
\frac{\partial \zeta_i}{\partial x_j} =  P_{ij} (x_1,\ldots,x_n,\zeta_1,\ldots,\zeta_i)  \quad\text{for} \quad i=1,\ldots,r, \quad j=1,\ldots, n,
\end{equation}
where $P_{ij}$ is a polynomial in $n+i$ variables. The crucial condition is that this system is triangular, in the sense that the derivative of $\zeta_i$  depends on the preceding functions in the chain $\zeta_1,\ldots,\zeta_i$, but not on $\zeta_{i+1},\ldots,\zeta_r$. Given such a chain, a Pfaffian function is a polynomial in the variables and the functions in the chain, i.e.~a function $f:U\to\bbR$ defined by
\begin{equation}\label{eq:Pfaffianfunction}
f(x_1,\ldots,x_n) = P(x_1,\ldots,x_n,\zeta_1,\ldots,\zeta_r) \,,
\end{equation}
where $P$ is a polynomial.
These functions satisfy the following theorem, which is of fundamental importance for the theory of sharp o-minimality \cite{Khovanskii}.

\begin{subbox}{Khovanskii's theorem}
Let $f_1,\ldots,f_n$ be Pfaffian functions on $\bbR^n$. Then the number of non-degenerate solutions of the system of equations $f_1=\ldots=f_n=0$ is finite and bounded by an explicit function of the defining data of the underlying Pfaffian chains.    
\end{subbox}
This theorem is a generalization of Bézout's theorem (subsection \ref{sec:geomcompAG}), which bounds the number of non-degenerate solutions of a polynomial system of equations. For this reason, Pfaffian functions are regarded as analytic functions which nevertheless behave like polynomials. \\

While we will not discuss the proof of this theorem, it is instructive to understand where the inherent finiteness of Pfaffian functions comes from. 
The underlying principle is quite simple: if $f:\bbR\to\bbR$ is a differentiable function taking the same value $f(a)=f(b)$ at two points $a$ and $b$, then there must be a point $c$ between $a$ and $b$ where $f'(c)=0$. Consequently, if $f$ has $N$ distinct isolated zeros, then its derivative $f'$ must have at least $N-1$ zeros. Reversing this principle, we can bound the number of zeros of $f$ in terms of the number of zeros of $f'$. Therefore, if $f$ satisfies a Pfaffian differential equation, its derivative is determined by a polynomial $P(x,f)$, meaning that the number of zeros of $f$ can be bounded in terms of the degree of $P$.  
A Pfaffian chain iterates this principle for every function that is added to the chain, thereby keeping algebraic control over the number of zeros.

\subsection{Pfaffian complexity}
The formulation of Khovanskii's theorem bounds the number of zeros of Pfaffian functions in terms of the defining data of the Pfaffian chain. This defining data 
consists of the following integer numbers:
\begin{itemize}
\item the number of variables $n$;
\item the number of functions $r$, also known as the \textit{order} of the Pfaffian chain;
\item the degree $\alpha = \text{max}(\deg{P_{ij}})$ of the underlying Pfaffian chain;
\item the degree $\beta=\deg P$ of the polynomial $P$ used to define $f$.
\end{itemize}
Together, these numbers define the \textit{Pfaffian complexity} of a Pfaffian function, and we denote this as
\begin{equation}\label{eq:Pcomplexity}
\cC_{\rm Pf}(f) = (n,r,\alpha,\beta)\,.
\end{equation}
The Pfaffian complexity of a Pfaffian function depends implicitly on the underlying Pfaffian chain, and since such a function may be represented by different chains, the Pfaffian complexity is not \textit{uniquely} defined. The notation $\cC_{\rm Pf}$ should therefore not be regarded as defining a single-valued quantity. In fact, given a Pfaffian function one can always make the underlying chain more complicated by adding unnecessary variables, auxiliary functions, and increased degrees. This observation already shows a resemblance between Pfaffian complexity and the filtration structure of sharp o-minimality: rather than assigning a single complexity, each Pfaffian function has a spectrum of complexities determined by all the Pfaffian chains which may represent it.

\subsubsection*{Examples of Pfaffian complexity}
Below we list some examples of Pfaffian functions and their Pfaffian complexities.
\begin{itemize}
\item \textbf{Polynomials.} An $n$-variable polynomial $Q$ of degree $\beta$ is a Pfaffian function of complexity $(n,0,1,\beta)$, as it can be defined using an empty Pfaffian chain.\footnote{Note that, even though there is no Pfaffian chain in this example, the degree $\alpha$ is defined to be greater than or equal to $1$ by convention.}
\item \textbf{The exponential.} The real exponential function $f(x)=e^{ax}$ for $a\in\bbR$ is a Pfaffian function of complexity $(1,1,1,1)$, since it fits into the Pfaffian chain 
\begin{equation}
    \frac{\partial \zeta}{\partial x} = a \zeta \,.
\end{equation}
\item \textbf{Rational functions.} The function $f(x)=1/x$, defined on the interval $(0,\infty)$, has complexity Pfaffian $(1,1,2,1)$, since it is a solution to the equation
\begin{equation}
    \frac{\partial \zeta}{\partial x} = - \zeta^2\,.
\end{equation}
\item \textbf{Restricted trigonometric functions.} The function $f(x)=\cos(x)$ defined on the interval $(-\pi,\pi)$ can be obtained as a Pfaffian function as follows. Consider the chain $\zeta$ formed by
\begin{align}
\frac{\partial \zeta_1}{\partial x} &=  \frac{1}{2}\big(1+ \zeta_1^2 \big)\,, \\
\frac{\partial \zeta_2}{\partial x} & = - \zeta_1\zeta_2\,, \nonumber
\end{align}
which has solutions $\zeta_1(x) = \tan(x/2)$ and $ \zeta_2(x) = \cos^2(x/2)$. Using trigonometric relations, we may then obtain $f$ as the Pfaffian function
\begin{equation}
f(x) = \cos(x) = 2\zeta_2(x) -1\,, 
\end{equation}
which shows that $\cC_{\rm Pf}(f) = (1,2,2,1)$. For any integer $m\geq 1$, the function $f_m(x)=\cos(mx)$ can be written as a polynomial of degree $m$ in $f(x)=\cos(x)$ using the Chebyshev polynomials, and therefore it has Pfaffian complexity $\cC_{\rm Pf}(f_m)=(1,2,2,m)$. This example is essential for applications to physics. 
\end{itemize}

To illustrate how the complexity of a Pfaffian function depends on the chosen chain, consider the mononomial $f(x)=x^d$. We examine two ways of constructing $f$ as a Pfaffian function. As in the examples discussed above, the most obvious construction is to let the Pfaffian chain be empty, in which case we have $\cC_{\rm Pf}(f) = (1,0,1,d)$. Alternatively, we can construct $f$ by means of differential equations, and consider the following Pfaffian chain $\zeta$ on the domain $U=(0,\infty)$,
\begin{align}
\frac{\partial \zeta_1}{\partial x} &=  - \zeta_1^2\,, \\
\frac{\partial \zeta_2}{\partial x} & =  d\, \zeta_1\zeta_2\,. \nonumber 
\end{align}
The solutions to these differential equations are $\zeta_1(x) = 1/x$ and $\zeta_2(x)=f(x)$, and the Pfaffian complexity of $f$ in this representation is given by $\cC_{\rm Pf}(f) = (1,2,2,1)$. More generally, if $f$ is a degree $d$ polynomial consisting of $M$ monomial terms, we can add each monomial to the set of differential equations to obtain a Pfaffian chain $\zeta$ of length $r=M+1$. The Pfaffian complexity of the polynomial $f$ is then
\begin{equation}
    \cC_{\rm Pf}(f) = (n,M+1,2,1)\, .
\end{equation}
If $f$ is a \textit{fewnomial}, i.e.~a polynomial function with $M\ll d$, then the latter representation may be favorable for minimizing the complexity. This idea underlies Khovanskii's fewnomial theory \cite{Fewnomials}.

\subsubsection*{Calculus of Pfaffian chains}
When performing manipulations on Pfaffian functions, it is useful to keep in mind that the class of Pfaffian functions is closed under several operations, such as taking sums, products, derivatives, and compositions. In fact, it is possible to explicitly track the Pfaffian complexity of the resulting functions:
\begin{itemize}
\item If $f_1$ and $f_2$ are Pfaffian with complexity $\cC_{\rm Pf}(f_i)=(n,r_i,\alpha_i,\beta_i)$ for $i=1,2$, then the sum $f_1+f_2$ and the product $f_1f_2$ are Pfaffian with complexity
\begin{equation}
\label{ComplexityRules}
\begin{aligned}
\cC_{\rm Pf}(f_1+f_2)&=(n,r_1+r_2,\max(\alpha_1,\alpha_2),\max(\beta_1,\beta_2))\,, \\
\cC_{\rm Pf}(f_1f_2)&=(n,r_1+r_2,\max(\alpha_1,\alpha_2),\beta_1+\beta_2)\,.
\end{aligned}
\end{equation}
The required Pfaffian chain is obtained by assembling the chains of $f_1$ and $f_2$ into a single chain. Note that if there is overlap in the two chains, then the order of the chain for the sum and product is less than $r_1+r_2$.

\item If $f$ is Pfaffian with complexity $\cC_{\rm Pf}(f)=(n,r,\alpha,\beta)$, then the partial derivative $\partial f/\partial x_j$ is Pfaffian in the same chain, with complexity 
\begin{equation}
 \cC_{\rm Pf}\Big(\frac{\partial f}{\partial x_j}\Big)=(n,r,\alpha,\alpha+\beta-1)  \,.
\end{equation}

\item If $f_1:U_1\to \bbR$ and $f_2:U_2\to \bbR$ are Pfaffian functions of complexity $\cC_{\zeta_i}(f_i)=(1,r_i,\alpha_i,\beta_i)$, where $U_1$ and $U_2$ are open intervals such that the image $f_1(U_1)$ is contained in $U_2$, 
then the composition $f_2\circ f_1$ is Pfaffian with complexity
\begin{equation}
\label{ComplexityRules2}
\cC_{\rm Pf}(f_2\circ f_1) = (1,r_1+r_2,\alpha_2\beta_1 +\alpha_1+\beta_1-1,\beta_2)\,.
\end{equation}
The required Pfaffian chain is obtained by combining the chain of $f_1$ with the composite of $f_1$ with all the functions in the chain of $f_2$. The resulting Pfaffian complexity is obtained by using the chain rule for partial derivatives. 
\end{itemize}
We will refer to these rules several times in this thesis.

\subsubsection*{Information-theoretic perspective on Pfaffian complexity} 
The complexity of a Pfaffian function can also be interpreted from the perspective of descriptive information, along the lines of the discussion in section~\ref{sec:introcomplexity}. Consider an analytic function on the real line with a local power series expansion. Generically, this power series is specified by infinitely many independent coefficients, signaling that an analytic function requires an infinite amount of information to be specified. The key idea for representing such a function with a finite amount of information is that there may be relations among the power series coefficients. For example, consider the analytic function $f(x) = e^x$. The power series $f(x)=\sum_{k=0}^\infty a_ k x^ k$ of this function has infinitely many non-zero coefficients, but they satisfy an infinite set of algebraic relations, namely $a_k=a_{k-1}/k$ for all $k$. Differential equations provide a method of efficiently encoding such algebraic relations. For the real exponential, the algebraic relations among the coefficients can be captured by a single logical statement, namely $\pd f /\pd x=f$. By this principle, the power series coefficients of Pfaffian functions can be encoded with a finite amount of information through the algebraic relations encoded in the differential chain. The Pfaffian complexity quantifies the complexity of the underlying algebraic relations, and in this way provides a measure of descriptive information of a Pfaffian function.

\subsection{Explicit bounds from Pfaffian complexity}
The main use of Pfaffian complexity is that it encodes sharp bounds on geometric objects constructed from Pfaffian functions. This provides a realization of the idea that descriptive complexity controls geometric complexity in the context of tame geometry, discussed in section~\ref{sec:introcomplexity}. In the following subsection we provide several examples of this idea.

\subsubsection*{Khovanskii's theorem}
As a first example of explicit bounds in the Pfaffian framework, let us revisit Khovanskii's theorem \cite{Khovanskii,Fewnomials}. Consider a set of $n$ Pfaffian functions $f_1,\ldots,f_n$ defined by a common Pfaffian chain on an  $n$-dimensional domain $U\subseteq \bbR^n$, and assume that $\cC_{\rm Pf}(f_i) = (n,r,\alpha,\beta_i)$. Then the number of non-degenerate solutions to the equation $f_1=\cdots=f_n=0$ is bounded by the following quantity:
\begin{subbox}{Khovanskii's bound}
\begin{equation}
2^{r(r-1)/2} \beta_1 \cdots \beta_n \big( \text{min}(n,r)\alpha + \beta_1 + \cdots + \beta_n -n+1\big)^{r}.
\end{equation}
\end{subbox}
The factor $\beta_1\cdots\beta_n$ is the product of the degrees of the Pfaffian functions, reflecting the algebraic Bézout bound. In addition,  there are several factors growing exponentially in the order $r$ of the chain. This bound is the foundation of the quantitative finiteness of Pfaffian functions.

\subsubsection*{Topological complexity}
Khovanskii's theorem extends to a much more general setting. Instead of focusing on systems of equations where the number of functions and variables is equal, we now consider sets defined by an arbitrary number of equations. In addition, we now also allow for sets defined by inequalities. An \textit{elementary semi-Pfaffian set} is a set in $\bbR^n$ which is defined by solutions to equations and inequalities of Pfaffian functions; more precisely, it is a set of the form
\begin{equation}
\{x\in U \, | \, f_1(x) =0, \,\ldots,\,f_I(x)=0, \,g_1(x)>0, \ldots,\, g_J(x) >0  \}\,,
\end{equation}
where the $f_i$ and $g_j$ are Pfaffian functions with a common Pfaffian chain defined on the domain $U$. A general semi-Pfaffian set is a finite union of elementary semi-Pfaffian sets, meaning that it can be written as 
\begin{equation} \label{eq:def-Xsemi}
    X = \bigcup_{1\leq i\leq M}\{x\in U\,|\,f_{i1}(x) =0, \,\ldots,\,f_{iI_i}(x)=0, \,g_{i1}(x)>0, \ldots,\, g_{iJ_i}(x) >0  \}\, .
\end{equation}
In what follows we will be interested in the topological complexity of semi-Pfaffian sets. For a semi-Pfaffian set $X\subseteq \bbR^n$, we define the topological complexity as the sum of its Betti numbers
\begin{equation}
b(X) = b_0(X) + \ldots + b_n(X)\,,
\end{equation}
where $b_i(X) = \text{dim}(H_i(X;\bbR))$. Suppose that $X$ is a semi-Pfaffian set defined using $M$ equalities or inequalities of Pfaffian functions as in \eqref{eq:def-Xsemi}, with each function defined in the same underlying Pfaffian chain and having Pfaffian complexity $(n,r,\alpha,\beta)$. Then, as a generalization of Khovanskii's theorem, it was shown in \cite{GabVor04} that there is a bound on the topological complexity given by 
\begin{equation}
    b(X) \leq M^2 2^{r(r-1)/2} O(\text{min}(n,r)\alpha + n\beta )^{n+r} \,. 
\end{equation}
It is worth noting how the various components of the complexity appear in this bound. It depends only polynomially on the degrees $\alpha$ and $\beta$, but exponentially on the number of variables $n$ and the length of the chain $r$.

\subsubsection*{Computational complexity of cell decomposition}
Recall from section~\ref{sec:tamenesstheorems} that the cell decomposition theorem states that every tame set admits a decomposition into finitely many cylindrical cells. Since every tame set can be constructed from these elementary cells, it is natural to ask if there exists an algorithm which explicitly finds the cells needed to construct a given tame set. Whereas for general tame sets such an algorithm is not available, there does exist an explicit algorithm for semi-Pfaffian sets definable in the o-minimal structure $\bbR_{\rm Pf}$ \cite{GabVor01}. Consider a semi-Pfaffian set $X\subseteq \bbR^n$ of dimension $d$, defined using $M$ equalities and inequalities of Pfaffian functions as in \eqref{eq:def-Xsemi}, all of which have Pfaffian complexity $(n,r,\alpha,\beta)$. In this setting, the algorithm which finds the cell decomposition of $X$ has a computational complexity given by\footnote{Note that the algorithms in question are based on \textit{real numbers machines}, which are formal computational devices capable of performing exact arithmetic computations on real numbers. This is in line with the assumption that a real number constitutes one unit of information.}
\begin{equation} \label{CompComplex}
M^{(r+n)^{O(d)}}(\alpha + \beta) ^{(r+n)^{O(d^2n)}}\,.
\end{equation}
The dimension $d$ of the semi-Pfaffian set plays a key role in this estimate. The degrees $\alpha$ and $\beta$ again appear polynomially, whereas the number of variables $n$ appear double exponentially. 
As a special case, the complexity of computing the zeros of a Pfaffian function of complexity $(n,r,\alpha,\beta)$ is estimated by
\begin{equation}
    (\alpha+\beta)^{(r+n)^{O(d^2n)}} \,.
\end{equation}
For more details on these algorithms we refer to \cite{GabVor01}.

\subsection{Pfaffian functions and sharp o-minimality}
The previous discussion shows that the finiteness of geometric features of objects constructed from Pfaffian functions can be quantitatively controlled by a descriptive measure of information associated to Pfaffian chains. This indicates that Pfaffian complexity may provide a way of defining a format-degree filtration on Pfaffian o-minimal structures. 

\subsubsection*{Format and degree in Pfaffian structures}
Given a Pfaffian function $f$ satisfying a Pfaffian chain, we define its format $\cF$ and degree $\cD$ as 
\begin{equation} \label{eq:degree-format-Pfaffalg}
    \cF = n+ r \,, \qquad \cD = \deg P + \sum_{i,j}\deg P_{ij}\, .
\end{equation}
In view of the bounds shown in the previous subsection, this definition reflects the exponential growth in $n$ and $r$ and the polynomial growth in the degrees of the chain. With this notion of format and degree, it is possible to construct a format-degree filtration which makes $\bbR_{\rm rPf}$ into a sharply o-minimal structure \cite{binyamini_tameness_2023,binyamini2022sharply,EffCell}. This construction is rather technical and will not play an explicit role in this thesis, and hence we refer to \cite{EffCell} for the details of this procedure. Note that, although this result is only shown for the structure generated by restricted Pfaffian functions, it is expected to hold for the structure $\bbR_{\rm Pf}$ generated by unrestricted Pfaffian functions as well \cite{BinyaminiTalk}. In this way, the structure of Pfaffian functions provides the prototype of sharp o-minimality beyond the algebraic setting.

\subsubsection*{On the complexity of Pfaffian functions}
In this thesis, we view sharp o-minimality as the fundamental notion of complexity in tame geometry. Yet, we will often prefer to use the Pfaffian complexity due to its explicitness and the direct availability of sharp bounds as shown in the previous subsection. Furthermore, in the applications to physics we will usually be interested in the tameness and complexity of functions themselves rather than geometric objects constructed from the graphs of these functions. In these settings we use both equation \eqref{eq:Pcomplexity} and equation \eqref{eq:degree-format-Pfaffalg} to characterize the complexity of a Pfaffian function. Here we keep in mind that both definitions in principle control the geometric complexity of sets in Pfaffian structures, and that the definition in terms of format and degree is ultimately generalizable to other o-minimal structures in tame geometry.

\subsection{Emergence of simplicity}\label{sec:emergentsimplicity}
Earlier in this chapter we have noted that the complexity spectrum of a tame set can be interpreted as encoding the possible representations of the set, similar to dualities in physics. A remarkable aspect of Pfaffian functions which underlines this is the idea of \textit{emergence of simplicity}. 
This is the phenomenon in which a sequence of tame functions $f_N$ has a complexity which seems to diverge as $N\to\infty$, while the limiting function $f = \lim_{N\to\infty}f_N$ has finite complexity due to the emergence of a simpler dual description.\\

As an example, consider the function $f(x)=\tanh(x)$ and its $N$th order Taylor expansion
\begin{equation}
    f_N(x) = \sum_{n=1}^N \frac{2^{2n}(2^{2n}-1)B_{2n} }{(2n)!}x^{2n-1}\,,
\end{equation}
where $B_{2n}$ is the $2n$th Bernoulli number. The function $f_N(x)$ is a polynomial with Pfaffian complexity $\cC_{\rm Pf}(f_N)=(1,0,1,N)$, which diverges as $N\to \infty$. However, in the $N\to\infty$ limit, the limiting function $f(x)$ has an emergent simpler description as a Pfaffian chain with finite complexity, since it satisfies $\pd f/\pd x=1-f^2$. Another example is given by Fourier series, consisting of a sum of increasingly complex functions which may ultimately sum to a simple function. \\

In some cases, the limiting emergent description can even be used to show that the complexity does not diverge, and remains uniformly bounded for every $N$. Let us illustrate this by a simple example. Consider the sequence of functions 
\begin{equation}
    f_N(x) =  \sum_{n=0}^N \frac{x^n}{n!} \,.
\end{equation}
As a polynomial, the function $f_N$ has Pfaffian complexity $\cC_{\rm Pf}(f_N)=(1,0,1,N)$, diverging as $N\to \infty$. The limiting function $f(x)=\lim_{N\to\infty}f_N(x)=e^x$ has the simple Pfaffian chain representation $\pd f/\pd x=f$. Inspired by this differential structure, one finds that $f_N$ in fact satisfies
\begin{equation}
    \frac{\pd f_N}{\pd x}= f_N -\frac{x^N}{N!}\,.
\end{equation}
While it at first seems that this Pfaffian chain has degree $N$, note that the second term is a monomial, which by the fewnomial representation has a description of lower complexity. This means that $f_N$ has a description with Pfaffian complexity $\cC_{\rm Pf}(f_N)=(1,3,2,1)$, which is independent of $N$. This idea relies on the fewnomial representation, and only works for special Pfaffian chains of low degree. \\

The emergence of simplicity is naturally embedded within the theory of sharp o-minimality when formulated in terms of format and degree. We summarize this observation as follows.

\begin{subbox}{Emergence of simplicity}
A sequence of tame functions whose complexity appears to diverge may have a description of finite complexity after the limit is taken. 
\end{subbox}
This phenomenon will be encountered several times in this thesis.

\subsection{Beyond Pfaffian: log-Noetherian functions}\label{sec:LNfunctions}
The setting of Pfaffian functions provide an essential example of sharp o-minimality, but in applications we may encounter functions which do not admit a Pfaffian description, necessitating the need for a sharply o-minimal structure beyond Pfaffian functions. 
To close this section, we highlight another class of functions defined by means of differential chains, the \textit{log-Noetherian functions} \cite{binyamini2024lognoetherianfunctions}. The motivation for introducing these here is twofold: they include many functions encountered in physics which are not Pfaffian, and they are conjectured to be sharply o-minimal. 

\subsubsection*{Local bounds and domain dependence}
First, let us emphasize that the significance of Pfaffian functions lies in the \textit{global} bounds which they satisfy, which makes them behave like polynomials despite being a fairly general class of functions. Generic analytic functions do not satisfy such bounds and may therefore have an infinite complexity. One way to nonetheless include, in addition to Pfaffian functions, more general analytic functions in the framework is to restrict these functions to bounded domains, and to rely on \textit{local} complexity bounds. However, it then becomes evident that the complexity of such a restricted analytic function should depend on the size of the domain. As a simple example, the function $f(x) = \sin(x)$ has infinitely many zeros on the real line, and when restricted to a bounded interval it has a finitely many zeros depending on the size of the interval.

\subsubsection*{Log-Noetherian chains}
The log-Noetherian functions introduced in \cite{binyamini2024lognoetherianfunctions} provide a class of functions which satisfies strong local complexity bounds. We will not give the complete definition of these functions, but rather describe an illustrative case that captures many of their core features. Consider a complex domain of the form 
\begin{equation}\label{eq:domainU}
U =D_\circ(\rho_1)\times \cdots \times D_\circ(\rho_N)  \subset \bbC^N \,,
\end{equation}
where $D_\circ(\rho)=\{z\in\bbC\,|\,|z|<\rho\}$ is the punctured disk of radius $\rho$.
A \textit{log-Noetherian} chain on $U$ consists of bounded holomorphic functions $\xi_1,\ldots,\xi_R: U \to \bbC$ satisfying a system of differential equations
\begin{equation}
    z_j \frac{\pd \xi_i}{\pd z_j} = P_{ij}(\xi_1,\ldots,\xi_R)\, .
\end{equation}
This resembles the definition of a Pfaffian chain, but there are a few crucial differences:
\begin{enumerate}
    \item[(i)] the triangularity requirement is dropped, and the polynomials $P_{ij}$ may depend on all functions $\xi_1,\ldots,\xi_R$. In this construction, the variables $z_1,\ldots,z_N$ themselves have to be defined as functions in the chain;
    \item[(ii)] the functions and variables are complex-valued, and instead of regular derivatives the chain uses logarithmic derivatives $z_j\frac{\pd}{\pd z_j}$;
    \item[(iii)] the domain $U$ is bounded and has to be of a specific cellular form, as explained in detail in \cite{binyamini2024lognoetherianfunctions}.\footnote{Essentially, the domains $U$ are constructed inductively by fibering disks, punctured disk and annuli over each other. The radii of these fibered disks and annuli are allowed to vary as a log-Noetherian function on the base space. The domain $U$ given in equation \eqref{eq:domainU} is a simple example of such a domain, in which the radii are constant.}
\end{enumerate}
A log-Noetherian function is then any function $f(z_1,\ldots,z_N) = P(\xi_1,\ldots,\xi_R)$ which depends polynomially on the functions in the chain. These functions satisfy controllable local complexity bounds, and it is shown in \cite{binyamini2024lognoetherianfunctions} that they generate an o-minimal structure $\bbR_{\rm LN}$, which is expected to be sharply o-minimal.
While it is presently not known how to precisely assign a format and degree to such a function and domain, it is suggestive\footnote{To be precise, it is shown in \cite{binyamini2024lognoetherianfunctions} that these functions are \textit{effectively} o-minimal, meaning that they satisfy effectively computable bounds characterized by a single number, forfeiting the polynomial control in the degree.} from  \cite{binyamini2024lognoetherianfunctions} that their sum should be given by 
\begin{align}\label{eq:formatLN}
    \cF_{\rm LN} + \cD_{\rm LN}  =  &\ \ R  +\deg P +  \sum_{i,j}\deg P_{ij} \\
    & \ \ + \norm{P} +  \sum_{i,j}  \norm{P_{ij}}  + \max_{i,z\in \overline{U}} |\xi_i (z)| + \cF(U) + \cD(U)\,. \nonumber
\end{align}
Note that the terms in the first line of this equation 
are also present in the case of Pfaffian functions, when taking into account that the variables are now considered as being part of the functions in the chain. In the log-Noetherian setting there are now several new terms capturing details of the functions and the domain on which they are defined. In particular, there are the norms $\norm{P}$, $\norm{P_{ij}}$ of the polynomials, which are defined as the sum of the absolute values of their coefficients. These are essential for obtaining the required local bounds on the function. The domain complexity $\cF(U) + \cD(U)$ given in \cite{binyamini2024lognoetherianfunctions} is given as a linear function in $\sum_{j}\rho_j$. In contrast to Pfaffian functions, the complexity of the domain complexity must be taken into account, as we will illustrate in the following example. \\

Consider the function $f(z)=\frac{e^{\lambda z}-1}{z}$, with $\lambda$ real and positive, on a punctured disk $D_\circ(\rho)\subseteq\bbC$ of radius $\rho$. This function fits in the log-Noetherian chain given by 
\begin{align}
    \xi_1(z)& = z \,, &&  z\frac{\pd\xi_1}{\pd z}=  \xi_1   \,, \\
    \xi_2(z)& = e^{ \lambda z}\,,  &&  z\frac{\pd\xi_2}{\pd z}=   \lambda \xi_1\xi_2   \,, \nonumber \\
    \xi_3(z)&= \frac{e^{ \lambda z}-1}{z} \,,    && z\frac{\pd\xi_3}{\pd z}=  \lambda \xi_2-\xi_3  \,. \nonumber
\end{align}
The numbers of zeros of $f$ on $\overline{D_\circ(\rho)}$ is given by $2\lfloor \frac{\lambda\rho}{2\pi} \rfloor$, and this should be reflected by the complexity of the function. Indeed, in the log-Noetherian chain $\lambda$ appears as a coefficient in the polynomials on the right-hand side, and the dependence on $\rho$ appears through the domain complexity $\cF(D_\circ(\rho)) + \cD(D_\circ(\rho))$. This shows that the terms in the second line of equation \eqref{eq:formatLN} are necessary.

\subsubsection*{The o-minimal structure \texorpdfstring{$\bbR_{\rm LN,Pf}$}{}} 
The proposal of \cite{binyamini2024lognoetherianfunctions} is to combine the log-Noetherian functions with Pfaffian functions, by considering the Pfaffian closure of the log-Noetherian functions. This class of functions, though requiring a somewhat complicated construction, turns out to include a wealth of functions, while still retaining controllable local and global complexity bounds. This has led Binyamini to conjecture the following \cite{binyamini2024lognoetherianfunctions}.

\begin{subbox}{Conjecture: sharp o-minimality of $\bbR_{\rm LN,Pf}$}
The o-minimal structure $\bbR_{\rm LN,Pf}$ obtained by taking the Pfaffian closure of the structure generated by log-Noetherian functions admits a format-degree filtration which makes it into a sharply o-minimal structure. 
\end{subbox}
Notably, this structure includes the period maps appearing in Hodge theory, which as explained in subsection~\ref{sec:hodge} play a major role in several physics settings. 

In the remainder of this thesis we will invoke this conjecture a number of times, to indicate that the notion of complexity in tame geometry conjecturally extends to these instances. For an analysis of log-Noetherian functions and their complexity in the physical context of Seiberg-Witten theory we refer to \cite{Carrascal:2025vsc}.

\section{Quantitative tameness theorems}\label{sec:tamenesstheorems2}
We have alluded to the fact that the complexity of a tame set, measured by its format and degree, makes the tameness theorems stronger by replacing finiteness by computable bounds. In this section we briefly illustrate some technical details of these quantitative tameness theorems.

\subsection{Bound on connected components}
Recall from section~\ref{sec:defsharp} that the complexity of the closure $\overline{X}$ of a set $X\subseteq \bbR^n$ of complexity $(\cF,\cD)$ has complexity $(\cF+n+2,\cD+2)$. The underlying mechanism for this is that the closure $\overline{X}$ can be constructed from $X$ with simple geometric operations, and that the change in complexity can be tracked in each step of the construction. By generalizing this mechanism to more advanced constructions, and resorting to the polynomials $\{P_\cF\}$ in the definition of a sharply o-minimal structure, we encounter a quantative tameness theorem which controls the number of connected components of a tame set \cite{binyamini2022sharply}.  

\begin{subbox}{Bound on connected components}
Let $\cS$ be a sharply o-minimal structure with format-degree filtration $\Omega$, and let $X\in\Omega_{\cF,\cD}$. Then $X$ has at most  $\text{poly}_{\cF}(\cD)$ connected components.   
\end{subbox}

In the statement of this theorem, the notation $\text{poly}_{\cF}(\cD)$ refers to an $\cF$-dependent polynomial in $\cD$, whose precise form is suppressed but can be explicitly computed in a given sharply o-minimal structure. The theorem generalizes the explicit boundedness of one-dimensional tame sets in terms of $P_\cF(\cD)$ to tame sets in any dimension. The underlying idea is that any tame set in a sharply o-minimal structure must satisfy the one-dimensional bound when projected down to the real line, and that these linear projections preserve the complexity by the axioms of sharp o-minimality.
The proof of this statement is already surprisingly technical, since it needs to account for higher-dimensional tame sets in complete generality \cite{binyamini2022sharply}.

\subsection{Sharp cell decomposition}
The cell decomposition is one of the fundamental theorems of tame geometry, and in sharp o-minimality there is a quantitative formulation of this theorem, proven by Binyamini, Novikov, and Zak \cite{binyamini2022sharply}. 

\begin{subbox}{Sharp cell decomposition theorem}
Let $\cS$ be a sharply o-minimal structure with format-degree filtration $\Omega$. Then $\Omega$ is reducible to a format-degree filtration $\widehat \Omega$, with respect to which the following holds: for any finite collection of $\cS$-definable sets  $X_1,\ldots,X_m\subseteq \bbR^n$, there exists a cell decomposition of $\bbR^n$ such that each $X_k$ is a union of cells, the total number of cells is bounded by $\text{poly}_{\cF}(n,\cD)$, and the cells have complexity $(O(\cF),\text{poly}_\cF(\cD))$.    
\end{subbox}

The theorem is a significantly stronger version of the cell decomposition, since the number of cells and the complexity of the cells can be explicitly bounded in terms of the underlying formats and degrees. It is this theorem which truly materializes the interpretation of format and degree as the geometric complexity of a tame set.
It was noted in \cite{binyamini2022sharply} that this statement does not hold in general for all format-degree filtrations, which necessitates the need for a reduction to another filtration $\widehat\Omega$. We will not explicitly use this theorem in this thesis, but we keep in mind that it provides the foundation for quantitative tame geometry. 

\subsection{Effective counting theorems for rational points}
As highlighted in subsection~\ref{sec:pilawilkie}, the Pila-Wilkie theorem on counting rational points in tame sets is an essential tool in mathematical applications of o-minimality. The original version of the theorem bounds the number of rational points of as a function of the height up to a multiplicative constant, and in a sharply o-minimal structure this constant can be estimated \cite{binyamini_tameness_2023}. 

\begin{subbox}{Sharp Pila-Wilkie theorem}
Let $\cS$ be a sharply o-minimal structure with format-degree filtration $\Omega$. Then for any $\eps>0$ and $\cF$, there is a polynomial $P_{\cF,\eps}$ such that for every tame set $X\in \Omega_{\cF,\cD}$ and height $H\geq 2$, we have 
\begin{equation}
  \#\{ X^{\rm tr}( \bbQ) \,|\,H(x)\leq H \} \leq P_{\cF,\eps}(\cD) \,H^\eps\,.
\end{equation}
\end{subbox}

This refinement of the Pila-Wilkie counting theorem shows that sharp o-minimality goes beyond only constraining the topology of tame sets, and opens the door for making many of the applications of tameness in algebraic geometry and number theory quantitative. Formulating effective counting theorems of this type is an active research topic \cite{Jones21,BinyaminiCluckers22,binyamini2023effective,BinyaminiCluckers24,BinyaminiCluckers25,BinyaminiFumiharu25,binyamini2026counting}.

\subsection{On the sharpness of bounds}
Before concluding this section, it is worthwhile to briefly reflect on the strength of the bounds provided by sharply o-minimal structures. The central premise of sharp o-minimality is that geometric features constrained by tameness can be explicitly bounded in terms of the complexity $(\cF,\cD)$, determined by a function which depends polynomially on the degree $\cD$ and typically super-polynomially on the format $\cF$. An apparent flaw of the framework is that when these bounds are applied to individual objects, one often finds that the bounds severely overestimate the actual geometric complexity of an object. In concrete examples, the true number of connected components, cells, or other topological data, may be far smaller than the general upper bounds predicted by the theory. 
The fundamental reason for this is that the complexity measure must simultaneously account for all tame sets in an o-minimal structure, which generally forms an extremely large and diverse collection of geometric objects. This is in fact where the strength of sharp o-minimality lies: it should not be regarded primarily as a tool to obtain optimal estimates, but rather as a universal framework for controlling complexity across an enormous range of sets and functions.

\section{Summary and discussion}
Reaching the end of part I of the thesis, let us summarize the main points which have been developed in these two chapters.
In chapter~\ref{ch:tameness}, we introduced o-minimal structures, which provide a framework of tame geometry founded on the finiteness assumption that tame sets have finitely many connected components. The definable sets in these structures can be thought of as being delimited by a selected class of sufficiently tame functions, and we discussed numerous examples of function classes that are consistent with o-minimality. The idea that these sets are tame is realized by the tameness theorems which they satisfy, of which the central one is the cell decomposition theorem. In the spirit of the introduction to this thesis, o-minimality provides a tame mathematical language in which the building blocks are tame sets.\\

In chapter~\ref{ch:complexity}, we discussed the quantitative refinement of o-minimality, defined through sharply o-minimal structures. We explained in general terms how the notion of format and degree defines a measure of complexity for tame sets. The strength of this framework lies in its universality, generality, and consistency under geometric and logical operations, which together ensure that the tameness theorems can be upgraded to quantitative statements. We discussed how the class of Pfaffian functions, with its associated theory of Pfaffian complexity, provides the main example of this idea. For the purposes of this thesis, the most important lesson from this chapter is that it is possible to define an intrinsic notion of complexity for functions, in way that is geometrically and logically meaningful.


\setpartpreamble[u][\textwidth]{
	\vspace*{1cm}
	\hrulefill 
	\vspace*{0.5cm}
	
The second part of this thesis is devoted to applications of tameness and complexity to quantum field theory. We begin in chapter~\ref{ch:classical} by discussing general aspects of o-minimality in physics, and study the first encounters in quantum mechanics and classical field theory. In chapter~\ref{ch:observables} we analyze the tameness of observables in quantum field theory, with a focus on non-perturbative correlation functions. Chapter~\ref{ch:CCC} specializes this analysis to cosmological correlators, the natural observables in cosmological quantum field theories. Finally, in chapter~\ref{ch:complexityQFT} we propose an intrinsic measure of complexity for quantum field theories based on sharp o-minimality.

	\vspace*{0.5cm}
	\hrulefill }

\part{Quantum field theory}\label{part3}
\chapter{Tameness and complexity in physical theories}
\label{ch:classical}
\setlength{\parindent}{0pt}
This chapter forms the starting point of our exploration of tameness and complexity in physical theories. The purpose of this chapter is to provide a general perspective on o-minimality in physics, and to consider aspects of quantum mechanics and classical field theory.

\section{General aspects of tameness and physics}\label{sec:tamenessphysics}
In the previous chapters we have given a definition of tameness in mathematical theories. Based on the general considerations of finiteness in physics presented in the introduction of this thesis, it is a natural expectation that the finiteness at the heart of tame geometry manifests itself to some degree in physical theories. Through the remainder of this thesis, we will be guided by the following questions:

\begin{subbox}{Guiding questions for o-minimality and physics}
\begin{itemize}
    \item Are the mathematical objects appearing in physical theories tame, in the sense that they are definable in an o-minimal structure?
    \item Can physical quantities be described with finite complexity, as measured by sharp o-minimality, and is there a physical interpretation of this notion of complexity?
    \item What physical conditions 
    impose tameness and finiteness of complexity on physical theories?
\end{itemize}
\end{subbox}
To address these questions, we will have to sharpen our view on the mathematical objects appearing in physics. While o-minimality is a geometric framework and geometry is prevalent in physics, the most natural starting point
is to consider the functions used in physical theories, which we will refer to as \textit{physical functions}. These functions generally depend on different classes of variables: spacetime coordinates, kinematic variables such as momentum and energy, and parameters of the theory.

\subsubsection*{Physical functions}
To structure our analysis, it will be helpful to conceptually organize the physical functions encountered in this thesis into various types. 
The first type we consider are the \textit{observables}, which describe measurable quantities of physical systems. In this class we also include indirectly observable quantities, such as correlation functions which form the building blocks of amplitudes in quantum field theory. Then there are the \textit{configurational functions} which encode the state of a system, typically as functions of space and time. In turn, these configurations arise as solutions to equations of motions, whose precise form is specified by functions such as the Lagrangian or Hamiltonian. These functions can be thought of as defining a \mbox{theory}, and will be called \textit{descriptive functions}. Finally, there are \textit{auxiliary functions}, which appear in calculations but may not always have a clear physical interpretation. \\

We remark that this list is not exhaustive, nor is there a well-defined boundary between these types. For example, the partition function of a quantum field theory can be viewed as an observable (the vacuum-to-vacuum transition amplitude), a descriptive function (the generating functional of the correlation functions), or an auxiliary function (the evaluation of the path integral). 
In this thesis we will investigate the realization of o-minimality across these types of physical functions, each of which reveals another layer of tameness in physics. Note that there is a certain hierarchy in the types outlined above: descriptive functions dictate the configurational functions, which in turn determine the observables. The propagation of tameness and finiteness of complexity through this hierarchy is one of the main themes of this thesis, and we will already see instances of this in quantum mechanics and classical field theory later in this chapter.

\subsubsection*{Oscillations, domain-dependence, and tameness}
The discussion of finiteness in physics presented in the introduction, together with the idea that tame geometry provides a precise mathematical framing of finiteness, may lead one to the ambitious conjecture that all physical functions are tame. However, the simplest physical systems already demonstrate that this is not true. Oscillations parametrized on unbounded domains are not definable in any 
o-minimal structure, but are omnipresent in physics. For example, any wave equation has solutions comprised of sine and cosine functions, which are not tame when spatial or temporal variables range over the entire real line.\footnote{As demonstrated in chapter \ref{ch:tameness}, the clearest way to see this is to note that the zero set of this function defines the set of integers.} 
In order to address these apparent counterexamples, we therefore have to refine the scope of tameness in physics. \\

At this point it is essential to note that the non-tameness of an oscillation is only due to the infinite size of the domain. When a wave is restricted to a finite interval in space and time, it is in fact a tame function, definable in structures such as $\bbR_{\rm an}$ or $\bbR_{\rm Pf}$. We could therefore regard the non-tameness of oscillations to be a descriptional artifact, since no physical system that we have complete access to is truly infinite in spatial or temporal extent. While we do not wish to enter too far into this philosophical direction, let us point out that this observation is consistent with Bekenstein's bound, which imposes finiteness of information only for systems which have a finite size. \\

There is, however, some conceptual tension between considering unbounded and restricted domains: 
since any restricted analytic function is tame, the restriction of physical functions to compact domains makes tameness trivial in many cases. Within this tension, the most meaningful cases to consider are therefore:
\begin{itemize}
    \item[(i)] Physical functions which are tame on an infinite domain.
    \item[(ii)] Non-analytic physical functions.
    \item[(iii)] Restricted physical functions whose complexity depends on the domain size.
\end{itemize}
We will be predominantly concerned with tameness and complexity in one of these three forms. With this general discussion in mind, let us proceed to the details.

\section{Wavefunctions in quantum mechanics}

In this section we consider the tameness and complexity of the spatial dependence of quantum-mechanical wavefunctions. We establish that, under mild assumptions, the wavefunctions of bound eigenstates are tame on an unbounded domain and have a complexity scaling with their energy eigenvalue. Our analysis is based on the structure of the one-dimensional time-independent Schr\"odinger equation
\begin{equation}\label{eq:1dSchrod}
    -\frac{1}{2m} \psi''(x) + \big(V(x)-E\big)\psi(x) =0\,.
\end{equation}
Here $\psi(x)$ denotes the wavefunction on the real line, $m$ is the mass, $V(x)$ is the external potential, and $E$ is the energy eigenvalue.
The mathematical properties of this equation can be understood in great detail using Sturm-Liouville theory \cite{Teschl,Teschl2}. 
The physical intuition behind the tameness of wavefunctions is that they oscillate in the classically allowed region where $E>V(x)$, whereas they decay exponentially in the tunneling region where $E<V(x)$. Therefore, for a bound state with a finite classically allowed region, there will be finitely many oscillations.

\subsection{The harmonic oscillator}
Let us start with an example and consider the wavefunctions of the eigenstates of the harmonic oscillator. The potential of this system is given by
\begin{equation}
    V(x) = \frac{1}{2}m\omega^2 x^2 \,,
\end{equation}
where $m$ is the mass of the particle, $\omega$ is the frequency of the oscillator. This theory is exactly solvable, and the solutions take the form 
\begin{equation}
\psi_n(x) = \frac{1}{\sqrt{2^n n!}} \left( \frac{m\omega}{\pi}\right)^{1/4} e^{-m\omega x^2/2} H_n(\sqrt{m \omega }x)
\end{equation}
where $H_n$ is the $n$th Hermite polynomial. The wavefunction $\psi_n$ corresponds to a state with energy given by $E_n = (n+\tfrac{1}{2})\omega $. \\

We now show that the functions $\psi_n$ are tame, by demonstrating that they can be described with a Pfaffian chain. Consider the following Pfaffian chain of length one: 
\begin{align}
    \zeta(x) &= e^{-m\omega  x^2/2}\,,   && \frac{\pd\zeta}{\pd x}=  -m\omega \,x\,\zeta \,.
\end{align}
This gives us a Pfaffian chain containing the Gaussian function, which has order 1 and degree 2. The Hermite polynomial $H_n$ has degree $n$, so it follows that, for each $n$, the function $\psi_n $ is a  polynomial of degree $n+1$ in in the variable $x$ and the function $\zeta$, and hence a Pfaffian function. The Pfaffian complexity of $\psi_n$ in this chain is given by 
\begin{equation}
    \cC_{\rm Pf}(\psi_n ) =  (1,1,2,n+1).
\end{equation}
The function $\psi_n$ describes a state $\ket{n}$ in the theory, so that we may define the complexity of the state $\ket{n}$ as the complexity of the underlying wavefunction $\psi_n$. This leads us to the simple observation that the complexity in this system scales with the energy of the eigenstates.

\subsection{Eigenstates in polynomial potentials}
Consider now a particle on the real line $\bbR$ in a general polynomial potential $V(x)$. For the stability of the system, we choose $V$ to be bounded from below, which is accomplished by taking the highest degree term of $V$ to have even degree $D$ and positive coefficient. In this setting, Sturm-Liouville theory tells us that the eigenstate wavefunctions satisfy two key properties \cite{Teschl,Teschl2}:
\begin{itemize}
    \item \textbf{Discreteness of spectrum.} The solutions $\psi_0,\psi_1,\psi_2,\ldots$ to the Schr\"odinger equation  have a discrete energy spectrum $E_0<E_1<E_2<\ldots$ with finite ground state energy $E_0$.
    \item \textbf{Node theorem.} The $n$th wavefunction $\psi_n$ has exactly $n$ zeros.
\end{itemize}
The second property will enable us to prove that the solutions to \eqref{eq:1dSchrod} are tame. \\

According to the node theorem, the ground state wavefunction $\psi_0$ is non-vanishing, which allows us to introduce the auxiliary function $\rho_0 = \psi_0'/\psi_0$ on the real line. The function $\rho_0$ satisfies the Riccati equation
\begin{equation}
    \rho_0'+ \rho_0^2 -2m\big( V(x)-E_0\big) =0
\end{equation}
This substitution thus reduces the Schr\"odinger equation to two coupled first-order differential equations, and we will refer to this procedure as the \textit{Riccati substitution}. It follows that $\psi_0$ fits into the Pfaffian chain given by
\begin{align}
    \zeta_1(x) &= \rho_0(x) \,,   &&  \frac{\pd\zeta_1}{\pd x}=  -\zeta_1^2+2m\big(V(x) - E_0 \big) \,, \\ 
    \zeta_2(x) &= \psi_0(x)\,,  &&  \frac{\pd\zeta_2 }{\pd x} =\zeta_1\zeta_2 \,. \nonumber 
\end{align}
which has order $2$ and degree $D$ specified by the degree of the potential $V(x)$. The Pfaffian complexity of the ground state wavefunction $\psi_0$ with respect to this chain is therefore
\begin{equation}
    \cC_{\rm Pf}(\psi_0) = (1,2,D,1) \,. 
\end{equation}
Recalling the discussion of fewnomials from section \ref{sec:Pfaff}, we note that if $V$ only has a small number of monomial terms $M$ compared to the degree $D$, then there exists a Pfaffian chain with a lower Pfaffian complexity. In this chain we include each of the monomials of $V$, and the resulting Pfaffian complexity of the ground state wavefunction is 
\begin{equation}
    \cC_{\rm Pf}(\psi_0) = (1,M+3,2,1) \,.
\end{equation}
This simple analysis already shows that in the Pfaffian framework, more complicated potentials lead to ground states of higher complexity. Remarkably, this complexity analysis can be performed without ever explicitly solving for the wavefunction. \\

The above analysis relied on the absence of zeros in the ground state wavefunction for performing the Riccati substitution. The wavefunctions of the excited states necessarily have zeros, but by the node theorem we know exactly how many there are, enabling us to circumvent them by a finite piecewise application of the Riccati substitution. For the $n$th wavefunction $\psi_n$, denote the zeros by $a_1<\ldots<a_n$ and set $a_0=-\infty$ and $a_{n+1}=\infty$. On the open intervals $(a_i,a_{i+1})$, the restriction $\psi_{n,i} = \psi_{n}|_{(a_i,a_{i+1})}$ is non-zero, allowing us to locally define the Riccati substitution $
    \rho_{n,i} = \psi'_{n,i}/ \psi_{n,i}$
for each $i=0,\ldots,n$. This makes the restriction $\psi_{n,i}$ a Pfaffian function on the interval $(a_i,a_{i+1})$ by means of the Pfaffian chain
\begin{align}\label{eq:piecewiseRicPchain}
    \zeta_{1,i}(x) &= \rho_{n,i}(x) \,,   &&  \frac{\pd\zeta_{1,i}}{\pd x}=  -\zeta_{1,i}^2+2m\big(V(x) - E_n \big) \,, \\ 
    \zeta_{2,i}(x) &= \psi_{n,i}(x)\,,  &&  \frac{\pd\zeta_{2,i} }{\pd x} =\zeta_{1,i}\zeta_{2,i} \,, \nonumber 
\end{align}
from which it follows that $\cC_{\rm Pf}(\psi_{n,i})=(1,2,D,1)$. This leads to the remarkable conclusion that, on each interval $(a_i,a_{i+1})$, the restricted wavefunction of the $n$th excited state $\psi_{n,i}$ has the same Pfaffian complexity as the ground state. \\

Geometrically, the full wavefunction $\psi_n$ can be obtained by taking the union of the graphs of the restrictions $\psi_{n,i}$, together with the $n$ isolated zeros $\psi_n(a_i)=0$:
\begin{equation}
    \Gamma(\psi_n) = \bigcup_{i=0}^n \big(\Gamma(\psi_{n,i}) \cup (a_i,0) \big) \,.
\end{equation}
Since this is a finite union of tame sets definable in $\bbR_{\rm Pf}$, we conclude the following:

\begin{subbox}{Tameness of eigenstates in polynomial potential} 
Let $V(x)$ be a polynomial potential which is bounded from below. Then any solution to the Schr\"odinger equation on the real line $\bbR$ is definable in the o-minimal structure $\bbR_{\rm Pf}$.
\end{subbox}

The axioms of sharp o-minimality tell us that the degrees are summed when the union of sets is taken, from which it furthermore follows that the complexity of excited states has a constant format $\cF$ and a degree $\cD$ which grows linearly with the level of the state.

\subsection{Bound states in tame potentials}
We now generalize the setting, and consider a particle on the real line $\bbR$ in a potential $V(x)$ which is bounded from below and definable in an o-minimal structure $\cS$. We assume that we have a solution $\psi(x)$ to the corresponding Schr\"odinger equation which describes a bound state, meaning that it is localized and has finitely many zeros. Since the potential $V(x)$ is assumed to be a tame function, it is piecewise differentiable with finitely many points of non-differentiability. Therefore, there exist finitely many points $a_i$ with $-\infty =a_0<a_1<\ldots a_n<a_{n+1}=\infty$ such that on each interval $(a_i,\ldots,a_{i+1})$, the restriction $\psi|_{(a_i,a_{i+1})}$ is non-zero and $V|_{(a_i,a_{i+1})}$ is differentiable. The potential is not assumed to be polynomial, so the equation
\begin{align}
    \frac{\pd\rho_{i}}{\pd x}=  -\rho_{i}^2+2m\big(V(x) - E \big) 
\end{align}
for the Riccati substitution $\rho_i$ on $(a_i,a_{i+1})$ no longer describes a Pfaffian chain. However, the differentiability and definability of $V$ on this interval is enough to guarantee that each $\rho_i$ is definable in the \textit{Pfaffian closure} of $\cS$, which defines an o-minimal structure $\cS_{\rm Pf}$ \cite{Speisegger99} (see subsection \ref{sec:Pfaffclosure}). Continuing the construction as for the polynomial potentials, assembling the finitely many pieces of $\psi(x)$, we find the following:

\begin{subbox}{Tameness of bound states in a definable potential} 
Let $V(x)$ be a potential which is bounded from below and definable in an o-minimal structure $\cS$. Then a bound state solution to the Schr\"odinger equation with finitely many zeros on the real line $\bbR$ is definable in the Pfaffian closure $\cS_{\rm Pf}$.
\end{subbox}

Recalling the discussion of the previous section, this is the first instance of the phenomenon that tameness of a descriptive function controls the tameness of a configurational function. In this case, the descriptive function in question is the potential $V(x)$, which defines the Hamiltonian of this system, and the configurational function is the wavefunction of a bound state in this system. If we lift this assumption by considering a particle in a non-tame potential, the tameness of the wavefunction is no longer guaranteed. 
To conclude, we note that wavefunctions in higher-dimensional quantum systems can be analyzed similarly, but require additional symmetry assumptions such as separability or rotational invariance which can be used to formulate the system in terms of one-dimensional differential equations. We will not study these cases further in this thesis.

\subsection{Comparison to quantum computational complexity}
The previous discussion indicates that many states in one-dimensional quantum mechanics admit a description of finite complexity, by capturing their wavefunction in a Pfaffian structure. Within the context of quantum-mechanical states, we mentioned in chapter \ref{ch:complexity} that there exists another established notion of complexity, namely quantum computational complexity (for a detailed review we refer to \cite{Chapman:2021jbh}). Since these quantities both describe the descriptive information required to specify a quantum state, it is natural to investigate how these two concepts are related. As we will see, the two measures of complexity differ in a number of fundamental aspects.\\

Quantum computational complexity counts the number of quantum gates needed to construct a  given target quantum state from a reference state. These gates represent physical operations which may be performed on the states, and form a finite set of elementary unitary operators which has to be selected by hand. 
The resulting notion of complexity initially only captures quantum systems with finite-dimensional Hilbert spaces. However, generalizations have been proposed which replace a discrete sequence of gates by a continuous unitary evolution of the reference state. Subsequently, these ideas have been applied to the harmonic oscillator, with the aim of assigning a quantum computational complexity to states in quantum field theories \cite{Jefferson:2017sdb}.\footnote{Similar ideas were implemented in the setting of conformal field theories in \cite{Chagnet:2021uvi}.}
In order to compare the two notions, we enumerate three essential features of quantum computational complexity.
\begin{itemize}
     \item[(i)] It is defined \textit{relative} to another quantum state, and without a fixed reference state there is no inherent notion of quantum computational complexity.
     \item[(ii)] Applications require careful generalizations from discrete to continuous complexity. Extending the framework in this way requires making additional choices, such as a distance measure on the space of unitary operators.
     \item[(iii)] It depends in a natural way on the parameters of the theory. For example, in \cite{Jefferson:2017sdb} it was shown that the complexity of the ground state $\psi(x)\sim e^{-m\omega x^2/2}$ of the harmonic oscillator relative to the ground state $\psi_0(x)\sim e^{-m\omega_0 x^2/2}$ of another harmonic oscillator depends on the frequency ratio $\omega/\omega_0$.
\end{itemize}
Let us now comment on the relation of these features to Pfaffian complexity, and point out a number of open problems in understanding this connection. \\

To address point (i), we note that Pfaffian complexity, or more generally the format and degree in sharp o-minimality, measures the complexity of physical quantities in an \textit{absolute} sense, encoding the amount of logical information required to describe a physical function without referencing an initial state. This means that this notion of complexity, while it still depends on the choice of representation and format-degree filtration, may provide intrinsic information about the state under consideration. In addition, this intrinsic information controls bounds on computational quantities associated to the physical functions, as explained in the previous chapter. \\

Turning to point (ii), we note that format and degree are fundamentally discrete quantities, which admit no immediate continuous generalization. In the context of quantum circuits, one attempt to implement this generalization could be to adopt the geometric perspective on circuit complexity as pioneered by Nielsen \cite{nielsen2005geometricapproachquantumcircuit}, replacing the search for an optimal discrete circuit by the optimization of a path in an operator space. The geometric nature of this approach resonates with tame geometry, and it is foreseeable that a quantum circuit can be reformulated within sharp o-minimality, so that the logical complexity of the circuit can be captured by a format and degree or a Pfaffian complexity. \\

Finally, addressing point (iii), we emphasize that the Pfaffian framework describes the complexity of the functional dependence on a set of variables. Therefore, the Pfaffian complexity itself cannot depend on these variables, and only on external parameters. Following the example above, consider the Pfaffian complexity of the function $\psi(x)=e^{-m\omega x^2/2}$ with respect to a Pfaffian chain containing the function $\psi_0(x) = e^{-m\omega_0 x^2/2}$. Since these are functions of the variable $x$, the complexity is independent of $x$. However, we may consider the dependence on the frequencies $\omega$ and $\omega_0$. The function $\psi$ is only Pfaffian in this chain in the special instance that the ratio $\omega/\omega_0$ is an integer, so that we have the algebraic relation $\psi=\psi_0^{\omega/\omega_0}$. In this case the complexity of $\psi$ within this fixed chain is given by $\cC_{\rm Pf}(\psi)=(1,1,2,\omega/\omega_0)$, and the dependence on physical parameters appears. Generically, $\omega/\omega_0$ is not an integer, and in order for the wavefunction $\psi$ to be Pfaffian, the chain must be extended, leading to an increase in the Pfaffian complexity independent of the frequency parameters $\omega$ and $\omega_0$. \\

In conclusion, there are a number of fundamental differences between quantum computational complexity and Pfaffian complexity, and therefore Pfaffian complexity may provide a complementary perspective on complexity in quantum systems.

\section{Field configurations in classical field theory}\label{sec:Fieldconfigurations}
In the following section we move from mechanics to field theory, and  analyze field configurations in simple relativistic classical field theories. 
Specifically, we will investigate the relation between complexity, energy, and spatial size, with the aim of finding a physical interpretation of the complexity defined by format and degree, exemplified through Pfaffian complexity. For simplicity we mostly focus on field theories in two spacetime dimensions, and comment on the higher-dimensional case towards the end of the section.

\subsection{Two-dimensional free massless scalar field}
Consider a classical field theory describing a non-interacting massless real scalar field in one spatial dimension and one time dimension. This theory is specified by the Lagrangian
\begin{equation}\label{eq:Lagfreescalar}
    \cL(\phi) = -\frac{1}{2}\pd_\mu \phi \,\pd^\mu \phi  \,, 
\end{equation}
where $x^\mu=(t,x)$. The equation of motion is the (1+1)-dimensional wave equation
\begin{equation}\label{eq:waveeq}
    \pd^2 \phi = 0 \,,
\end{equation}
and the simplest solutions are plane waves. In view of the discussion of section \ref{sec:tamenessphysics}, let us consider the complexity of field configurations in this theory on a finite space interval $[0,R]$ at fixed time $t=0$. For simplicity, let us in addition impose vanishing boundary conditions $\phi(0)=\phi(R)=0$. The elementary solutions are the Fourier modes
\begin{equation}
\phi_n(x) = \alpha \sin(\frac{n\pi  x}{R}) \,,
\end{equation}
where $\alpha$ is a dimensionless amplitude.
These functions are definable in the o-minimal structure $\bbR_{\rm Pf}$ as follows. Consider the Pfaffian chain on $(0,R)$ given by 
\begin{align}
    \zeta_1(x)& = \tan(\frac{\pi x}{2R} -\frac{\pi}{4} ) \,, &&  \frac{\pd\zeta_1}{\pd x}=\frac{\pi}{2R}\big(1+ \zeta_1^2 \big) \,, \\
    \zeta_2(x)&=  \sin\Big(\frac{\pi x}{R} \Big) \,,   && \frac{\pd\zeta_2}{\pd x}= -\frac{\pi}{R}\zeta_1(\zeta_2+1)\,. \nonumber 
\end{align}
For $n$ odd, the $n$th Fourier mode can then be expressed as $\phi_n(x) = \alpha\, T_n( \zeta_2(x) ) $, where $T_n$ is the Chebyshev polynomial of degree $n$. As a result, the field configuration $\phi_n$ is a tame function with Pfaffian complexity 
\begin{equation}\label{eq:1dfourierPcomp}
    \cC_{\rm Pf}(\phi_n) = (1,2,2,n )\,. 
\end{equation}
A slightly modified Pfaffian chain can be used to define the even Fourier modes with the same complexity. The fourth component of the Pfaffian complexity, which indicates the degree of the Pfaffian function, increases linearly with the wavenumber. In terms of the format and degree, the complexity of the Fourier modes thus scale as $\cF \sim 1$, $\cD \sim n$.

\subsection{Energy, radius, and complexity}
The Bekenstein bound states that the entropy $S$ of a physical configuration of radius $R$ and energy $E$ is bounded by\footnote{Here we have temporarily reinstated the constants $\hbar$ and $c$ to emphasize the physical origin of the bound.}
\begin{equation}
    S \leq \frac{2\pi k_{\rm B}}{\hbar c}RE \,.
\end{equation}
The bound imposes a finiteness condition on the  information density of physical systems. 
In the spirit of this bound, it is compelling to consider the relation between the complexity of the classical field configuration and the energy and size of the system. The energy of the configuration $\phi_n$ is calculated by integrating the Hamiltonian density, which yields
\begin{equation}\label{eq:1denergy}
E_n = \frac{1}{2} \int_{0}^{R} \dd x \, (\pd_x \phi_n)^2  = \frac{\alpha^2  \pi^2 n^2 }{4R } \,.
\end{equation}
By combining equations \eqref{eq:1dfourierPcomp} and \eqref{eq:1denergy}, we observe the following.

\begin{subbox}{Complexity of a two-dimensional free massless scalar}
The format and degree of a plane wave field configuration in a two-dimensional massless real scalar field theory scale as 
\begin{equation}
    \cF \sim 1, \quad \cD \sim \frac{1}{\alpha}\sqrt{R E} \,.
\end{equation}
\end{subbox}

This relation, although it relies on numerous simplifications, reveals a connection between the complexity measure of sharp o-minimality and physical quantities. 
An essential observation is that the complexity is not strictly bounded in terms of $R$ and $E$, since it also depends on the amplitude $\alpha$, which is a free parameter. By taking $\alpha\to 0$, we can generate arbitrarily high complexity at fixed $R$ and $E$. From a geometric perspective this makes sense, since this notion of complexity does not see the size of an object, but only the structure of its geometric features. 

For example, a wave with a very small amplitude but a high frequency can achieve a large complexity at the expensive of little energy. This is a consequence of the classical nature of the setting; after quantization, we cannot expect to make the parameter $\alpha$ arbitrarily small. This resonates with the quantum-mechanical nature of the Bekenstein bound.

\subsection{Superpositions of waves}
We now generalize the discussion to a finite superposition of waves, and consider a field configuration of the form
\begin{equation}
    \phi(x) = \sum_{n=1}^N \alpha_n \sin(\frac{n\pi  x}{R} )\,.
\end{equation}
Since the Fourier modes arise from the same Pfaffian chain, the complexity of this superposition scales as the complexity of the highest-degree mode, i.e.~$\cF\sim 1$ and $\cD\sim N$. Due to the orthogonality of the modes, the energy of this configuration takes the simple form
\begin{equation}
    E = \frac{\pi^2}{4R} \sum_{n=1}^N \alpha_n^2 n^2  \,.
\end{equation}
For this simple generalization, the connection between the quantities $R$, $E$, and $\cD$ is already obscured by the free parameters $\alpha_1,\ldots,\alpha_n$. Instead of this precise relation, we can bound the energy as 
\begin{equation}
    E \leq \frac{\alpha^2 \pi^2 N^2}{4R}   \,,
\end{equation}
where $\alpha$ is now defined by $\alpha^2=\sum_{n=1}^N\alpha_n^2$. Perhaps surprisingly, this leads to a \textit{lower} bound on complexity,
\begin{equation}
    \cD \gtrsim \frac{1}{\alpha}\sqrt{RE}\,.
\end{equation}
Upon keeping the amplitude $\alpha$ fixed, this expression suggests that a classical field configuration of size $R$ and energy $E$ requires \textit{at least} a certain amount of geometric complexity, measured through the degree $\cD$. Generalizing the present setting by removing the simplifying assumptions will only increase the complexity of the description of the field configuration, indicating that this lower bound generalizes, unless a special description of lower complexity exists. 

\subsection{Emergence of simplicity in field configurations}
A situation in which a special description may emerge is when the coefficients of the plane waves satisfy algebraic relations. This occurs for instance when the amplitudes describe the Fourier coefficients of a simple function. As an explicit example, consider the function $\phi:[0,R]\to\bbR$ defined by $\phi(x)=x(R-x)$. This function is algebraic and has a low complexity, and it can also be expressed in terms of the Fourier series
\begin{equation}\label{eq:xR-xsolution}
\phi(x) = \sum_{n=1}^\infty \frac{8R^2}{\pi^3(2n-1)^3} \sin(\frac{(2n-1)\pi  x}{R} )  \,.
\end{equation}
This example shows that field configurations can have a finite complexity despite being a superposition of infinitely many plane waves of increasing complexity, demonstrating a physical instance of the phenomenon of \textit{emergence of simplicity} discussed in chapter~\ref{ch:complexity}. 

\subsection{Complexity and time-dependence}
Let us remark that so far we have only considered static field configurations at a fixed time $t=0$. When these solutions are evolved in time according to the wave equation \eqref{eq:waveeq}, we have to account for the complexity of the time-dependence. This can be done in two ways: either we regard time as a parameter and analyze how the complexity changes with time, or we view time as a variable and quantify the complexity of the field configuration as a function on spacetime. \\

Applying the first approach to the example of a field on an interval $[0,R]$ studied above, we note that the time-dependent solutions are standing waves, with each wave component receiving a time-dependent factor that leaves its complexity unchanged. The only time-dependence of the complexity then arises from special points in time at which a simpler description emerges. For example, in the example above the time-dependent solution takes the form
\begin{equation}
\phi(x,t) = \sum_{n=1}^\infty \frac{8R^2}{\pi^3(2n-1)^3}\cos(\frac{(2n-1)\pi  t}{R} ) \sin(\frac{(2n-1)\pi  x}{R} )  \,.
\end{equation}
There is a low-complexity description at $t=0$, where the solution reduces to equation~\eqref{eq:xR-xsolution}, which disappears as soon as the modes start to oscillate independently. This simpler description reappears whenever $t$ is an integer multiple of $R$.\\

\newpage
To implement the second point of view, we must construct a Pfaffian chain which describes the oscillating time-dependence of the solution. Given the relativistic nature of the theory, it is evident that this can be done in the same way as for the spatial variable $x$, and that the resulting complexity depends on the size of the time interval. We will thus not explore this perspective further, and focus on the complexity of the spatial dependence.

\subsection{Field configurations in potentials}
Let us now generalize the setting by considering a two-dimensional real scalar theory with a general potential, described by a Lagrangian of the form 
\begin{equation}
    \cL(\phi) = -\frac{1}{2}\pd_\mu \phi \,\pd^\mu \phi -V(\phi) \,. 
\end{equation}
Unless the potential only consists of a mass term $V(\phi)=\frac{1}{2}m^2\phi^2$, the field configurations in this theory will no longer be described by a simple wave equation. The general form of the equation of motion is 
\begin{equation}
    \pd^2\phi -  V'(\phi) = 0\,.
\end{equation}
While it is generally no longer possible to explicitly solve for the solutions to this equation, we can proceed along the same lines as for the time-independent Schr\"odinger equation and deduce the complexity of solutions to this equation using the Riccati substitution. As before, the complexity depends on the size of the spacetime domain on which the field is considered. In addition, if the complexity of $\phi$ is captured by means of a differential chain, it will directly depend on the complexity of the potential $V(\phi)$.
\begin{subbox}{Complexity in field theory}
The complexity of a classical field configuration depends on the complexity of the underlying theory, measured by the complexity of the Lagrangian.     
\end{subbox}
This observation foreshadows that it might be possible to use sharp o-minimality to define the complexity of a field theory itself, beyond merely considering the physical functions appearing in calculations.  \\

\subsection{Complexity in higher-dimensional field configurations}
To conclude this section, let us briefly comment on the generalization to higher-dimensional field theories. 
We consider a $d$-dimensional free massless scalar field, described by the Lagrangian of equation~\eqref{eq:Lagfreescalar} with $\mu=0,\ldots,d-1$. The simplest solutions to the equation of motion are plane waves, and as before we consider a physical configuration confined to a bounded spatial region, which for simplicity we assume to have the shape of a box $[0,R]^{d-1}\subseteq \bbR^{d-1}$. By aligning the momentum with the coordinates of the box, a  plane wave solution takes the form
\begin{equation}
    \phi(\vec x) = \alpha \sin(\frac{n\pi x}{R})\,,
\end{equation}
where $\vec x =(x,\ldots)$ and $\alpha$ is a dimensionful parameter with dimension $[\alpha]=(d-2)/2$. This configuration can be described with the same Pfaffian chain as the $d=2$ case, and the complexity scales in the same way with the size of the system. In particular, the additional dimensions do not contribute to  $(\cF,\cD)$, essentially because the plane wave oscillates only in one spatial direction. The energy of the field configuration is calculated as 
\begin{equation}
    E = \frac{1}{2}\int_{[0,R]^{d-1}} \!\!\! \dd^{d-1} \vec x \, (\nabla \phi)^2 = \frac{1}{4}\alpha^2\pi^2 n^2 R^{d-3}\,.
\end{equation}
The difference with the two-dimensional case is the dimensionality of the parameter $\alpha$, and the contribution of the volume of the box, which modifies how $E$ scales with $R$. The parameter $\alpha$ can be made dimensionless by dividing by an appropriate energy scale. The natural scale in the system is set by the length $R$, and upon rescaling $\alpha\to R^{-(d-2)/2}\alpha$, the relation between the energy, the radius, and the complexity of the configuration becomes
\begin{equation}
    \cF \sim d, \quad \cD \sim \frac{1}{\alpha}\sqrt{R E} \,.
\end{equation}
While this relation again relies on numerous  simplifying assumptions, it is intriguing to note that the dependence of the degree $\cD$ on $R$ and $E$ is independent of the spacetime dimension $d$, similar to the Bekenstein bound. Again, the classicality of the theory permits an unbounded complexity by taking $\alpha\to 0$. This signals that, although classical physics has the potential to be governed by tame functions, there is no uniform bound on the information contained in classical configurations. \\

In the next chapter we turn from quantum mechanics and classical field theory to tameness and complexity in quantum field theory. 

\chapter{Tameness of non-perturbative observables in quantum field theory}\label{ch:observables}
\setlength{\parindent}{0pt}

The observables of quantum field theories are typically defined in terms of path integrals over field configurations and generally depend in a non-trivial way on kinematic variables and parameters of the Lagrangian of the theory. 
While these physical functions are often extremely hard to compute, it appears
reasonable to presume that they are not arbitrarily complex and have many regularity features. Specifically, one may have the hope that they are specified by inputting a finite amount of information that can be matched with experimental data.
With this motivation in mind, we set out to analyze the tameness of observables in quantum field theories, including amplitudes, partition functions, and correlation functions. We begin by briefly reviewing the case of perturbative amplitudes, whose tameness is established in \cite{Douglas:2022ynw}. 
The focus of this chapter lies on non-perturbative observables, which are accessible only in exceptional cases. We mostly analyze the setting of zero-dimensional quantum field theories, where exact calculations can be performed. Using two distinct approaches based on differential chains and the theory of Borel resummation, we show that tameness holds in a general class of zero-dimensional models, despite the non-analytic nature of the weak coupling limits of observables.

\section{Perturbative amplitudes}
We begin this chapter by briefly reviewing the tameness of perturbative amplitudes, analyzed in detail in \cite{Douglas:2022ynw}. Amplitudes in quantum field theory encode the probability density associated to a physical process. The physical processes under consideration describe the scattering of particles, and the corresponding amplitudes are functions $\cA$ of kinematic variables (such as positions or momenta of the particles)  and parameters of the theory (such as coupling constants describing the strengths of the interactions). Typically, the dependence on the parameters is extremely difficult to determine, and one resorts to approximations via perturbation theory. 
In this procedure, the amplitude is expanded as a formal power series in a parameter $\lambda$.
This results in a perturbative expansion of the form
\begin{equation}
    \cA(\kappa,\lambda) = \sum_{\ell=0}^\infty \lambda^\ell \cA_\ell(\kappa)\,.
\end{equation}
Perturbative quantum field theory is concerned with calculating the functions $\cA_\ell(\kappa)$, which depend on the kinematic variables $\kappa$ as well as the remaining parameters of the theory. This perturbative expansion can be combinatorially organized in terms of Feynman graphs, and the rules for evaluating them is determined by the descriptive functions of theory. The function $\cA_\ell(\kappa)$ is then obtained by evaluating all Feynman graphs with $\ell$ loops, and calculations become systematically more difficult as $\ell$ increases. In this setting, the following result was shown in \cite{Douglas:2022ynw}. 

\begin{subbox}{Tameness of perturbative amplitudes}
For any renormalizable quantum field theory with finitely many particles and interactions, all perturbative amplitudes with finitely many loops and external edges are definable in the o-minimal structure $\bbR_{\rm an,exp}$ as functions of the masses, external momenta, and coupling constants.    
\end{subbox}

The proof of this statement proceeds along the following lines.
\begin{enumerate}
    \item[(i)] Under the assumption of finitely many particles and interactions, a finite-loop amplitude is obtained by summing over finitely many Feynman graphs and evaluating the corresponding Feynman integrals.
    \item[(ii)] The individual Feynman integrals are cast into period integrals on an auxiliary geometry. Potential divergences are controlled by dimensional regularization.
    \item[(iii)] The period integrals are definable in $\bbR_{\rm an,exp}$ by the results of \cite{BKT,BakkerMullane}.
\end{enumerate}
The key point in this proof is that each perturbative amplitude is calculated from finitely many contributions, and that each contribution has a tame geometric interpretation. The universality of this result is striking: regardless of the precise form of the Lagrangian of the quantum field theory and the precise nature of the physical process, the perturbative amplitude can be captured by an o-minimal structure. The assumption that the theory is renormalizable is essential in order to keep the renormalization process under control, ensuring that only finitely many counterterms have to be evaluated to obtain a finite amplitude. Note that the statement mainly concerns the dependence on the external momenta and the mass parameters, since the dependence on the parameter $\lambda$ is trivialized to a polynomial at finite loop order.

\section{Non-perturbative observables via differential chains}\label{sec:0dQFTdiff}
While the evaluation of perturbative amplitudes lies at the heart of quantum field theory, they are only approximations of the exact physical amplitude. The non-perturbative observable in principle receives contributions from infinitely many Feynman graphs, and although the previous section shows that each of these contributions is tame, it is a delicate question whether they combine into a tame function of finite complexity. Nonetheless, since the parameters of the Lagrangian determine the description of the theory in a relatively simple way, it is a reasonable hope that non-perturbative observables realize an emergence of simplicity. 
In this first section on non-perturbative observables, we use differential chains to study correlation functions in zero-dimensional quantum field theories, where exact calculations can be performed. The differential chain approach will allow us to not only demonstrate the tameness of observables, but also to capture their complexity.

\subsection{Zero-dimensional QFTs and Noetherian chains}
We begin by considering a zero-dimensional theory consisting of a scalar field on a finite lattice with a polynomial Lagrangian. Our aim is to demonstrate, with the help of the results of \cite{Weinzierl:2020nhw,Weinzierl:2022mmp}, that there exists a general systematic procedure for constructing a Noetherian chain containing the correlation functions of the theory. To precisely specify our setting, we fix a $d$-dimensional Euclidean lattice $\Lambda$ consisting of $N$ points, and consider a theory described by the Lagrangian
\begin{equation}\label{eq:latticeaction}
    \cL(\phi) =\sum_{x\in \Lambda}\left(-\sum_{\mu=0}^{d-1}\phi_x\phi_{x+  e_\mu}+d\phi_x^2+\sum_{j=2}^{D}\frac{\lambda_j}{j!}\phi_x^j\right)\,.
\end{equation}
Here the $x$ labels the lattice sites, $\phi_x$ is the value of the scalar field on the site $x$, and $e_\mu$ are basis vectors for the lattice $\Lambda$. The Lagrangian includes the most general polynomial interaction term, and $D$ indicates the highest-order interaction. Note that the nearest-neighbor interactions of the form $\phi_x \phi_{x+e_\mu}$ encode the kinetic term of the scalar. The correlation functions of the theory now take the form of the following Euclidean lattice path integral,\footnote{The domain of integration is $\gamma^N$, where $\gamma$ is a curve in the complex plane chosen such that the integrand vanishes at the boundary; this choice of contour enables us to integrate by parts.} 
\begin{equation}\label{Inu}
    I_{\nu_1\cdots \nu_N}(\lambda_2,\ldots,\lambda_D)=\int\dd^N \phi \bigg(\prod_{k=1}^{N}\phi_{x_k}^{\nu_k}\bigg)e^{-\cL(\phi)} \,.
\end{equation}
Crucially, these integrals are related among each other through a collection of integration-by-parts (IBP) identities 
\begin{equation}
    \int\ \dd^N \phi  \, \frac{\pd}{\pd \phi_x}\bigg[\bigg(\prod_{k=1}^{N}\phi_{x_k}^{\nu_k} \bigg)e^{-\cL(\phi)}  \bigg] =0 \,.
\end{equation}
Using techniques of twisted cohomology which formalize these identities, it was argued in \cite{Weinzierl:2020nhw,Weinzierl:2022mmp} that there exists a finite basis $I_1,\ldots,I_{n}$ of correlation functions, such that every correlation function can be written as a linear combination of this basis, with coefficients given by Laurent polynomials in the couplings $\lambda_2,\ldots,\lambda_D$. Moreover, the number of basis vectors is at most $N_{\rm F} = (D-1)^N$. Meanwhile, differentiating a correlation function with respect to a coupling $\lambda_j$ yields
\begin{equation}
    \frac{\pd }{\pd \lambda_j } I_{\nu_1\cdots \nu_N} = -\frac{1}{j!} \sum_{i=1}^N I_{\nu_1\cdots (\nu_i +j) \cdots \nu_N} \,.
\end{equation}
Combining this differential relation with the existence of a finite basis, it was then shown that the basis integrals satisfy a system of differential equations of the form 
\begin{equation}\label{eq:Noether}
    \frac{\pd}{\pd \lambda_j}I_i=\sum_{k=1}^{N_{\rm F}}A_{ik} I_k,
\end{equation}
for some $N_{\rm F}\times N_{\rm F}$ square matrix $A$. The entries of $A$ are Laurent polynomials in the couplings. This essential fact about the structure of correlation functions in lattice QFTs can now be used to construct a Noetherian chain for which all correlation functions are Noetherian. We start with the functions $f_2,\ldots,f_D$ defined by 
\begin{equation}
    f_j(\lambda_2,\ldots,\lambda_D)  = \frac{1}{\lambda_j} \,,
\end{equation}
which are defined on the domain $\lambda_j\in(0,\infty)$ and satisfy the differential equation
\begin{equation}
     \frac{\pd}{ \pd \lambda _j} f_j = -f_j^2 \,.
\end{equation}
The entries of the matrix $A$ are now polynomial in the couplings $\lambda_2,\ldots,\lambda_D$ and the functions $f_2,\ldots,f_D$, meaning that the system of differential equations in \eqref{eq:Noether} forms a Noetherian chain. In fact, since $I_1,\ldots,I_n$ form a basis, \textit{all} correlation functions in the theory are Noetherian with respect to this chain. Though Noetherian functions satisfy strong local finiteness properties, the argument above is unable to show that general lattice correlation functions are tame on their complete domain. \\

If the matrix $A$ were triangular, this chain would have been Pfaffian, demonstrating the tameness of the correlation functions on the full domain of the couplings. In general, $A$ is rarely triangular or triangularizable, which means that the Pfaffian framework cannot be implemented there unless another representation is found. \\

A simple way to nonetheless characterize the local tameness of these correlation functions is to restrict the space of couplings $\lambda_2,\ldots,\lambda_D$ by imposing finite lower and upper bounds on each $\lambda_j$, so that equation \eqref{eq:Noether} defines a restricted Noetherian chain. On this domain, the correlation functions are then definable in the o-minimal structure $\bbR_{\text{rN}}$, which is conjectured to be sharply o-minimal \cite{binyamini_tameness_2023}. If this conjecture holds, then the sharply o-minimal complexity measure can be implemented for any lattice QFT described by a Lagrangian of the form in equation \eqref{eq:latticeaction}. In particular, all correlation functions in these theories would then carry a format $\cF$ and degree $\cD$, whose dependence on the parameters of the theory such as the dimension $d$, the size of the lattice, and the highest interaction $D$ could be analyzed. Let us emphasize that the preceding discussion is only valid upon restricting the space of couplings. This means that, in the restricted Noetherian setting, one loses the ability to probe potentially interesting zero and infinite coupling limits. In contrast, the Pfaffian framework is globally valid, but can only be implemented in special cases, as we will see in the next subsection.

\subsection{Zero-dimensional QFTs and Pfaffian chains}
We now turn to a significant simplification of the setting of the previous subsection, focusing on a scalar field $\phi$ on a one-point spacetime. The simplicity of this setting yields strong computational control, which will allow us to perform a detailed analysis of the non-perturbative correlation functions. These functions are single-variable integrals 
\begin{equation}
    I_{\nu}=\int \dd\phi \,\phi^\nu e^{-\cL(\phi)}\, ,
\end{equation}
where $\cL(\phi)$ is a polynomial Lagrangian. Following the discussion above, we now know that the IBP identity
\begin{equation}
  I_\nu = \int \dd \phi \, \left( \frac{\pd }{\pd \phi}\frac{1}{\nu+1}\phi^{\nu+1} \right) e^{-\cL(\phi)} =  \frac{1}{\nu+1}\int \dd \phi \,  \left(\frac{\pd }{\pd \phi}\cL(\phi) \right) \phi^{\nu+1}  e^{-\cL(\phi)} 
\end{equation}
implies that there exists a finite basis of correlation functions, with the size of the basis depending on the degree of the Lagrangian.

\subsubsection*{$\phi^4$-theory on a point}
We first study a massive scalar field with a single $\phi^4$-interaction. The Lagrangian of this theory, in the standard convention, is given by
\begin{equation} \label{phi4action}
\cL(\phi) = \frac{m^2}{2}\phi^2 + \frac{\lambda}{4!}\phi^4 \,. 
\end{equation}
The parameters of the theory are the mass $m$ and the coupling $\lambda$, but since we can freely rescale the field $\phi$ without changing the partition function, the theory can be fully specified by a single parameter. In fact, reparametrizing the theory may also lead to a reduction in the complexity of computable quantities. This motivates us to rescale the field $\phi$ and introduce a new coupling parameter as follows
\begin{equation*}
    \phi\rightarrow\sqrt{\frac{3}{2 \lambda}} m \phi, \qquad g= \frac{3m^4}{4\lambda} \,.
\end{equation*}
In these variables, the Lagrangian takes the simpler form
\begin{equation}\label{eq:Lphi4g}
    \cL(\phi) =g \phi^2 +\frac{g}{8} \phi^4 \,,
\end{equation}
and the coupling $g$ is now an overall multiplicative parameter. The general theory mentioned above tells us that a spanning set of correlation functions is given by $\{I_0,I_1,I_2\}$. However, since $I_1=0$ by symmetry, the integrals $I_0$ and $I_2$ suffice. The explicit form of these basis integrals is  
\begin{equation}
     I_0=\sqrt{2} e^{g} K_{1/4}(g)\,,\qquad  
    I_2=-2\sqrt{2} e^{g}(K_{1/4}(g)-K_{3/4}(g)) \,,
\end{equation}
where $K_\alpha$ is the modified Bessel function of the second kind. \\

The IBP identity relating the correlation functions is given by
\begin{equation}\label{eq:IBPpoint}
    I_\nu = \frac{2(\nu-3)}{g}I_{\nu-4}-4 I_{\nu-2} \,. 
\end{equation}
By iteratively invoking this identity, it is possible to write any $I_{\nu}$ as a linear combination of the two basis functions $I_0$ and $I_2$, with coefficients given by polynomials in $1/g$. The degree of these polynomials is determined by the number of times the IBP identity has to be used. For example, we have
\begin{align*}
    I_4 &= \frac{6}{g}I_0 -4I_2\, ,\\ 
    I_6 &= -\frac{24}{g}I_0 + \left(16 +\frac{18}{g} \right)I_2\, , \\
    I_8 & = \left(\frac{96}{g} +\frac{180}{g^2} \right)I_0 + \left(-64 -\frac{192}{g} \right)I_2\, .
\end{align*}
In general, the degree of the polynomials appearing in the expansion of $I_\nu$ is bounded above by $\nu/4$. \\

We now claim that all the correlation functions $I_\nu$ in this theory are Pfaffian, which implies that they are tame, and in addition enables us to compute their complexity. To show this, we construct an explicit Pfaffian chain for $I_0$ and $I_2$. First, note that the derivatives of the correlators are given by
\begin{equation}
\label{derivative}
   \frac{\pd I_\nu}{\pd g }= -I_{\nu+2}-\frac{1}{8} I_{\nu+4}\,.
\end{equation}
Together with the IBP relation of equation \eqref{eq:IBPpoint} this allows us to write down a second-order differential equation for $I_0$, namely
\begin{equation}
\label{phi4equation}
    \frac{\pd^2 I_0}{\pd g^2} -\left(2-\frac{1}{g}\right)\frac{\pd I_0}{\pd g}-\left(\frac{1}{g} + \frac{1}{16g^2}\right)I_0 =0 \,.
\end{equation}
Making use of the fact that $I_0$ is non-vanishing, it is possible to perform a Riccati substitution by introducing the auxiliary function
\begin{equation}
    h(g) = -\frac{1}{I_0} \frac{\pd I_0}{\pd g} \,,
\end{equation}
which fulfills the Riccati equation
\begin{equation}
  \frac{\pd h}{\pd g}=h(g)^2+\left(2-\frac{1}{g}\right) h(g)-\frac{1}{16 g^2}-\frac{1}{g}\,.  
\end{equation}
These functions fit together into a Pfaffian chain as follows
\begin{align}
    \zeta_1(g)& = \frac{1}{g}\,,  &&  \frac{\pd\zeta_1}{\pd g}=- \zeta_1^2 \,, \\
    \zeta_2(g)&=h(g)\,,   && \frac{\pd\zeta_2}{\pd g}= \zeta_2^2+\left(2-\zeta_1\right) \zeta_2-\frac{1}{16 }\zeta_1^2-\zeta_1\,, \nonumber\\ 
    \zeta_3(g)&=I_0(g) \,, && \frac{\pd\zeta_3 }{\pd g} =-\zeta_2\zeta_3 \,. \nonumber
\end{align}
Note that the domain of these functions is the interval $(0,\infty)$. This  excludes the value $g=0$ of the coupling, on which we will comment later. Within this chain, $I_2$ can be expressed as a Pfaffian function  
\begin{equation}\label{eq:I2Pfaff}
    I_2=2\zeta_2\zeta_3-\frac{1}{2}\zeta_1\zeta_3\, , 
\end{equation}
which follows from the IBP relation and the derivative relation in equation \eqref{derivative}. Since any correlation function $I_\nu$ can be written as a linear combination of $I_0$ and $I_2$, with coefficients given by polynomials in $\zeta_1$, it follows that all correlation functions in this theory are Pfaffian functions. This forms our first example of tameness on an unbounded domain for non-perturbative observables in QFT. \\

The Pfaffian complexity of the correlation functions is now determined by this Pfaffian chain. First, consider the partition function $Z(g)=I_0(g)$ of the theory. Its Pfaffian complexity can be read off from the chain as
\begin{equation}
    \cC_{\rm Pf}(Z) = (1,3,2,1)\, .
\end{equation}
Next, the correlator $I_2$ is a Pfaffian function of degree 2 with respect to the given chain as shown by equation \eqref{eq:I2Pfaff}, so its Pfaffian complexity is $\cC(I_2) = (1,3,2,2)$. For general correlation functions, we use the fact that the expansion in terms of $I_0$ and $I_2$ has coefficients which are polynomial in the function $\zeta_1(g)=1/g$. By counting the degrees in this expansion, we infer the following.

\begin{subbox}{Complexity of correlation functions in 0d $\phi^4$-theory}
The Pfaffian complexity of the correlation function $I_\nu$ is given by 
\begin{equation}
\cC_{\rm Pf}(I_\nu) = \left(1,3,2, \left\lceil  \frac{\nu}{4} \right\rceil +1 \right) \qquad (\nu \text{  even})\, .
\end{equation}
\end{subbox}

Here $\lceil\cdot\rceil$ denotes the ceiling function. It is natural to expect that the Pfaffian complexity of the correlation function grows with the number of field insertions $\nu$, and our calculation shows that this growth is (stepwise) linear. Moreover, the degree $\beta$ is the only complexity parameter that grows, whereas the other parameters appear to be fixed by the theory.\footnote{This includes the parametrization of the theory; in a different description, the complexity could take a different value.}

\subsubsection*{$\phi^6$-theory on a point}
A natural continuation is to study the $\phi^6$-theory on a point. For this theory, we choose the following parametrization of the Lagrangian:
\begin{equation}
\label{eq:actionphi6}
    \cL(\phi )=-\frac{g}{2}\phi^2+\frac{1}{96}\phi^6\,.
\end{equation}
The structure of the theory is similar to the $\phi^4$-theory, and the main difference is that the set of basis integrals now consists of the three functions $I_0$, $I_2$, and $I_4$. Note that, for technical reasons on which we will comment later, the mass term now carries a different sign than in the previous example. With this normalization, the IBP relation takes the form
\begin{equation}
   I_\nu= 16 \left((\nu -5) I_{\nu -6} +g I_{\nu -4}\right)\,. 
\end{equation}
The three basis correlation functions are explicitly given by
\begin{align}\label{eq:I0I2I4}
    I_0(g) &=\sqrt{2} \pi ^{3/2} \left(\text{Ai}(g)^2+\text{Bi}(g)^2\right) \,,\\
    I_2(g) &=4 \sqrt{2} \pi ^{3/2} (\text{Ai}(g) \text{Ai}'(g)+\text{Bi}(g) \text{Bi}'(g)) \,, \\
    I_4(g) &=8 \sqrt{2} \pi ^{3/2} \left(\text{Ai}'(g)^2+\text{Bi}'(g)^2+g \left(\text{Ai}(g)^2+\text{Bi}(g)^2\right)\right) \,,
\end{align}
where we use a prime to denote the first derivative, and where Ai and Bi are the Airy functions fulfilling the Airy equation
\begin{equation}
    \frac{\pd^2 f }{ \pd x^2}- x f(x)=0\, .
\end{equation}
We now claim that $I_0$, $I_2$ and $I_4$ are Pfaffian, and our strategy to show this will be to argue that the Airy functions are Pfaffian. To this end, we introduce the logarithmic derivatives $h_A=\text{Ai}'/\text{Ai}$ and $h_B=\text{Bi}'/\text{Bi}$ which satisfy
\begin{equation}
    \frac{\pd h_A }{\pd g}=h_A^2+g\,, \quad \frac{\pd h_B }{\pd g}=h_B^2+g\,.
\end{equation}
Note that these are only defined on a domain on which the Airy functions are non-vanishing. Since these are first-order, we are now able to write down the following Pfaffian chain: 
\begin{align}
    \zeta_1(g) &= h_A(g) \,,  && \frac{\pd\zeta_1}{\pd g}= \zeta_1^2 +g \,, \\ 
    \zeta_2(g) &= \text{Ai}(g) \,,&& \frac{\pd\zeta_2 }{\pd g} =\zeta_1\zeta_2 \,, \nonumber \\
    \zeta_3(g) &= h_B(g)  \,, && \frac{\pd\zeta_3}{\pd g}= \zeta_3^2 +g \,, \nonumber\\ 
    \zeta_4(g) &= \text{Bi}(g) \,, && \frac{\pd\zeta_4 }{\pd g} =\zeta_3\zeta_4 \,. \nonumber
\end{align}

It follows that the Airy functions 
are Pfaffian on a domain in which the logarithmic derivatives can be defined. Since the Airy functions are non-vanishing for $g>0$ and have infinitely many zeros\footnote{It is interesting to note that the presence of infinitely many zeros means that the Airy functions are not tame on the real line, but that the combinations appearing in the correlation functions $I_0$, $I_2$, and $I_4$ in equation~\eqref{eq:I0I2I4} are tame.} for $g<0$, we restrict to positive $g$.\footnote{This explains the choice of the sign of the mass term in equation \eqref{eq:actionphi6}.}\\

Having set up the Pfaffian chain for the Airy functions, let us calculate the Pfaffian complexity of the correlation functions, starting with the three basis correlators. Firstly, the partition function $Z(g)=I_0(g)$ is quadratic in $\zeta_1$, so its Pfaffian complexity is 
\begin{equation}
    \cC_{\rm Pf}(Z) = (1,4,2,2)\,.
\end{equation}
Recalling that the partition function of the $\phi^4$-theory has Pfaffian complexity $(1,3,2,1)$, we can already observe that the complexity has increased as a consequence of the higher-order interaction term. The correlators $I_2$ and $I_4$ involve the derivatives of the Airy functions, which can be written with degree $2$ as $\zeta_1\zeta_2$ and $\zeta_3 \zeta_4$, respectively. It follows that 
\begin{equation}
    \cC_{\rm Pf}(I_2) = (1,4,2,3 ) \,, \quad  \cC_{\rm Pf}(I_4) = (1,4,2,4) \,.
\end{equation}
For the higher correlation functions $I_\nu$, we use that they can be written as a linear combination of $I_0$, $I_2$ and $I_4$ with polynomial coefficients in $g$. As in the $\phi^4$-theory, the degrees of these polynomials depend stepwise linearly on $\nu$. 

\begin{subbox}{Complexity of correlation functions in 0d $\phi^6$-theory}
The Pfaffian complexity of the correlation function $I_\nu$ is given by 
\begin{equation}
\cC_{\rm Pf}(I_\nu) = \left(1,4,2, \left\lfloor \frac{\nu}{4} \right\rfloor+3 \right) \qquad (\nu \text{  even and  } \nu\neq  0)\, .
\end{equation}
\end{subbox}
Here $\lfloor\cdot\rfloor$ denotes the floor function. The degree of the Pfaffian complexity again grows linearly with the number of field insertions $\nu$. It is interesting to note that the growth of the degree has the same rate as the complexity of the correlation functions in $\phi^4$-theory, whereas the `initial' value of the complexity is higher. Let us also comment on the fact that the order $r$ and degree $\alpha$ are independent of $\nu$. The reason for this is that all correlation functions can be expressed using the same Pfaffian chain, which is a remarkable consequence of the algebraic relations between the correlation functions. In general, one expects that more complicated physical quantities also require more involved Pfaffian chains with increased $r$ and $\alpha$. \\

While this analysis shows that non-perturbative observables in simple theories can be shown to be tame when the coupling is restricted to the infinite half-line $(0,\infty)$, it misses the zero-coupling limit, where new subtleties emerge. \\

\section{Non-perturbative observables via resummation}
We now continue with an alternative approach for calculating non-perturbative observables, which proceeds by an analysis of the perturbative expansion.
As the factorial growth of the number of Feynman graphs in a perturbative expansion shows, the resulting series often diverges and is only asymptotic to the exact observable. This prevalent feature of QFTs has been 
recognized since the seminal work of Dyson \cite{PhysRev.85.631}. 
This behavior manifests itself in the non-analyticity of observables as a coupling $\lambda$ approaches zero, posing a significant challenge to using perturbation theory for testing the tameness of the full quantum observable. \\

A powerful method of addressing the divergence of the perturbative expansions in QFT is Borel resummation, which is a procedure that transforms the asymptotic series arising from perturbation theory into a well-defined function of the coupling. In this section we will focus on perturbative expansions that are Borel summable, and prove that the partition functions and correlation functions in certain QFTs are definable functions in the o-minimal structure generated by the Gevrey functions, known as $\bbR_{\rm G}$ \cite{DRIES_SPEISSEGGER_2000}. Before we begin the analysis, we briefly review some preliminaries.

\subsection{Introduction to non-analyticity and Borel resummation}\label{sec:Borel-summability-section}

In this subsection we introduce the phenomena of non-analyticity, 
and the theory of Borel resummation by which one can turn the associated divergent series into a  function which agrees with the original function in the absence of resurgence \cite{Sauzin:2014qzt,balser1994divergent,Hardy-book,LodayRichaud2014DivergentSA}.

\subsubsection*{Non-analytic physical functions}
There are two major sources for non-analyticity in physical functions. The first arises from the prototypical example of a function which is smooth but non-analytic,
\begin{equation}\label{f-non-an}
f(x) = 
    \begin{cases}
        e^{-1/x} & x > 0\ , \\
        0 & x=0\ .
    \end{cases}
\end{equation}
Every derivative of $f$ vanishes at $x=0$, while $f$ itself is not constant on any neighbourhood of $x=0$. Therefore, the Taylor series of $f$,
although it has an infinite radius of convergence, is identically zero. This type of non-analyticity is encountered in the presence of instanton corrections, which are genuinely invisible in perturbation theory.\\

The second source of non-analyticity comes from a rather different reason, namely when the function has a diverging Taylor series with zero radius of convergence. This form of non-analyticity is encountered ubiquitously in quantum field theory, due to the factorial growth of the number of Feynman graphs in perturbative expansions. This is the type of non-analyticity with which we will be concerned in the remainder of this chapter.\\

As an example, consider the zero-dimensional $\phi^4$-theory analyzed in the previous section, with partition function 
\begin{equation}\label{eq:Z(g)}
    Z(g) =\int_{-\infty}^{\infty}\dd\phi \, e^{-\cL(\phi)} =\sqrt{2}e^g K_{1/4}(g) \, .
\end{equation}
Recall that the auxiliary coupling $g$ is inversely related to the original $\phi^4$-coupling $\lambda$ via $g=3m^4/(4\lambda)$. The Bessel function $K_{1/4}$, which can be represented as 
\begin{equation}\label{eq:K1/4}
    K_{1/4}(z) = \int_{0}^{\infty}\dd t\, \cosh\left(\tfrac{t}{4}\right)e^{-z\cosh(t)} \qquad \text{with}\quad |\arg(z)| < \frac{\pi}{2} \, ,
\end{equation}
has an essential singularity at $z=\infty$, implying that $Z(\lambda)$ is non-analytic in the $\lambda\to 0$ limit. Note however that the limit $\lim_{g\rightarrow \infty}Z(g)$ exists and takes the value $0$, so that $Z(\lambda)$ is smooth at $\lambda= 0$. The non-analyticity can be seen more explicitly from the rapid growth of the coefficients in the perturbative expansion, which we analyze in detail later in this section. Nonetheless, a physical meaning can be extracted from the divergent power series using the technique of Borel resummation.

\subsubsection*{Borel resummation}
We now briefly introduce the basic concepts of Borel summability, referring to \cite{Sauzin:2014qzt,balser1994divergent,Hardy-book,LodayRichaud2014DivergentSA} for more detailed expositions. Consider a divergent (formal) power series 
\begin{equation}\label{eq:formal-power}
    \widetilde{\varphi} = \sum_{n=0}^{\infty} a_n z^{n+1} \,.
\end{equation}
By divergent, it is meant that the power series has zero radius of convergence in the complex plane. The power series $\widetilde{\varphi}$ is said to belong to the \textit{Gevrey class} $1/p$ if there exist constants $A, B$ such that
 \begin{equation}
     |a_n| \leq A B^n (n!)^p \,,
 \end{equation}
uniformly in $n$. In this case, the Borel operator $\mathcal{B}_p$ is defined as
\begin{equation}
   \widehat{\varphi}(\zeta)  = \mathcal{B}_p\widetilde\varphi = \sum_{n=0}^{\infty} \frac{a_n}{\Gamma(np +1)} \zeta^{n} \,.
\end{equation}
This operation produces a new series $\widehat \varphi(\zeta)$ with a finite radius of convergence, and is known as the (formal) Borel transform. The following definition is central.

\vspace{0.5cm}

\begin{subbox}{Definition: Borel summability}
A formal power series $\widetilde\varphi$ of Gevrey class $1/p$ is said to be $p$-\textit{summable} (or \textit{Borel-summable}), if its Borel transform $\mathcal{B}_p\widetilde\varphi = \widehat\varphi$ satisfies the following conditions:
    \begin{itemize}
        \item It admits an analytic continuation to a horizontal strip 
        \begin{equation}
            S_{\delta} = \big\{ \zeta \in \bbC \,\big|\,  |\Im{\zeta^{1/p}}| < \delta \big\}\,.
        \end{equation}
        \item The analytic continuation has exponential size no larger than $1/p$, meaning that
\begin{equation}\label{eq:exp-size}
    |\widehat\varphi(\zeta)| \leq C \exp\big(c |\zeta|^{1/p}\big)
\end{equation}
for some constants $C$ and $c$, and for sufficiently large $\zeta \in S_\delta$.
    \end{itemize}
\end{subbox}

\vspace{0.5cm}

\begin{figure}[h]
    \centering
    \begin{subfigure}[t]{.5\textwidth}
  \centering %
 \begin{tikzpicture}[scale=0.8]
  \draw[fill=black!10] (1,0) circle [radius=1cm];
      \draw[->] (-3,0) -- (3,0) node[right] {$\Re (z)$};
  \draw[->] (0,-2) -- (0,2) node[above] {$\Im(z)$};
 
\end{tikzpicture}
\end{subfigure}%
\hfill
\begin{subfigure}[t]{.5\textwidth}
  \centering %
 \begin{tikzpicture}[scale=0.8]
    \draw[smooth,fill=black!10] plot[domain=0:540,samples=200] (\x:{2*cos(\x/2)^2});
  \draw[->] (-3,0) -- (3,0) node[right] {$\text{Re}(z)$};
  \draw[->] (0,-2) -- (0,2) node[above] {$\text{Im}(z)$};
 
\end{tikzpicture}
\end{subfigure}
\caption{Sokal disks for $p=1$ (left), and $p=2$, also known as a cardioid (right).}
    \label{fig:Sokal_disks}
\end{figure}
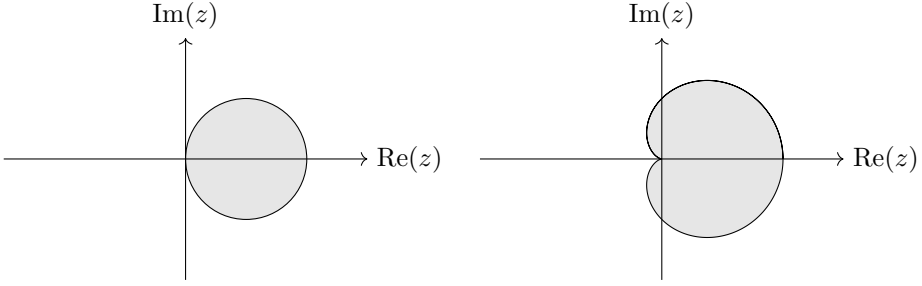

By a slight abuse of notation, we denote the analytic continuation of $\widehat\varphi(\zeta)$ by the same symbol. In case the above conditions are not satisfied, the positive real line $\mathbb{R}^+$ is said to be a \textit{Stokes line} for $\widetilde\varphi$. If a formal power series $\widetilde\varphi$ is $p$-summable, then the Laplace operator $ \mathcal{L}_p$ can be applied to $\widehat\varphi$ as
\begin{equation}\label{eq:Laplacetransform}
   \varphi(z)=\mathcal{L}_p\widehat{\varphi} = \frac{1}{p}\int_{0}^{\infty}\dd\zeta\left(\frac{\zeta}{z}\right)^{\frac{1}{p}-1}e^{-\left(\frac{\zeta}{z}\right)^{\frac{1}{p}}}\widehat{\varphi}(\zeta)\ .
\end{equation}
The resulting function $\varphi(z)$ is called the \textit{Borel sum} of $\widetilde\varphi$, and the procedure described to obtain it is called Borel resummation. One can then prove that the resulting function $\varphi(z)$ must be analytic in a domain $D^p_c = \{z \in \bbC \,|\, \textup{Re}(z^{-1/p}) > c\}$, where $c$ is the constant appearing in (\ref{eq:exp-size}). This domain is called a Sokal disk, and it is depicted for $p=1$ and for $p=2$ in figure~\ref{fig:Sokal_disks}. Note that the Sokal disk is open and does not include the origin: although $\varphi(0)$ can be defined as the limit for $z \rightarrow 0$ of $\varphi(z)$, the function thus extended will not be analytic at $z=0$, unless the original formal power series \eqref{eq:formal-power} was actually convergent. The limit $\lim_{z \rightarrow 0}\varphi(z)$ must exist in the Sokal disk if $\widetilde\varphi$ is Borel-summable, and it is actually $0$ if we assume the form \eqref{eq:formal-power} for the initial formal power series.\\

\newpage 
The interest in the Borel sum $\varphi(z)$ is due to the Nevanlinna-Sokal theorem \cite{Sokal}. 
This theorem guarantees that, if $\varphi(z)$ is the Borel sum of a Borel-summable formal power series $\widetilde\varphi$ of Gevrey class $1/p$, then $\varphi(z)$ admits $\widetilde\varphi$ as asymptotic expansion of order $p$ (denoted by $\varphi(z) \sim_p \widetilde{\varphi}$). This means that
\begin{equation}\label{asymptotic-expansion}
    \left|\varphi(z) - \sum_{n=0}^{N-1}a_n z^{n+1}\right| \leq A \delta^{-N}(pn!) z^{N+1}
\end{equation}
for every $z$ in the Sokal disk $D^p_c$.\footnote{Note that we can always substitute $(pn!)$ with $(n!)^p$ with a due rescaling of the constants $A$ and $\delta$ without altering the Gevrey class of the asymptotic expansion.} Observe how the exponential growth of the upper bound depends inversely on the width of the sector in which the Borel transform admits analytic continuation with exponential size at most $1/p$.\\

Crucially for our purposes, the theorem also holds in reverse. That is, the existence of a function $\varphi(z)$ which is analytic on the Sokal disk $D^p_c$ and admits a uniform asymptotic expansion as (\ref{asymptotic-expansion}) on $D^p_c$, suffices to conclude that the formal power series $\widetilde\varphi$ is Borel-summable, and that $\varphi(z)$ is its unique Borel sum. The combination of the two statements of the Nevanlinna-Sokal theorem ensures that there is a one-to-one correspondence between the algebra of functions holomorphic on a Sokal disk and the algebra of their asymptotic expansions. In other words, the Taylor map on this algebra of functions is injective, a feature which is known as \textit{quasi-analyticity}. The Nevanlinna-Sokal theorem is at the core of the results shown in this section: in particular, we will use the converse statement to prove the Borel summability of a path integral with a polynomial interaction.

\subsection{Gevrey functions and the o-minimal structure \texorpdfstring{$\bbR_{\rm G}$}{}}\label{sec:Rg-section}
In order to show that tameness of non-perturbative physical functions persists in the presence of non-analyticity, we need an o-minimal structure which manifestly hosts non-analytic functions of the type encountered above.
The required functions are the \textit{Gevrey functions}, which were shown to generate an o-minimal structure $\bbR_{\rm G}$ by Van den Dries and Speisegger  \cite{DRIES_SPEISSEGGER_2000}. 
Since the complete definition of Gevrey functions is quite technical, and because this chapter only considers one-parameter limits in quantum field theory, we focus on one-variable Gevrey functions. These functions are defined on domains of the form 
\begin{equation}
    S(R,\phi, p) = \big\{ z \in \bbC \, \big|\, 0< |z| < R ,\, |\text{arg}(z)| < p \phi\big \}\,,
\end{equation}
where $R \in (0, \infty)$, $0< p \leq 1$ and, most importantly, $\phi \in (\frac{\pi}{2}, \pi)$.
On these sectors of the complex plane we consider holomorphic functions $f(z)$ such that:
\begin{enumerate}
    \item[(i)] The limit $\textup{lim}_{z \rightarrow 0}f^{(n)}(z)$ when taken in $ S(R,\phi, p)$ exists for every $n \in \bbN$.
    \item[(ii)] The function $f$ satisfies the Gevrey condition: there exist constants $A, B$ depending on $f$, such that
\begin{equation}\label{GevreyvdD}
    \left|\frac{f^{(n)}(z)}{n!}\right| \leq A B^n (n!)^{p}
\end{equation}
for all $n\in \bbN$ and $z\in S(R,\phi,p)$\,.
\end{enumerate}
The first condition ensures that $f(z)$ admits an asymptotic expansion on $S(R, \phi, p)$, and the second ensures that the asymptotic expansion is of Gevrey class $1/p$  \cite{LodayRichaud2014DivergentSA}. Furthermore, the condition $\phi \in (\frac{\pi}{2}, \pi)$ ensures that the Nevanlinna-Sokal theorem holds, and therefore $f(z)$ is equal to the Borel sum of its asymptotic expansion. Although the sector $S(R, \phi, p)$ is larger than a Sokal disk $D^p_R$, the results of \cite{balser1994divergent} 
show that the above conditions are not more restrictive: the analyticity domain where the Gevrey condition holds can be extended beyond the Sokal disk by rotating the line of integration of the Laplace transform by an angle $\varepsilon$, which is always possible for $\varepsilon$ small enough provided that Borel summability holds.\\

Since the limit for $z\rightarrow 0$ must exist, we are enabled to extend every such function $f$ to $S \cup \{0\}$ by setting $f(0) = \textup{lim}_{z \rightarrow 0}f(z)$. The Gevrey functions of one variable are then defined by restricting these functions to the real closed interval $[0,R]$, including the limit point $0$. 
This means that these one-variable Gevrey functions are Borel sums, non-analytic in the weak-coupling limit $z=0$. \\

\newpage 
Note that whenever $p>1$, it is possible by a simple change of variable $z \rightarrow z^p$ to turn the Gevrey class of a formal power series from  $1/p$ to  $1$ while preserving Borel summability \cite{LodayRichaud2014DivergentSA}. Multi-variable Gevrey functions are defined in a similar spirit, but require significantly more technicalities, for which we refer to \cite{DRIES_SPEISSEGGER_2000}. The class of all Gevrey functions then generates the structure $\bbR_{\rm G}$. 
For our purposes the one-variable setting suffices, and the essential conclusion is as follows.

\begin{subbox}{Tameness of Borel sums}
The Borel sums of Borel-summable asymptotic series, defined on a suitable closed interval on the real line which includes the origin, are tame functions definable in the o-minimal structure $\bbR_{\rm G}$. 
\end{subbox}

This fact provides a strategy for proving the tameness of non-perturbative partition functions and correlation functions, as a function of a coupling $\lambda$ including its weak coupling limit $\lambda\to0$. 
The first step is to perform the complete perturbative expansion, obtaining an asymptotic series $\widetilde \varphi$. Next, we compute the Borel transform $\widehat\varphi = \mathcal{B}_p\tilde\varphi$ and then use the fact that if $\widehat\varphi(\zeta)$ has no poles on the positive real line $\bbR^+$ and exponential size less than $1/p$, then the series $\widetilde\varphi$ is $p$-summable along $\bbR^+$. Consequently, the resulting Borel sum $\varphi(\lambda)$ is definable in the structure $\bbR_{\rm G}$, and hence a tame function. Note that the structure of restricted analytic functions $\bbR_{\rm an}$ is contained in $\bbR_{\rm G}$, so that the definability can be extended to compact intervals on which the partition function is analytic. Specifically, given an $\bbR_{\rm G}$-definable function on $[0,R]$, its analytic continuation to a larger interval $[0,R']$ remains tame as long as $R>0$ and $R'<\infty$.

\subsection{Zero-dimensional QFTs and Borel resummation}\label{sec:0dQFTborel}
In this subsection we discuss partition functions of various interacting quantum field theories on a point spacetime and prove that they are tame functions of $\bbR_{\rm G}$. We begin by revisiting the $\phi^4$-theory, and afterwards we continue to discuss real and complex scalar theories with a higher-order monomial interaction term. We then proceed with an analysis of complex matrix theories. Finally, we prove the tameness of partition functions for a scalar field theory with a general polynomial interaction. Throughout this discussion, we use the symbol $\widetilde\varphi$ for asymptotic series, $\widehat\varphi$ for the corresponding Borel transforms, and $\varphi$ for the Borel-resummed functions. 

\subsubsection*{$\phi^4$-theory on a point revisited}\label{phi4}
To begin with, we return to our initial example defined by the Lagrangian of equation \eqref{eq:Lphi4g} and provide a new perspective. Let us proceed perturbatively by expanding the quartic interaction, which yields the asymptotic expansion
\begin{align} 
\label{eq:Z-expansion}
    Z(g) &= \int_{-\infty}^{\infty}\dd\phi \,e^{-g\phi^2}\sum_{n=0}^{\infty}\frac{(-1)^n}{n!}\left(\frac{g\phi^4}{8}\right)^n \\
    & \sim_1\, \sum_{n=0}^{\infty}\frac{(-1)^n}{n!}\left(\frac{g}{8}\right)^n \int_{-\infty}^{\infty}\dd\phi\, e^{-g\phi^2}\phi^{4n} \,.\nonumber
\end{align}
It is of great importance to highlight that in passing from the first to the second infinite sum we have changed the order of the integral with the sum, and hence we can no longer claim that these two expansions are equal. Instead, we can only claim that it is its asymptotic expansion and therefore use the symbol $\sim$. The Gaussian integral can be performed explicitly in terms of a Gamma function, yielding the formal power series
    \begin{equation}\label{eq:Z-perturb2}
   Z(g) \sim_1 \sum_{n=0}^{\infty}\frac{(-1)^n}{n!}\frac{1}{8^ng^{n+1/2}}\Gamma\left(2n +\tfrac{1}{2}\right)=\sqrt{\pi g}\widetilde{\varphi}(g) \, .
\end{equation}
Here we have extracted a factor $\sqrt{\pi g}$ for convenience, thus defining the asymptotic series $\widetilde\varphi  = \sum_{n=0}^{\infty}a_ng^{-n-1}$. Recalling that  $\Gamma(n/2) = \sqrt{\pi}2^{-(n-1)/2}(n-2)!!$ for any odd $n$, one finds that the coefficients $a_n$ are given by
\begin{equation}
    a_n = \frac{(-1)^n}{n!}\frac{(4n-1)!!}{32^n}\,.
\end{equation}
These coefficients diverge factorially and therefore the Gevrey class of $\widetilde\varphi$ is $1$. Note that $(4n-1)!!$ is the number of Wick contractions among $4n$ fields under Gaussian integration, corresponding to the number of Feynman graphs; it can now be seen explicitly that they their number indeed grows faster than $n!$ as claimed before. The Borel transform of $\widetilde{\varphi}$ is\footnote{We are now working in the large $g$ limit rather than the small $\lambda$ limit, but this does not change the previous discussion about Borel summability.}
\begin{equation}
        \widehat{\varphi}(\zeta) = \sum_{n = 0}^{\infty}\frac{(-1)^n}{32^n}\frac{(4n-1)!!}{(n!)^2}\zeta^n = \frac{2}{\pi}\frac{1}{(1+\zeta/2)^{1/4}} K\Big(\tfrac{1}{2} -\tfrac{1}{2\sqrt{1+\zeta/2}}\Big)\ ,
\end{equation}
where $K(z)$ is the complete elliptic integral of the first kind. \\

The only pole of $\widehat\varphi(\zeta)$ is $\zeta = -2$, and the reason that it lies on the negative real axis is that the coefficients of the asymptotic series $\widetilde\varphi$ has alternating signs. Furthermore, it can be checked that $\widehat{\varphi}(\zeta)$ decays in the large $\zeta$ limit. Since $\widehat\varphi(\zeta)$ has no poles on $\bbR^+$, we conclude that $\widetilde\varphi$ is $1$-summable, and the asymptotic expansion is of order $1$, i.e.~$Z(g) \sim_1 \sqrt{\pi g}\widetilde\varphi(g)$ holds uniformly on every closed subsector of $\bbC \backslash \bbR^-$. The Borel sum 
\begin{equation}
    \varphi(g) = \int_0^{\infty}\dd\zeta e^{-g\zeta} \,\widehat\varphi(\zeta)
\end{equation}
is then definable in the structure $\bbR_{\rm G}$, from which we conclude that $Z(g) = \sqrt{\pi g}\varphi(g)$ is also definable in $\bbR_{\rm G}$. The resummed function can be calculated explicitly and yields $\sqrt{\pi g}\varphi(g) = \sqrt{2}e^g K_{1/4}(g)$, recovering the same expression as found by direct integration. This shows that the partition function of this theory is tame even when the non-analytic weak coupling limit $\lambda\to0$ is included in the domain.

\subsubsection*{$\phi^{2p}$-theory on a point}\label{sec:phi-2p}
The results obtained above can be extended to a more general potential. Let us consider the Lagrangian
\begin{equation}\label{2paction-real}
    \cL(\phi) = \frac{m^2}{2}\phi^2 + \lambda \phi^{2p}
\end{equation}
where $p>1$ is an integer. Upon setting $y^2 = \frac{m^2}{2}\phi^2$  and redefining the coupling by the rescaling $ \lambda \left(\frac{2}{m^2}\right)^{p}\rightarrow \lambda$, the partition function depends only on $\lambda$ and becomes
\begin{equation}
    Z(\lambda) = 
    \frac{\sqrt{2}}{m} \int_{-\infty}^{\infty} \dd y \, e^{-y^2-\lambda y^{2p}}\, .
\end{equation}
Again proceeding perturbatively, in analogy to the previous case, we find the asymptotic expansion
\begin{equation} \label{eq:def-tildephi}
        Z(\lambda) \sim_{p-1} \frac{\sqrt{2}}{m}\sum_{n=0}^{\infty}\frac{(-1)^n}{n!}\lambda^n\Gamma\left( np + \frac{1}{2}\right) =\frac{\sqrt{2}}{\lambda m}\widetilde\varphi(\lambda)
\end{equation}
where $\widetilde\varphi(\lambda)$ is now a formal power series of Gevrey class $1/(p-1)$. \\

To show that $Z(\lambda)$ is definable in $\bbR_{\rm G}$, we compute the Borel transform of \eqref{eq:def-tildephi}. As detailed in appendix \ref{app:Borel-explicit}, the Borel transform of $\widetilde\varphi(\lambda)$, $\mathcal{B}_{p-1}\widetilde\varphi = \widehat\varphi(\zeta)$ can be computed explicitly in terms of the generalized hypergeometric function ${}_a F_b \big[{\vec{a} \atop \vec{b}} \big| z\big ]$. Explicitly, we find
\begin{equation}\label{eq:hyper-sol}
    \widehat{\varphi}(\zeta) = \sqrt{\pi}\,_aF_b\Big[ {\vec{a} \atop \vec{b}} \Big | -\frac{p^p}{(p-1)^{p-1}}\zeta \Big] \,,
\end{equation}
where $a = |\vec{a}|$, $b = |\vec{b}|$ with $a=p$, $b = p-1$ and
\begin{equation}\label{eq:abvectors}
    \vec{a} = \frac{1}{2p}\left(1, 3,\ldots, 2p-1\right)  \,, \qquad 
    \vec{b} = \frac{1}{p-1}\left(1, 2,\ldots,  p-1\right) \,.
\end{equation}
The poles of the hypergeometric function ${}_a F_b \big[{\vec{a} \atop \vec{b}} \big| z\big]$ lie at $z =1$ whenever $a = b+1$ \cite{Douglas:2022ynw}. Therefore the unique pole in the Borel plane of $\widehat\varphi(\zeta)$ lies at $\zeta = -(p-1)^{p-1}/p^p$, and hence it is analytic in a horizontal strip $S_\delta$. Additionally, it can be argued using the differential equation satisfied by the generalized hypergeometric functions that it has no exponential growth on this strip \cite{Smith39}. 
The asymptotic expansion $\widetilde\varphi$ is then $(p-1)$-summable and its Borel sum $\varphi(\lambda) = \mathcal{L}_{p-1}\widehat\varphi$ is definable in $\bbR_{\rm G}$. We finally conclude that the partition function for the Lagrangian of equation \eqref{2paction-real}, $Z(\lambda) = \frac{\sqrt{2}}{m\lambda}\varphi(\lambda)$, is definable in $\bbR_{\rm G}$ for any integer $p$.\\

The correlation functions in this theory can be shown to be Borel-summable in a similar manner. For $\nu\in\bbN$, the $2\nu$-point correlation function is given by
\begin{equation}
    I_{2\nu}(\lambda) = \int_{-\infty}^{\infty}\dd\phi \, \phi^{2\nu}e^{-\phi^2-\lambda\phi^{2p}}.
\end{equation}
As before, expanding the exponential yields a formal power series of Gevrey class $1/(p-1)$ which takes the form
\begin{equation}
    \frac{1}{\lambda}\widetilde{\varphi}_\nu=\sum_{n=0}^{\infty}\frac{(-\lambda)^n}{n!}\Gamma\left(np+\nu+\tfrac{1}{2}\right).
\end{equation}
The Borel transform is then computed to be
\begin{equation}\label{eq:hyper-sol-j}
    \widehat{\varphi}_\nu(\zeta) = \mathcal{B}_{p-1}\widetilde\varphi_\nu = \Gamma\left(\tfrac{1}{2}+\nu\right) \;_aF_b\Big[ {\vec{a} \atop \vec{b}} \Big | -\frac{p^p}{(p-1)^{p-1}}\zeta \Big]\,,
\end{equation}
where again $a = |\vec{a}|$, $b = |\vec{b}|$ with $a=p$, $b = p-1$ and
\begin{equation}\label{eq:abvectors2}
    \vec{a} = \frac{1}{2p}\left(1+2\nu, 3+2\nu,\ldots, 2p-1+2\nu\right), \qquad
    \vec{b} = \frac{1}{p-1}\left(1, 2,\ldots,  p-1\right). 
\end{equation}
The remainder of the proof is similar to the previous case and is discussed in appendix \ref{app:Borel-explicit}. The location of the pole in the Borel plane is independent of $\nu$, and Borel summability follows from the properties of the hypergeometric function. Thus, the function $\varphi_\nu(\lambda) = \mathcal{L}_{p-1}\widehat\varphi_j$ is definable in $\bbR_{\rm G}$ and so is the correlation function moment $I_{2\nu}(\lambda) =\tfrac{1}{\lambda }\varphi_\nu$, as a function of $\lambda$ on a suitable domain.

\subsubsection*{Complex $(\phi\bar\phi)^{p}$-theory on a point }
Let us now consider again a monomial potential, but for a complex scalar field $\phi$. The zero-dimensional partition function for such a field is 
\begin{equation}
    Z(\lambda) = \int \frac{\dd\phi \, \dd\bar{\phi}}{2\pi i}e^{-\phi\bar\phi-\lambda(\phi\bar\phi)^p} \,,
\end{equation}
where we have already removed the dependence on a mass parameter by rescaling the field. In polar coordinates, this partition function becomes
\begin{equation}
   Z(\lambda) = \int_0^{\infty}\dd r \,e^{-r-\lambda r^{p}}.
\end{equation}
We can now perform the asymptotic expansion in $\lambda$ which yields
\begin{equation}
    \frac{1}{\lambda}\widetilde\varphi(\lambda) = \sum_{n=0}^{\infty} \frac{(-1)^n\lambda^n}{n!}\int_{0}^{\infty}\dd r\, e^{-r}r^{pn} = \frac{1}{\lambda}\sum_{n=0}^{\infty} (-1)^n\lambda^{n+1} \frac{(pn)!}{n!}\, ,
\end{equation}
which is of Gevrey class $1/(p-1)$ and thus $Z(\lambda) \sim_{p-1} \frac{1}{\lambda}\widetilde\varphi(\lambda)$. The Borel transform of the above sum is then
\begin{equation}\label{eq:Borel-trans- cplx}
    \widehat{\varphi}(\zeta) = \sum_{n=0}^{\infty}(-1)^n\zeta^n\frac{pn!}{n!((p-1)n)!} = \sum_{n=0}^{\infty}(-1)^n\zeta^n\binom{pn}{n}\, .
\end{equation}
This particular power series was also studied in \cite{Rivasseau_2017}, where it was observed that the binomial coefficient $\binom{pn}{n}$ is closely related to the $n$th Fuss-Catalan number
\begin{equation}
    C^{(p)}_n = \frac{1}{pn+1}\binom{pn+1}{n} = \frac{1}{(p-1)n+1}\binom{pn}{n}\, .
\end{equation}
These Fuss-Catalan numbers arise as coefficients in the generating function 
\begin{equation}
  T_p(z) = \sum_{n=0}^{\infty}C^{(p)}_nz^n\,,  
\end{equation}
which satisfies the algebraic relation $zT_p^p(z)+1 = T_p(z)$. From this we derive
\begin{equation}
    \widehat\varphi(\zeta) = \frac{1}{1+p\zeta T_p^{p-1}(-\zeta)}\,.
\end{equation}
Furthermore, in \cite{Rivasseau_2017} it was noted that the radius of convergence of the Borel transform $\widehat\varphi(\zeta)$ is the same of that of the Fuss-Catalan generating function $T_p(z)$, which is $R_p = (p-1)^{(p-1)}/p^p$. As it might have been expected, this is the same radius of convergence of the Borel transform as the case of the real field dealt with earlier in this section. As shown in \cite{Rivasseau_2017}, the function $\widehat\varphi(\zeta)$ is holomorphic on $\bbC\setminus (-\infty,-R_p]$ and its growth for $\Re(\zeta) \rightarrow \infty$ is polynomially bounded. We thus conclude that $Z(\lambda)$ is $(p-1)$-summable and hence equal to the Borel sum of its asymptotic expansion. As in the previous cases, Borel summability is owed essentially to the alternate signs in the Borel transform \eqref{eq:Borel-trans- cplx}, which cause the pole to lie on the negative real axis. We can then conclude that $Z(\lambda)$ is a tame function on a closed interval containing $\lambda=0$, for all positive integers $p$.

\subsubsection*{Complex matrix theory on a point}
The previous example can be generalized to a complex matrix theory by relying on results from constructive field theory \cite{Rivasseau_2007,Rivasseau_2009,Rivasseau_2017}. In essence, this theory provides an alternative to perturbative field theory, avoiding the issue of divergence. While perturbative field theory produces a divergent power series in powers of the coupling $\lambda$, whose coefficients are computed by sums over Feynman graphs, constructive field theory instead reorganizes the combinatorial expansion in terms of forests and spanning trees, which encode the same information but without loop diagrams \cite{abdesselam1995trees,Rivasseau_2013}. \\

We consider an $N\times N$ complex matrix theory with a partition function of the form
\begin{equation}
    Z(\lambda) = \int \dd M \dd M^{\dagger} \exp\Big(-\textup{Tr}(MM^{\dagger})-\frac{\lambda}{N^{p-1}}\textup{Tr}(MM^{\dagger})^p\Big) \,,
\end{equation}
and as in the previous examples we focus on summability in the coupling $\lambda$. This model was considered in the context of constructive field theory in \cite{Rivasseau_2009} for a quartic interaction and in \cite{Lionni:2016ush,Krajewski:2017thd, Krajewski:2019tsi} for higher-order interactions with $p>2$. Specifically, it was shown in \cite{Lionni:2016ush} that the partition function is $(p-1)$-summable for each $N$. The precise statement is that it is analytic and satisfies Gevrey bounds on a Sokal disk
\begin{equation}
    D^{p-1}_{\rho(p,N)} = \big\{\lambda \in \bbC \big| \Re \lambda^{-1/(p-1)} >  \rho(p,N) \big\}\,,
\end{equation}
where the dependence on $N$ is given by $\rho(p,N) = \rho_pN^{1+2/(p-1)}$. This disk shrinks as the matrix size $N$ increases, but it was proved in  \cite{Krajewski:2017thd, Krajewski:2019tsi} that the analyticity domain can be extended to a domain $P(\varepsilon,\eta) = \big\{ \lambda \in \bbC \,\big|\, 0 < |\lambda| < \eta \;, |\arg{\lambda}| < \pi -\varepsilon\big\}$ which is uniform in $N$. 
We conclude that $Z(\lambda)$, once restricted to an interval including the origin, is definable in $\bbR_{\rm G}$ for every $N$ and $p$. \\

\newpage

This result can be extended to correlation functions in the theory by considering a partition function with sources $Z(\lambda,J,J^\dagger)$ obtained by adding the terms
\begin{equation}
      \sqrt{N} \textup{Tr}(JM^{\dagger}) + \sqrt{N} \textup{Tr}( MJ^{\dagger})
\end{equation}
to the Lagrangian. The matrix-valued $2\nu$-point correlation function is computed as
\begin{equation}
    I_{2\nu}(\lambda)= \frac{\pd^{2\nu}}{\pd J^\nu \pd (J^\dagger)^\nu } \log Z(\lambda, J, J^{\dagger}) \big|_{J = J^{\dagger} = 0} \,.
\end{equation}
The constructive field theory analysis of \cite{Rivasseau:2023vba} then shows that these functions are analytic on the cardioid-shaped Sokal disk $D^{p-1}_{c(p)}$ with $c= (2(p-1))^{p-1}$, uniformly in $N$. Moreover, it is shown that the $2k$-point functions admit an asymptotic expansion of Gevrey class $1/(p-1)$, so it follows that they are Borel-summable. As a result, we conclude that the non-perturbative $2k$-point functions are also definable in $\bbR_{\rm G}$ as functions of the coupling $\lambda$ on a suitable domain.

\subsubsection*{Scalar with polynomial potential on a point}
Up until this point we have analyzed zero-dimensional quantum field theories with a single monomial interaction term. We now continue with a scalar theory on a point, but with a general polynomial potential. In this case, the explicit computation of the Borel transform is not possible, and we must rely on the Nevanlinna-Sokal theorem to prove Borel summability. We consider the partition function
\begin{equation}\label{eq:Z-def-poly}
    Z(\lambda) = \int_{-\infty}^{\infty}\dd\phi \,e^{-\phi^2 -\lambda V(\phi)}\, ,
\end{equation}
where we assume that $V(\phi)$ is a polynomial of degree $2p$, given by 
\begin{equation}
    V(\phi) = a_{2p}\phi^{2p}+ a_{2p-1}\phi^{2p-1}+\ldots+a_3\phi^3\ .
\end{equation}
We further assume that the leading coefficient satisfies $a_{2p} > 0$ so that the integral is well-defined for a positive coupling $\lambda$. Moreover, under our assumptions, $V(0)=0$. The main result of this section is that this partition function is non-perturbatively tame, on a domain including the weak coupling limit $\lambda\to0$. We do this by showing that $Z(\lambda)$ is $(p-1)$-summable using the Nevanlinna-Sokal theorem. This requires two steps: 
\begin{enumerate}
    \item[(i)] Finding an analyticity domain for $Z(\lambda)$ which contains a Sokal disk $D^{p-1}_0$.
    \item[(ii)] Proving that $Z(\lambda)$ is asymptotic to a $1/(p-1)$-Gevrey formal power series on that domain. 
\end{enumerate}
By following these steps, we establish the summability and hence the tameness of the partition function, and a similar argument shows the tameness of correlation functions. The technicalities of the proof are deferred to appendix~\ref{app:analyticityproof}, and we summarize the result as follows. 
\begin{subbox}{Tameness of correlation functions in 0d theory with polynomial potential}
The non-perturbative $\nu$-point correlation functions
\begin{equation}
    I_\nu(\lambda) =  \int_{-\infty}^{\infty} \dd \phi \,\phi^\nu e^{-\phi^2-\lambda V(\phi)} 
\end{equation}
in a zero-dimensional scalar theory with polynomial potential $V(\phi)$ are tame as a function of $\lambda$ on a closed real interval containing the origin $\lambda=0$.
\end{subbox}

\subsection{On generalizations to higher-dimensional QFTs}
In the previous section we have used zero-dimensional quantum field theories as a laboratory to gain insight into how the tameness of partition functions and correlation functions persists in the non-analytic weak coupling limits. The power of this setting comes from the availability of non-perturbative results and detailed exact computations. The aim of this subsection is to discuss extensions of these observations to quantum field theories formulated in non-zero dimension, and comment on challenges that arise.

\subsubsection*{$\phi^4$-theory in four dimensions}
Let us proceed by again considering $\phi^4$-theory, but now in four Euclidean spacetime dimensions, in which the theory is renormalizable. Aside from computational difficulties in evaluating path integrals, the main challenge arises from divergences in the UV and the IR which require regularization. If we were to formulate the theory on a lattice as a regularization method, we would in fact be studying a matrix model in zero dimensions which has already been discussed earlier in this  section. Another way to implement UV and IR cutoffs, adopted in the constructive field theory literature \cite{Magnen_2008}, is as follows.  We restrict the interaction to a bounded spacetime region $U\subseteq \bbR^4$, and consider the Schwinger proper time expression for the free propagator
\begin{equation}
    \Delta(p) = \int_0^\infty \dd\tau \, e^{-\tau(p^2+m^2)}\,.
\end{equation}
We then express this propagator in terms of renormalization group slices indexed by integers $j$,
\begin{equation}
    \Delta_j(p) = \int_{M^{-2j}}^{M^{-2j+2}} \dd\tau \, e^{-\tau(p^2+m^2)}\,,
\end{equation}
where $M$ parametrizes the width of the slices, and the full propagator is recovered via $\Delta=\sum_j\Delta_j$. A regularized expression for the partition function may then be written as 
\begin{equation}
    Z(\lambda;U) = \int D\phi \, \exp( -\frac{1}{2}\int \dd^4 x\, \phi \Delta^{-1}_j\phi -\lambda \int_U \dd^4 x \,\phi^4)\,.
\end{equation}
Note that only the interaction is localized on the region $U$, since the quadratic term already has an IR cutoff implemented. It was proved in \cite{Magnen_2008} that the connected correlation functions
\begin{equation}
    I^{\rm c}_{2\nu}(x_1,\ldots,x_{2\nu}; \lambda) = \lim_{U \rightarrow \mathbb{R}^4}  \frac{1}{Z(\lambda; U) }\int \! D\phi\,\phi(x_1)\cdots\phi(x_{2\nu}) e^{-\tfrac{1}{2}\int \dd^4 x \phi \Delta^{\!-\!1}_j\phi-\lambda\int_U\!\dd^4x\,\phi^4}
\end{equation}
are Borel-summable, uniformly in the slice index $j$. From the techniques of this section it then follows that $I^{\rm c}_{2\nu}(x_1,\ldots,x_{2\nu}; \lambda)$ is a well-defined regularized physical quantity which is definable in the o-minimal structure $\bbR_{\rm G}$, despite being non-analytic in the $\lambda\to0$ limit. 

\subsubsection*{Vector model in two dimensions}
Going beyond theories of a single scalar field, we now consider a theory with $N$ scalars coupled by a general quartic interaction. The Borel summability of such a model was analyzed in \cite{Erbin_2021} for $d=2$, and we will build on these results to argue for the tameness of the partition function. The partition function for the model in question is given by
\begin{equation}
    Z(\lambda) = \int D\phi_1\cdots D\phi_n \, \exp\Bigg( -\int \dd^2 x \bigg(\frac{1}{2}\sum_{i=1}^{N} \phi_i \Delta^{-1}\phi_i + \frac{\lambda}{4!} \sum_{i,j,k,l}^{N}\mathcal{W}_{ijkl}\phi_i\phi_j\phi_k\phi_l\bigg)\Bigg)\,,
\end{equation}
where $\mathcal{W}_{ijkl}$ is a completely symmetric tensor, and $\Delta^{-1}$ is the inverse propagator, which in this case is regularized by writing it as
\begin{equation}
    \Delta(p) = \int_{M}^\infty \dd\tau \, e^{-\tau(p^2+m^2)}\,,
\end{equation}
where $M$ imposes a UV cutoff. The renormalization of the theory can be implemented by using the multiscale loop vertex expansion, as detailed in \cite{Gurau:2013oqa}. 

The authors of \cite{Erbin_2021} proved, under the assumption that $\mathcal{W}_{ijkl}$ has only positive eigenvalues, that the free energy $F(\lambda) = \log Z(\lambda)$ is analytic and Borel-summable in a cardioid domain defined by
\begin{equation}
    |\lambda| \leq O(1)\frac{1}{N w_0^2}\cos^2(\theta/2),
\end{equation}
where $\lambda = |\lambda| e^{i\theta}$, $w_0$ is the largest eigenvalue of $\mathcal{W}$, and $O(1)$ indicates the dependence on constants of the specific model. Once more, Borel summability implies that the free energy $F(\lambda)$ is tame and definable in the structure $\bbR_{\rm G}$. By composing with the exponential function, we then conclude that the partition function $Z(\lambda)$ is definable in $\bbR_{\rm G}$ on a domain including $\lambda=0$.

\section{Outlook on non-perturbative tameness}
In this chapter we have shown that various non-perturbative observables in quantum field theory are tame functions of the parameters of the theory. Our analysis was motivated by the idea that physical functions, under suitable assumptions, should be representable with a finite amount of information, as made precise by o-minimality. Our results are mostly formulated in zero-dimensional settings, where a rigorous mathematical analysis of non-perturbative observables can be performed. In specific cases we were able to formulate a differential chain for these observables, enabling us to calculate their complexity in the framework of sharp o-minimality. These chains, being analytic in nature, miss the weak coupling limit at which the observables are generally non-analytic. To capture these limits, we utilized the class of Gevrey functions underlying the o-minimal structure $\bbR_{\rm G}$, and demonstrated that non-perturbative tameness persists provided that the divergent perturbative expansion is Borel-summable. The settings in which this holds include real scalar theories with a general polynomial potential, as well as complex scalar theories and matrix theories with a monomial potential. The generalization to higher-dimensional quantum field theories is challenged by UV and IR divergences, and we discussed examples of regularized theories in which summability persists.
These conclusions reveal the following general expectation.
\begin{subbox}{Non-perturbative tameness}

The tameness of non-perturbative observables in quantum field theory requires definability in o-minimal structures generated by non-analytic functions.     
\end{subbox}

We identify two directions in which the program of analyzing the tameness of exact observables in quantum field theory can be extended further. 
\begin{itemize}
    \item \textbf{Resurgence.} The methodology in this chapter assumes the absence of Stokes phenomena due to singularities in the Borel plane. This structure is present in many observables in quantum field theory; in these cases Borel resummation is not sufficient to recover the observable from its perturbative expansion and a resurgence analysis is needed \cite{Sauzin:2014qzt}. To establish the tameness of these functions one therefore requires another o-minimal structure, and we expect that the classes of o-minimal structures considered in \cite{rolin2007quasi}, generated by quasi-analytic solutions to  differential equations, form a natural candidate.
    \item \textbf{Non-perturbative complexity.} The o-minimal structures generated by quasi-analytic function classes extend the structure of restricted analytic functions, which by the discussion of chapter~\ref{ch:complexity} implies that they cannot admit a format-degree filtration enhancing them to a sharply o-minimal structure. The results of this chapter thus motivate addressing the open problem of establishing a sharply o-minimal structure generated by non-analytic functions. Defining such a structure would greatly extend the scope of the applications of the complexity framework of tame geometry to quantum field theory.
\end{itemize}

With this outlook in mind, we conclude that this chapter takes a rigorous step towards uncovering the tame mathematical structures in quantum field theory. 

\begin{subappendices}

\section{Borel transforms as hypergeometric functions} \label{app:Borel-explicit}
In this appendix we prove the identities \eqref{eq:hyper-sol} and \eqref{eq:hyper-sol-j}, which were used in section~\ref{sec:0dQFTborel}. We begin by recalling the most general definition of the hypergeometric function $_aF_b\big[ {\vec{a} \atop \vec{b}} \big |x\big]$, where $a = |\vec{a}|$, $b = |\vec{b}|$:
\begin{equation}\label{eq:hypergeometric}
  _aF_b\Big[ {\vec{a} \atop \vec{b}} \Big |x\Big] = \dfrac{\prod_{i=1}^{b}\Gamma(b_i)}{\prod_{i=1}^{a}\Gamma(a_i)} \sum_{n=0}^{\infty} \dfrac{\prod_{i=1}^{a}\Gamma(a_i+n)}{\prod_{i=1}^{b}\Gamma(b_i+n)}\frac{x^n}{\Gamma{(n+1)}}\,.
\end{equation}
We then observe that, for any integer $n$, recalling that $x\Gamma(x) = \Gamma(x+1)$,
\begin{align}\label{eq:generalised-recurrence}
  \frac{\Gamma(x+n)}{\Gamma(x)} &=  \frac{\Gamma(x+n)}{\Gamma(x+(n-1))} \frac{\Gamma(x+n-1)}{\Gamma(x+(n-2))} \cdot\cdot\cdot \frac{\Gamma(x+1)}{\Gamma(x)} \\
   \nonumber
   &= (x+n-1)(x+n-2) \cdots x \,.
\end{align}
By comparing (\ref{eq:hyper-sol}) with (\ref{eq:hypergeometric}), we realize that it suffices to prove that
\begin{equation}
  \frac{\Gamma\left( np + \frac{1}{2}\right) }{\Gamma\left(n(p-1)+1\right)} = \sqrt{\pi}\prod_{i=1}^{a}\dfrac{\Gamma(a_i+n)}{\Gamma(a_i)} \left(\prod_{i=1}^{b}\dfrac{\Gamma(b_i+n)}{\Gamma(b_i)}\right)^{-1}\left(\frac{(p-1)^{p-1}}{{p^p}}\right)^n \,.
\end{equation}
Let us focus on the first product. Plugging in the values in (\ref{eq:abvectors}) and resorting to (\ref{eq:generalised-recurrence}) we have
\begin{alignat}{4} 
\prod_{i=1}^{a}\dfrac{\Gamma(a_i+n)}{\Gamma(a_i)} = \quad
&\big(\tfrac{1}{2p} + n-1\big) && \quad\cdots\quad &&\big(\tfrac{1}{2p}
+1\big) &&\big(\tfrac{1}{2p}\big) \nonumber\\
&\big(\tfrac{3}{2p} + n-1\big) && \quad\cdots\quad && \quad && \big(\tfrac{3}{2p}\big) \nonumber\\
&\quad\quad\quad\vdots && && &&\quad\vdots \nonumber \\
&\big(\tfrac{2p-1}{2p} + n-1\big) && \quad\cdots\quad && \quad && \big(\tfrac{2p-1}{2p}\big) \,.
\end{alignat}
The rows have $n$ terms and the columns have $p$ terms. By taking products starting from the top right corner and moving first down the column and then moving to the column to the left, we have
\begin{equation}\label{eq:earlier-Gamma-hyp}
    \prod_{i=1}^{a}\dfrac{\Gamma(a_i+n)}{\Gamma(a_i)} = \frac{(2np-1)!!}{(2p)^{np}} = \frac{1}{\sqrt{\pi}}\Gamma\left(\frac{2np+1}{2}\right)\frac{1}{p^{np}}\,,
\end{equation}
where we recall that $\Gamma(n/2) = \sqrt{\pi}2^{-(n-1)/2}(n-2)!!$ for any odd $n$. Similarly, we can derive that
\begin{alignat}{4} 
\prod_{i=1}^{b}\dfrac{\Gamma(b_i+n)}{\Gamma(b_i)} = \quad
&\big(\tfrac{1}{p-1} + n-1\big) && \quad\cdots\quad &&\big(\tfrac{1}{p-1}
+1\big) &&\big(\tfrac{1}{p-1}\big) \\
&\big(\tfrac{2}{p-1} + n-1\big) && \quad\cdots\quad && \quad && \big(\tfrac{2}{p-1}\big)\nonumber \\
&\quad\quad \quad\vdots && && &&\quad\vdots \nonumber \\
&\big(\tfrac{p-1}{p-1} + n-1\big) && \quad\cdots\quad && \quad && \big(\tfrac{p-1}{p-1}\big) \nonumber
\\[.2cm]
=& \frac{(n(p-1))!}{(p-1)^{n(p-1)}} \nonumber\\
=& \frac{\Gamma(n(p-1)+1)}{(p-1)^{n(p-1)}} \,,\nonumber
\end{alignat}
which together with the earlier (\ref{eq:earlier-Gamma-hyp}) yields the desired identity. The proof of (\ref{eq:hyper-sol-j}) is analogous. The only difference is that, in this case, we have to show that
\begin{equation}
    \prod_{i=1}^{a}\frac{\Gamma(a_i+n)}{\Gamma(a_i)} = \frac{\Gamma\left(n(p-1)+j+\tfrac{1}{2}\right)}{\Gamma\left(\tfrac{1}{2}+j\right)} \,.
\end{equation}
The left-hand side can be expanded as before, yielding the $np$ products
\begin{alignat}{4} 
\prod_{i=1}^{a}\dfrac{\Gamma(a_i+n)}{\Gamma(a_i)} = \quad
&\big(\tfrac{1+2j}{2p} + n-1\big) && \quad...\quad &&\big(\tfrac{1+2j}{2p}
+1\big) &&\big(\tfrac{1+2j}{2p}\big) \nonumber \\
&\big(\tfrac{3+2j}{2p} + n-1\big) && \quad...\quad && \quad && \big(\tfrac{3+2j}{2p}\big) \nonumber\\
&\quad\vdots && && &&\quad\vdots  \nonumber\\
&\big(\tfrac{2p-1+2j}{2p} + n-1\big) && \quad...\quad && \quad && \big(\tfrac{2p-1+2j}{2p}\big) \,.
\end{alignat}
Reading the products from right to left and from top to bottom, we realize that the products at the numerator run from $1+2j$ up to $2(np+j)-1$, where the product runs over the odd integers. We conclude that the above product gives the desired result
\begin{equation}
\prod_{i=1}^{a}\dfrac{\Gamma(a_i+n)}{\Gamma(a_i)} 
= \frac{(2(np+j)-1)!!}{2^{np+j}}\frac{2^j}{(2j-1)!!} = \frac{\Gamma\left(np+j+\tfrac{1}{2}\right)}{\Gamma\left(\tfrac{1}{2}+j\right)}\,.
\end{equation}

\section{Partition function with polynomial potential}\label{app:analyticityproof}
In this appendix we prove the tameness of non-perturbative observables for a scalar theory with a general polynomial potential on a point, whose partition function is given by
\begin{equation}\label{eq:Z-def-poly2}
    Z(\lambda) = \int_{-\infty}^{\infty}\dd\phi \,e^{-\phi^2 -\lambda V(\phi)}\, ,
\end{equation}
following the steps outlined in the end of section~\ref{sec:0dQFTborel}.

\subsubsection*{Step (i): analyticity domain}
Let us commence with part (i) of the proof. The integral representation of equation \eqref{eq:Z-def-poly2} is analytic on $\{\lambda \in \bbC \,|\, \Re\lambda > 0\}$, but this domain can be extended by means of a change of variable. We let $\phi^2 = \lambda^{-\frac{q-1}{qp}}x^2$ for a positive integer $q$. Since we are eventually interested in a real-valued coupling $\lambda$, we implement the change of variable taking $\lambda$ to be real, and subsequently perform the analytic continuation to complex values. Writing $V(\phi) =\phi^{2p}+ Q(\phi) $ for a lower-degree polynomial $Q(\phi)$, and setting $a_{2p} =1$ for simplicity, we then have
\begin{equation}\label{eq:Z-change-of-variable}
     Z(\lambda) = \int_{-\infty}^{\infty}\dd x\,\lambda^{-\frac{q-1}{2qp}}\exp\Big(-\lambda^{-\frac{q-1}{qp}}x^2-\lambda^{\frac{1}{q}}x^{2p}-\lambda Q\big( \lambda^{-\frac{q-1}{2qp}}x\big)\Big)\,.
\end{equation}
The function $Z(\lambda)$ is now analytic on the Sokal disk\footnote{To be more precise, as the $Z(\lambda)$ has a branch cut on $\bbR^-$, Sokal disks with $D^q_0$ with $q>2$ should be viewed as subsets of the Riemann surface of the logarithm (see e.g.~\cite{Sauzin:2014qzt}).}
\begin{equation}
  D_{0}^q = \big\{ \lambda \in \bbC \,\big|\, \Re \lambda^{\frac{1}{q}} > 0\big\}=\left\{\lambda \in \bbC \,\middle|\,|\arg \lambda | < q \frac{\pi}{2} \right\} \,.  
\end{equation}
The analyticity domain can then be extended indefinitely by increasing $q$. However, we do not merely seek an analyticity domain, but an analyticity domain on which the function $Z(\lambda)$ admits an asymptotic expansion. As stated in \cite{LodayRichaud2014DivergentSA}, this is true if and only if the limit for $|\lambda| \rightarrow 0$  of $Z(\lambda)$ and all its derivatives exists along every direction of $D_0^q$. If we change variable again to $y^2 = |\lambda|^{-(q-1)/(qp)}x^2$ we find
\begin{equation}
    Z(\lambda) = \int_{-\infty}^\infty \dd y \,e^{-i  \frac{q-1}{2qp} \theta}\exp\left(-y^2e^{-i  \frac{q-1}{qp} \theta} - |\lambda| \left(e^{-i\frac{\theta}{q}}y^{2p} + Q\big(e^{-i\frac{q-1}{2qp}\theta}y\big)\right)\right)
\end{equation}
where $\lambda = |\lambda| e^{i\theta}$. From this we can infer that the limit for $|\lambda| \rightarrow 0$, with $\lambda \in D^{q}_0$, will only exist if  $q \leq p+1$. The largest domain of analyticity on which $Z(\lambda)$ admits an asymptotic expansion is then
\begin{equation}\label{eq:opening-sector}
    -(p+1)\frac{\pi}{2} < \theta < (p+1)\frac{\pi}{2} \,.
\end{equation}
This agrees with the expectation from our analysis of the monomial potential $V(\phi) = \phi^{2p}$, where the asymptotic expansion of $Z(\lambda)$ is of Gevrey class $1/(p-1)$, and the associated Borel transform $\widehat{\varphi}(\zeta)$ has a unique pole on the negative real line $\bbR^{-}$. Since $\widehat\varphi(\zeta)$ has the correct exponential size, the Laplace transform $\mathcal{L}_{p-1}\hat\varphi$ is analytic on a sector of opening $(p-1)\pi$ centered on $\bbR^+$, and it can be analytically continued to a sector of angle $2\pi + (p-1)\pi $ by rotating the integration line of the Laplace transform.
We thus conclude that $Z(\lambda)$ is analytic on the Sokal disk of infinite radius $D^{p-1}_0=\big\{\lambda \in \bbC\,\big|\, \Re \lambda ^{-\frac{1}{p-1}} > 0 \big\}$. The first assumption of the Nevanlinna-Sokal theorem is then satisfied.

\subsubsection*{Step (ii): asymptotic expansion}
We now turn to the second step of the proof. To apply the Nevanlinna-Sokal theorem we also need to show that $Z(\lambda)$ admits an asymptotic expansion of Gevrey class $1/(p-1)$. To do so, we use a Taylor expansion with integral remainder. Given a smooth function $f(x)$ differentiable infinitely many times at $0$, we can write
\begin{equation}
    f(x) = \sum_{k=0}^n\frac{f^{(k)}(0)}{k!}x^k + \frac{x^{n+1}}{n!}\int_0^1 \dd u (1-u)^n f^{(n+1)}(xu) \,.
\end{equation}
Hence, after setting $ k! c_k = \frac{\dd^k}{\dd\lambda^k}Z(0)$ we have
\begin{equation}\label{eq:Gevrey-remainder}
    \left|Z(\lambda) - \sum_{k=0}^{n}c_k\lambda^k \right| = \left| \frac{\lambda^{n+1}}{n!}\int_0^1\dd u (1-u)^n\int_{-\infty}^{\infty}\dd\phi \,\big(V(\phi)\big)^{n+1} e^{-\phi^2-\lambda u V(\phi)}\right| \,.
\end{equation}
We now have to prove that the right-hand side of this equation can by bounded by $\lambda^{n+1}AB^{n+1}((n+1)!)^{p-1}$ for two constants $A$ and $B$ independent of $\lambda$ and $n$. For the purposes of tameness, it suffices to show this uniformly in $\lambda$ on a finite-radius Sokal disk $D_\rho^{p-1}$ with $\rho>0$. We perform a change of variable 
\begin{equation}
    \phi = \left(\frac{\lambda}{|\lambda|}\right)^{-\frac{p+1}{2p^2}} y = e^{-i\frac{p+1}{2p^2}\theta}y \,,
\end{equation}
where $\lambda = |\lambda| e^{i\theta}$. The second integral in equation \eqref{eq:Gevrey-remainder} can then be expressed as
\begin{align}
I(\lambda)= \int_{-\infty}^{\infty}\!\!\dd y & \,e^{-i \frac{p+1}{2p^2} \theta} \big(V\big(e^{-i\frac{p+1}{2p^2}\theta}y\big)\big)^{n+1} \! \exp\!\left(-e^{-i\frac{p+1}{p^2}\theta}y^2-u|\lambda|\big(e^{-i\frac{\theta}{p}}y^{2p}+ \!\ldots\! \big) \right) \,.
\end{align}
On the Sokal disk $D^{p-1}_\rho$, we have $|\theta| < (p-1)\frac{\pi}{2}$, so the integral is well-defined because $e^{i \theta/p}$ has positive real part for every value of $\theta$. Due to the boundedness of the Sokal disk, the exponential in the integrand can be dominated by a Gaussian, i.e.~there are constants $C>0$ and $c>0$ such that
\begin{equation}
    \exp(\Re(-e^{-i\frac{p+1}{p^2}\theta}y^2-u|\lambda|\big(e^{-i\frac{\theta}{p}}y^{2p}+ \ldots))) \leq C e^{-cy^2}
\end{equation}
for all $\lambda\in D^{p-1}_\rho$. Consequently, we have
\begin{equation}
    |I(\lambda)| \leq C \int_{-\infty}^{\infty} \dd y  \big|V\big(e^{-i\frac{p+1}{2p^2}\theta}y\big)\big|^{n+1} \,e^{-c y^2} \,.
\end{equation}
By estimating the potential by its leading term, we can then derive 
\begin{equation}\label{final-bound-I}
    |I(\lambda)| \leq C(2Ap)^{n+1} \int_{-\infty}^{\infty} \dd y \, e^{-c y^2}y^{2p(n+1)} = \frac{C}{\sqrt{c}}\left(\frac{2Ap}{c^p}\right)^{n+1}\Gamma\left(p(n+1)+\tfrac{1}{2}\right) \,,
\end{equation}
 where $A$ is the maximal absolute value of the coefficients of the potential. The integral over $u$ in \eqref{eq:Gevrey-remainder} is now trivial and we finally reach a uniform asymptotic expansion is of Gevrey class $1/(p-1)$, given by
\begin{equation}
     \left|Z(\lambda) - \sum_{k=0}^{n}c_k\lambda^k \right| \leq \frac{\lambda^{n+1}}{(n+1)!} \frac{C}{\sqrt{c}}\left(\frac{2Ap}{c^p}\right)^{n+1}\Gamma\left(p(n+1)+\tfrac{1}{2}\right)
\end{equation}
for every $\lambda \in D^{p-1}_\rho$. Thus, the Nevanlinna-Sokal theorem is satisfied and $Z(\lambda)$ is Borel-summable, and more precisely $(p-1)$-summable. 
We conclude that $Z(\lambda)$, when restricted on a finite interval $\lambda \in [0, R] \subset \bbR$,  is a tame function, definable in the o-minimal structure $\bbR_{\rm G}$.\\

By means of a small modification in the proof, the correlation function
\begin{equation}
    I_\nu(\lambda) = \int_{-\infty}^{\infty} \dd \phi \, \phi^\nu e^{-\phi^2-\lambda V(\phi)}
\end{equation}
with $V(\phi)$ a polynomial as above, can be proved to be Borel-summable following the same steps as before. The only difference is that, instead of (\ref{final-bound-I}), one finds
\begin{align}
    |I| &\leq C(2Ap)^{n+1} \int_{-\infty}^{\infty} \dd y\, e^{-c\varepsilon \;y^2}y^{2p(n+1) +j} \\
    & = \frac{C}{\sqrt{(c\varepsilon)^{j+1}}}\left(\frac{2Ap}{(c\varepsilon)^p}\right)^{n+1}\Gamma\left(p(n+1)+\tfrac{j+1}{2}\right) \,, \nonumber
\end{align}
from which uniform Gevrey bounds on a finite-radius Sokal disk $D^{p-1}_\rho$ follow. Thus, $I_\nu(\lambda)$ is also $(p-1)$-summable and a tame function in $\bbR_{\rm G}$ when restricted to a suitable real closed interval.

\end{subappendices}

\chapter{Complexity of cosmological correlators}
\label{ch:CCC}
\setlength{\parindent}{0pt}

The following chapter is devoted to an analysis of the tameness and complexity of \textit{cosmological correlators}, which are the natural observables of cosmological quantum field theories. We begin with a brief introduction to the setting, reviewing aspects of cosmology and quantum field theory in expanding spacetimes. We then analyze tree-level wavefunction coefficients in a class of cosmological models and show that they are tame functions of the kinematic variables.  Specifically, we build on recent advances on differential equation techniques for cosmological correlators, captured by a procedure called \textit{kinematic flow} \cite{Arkani-Hamed:2023kig,Arkani-Hamed:2023bsv}, to show that these functions can be captured by a Pfaffian chain. This Pfaffian representation enables us  to calculate their complexity explicitly. We discuss a link between complexity and emergent time in kinematic flow, examine the physical interpretation of complexity in this context, and comment on how the complexity can be reduced.

\section{Introduction to cosmological correlators}
\subsection{Cosmology, inflation, and fluctuations}
Cosmology is concerned with understanding the universe at the largest scales. On these scales, the universe appears to be remarkably homogeneous and isotropic. Observations of the large-scale structure of the cosmos, such as surveys of the distribution of galaxies and measurements of the cosmic microwave background, allow us to probe the small deviations from homogeneity and isotropy. Remarkably,  observations suggest that these large-scale deviations originate from primordial quantum fluctuations in the earliest phases of the universe, which are amplified to cosmic scales during inflation, a phase of near-exponential expansion. This hypothesis provides a link between correlation functions in the early universe and observable cosmological data.
For this reason, the calculation of correlation functions of quantum field theory in an expanding universe is central to cosmology \cite{Lee:2024sks,baumann_lectures_}.

\subsection{Quantum field theory in expanding spacetimes}
The type of spacetime which we will be concerned with in this chapter is the four-dimensional Friedmann-Lema\^itre-Robertson-Walker (FLRW) metric, given by
\begin{equation}\label{eq:FLRW}
    g_{\mu\nu}\dd x^\mu \dd x^\nu = -\dd t^2 + a^2(t)g_{ij} \dd x^i \dd x^j \,.
\end{equation}
Here $g_{ij}$ is the metric on the three-dimensional spatial slices, which is maximally symmetric as a consequence of homogeneity and isotropy. The time-dependence of the metric is completely encoded in the function $a(t)$, called the scale factor. The dynamics of the scale factor are governed by the Einstein equation. It is customary to express the scale factor in terms of a \textit{conformal time} variable $\eta$, which is defined through $\dd \eta =\dd t/a(t)$. In simple cosmological models the scale factor takes the form of a power-law 
\begin{equation}\label{eq:powerlaw}
a(\eta) = \left(\frac{\eta}{\eta_0}\right)^{-(1+\varepsilon)}\,,
\end{equation}
where $\eta$ is determined from the dominant energy source in the universe. Particularly, $\varepsilon=0$ corresponds to de Sitter space, and $\varepsilon\approx 0$ during inflation. \\

Whereas conventional observables in quantum field theory are defined by means of transition amplitudes between asymptotic states in the infinite past and infinite future, the observables in cosmology take a fundamentally different form. Instead, these cosmological correlators are defined as statistical distributions of field operators generated by the time-evolution of an initial state on a specified spatial slice. There are several approaches to calculate these observables, and we will employ the formalism of the \textit{wavefunction of the universe} 
\cite{Maldacena:2002vr,Maldacena:2011nz,Benincasa:2018ssx,Benincasa:2022omn,Albayrak:2023hie,Creminelli:2024cge,Lee:2024sks,Stefanyszyn:2024msm}.
For a given state $\ket{\Psi}$, this wavefunction is a functional $\Psi[\phi]=\bra{\phi}\ket{\Psi}$ which quantifies the overlap with a spatial field configuration $\ket{\phi}$. This wavefunction admits an expansion of the form  
\begin{equation}
    \Psi[\phi] = \exp\Bigg(\sum_{n\geq 2} \int \dd^3 \vec k_1\cdots \dd^3 \vec k_n \,\psi_n(\vec k_1,\ldots,\vec k_n) \,\phi_{k_1}\cdots\phi_{k_n}  \Bigg) \,.
\end{equation}
Here the $\phi_{k_i}$ are the Fourier modes of the field configuration $\ket{\phi}$, and the wavefunction $\Psi[\phi]$ is hereby encoded in the \textit{wavefunction coefficients} $\psi_n(\vec k_1,\ldots,\vec k_n)$. These wavefunction coefficients, which are directly related to the cosmological correlators of interest, may be calculated perturbatively by using special Feynman rules that account for the expanding spacetime \cite{arkani-hamed_cosmological_2017}.\\

\subsection{Conformally coupled scalar in power-law cosmology}
The cosmological model whose observables we will analyze is a scalar field with a general polynomial potential which is conformally coupled to gravity, in an FLRW spacetime with a power-law scale factor. This theory is thus described by the action
\begin{equation}\label{eq:cosmoaction}
    S[\phi] = \int \dd^4 x \,\sqrt{-g} \bigg(-\frac{1}{2}\pd_\mu \phi \,\pd^\mu \phi - \frac{1}{12}\cR\phi^2 -\sum_{k=3}^D \frac{\lambda_k}{k!}\phi^k\bigg) \,,
\end{equation}
where the parameters $\lambda_k$ are the couplings of the various interactions, and $D$ is the power of the highest-order interaction. Here $\cR$ is the Ricci scalar, and the metric is given by $\dd s^2 = a(\eta)(-\dd \eta^2 +  \dd \vec x^2 ) $, with the scale factor $a(\eta)$ given by equation~\eqref{eq:powerlaw}. This model has been studied extensively \cite{Arkani-Hamed:2015bza,Arkani-Hamed:2018bjr,Benincasa:2019vqr,Kuhne:2022wze,Arkani-Hamed:2023bsv,Arkani-Hamed:2023kig,De:2023xue,
Benincasa:2024leu,Benincasa:2024lxe,Glew:2025otn}, and it has the advantage of being fairly general while still having a tractable mathematical structure in its observables. By implementing a conformal rescaling, the evaluation of a wavefunction coefficient can be recast in terms of a time integral over a flat space wavefunction coefficient with time-dependent couplings.\footnote{In what follows, we will also use the term `wavefunction coefficient' to indicate the contribution coming from a single Feynman graph, as well as to refer to the cosmological correlators.}\\

In this chapter, we focus on tree-level Feynman graphs. 
Although the associated functions can depend on all external momenta $\vec k_1,\ldots,\vec k_n$, their dependence is actually restricted to specific combinations of these momenta. In the following, we refer to these combinations as the \textit{kinematic variables}, and they are defined as follows. Every vertex with label $v$ in the diagram comes with a vertex energy $X_v = \sum_{i} |\vec k_i|$,
where the sum runs over all external propagators attached to the vertex. Meanwhile, every internal propagator is associated to an internal energy variable $Y$ given by the energy flowing over that edge, which by momentum conservation can be written in terms of the external momenta $\vec k_i$. For example, consider the tree-level single-exchange graph
\begin{center}
\includegraphics[width=0.30\linewidth]{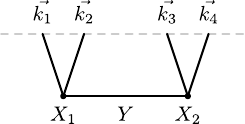} 
\end{center}
Here the dashed gray line indicates the spatial slice of  the state $\ket{\Psi}$ on which the external propagators are anchored, and the kinematic variables are given by $X_1=|\vec k_1|+|\vec k_2|$, $X_2=|\vec k_3|+|\vec k_4|$, and $Y = |\vec k_1 + \vec k_2 |  = |\vec k_3 +k_4|$. \\

\section{Pfaffian chain for wavefunction coefficients}
In the following section, we describe a procedure by which tree-level wavefunction coefficients of the model described by the action of equation~\eqref{eq:cosmoaction} can be defined in a Pfaffian chain. This shows that they are tame functions of the kinematic variables, and enables us to calculate their complexity. 

\subsection{Basis functions for wavefunction coefficients}
Consider a fixed tree-level Feynman graph generated by the vertices of this theory, and its corresponding value $\psi$ as a function of the kinematic variables. Our goal is to study these functions as generally as possible, without assumptions on the precise details of the underlying Feynman graph, the choice of interaction potential, or the parameter $\varepsilon$. A general strategy proposed in \cite{Arkani-Hamed:2023kig,Arkani-Hamed:2023bsv} is to construct a system of differential equations of the form 
\begin{equation}
    \dd I = A\, I \, ,
\end{equation}
where $I=(\psi,f_{2},\ldots,f_{N_{\rm F}})$ is a vector consisting of the function $\psi$ together with auxiliary functions $f_2,\ldots,f_{N_{\rm F} }$, and $A$ is an $N_{\rm F}\times N_{\rm F}$ matrix of one-forms. The precise form of the matrix $A$ can then be obtained by performing a combinatorial procedure on the graph, which is captured by an algorithm called \textit{kinematic flow}. In order to analyze the tameness and complexity of the wavefunction coefficients, we now provide a brief review of the algorithm. For a more detailed overview we refer to \cite{Arkani-Hamed:2023bsv}, and for an in-depth description and derivation we refer to \cite{Arkani-Hamed:2023kig}.

\subsection{Kinematic flow algorithm}
Let us first discuss the set-up of the algorithm, for which we have to introduce a diagrammatic notation. Starting with the Feynman graph under consideration, we first remove the external propagators, and mark the remaining edges of the graph with crosses. On the resulting marked graphs, we consider \textit{kinematic tubings}, which are
clusters of adjacent vertices and crosses including at least one vertex.  A kinematic tubing is complete if all vertices of the graph are enclosed by a tube. These kinematic tubings graphically encode certain functions of the kinematic variables in the following way. For a connected kinematic tubing, the corresponding function is given by the sum of the vertex energies $X_i$ in the tube and the internal energies $Y_j$ of the edges that enter the tube. If the tube includes the cross on such an edge, the sign of the internal energy is flipped. For example, for the single-exchange graph these functions may be represented by the following kinematic tubings

\begin{center}
\includegraphics{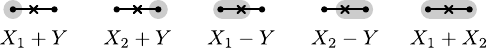} 
\end{center}

These particular linear combinations of the kinematic variables are also known as the \textit{letters} of the differential equation, which can be thought of as elementary building blocks encoding the singularity structure of the wavefunction coefficient $\psi$. Below we now review the steps of the algorithm \cite{Arkani-Hamed:2023bsv,Arkani-Hamed:2023kig}.

\begin{subbox}{Kinematic flow algorithm}
\begin{enumerate}
    \item[(i)] Select a complete kinematic tubing of the graph, for which the differential $\dd f$ of the corresponding function $f$ is to be computed.
    \item[(ii)] Write down a `kinematic flow tree' according to the following steps:
    \begin{enumerate}
        \item[1.] \textbf{Activation}. For each tube of the kinematic tubing, write down a descendant in which that tube is `activated' (indicated by a coloring).
        \item[2.] \textbf{Growth}. For each activated tube which contains no crosses, descendants are generated by `growing' the tube by including adjacent crosses in all possible combinations.
        \item[3.] \textbf{Merger}. If the tube resulting from the previous step intersects another tube, the two tubes merge and the union becomes activated. 
        \item[4.] \textbf{Absorption}. If a cross contained in an activated tube is adjacent to another tube containing a cross, the other tube is `absorbed' whereupon the union becomes activated, generating another descendant. 
    \end{enumerate}

    \item[(iii)] The expression for $\dd f$ is read off from the kinematic flow tree as follows: for each graph, include a term $ \dd \log \Phi$ where $\Phi$ is the letter of the active tube, and multiply this term by the  function associated to the graph minus the functions corresponding to the direct descendant graphs. Finally, multiply all terms by an overall factor $\varepsilon$ and the number of vertices included in the tube upon activation.
\end{enumerate}
Applying these steps sequentially to all functions $\psi,f_2,\ldots,f_{N_{\rm F}}$, the algorithm terminates and the matrix of one-forms $A$ is obtained. 
\end{subbox}

\subsubsection*{Example: kinematic flow for single-exchange graph}
To illustrate the abstract steps of the algorithm, let us consider a simple example, namely the single-exchange graph. There are four complete kinematic tubings, and hence four corresponding functions

\begin{center}
\includegraphics{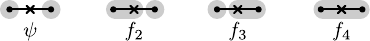} 
\end{center}

We start by computing $\dd \psi$, and write down the required kinematic flow tree below.
\begin{center}
\includegraphics{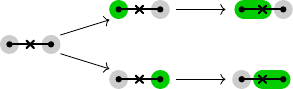} 
\end{center}

There are two tubes, and hence the activation step yields two descendant graphs. The next layer is generated by the growth of the active tubes to include the adjacent crosses. The kinematic flow terminates here, since the conditions for further growth, merger, or absorption are not fulfilled. Using the letters for the single-exchange graph given above, it follows from step (iii) that the differential equation for $\psi$ is 
\begin{align}
    \dd \psi = \varepsilon \big[ &(\psi-f_2) \dd \log(X_1+Y) + f_2 \dd\log(X_1-Y)\\ + &(\psi-f_3) \dd \log(X_2+Y) + f_3 \dd\log(X_2-Y)      \big] \,.\nonumber
\end{align}
Next, the kinematic flow tree for $f_2$ is given by

\begin{center}
\includegraphics{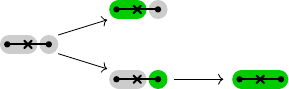} 
\end{center}

In the second layer the two tubes become active. The top channel terminates already at this step. The bottom channel shows the merger step, in which the entire graph is activated. The algorithm then tells us that
\begin{align}
    \dd f_2 = \varepsilon \big[ f_2 &\dd \log(X_1-Y) + (f_2-f_4)\dd \log(X_2+Y) \\ + f_4 &\dd \log(X_1+X_2)   \big] \,.\nonumber
\end{align}
The equation for $f_3$ follows by symmetry. Lastly, the kinematic flow for the final function $f_4$ is simply 
\begin{center}
\includegraphics{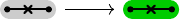} 
\end{center}

which gives (noting the factor of $2$ coming from the two vertices included in the tube upon activation)
\begin{equation}
    \dd f_4 = 2\varepsilon f_4 \dd \log(X_1+X_2) \,.
\end{equation}
The algorithm is now complete, and combining these equations we find the matrix $A$ which encodes the system of differential equations satisfied by the wavefunction coefficient and the auxiliary functions. 

\subsection{Constructing a Pfaffian chain}
\subsubsection*{Pfaffian chain for letters}
We will now argue that the wavefunction coefficient $\psi$ and the auxiliary functions discussed in the previous section form a Pfaffian chain. As a preparation, we need to include the $\dd\log$ terms of the letters into the chain. For each letter $\Phi_i(Z_I)= \sum_{I} c_i^I Z_I $, with $Z_I$ collectively denoting the kinematic variables, we have
\begin{equation}
    \dd\log \Phi_i = \frac{1}{\Phi_i}\sum_I c_i^I\dd Z_I \,.
\end{equation}
The coefficient $\ell_i=1/\Phi_i$ satisfies the differential equation
\begin{equation}\label{eq:letterdiffeq}
    \dd \ell_i = -\sum_{I}\ell_i^2 c_i^I \dd Z_I \,.
\end{equation}
In order to define these coefficients in a Pfaffian chain, we introduce the function
\begin{equation}
    L = \prod_i \ell_i \,,
\end{equation}
which satisfies the Pfaffian differential equation\footnote{Note that due to the large number of variables, it is convenient to express Pfaffian chains in terms of total derivatives in this discussion.}
\begin{equation}
    \dd L = -\sum_I \sum_i \bigg(\prod_{j\neq i} \Phi_j\bigg) L^2 c_i^I \dd Z_I \,, 
\end{equation}
which has degree $N_{\rm L}+1$, where $N_{\rm L}$ is the number of letters of the equation, depending on the Feynman graph under consideration. The inverse letters $\ell_i$ can then be expressed polynomially as
\begin{equation}\label{eq:letterpoly}
    \ell_i = \bigg(\prod_{j\neq i} \Phi_j\bigg) L\,.
\end{equation}
This enables us to use the letters as building blocks in subsequent functions in the Pfaffian chain. An alternative representation is to use equation~\eqref{eq:letterdiffeq} directly, which increases the length of the chain but lowers its degree. \\

\subsubsection*{Pfaffian chain for kinematic flow}
To set up a chain for the basis functions in the kinematic flow, we first organize them by the number of \textit{cuts} (line segments between vertices and crosses) appearing in the complete kinematic tubing. Complete kinematic tubings of a graph may then be uniquely described by the placement of these cuts. \\

The crucial observation is that, as a consequence of the rules of the algorithm, the differential of every basis function only depends on the function itself and basis functions corresponding to complete kinematic tubings with \textit{strictly fewer} cuts. In particular, the growing of the tubes in step 2.-4.~always implies a reduction in the number of cuts. Meanwhile, there is a single basis function $\zeta_0$ which corresponds to zero cuts, i.e.~the complete kinematic tubing with one tube. Taking this as the first basis function in the chain we can iteratively add the functions with an increasing number of cuts, thereby obtaining a Pfaffian chain. To be precise, let us denote the basis functions by $\zeta_{i,j}$, where $i$ is the number of cuts and $j=1,\ldots,n_i$ enumerates the basis functions with $i$ cuts. By the above argument we then have
\begin{equation}
    \dd \zeta_{i,j} = \sum_{I}  P_I (\ell_1,\ldots,\ell_{N_{\rm L}},\zeta_{i,j},\zeta_{i-1,1},\ldots)\dd Z_I \, ,
\end{equation}
which precisely fits the defining condition of a Pfaffian chain. \\

\subsubsection*{Example: Pfaffian chain for single-exchange graph}
Let us again turn to the single-exchange graph to exemplify our argument. The first function is given by
\begin{equation}
    L = \frac{1}{(X_1+Y)(X_2+Y)(X_1-Y)(X_2-Y)(X_1+X_2)}\,, 
\end{equation}
which satisfies the differential equation
\begin{equation}
    \dd L  =  -L (\ell_1+\ell_3+\ell_5)\dd X_1-L (\ell_2+\ell_4+\ell_5)\dd X_2 -L(\ell_1+\ell_2-\ell_3-\ell_4)\dd Y\,,
\end{equation}
where the letters $\ell_i$ corresponds to the kinematic tubings drawn earlier. Equation~\eqref{eq:letterpoly} shows that this differential equation can be expressed polynomially in $L$ and the kinematic variables $X_1,X_2,Y$, and thus constitutes the first level of the Pfaffian chain. 
Now we are able to write down the differential equations satisfied by the basis functions $(\psi,f_2,f_3,f_4)\equiv(\psi,\zeta_{1,1},\zeta_{1,2},\zeta_0)$ in terms of a Pfaffian chain, starting with $f_4$, which satisfies
\begin{align}
    \dd \zeta_0 &= 2\varepsilon \ell_5 \zeta_0 (\dd X_1 +\dd Y) \,.
\end{align}
Next, we define the basis functions belonging to graphs with one cut, namely $\zeta_{1,1}$ and $\zeta_{1,2}$. Employing the algorithm, one finds that they satisfy the Pfaffian equations
\begin{align}
     \dd\zeta_{1,1} =& \,\varepsilon \ell_2 \zeta_{1,1}(\dd X_1-\dd Y) +\varepsilon \ell_5 \zeta_0(\dd X_1 +\dd X_2) \\
     &+ \varepsilon \ell_3(\zeta_{1,1}-\zeta_0)(\dd X_2 +\dd Y)\,, \nonumber\\ 
     \dd\zeta_{1,2} =& \,\varepsilon \ell_4 \zeta_{1,2}(\dd X_2-\dd Y) +\varepsilon \ell_5 \zeta_0(\dd X_1 +\dd X_2) \\
     &+ \varepsilon \ell_3(\zeta_{1,2}-\zeta_0)(\dd X_1 +\dd Y)\,. \nonumber
\end{align}
Finally, the wavefunction itself is recovered in the Pfaffian chain as
\begin{align}
\dd \psi=\,\, &\varepsilon \ell_1(\psi-\zeta_{1,1})(\dd X_1 + \dd Y) + \ell_2 \zeta_{1,1} (\dd X_1-\dd Y)   \\
 + &\varepsilon \ell_3(\psi-\zeta_{1,2})(\dd X_2 + \dd Y) + \ell_4 \zeta_{1,2} (\dd X_2-\dd Y) \,. \nonumber
\end{align}
In this representation, $\psi$ has Pfaffian complexity $\cC_{\rm Pf}(\psi)=(3,5,6,1)$.

\subsection{Complexity of wavefunction coefficients}
\subsubsection*{Pfaffian complexity bounds}
The results from the previous section allow us to estimate the complexity of a general tree-level wavefunction coefficient in terms of the number of vertices of the graph, $N_{\rm V}$. The number of variables in the chain, i.e.~the number of kinematic variables, is given by $n=2N_{\rm V}-1$, since there is one variable for each vertex and edge. The order of the chain, $r$, is the component of the Pfaffian complexity which is most strongly dependent on the number of vertices. For a given graph, the Pfaffian chain consists of $N_{\rm L}$ letters and $N_{\rm F}$ basis functions. The number of letters strongly depends on the topology of the graph. For instance, we have
\begin{align} \label{eq:N-bounds}
    N_{\rm L}^{\rm chain}(N_{\rm V})&= 2N_{\rm V}^2 -2N_{\rm V}+1 \,, \\
    N_{\rm L}^{\rm star}(N_{\rm V}) &= 3^{N_{\rm V}-1} +2N_{\rm V}-3 \,.    \nonumber
\end{align}
It is worth noting that for fixed $N_{\rm V}$, the chain graph has the least letters and the star graph has the most letters, that is, for general $N_{\rm L}$ we always have 
\begin{equation} \label{eq:N_Lbound}
   N_{\rm L}^{\rm chain}(N_{\rm V}) \leq N_{\rm L}\leq N_{\rm L}^{\rm star}(N_V)\, . 
\end{equation}
Note that a star graph with $k$ legs is only present when the theory under consideration contains a $\phi^k$ interaction. In particular, in the notation of equation \eqref{eq:cosmoaction} this requires $k\leq D$ and $\lambda_k\neq 0$. Meanwhile, counting the number of basis functions one finds that
\begin{equation}
    N_{\rm F} = 4^{N_{\rm V}-1}\,.
\end{equation}
Therefore, for large graphs, we have $r\sim 4^{N_{\rm V}}$. The precise value of the Pfaffian complexity will now depend on whether we use the representation with the auxiliary function $L$ or the representation which defines the letters $\ell_1,\ldots,\ell_{N_{\rm L}}$ independently.

\begin{subbox}{Complexity of tree-level wavefunction coefficients}
In the representation in which the letters are defined by an auxiliary function, the Pfaffian complexity of $\psi$ is given by 
\begin{equation}\label{eq:wavefunctioncomplexity1}
    \cC_{\rm Pf}(\psi) = (2N_{\rm V}-1, 4^{N_{\rm V}-1} + 1,N_{\rm L}+1,1) \,.
\end{equation}
Alternatively, in the representation in which the letters are defined independently, the Pfaffian complexity of $\psi$ is given by
\begin{equation}\label{eq:wavefunctioncomplexity2}
    \cC_{\rm Pf}(\psi) = (2N_{\rm V}-1, 4^{N_{\rm V}-1} + N_{\rm L},2,1) \,.
\end{equation}
\end{subbox}
This result shows that the complexity of the letters can be either added to the length of the chain or to the degree of the chain.
In the language of sharp \mbox{o-minimality}, the former representation minimizes the format $\cF=n+r$, while the latter representation uniformly reduces the degree $\cD$ to a constant value at the price of adding an exponential contribution to the format.
The dependence of the complexity as a function of $N_{\rm V}$ is most apparent in the order $r$ of the chain, which shows an exponential growth. Also, note that the order $r$ increases by one for every function added to the Pfaffian chain. Since the steps of the kinematic flow algorithm generate more functions to be added to the chain, we note that \textit{complexity grows along the kinematic flow in a quantifiable way}. The remainder of this section is devoted to the interpretation of these observations.\\

\vspace{1cm}

Let us first note that the preceding discussion refers to the contribution to the wavefunction coefficient coming from a single Feynman graph viewed as a function of the kinematic variables $Z_I$. In general, the full wavefunction coefficients $\psi_n$ arises from a sum of graphs, and it will be a function of the couplings and kinematic variables $\psi_n(Z_I,\lambda_k)$. However, at tree level the sum over graphs is finite, implying that the tame structure in $Z_I$ is preserved while the coupling dependence is now a polynomial in the $\lambda_k$ with $p\leq n$. The complexity of the full tree-level wavefunction coefficient viewed as Pfaffian function in $Z_I$ and $\lambda_k$ can then be obtained through the complexities of the underlying graphs, which are determined by the interaction terms in the Lagrangian of the theory. \\ 

\vspace{-3mm}
\subsubsection*{Topological bounds and complexity reduction}
One of the essential features of Pfaffian complexity is its ability to encode other notions of complexity \cite{GabVor04}. Specifically, as explained in chapter~\ref{ch:complexity}, Pfaffian complexity encodes bounds on topological and computational measures of complexity. As a basic example, recall that a Pfaffian function $f$ with complexity $\cC_{\rm Pf}(f)=(1,r,\alpha,\beta)$ obeys bounds such as
\begin{equation}\label{eq:cosmoBezout}
    \text {number of isolated zeros} \leq 2^{r(r-1)} \beta \big(\alpha+\beta\big)^r \,,
\end{equation}
which may be interpreted as a generalized B\'ezout bound~\cite{GabVor04}. Since it counts the number of connected components of the solution set of an equation, it provides a coarse measure of topological complexity. 
Consequently, the optimal representation for a wavefunction coefficient could be one which minimizes the bound on one of the derived notions of complexity, such as the topological complexity given above. In sharp o-minimality, these bounds generically grow at least exponentially in the format $\cF$, which is in part determined by the order of the chain $r$. In our analysis of cosmological correlators, we find that $r$ itself grows exponentially in the number of vertices $N_{\rm V}$ of the Feynman graph, suggesting a doubly exponential growth in the topological or computational complexity of cosmological correlators. The bound given in equation \eqref{eq:cosmoBezout} applied to the wavefunction coefficient of an $N_{\rm V}$-vertex graph as a function of the kinematic variables becomes
\begin{equation} \label{BezoutCC}
    2^{(4^{N_{\text{V}}-1} +N_{\rm L})(4^{N_{\rm V}-1} +N_{\rm L}-1) }\big(4N_{\rm V}-1\big)^{4^{N_{\rm V}-1} +N_{\rm L}}  \,.
\end{equation}
Note that the number of letters $N_{\rm L}$ is bounded in terms of $N_{\rm V}$ as in \eqref{eq:N_Lbound}, with \eqref{eq:N-bounds} implying  doubly exponential growth in $N_{\rm V}$. However, from integral representation of  cosmological correlators one expects much slower growth. \\

This suggests that the complexity of the Pfaffian representations constructed in this section are far from optimal. In addition, there are physical arguments for the existence of a representation of reduced complexity. While a general solution to the differential equations may have a complicated singularity structure, many of these are not singularities of the wavefunction coefficient \cite{arkani-hamed_cosmological_2017,Armstrong:2023phb,salcedo_analytic_2023,Caloro:2023cep,lee_amplitudes_2024,fan_cosmological_2024}. Likewise, a further reduction is observed when passing from the wavefunction coefficients to cosmological correlators \cite{Chowdhury:2023arc}. 
A guiding principle for finding a simpler representation  arises from the locality of the physical theory. It was shown in \cite{Arkani-Hamed:2023kig} that, for some specific examples, locality implies the existence of simpler sets of differential equations for the wavefunction coefficients. Using the theory of GKZ systems and an associated notion of reducibility, it was indeed shown that the Pfaffian chain underlying these wavefunction coefficients can be simplified, leading to a significant reduction in complexity \cite{Grimm:2025zhv}. The reduction theory of GKZ systems and applications to cosmological correlators are explained in great detail in \cite{Hoefnagels:2025cnt}. 

\subsubsection*{Complexity and emergence of time}
The notion of complexity in this setting admits a speculative physical interpretation.
The authors of \cite{Arkani-Hamed:2023bsv,Arkani-Hamed:2023kig} argue that the kinematic flow algorithm is a boundary manifestation of the cosmological time evolution in the bulk. In this static description, time arises as an emergent concept. 
Through the results of \cite{GabVor04} explained in chapter~\ref{ch:complexity}, Pfaffian complexity can be used to control bounds on the computational complexity of algorithms involving Pfaffian functions. 
The results of this section thus show that there is a quantifiable growth of computational complexity along the kinematic flow. It is therefore tempting to speculate on the connection between time and complexity, and the idea that complexity may provide an emergent description of time. In physics there are many hints that time and complexity are related, for instance through entropy, or more recently, the idea of holographic complexity and time evolution in the bulk spacetime \cite{Susskind:2014rva,Carmi:2017jqz,Susskind:2018pmk}. The notion of complexity defined in sharp o-minimality is somewhat different from the one used in these works, and thereby provides a complementary perspective. Building further upon the idea of emergent time through kinematic flow, our findings provide a possible step in giving a quantitative connection between complexity and time, through the computational complexity of wavefunction coefficients. It is a compelling question whether the aforementioned notion of computational complexity has a physical interpretation in which the field theory formulated in the cosmological spacetime considered here acts as a computational device. We emphasize that these are speculative comments, and leave a further exploration of these ideas open to future work.  \\

\subsubsection*{Generalizations beyond tree-level graphs}
In this chapter we have focused on the complexity associated to an individual graph. An immediate follow-up question is to consider the complexity of the full tree-level wavefunction coefficient, 
both as a function of the kinematic variables and coupling constants. 
This complexity will depend on the properties of the Lagrangian, such as the number of scalar fields, the number of non-vanishing coupling constants, and the precise form of  the interaction polynomial.  The Pfaffian framework allows us to quantify this dependence for tree-level amplitudes.
Furthermore, this complexity measure allows us to quantify how simplicity emerges in amplitudes, since it accounts for possible algebraic relations among functions and gives a complexity reduction in the presence of a symmetry. An example of this arises
in \cite{Arkani-Hamed:2023kig} when considering multiple fields, for which the basis functions are shared between different channels. Due to these relations, the total complexity will be lower than expected based on an analysis of the individual diagrams. Investigating these possibilities is a key application of this notion of complexity, especially when eventually considering the full expression of the cosmological wavefunction.  
It was shown in \cite{Baumann:2024mvm} that kinematic flow can be generalized to loop integrands in the perturbative expansion of this model, which suggests that the results of this chapter can be extended to quantum corrections to cosmological correlators.\\

\chapter{On the complexity of quantum field theory}
\label{ch:complexityQFT}
\setlength{\parindent}{0pt}

The vast framework of quantum field theory remains one of the cornerstones of theoretical physics, describing an enormous variety of physical settings with remarkable precision. While many aspects of the framework are understood, a complete classification of quantum field theories remains elusive. Ultimately, one could envision a \textit{space of quantum field theories}, in which all quantum field theories and their observables can be systematically organized and compared according to various metrics \cite{Douglas:2010ic}. An essential part of such a construction is to understand what data fundamentally defines a quantum field theory, and how much information this data contains. Information-theoretic methods have been widely investigated in the context of quantum field theory, but currently a precise notion of information content underlying a theory does not exist. The previous chapters show that tame geometry appears naturally in various physical functions in quantum field theory. The notion of complexity provided by sharply o-minimal structures may therefore provide a systematic method of measuring information in this setting. \\

In this chapter we aim to explore whether sharp o-minimality can be used to define a complexity measure on the hypothetical space of quantum field theories. This notion of \textit{QFT complexity} should measure the information contained within a theory in a physically meaningful way, thereby providing a quantitative tool for analyzing the structure of the quantum field theories and their observables. We take two complementary perspectives: a top-down perspective based on the complexity of the descriptive functions of the theory, and a bottom-up perspective which views the observables as the fundamental starting point. After discussing ideas to define QFT complexity, we investigate the physical meaning of our proposals. We do this by commenting on how it relates to various standard aspects of quantum field theory, such as symmetries, perturbation theory, and renormalization group flow. In addition, our framework allows us to compare the complexity of the descriptive functions of a theory to the complexity of its observables, following the ideas laid out in chapter~\ref{ch:classical}. The analysis of this chapter will be exploratory in nature, initiating the first steps but leaving many questions open for future research.

\section{Complexity of Lagrangians}\label{sec:QFTactions}
The purpose of this section is to investigate how to assign a complexity to QFTs by considering the descriptive functions specifying them. In particular, we will focus on theories which can be described by quantizing a classical field theory, so that the data defining the theory consists of a collection of fields $\Phi$ and a Lagrangian $\cL(\Phi)$. 
Instead of implementing sharp o-minimality directly, we start this chapter from a perspective which is somewhat agnostic to the notion of complexity developed in chapter~\ref{ch:complexity}. By studying progressively more complicated theories, we then argue that we are led towards sharp o-minimality in a fairly natural way. 

\subsection{Complexity of scalar Lagrangians}

\subsubsection*{Single scalar with algebraic potential} 
Let us start with a simple example, namely a $d$-dimensional theory consisting of a single real scalar field $\phi$ with polynomial self-interactions. The Lagrangian then takes the form 
\begin{equation}
    \cL(\phi) = -\frac{1}{2}\pd_\mu \phi \, \pd^ \mu \phi -V(\phi) \,,
\end{equation}
where $V(\phi)= \lambda_D \phi^ D + \lambda_{D-1}\phi^{D-1}+ \ldots + \lambda_0$ is a polynomial. What is the complexity of this Lagrangian? Assuming that the field $\phi$ is canonically normalized, the Lagrangian is specified completely by the polynomial $V(\phi)$, and the question reduces to measuring the complexity of a polynomial in one variable. A natural candidate is the degree $D$ of $V$, since this Lagrangian requires $D$ real numbers to be uniquely specified, namely the coupling constants $\lambda_D,\ldots,\lambda_0$. The complexity of this set of QFTs would then be indexed by a single integer $D$. \\

However, this measure is not sufficiently refined to accurately reflect the complexity of a polynomial potential, since it only sees the highest-power interaction term. There may be special algebraic relations among the coefficients $\lambda_0,\ldots,\lambda_D$ which simplify the potential significantly. For instance, consider the simple algebraic relation $\lambda_0=\ldots=\lambda_{D-1}=0$, so that the potential reduces to a monomial $V(\phi)=\lambda_D\phi^D$. Naturally, the QFT defined by this polynomial potential should have a lower complexity than the one for which all coefficients $\lambda_0,\ldots,\lambda_D$ are free. This example already calls for a more refined notion of complexity. 

\subsubsection*{Multiple scalars with algebraic potential}
A natural extension of the previous example is to increase the number of degrees of freedom.\footnote{By number of degrees of freedom, we refer to the number of fields in the theory throughout this chapter.} Hence, we now consider a theory with $N$ real canonically normalized scalar fields $\phi_1,\ldots,\phi_N$ described by a Lagrangian
\begin{equation}\label{eq:Lscalars}
    \cL(\phi_1,\ldots,\phi_N) = -\frac{1}{2}\sum_{k=1}^N \pd_\mu \phi_k \, \pd^ \mu \phi_k -V(\phi_1,\ldots,\phi_N)\, .
\end{equation}
The potential $V$ is now taken to be a general polynomial of degree $D$
\begin{equation}
    V(\phi_1,\ldots,\phi_N) = \sum_{I,\,|I|\leq D} \lambda_I \phi^I\, .
\end{equation}
In this notation, $I=(I_1,\ldots,I_N)$ is a multi-index, and we define
\begin{equation}
    |I|=I_1+\cdots+I_N, \quad \lambda_I = \lambda_{I_1\cdots I_N}, \quad \phi^I = \phi_1^{I_1} \cdots \phi_N^{I_N} \, .
\end{equation}

As in the previous example, the complexity of this QFT completely lies in the polynomial potential $V(\phi_1,\ldots,\phi_N)$. In addition to the degree $D$, the complexity of the theory should now also depend on the number of fields $N$. A first natural attempt at  quantifying the complexity of this theory would be to define it as a number $\cC$ which depends on $N$ and $D$, e.g.~the number of independent coupling constants $\lambda_I$:
\begin{equation}
    \cC =  \frac{(N+D)!}{N!D!} \,.
\end{equation}
However, as explained in section~\ref{sec:introcomplexity}, there is a fundamental problem with this approach. From the perspective of computational complexity, $N$ and $D$ play a radically different role. This difference is captured by Bézout's bound, which states that the number of common zeros of $N$ polynomials of degree $D$ is bounded above by $D^N$. In particular, it implies that the computational complexity of solving polynomial equations satisfies a lower bound depending polynomially on the degree $D$, while depending \textit{exponentially} on the number of variables $N$. With this in mind, the complexity of this QFT should reflect the difference in the dependence on $N$ and $D$ in order to be meaningful for computational applications.  \\

This observation suggests that it may be natural to characterize the complexity of a QFT with \textit{two numbers}, in line with sharp o-minimality. 
Besides being stronger for computational applications, from a physical perspective it reflects the fact that a QFT depends on the number of degrees of freedom and the interactions among them in a rather different way. For the class of QFTs described by the Lagrangian of equation \eqref{eq:Lscalars}, we could measure the complexity by the pair of integers $(N,D)$. Let us note that, although this is a good starting point, these integers are not yet refined enough to capture certain details of the QFT, such as the geometry of the field space and the strength of the coupling constants. We discuss these aspects more carefully later.

\subsubsection*{Scalars with analytic potential}
Let us now generalize the theory described by the Lagrangian of equation \eqref{eq:Lscalars} one step further, and assume that $V(\phi_1,\ldots,\phi_N)$ is a general analytic function of the fields. We then seem to run into an obstacle, since the potential $V$ could now contain an infinite amount of information. For instance, if we fix a vacuum and expand the potential in a power series, we will need infinitely many coefficients $\lambda_I$ to specify the Lagrangian. Therefore, this set of QFTs already contains theories with infinite information content in their description, for which a naive algebraic notion of complexity seems to break down. \\

From the previous chapters in this thesis, we know how to amend this situation: among these QFTs with analytic potentials, we consider those whose potential can be described with a finite amount of information in a sharply o-minimal structure. For instance, in case the potential is a Pfaffian function, the underlying Pfaffian chain encodes algebraic relations among the coefficients $\{\lambda_I\}$ in the power series expansion. This suggests that we that we measure QFT complexity in a general scalar theory by assigning a format and degree $(\cF,\cD)$ to the Lagrangian. The advantages of this prescription are that it reflects the number of degrees of freedom and the complexity of the interactions, and encompasses theories whose Lagrangians are formulated with complicated transcendental functions.\footnote{When focusing on renormalizable theories, it is often sufficient to consider operators of finite dimension in the Lagrangian, but in effective theories much more complicated functions may arise. For example, as we will see in part III, in the effective actions coming from string theory compactifications one frequently encounters coupling functions coming from period integrals.} The number of degrees of freedom is then counted by $\cF$, defined as the sum of the number of scalars $N$ and the number of auxiliary functions $r$ appearing in the Lagrangian. The degree $\cD$ encodes the complexity of the interactions and relations among the couplings. To gain some intuition for this prescription, let us discuss a few examples.

\subsubsection*{Example: polynomial and fewnomial interactions}
We first compare two theories of a single real scalar $\phi$ with Lagrangians $\cL_{\rm mono}$ and $\cL_{\rm poly}$, given by
\begin{equation}\label{eq:Lpoly}
    \cL_{\rm poly} =  -\frac{1}{2}\pd_\mu \phi \, \pd^ \mu \phi -\sum_{k=0}^D\lambda_k \phi^k
     \,, \qquad   
     \cL_{\rm mono} =  -\frac{1}{2}\pd_\mu \phi \, \pd^ \mu \phi -\lambda \phi^D \,.
\end{equation}
Even though these two Lagrangians have the same polynomial degree $D$, the theory described by $\cL_{\rm mono}$ should have a lower complexity, since it only contains a single interaction term. This is indeed discerned by format and degree by means of the fewnomial representation discussed in section~\ref{sec:Pfaff}. The format and degree of $\cL_{\rm poly}$ is simply given by $(\cF,\cD)=(1,D)$, since it can be represented as a Pfaffian function with a Pfaffian chain of length zero. However, for $\cL_{\rm mono}$ we can implement the more information-efficient fewnomial representation by considering the Pfaffian chain
\begin{align}
    \zeta_1(\phi)& = \frac{1}{\phi} \, , &&  \frac{\pd\zeta_1}{\pd \phi}=- \zeta_1^2 \,, \\
    \zeta_2(\phi)&=  \phi ^D  \,,  && \frac{\pd\zeta_2}{\pd \phi}= D\zeta_1\,\zeta_2, \nonumber
\end{align}
so that $\cL_{\rm mono}(\phi)=-\frac{1}{2}\pd_\mu \phi \,\pd^\mu\phi +\lambda \, \zeta_2(\phi)$, which has format and degree given by $(\cF,\cD)=(3,5)$. At the price of introducing two auxiliary field variables $\zeta_1$ and $\zeta_2$, we have effectively reduced the complexity of the representation and removed the dependence on the power of the interaction $D$.

\subsubsection*{Example: cosine potential}
As another instructive example, consider a theory with a real massive scalar field $\phi$ whose field space is the finite interval $(-L,L)$, described by the Lagrangian
\begin{equation}
    \cL_{\rm cos}(\phi) = -\frac{1}{2}\pd_\mu \phi \,\pd^\mu \phi - \frac{1}{2}m^2\phi^2 + \lambda \cos (k\pi \phi/L) \,,
\end{equation}
where $k$ is a positive integer. Oscillating potentials of this form appear in theories such as the sine-Gordon model. Expanded in a power series, the potential has infinitely many interaction terms. However, this Lagrangian can be expressed by means of the Pfaffian chain 
\begin{align}
    \zeta_1(\phi)& = \tan(\frac{\pi \phi}{2L}) \,, &&  \frac{\pd\zeta_1}{\pd \phi}=\frac{\pi}{2L}\big(1+ \zeta_1^2 \big) \,, \\
    \zeta_2(\phi)&=  \cos\left(\frac{\pi\phi}{L}\right)  \,,  && \frac{\pd\zeta_2}{\pd \phi}= -\frac{\pi}{L}\zeta_1(\zeta_2+1)\,, \nonumber
\end{align}
which has made an appearance earlier in section~\ref{sec:Fieldconfigurations}. Subsequently, we can write $\cL_{\rm cos}$ as a Pfaffian function, 
\begin{equation}
    \cL_{\rm cos}(\phi) = -\frac{1}{2}\pd_\mu \phi \,\pd^\mu \phi - \frac{1}{2}m^2\phi^2 + \lambda \, T_k(\zeta_2(\phi)) \,,
\end{equation}
where $T_k$ is the Chebyshev polynomial of degree $k$. The theory now has a finite complexity, measured by the format and degree $(\cF,\cD)=(3,4+k)$. As a function of $k$, the format of the theory stays fixed, but the degree grows linearly. A curious observation is that the number of minima of the potential, i.e.~the number of vacua, also grows linearly with $k$. This exemplifies how format and degree of encode precise bounds for geometric and topological properties of sets constructed from Pfaffian functions, e.g.~the number of zeros of a Pfaffian function. Applied to QFTs, this means that our notion of QFT complexity for actions is able to accurately probe the vacuum structure of the theory. We discuss this further in section~\ref{sec:vac}.

\subsubsection*{Field redefinitions and minimal representations}
The format and degree of a Lagrangian depend on the representation, and a field redefinition may result in a change in the complexity. As an example, consider the Hubbard-Stratonovich transformation in $\phi^4$ theory. Starting with the Lagrangian 
\begin{equation}
\cL(\phi)=-\frac{1}{2}\pd_\mu \phi\,\pd^\mu \phi -\lambda \phi^4 \,,
\end{equation}
we introduce an non-dynamical auxiliary scalar $\sigma$ and use the path integral identity
\begin{equation}
     e^{- \lambda \phi^4}  =  \int D\sigma \, e^{-\frac{1}{2}m^2\sigma^2 +\sqrt{2\lambda}\,m\,\sigma\phi^2}\,.
\end{equation}
The resulting transformed Lagrangian is 
\begin{equation}
    \cL(\phi,\sigma) = -  \frac{1}{2}\pd_\mu \phi\,\pd^\mu \phi -\frac{1}{2}m^2\sigma^2 +\sqrt{2\lambda}\,m \,\sigma\phi^2 \,.
\end{equation}
This Lagrangian describes the same QFT, but its format $\cF$ has increased by one and its degree $\cD$ has decreased by one.\\

The fact that a given QFT can be represented by different Lagrangians allows one to ask, in a quantitative way, whether a given QFT admits a `minimal representation', represented by an extremal point in the complexity spectrum. The complexity of this minimal representation would then measure the minimal amount of information needed to specify a theory. For example, it is known that dualities such as AdS/CFT may lead to significantly simpler descriptions of the same theory. A notion of QFT complexity formalizes this idea, and we will revisit this idea in chapter~\ref{ch:EFTcomplexity}.

\subsection{Complexity of general Lagrangians} 
\label{sec:CoA-general}
For theories with only scalar fields, we have found that the format and degree of the Lagrangian, defined by means of sharp o-minimality, may provide a reasonable measure of QFT complexity. We now extend this discussion to more general theories, including gauge fields, fermions, and non-trivial target spaces. This generalization will lead to a number of conceptual challenges for complexity, and along the way we will discuss how some of these challenges may be resolved.

\vspace{-1mm}

\subsubsection*{Abelian gauge theories} 
Continuing with bosonic theories for now, let us start by discussing abelian gauge theories. Consider a $d$-dimensional $\text{U}(1)$ gauge field $A_\mu$, described by the Lagrangian 
\begin{equation}\label{eq:U1}
    \cL_{\text{U(1)}} = -\frac{1}{4g^2}F_{\mu\nu} F^ {\mu\nu} \, ,
\end{equation}
with $F_{\mu\nu} = \pd_\mu A_{\nu} - \pd_\nu A_\mu$. Compared to Lagrangians with only scalar fields, we encounter three new challenges for complexity: (i) $A_\mu$ is a 1-form instead of a scalar; (ii) there is a gauge symmetry which renders some of the degrees of freedom unphysical; and (iii) the Lagrangian only contains derivatives of the fields.
How should we assign a format and degree to this Lagrangian, with these points in mind? To address point (i), we could treat the components $A_0,\ldots, A_{d-1}$ as $d$ real field variables. For the moment, we thereby disregard non-trivial topological and geometric features of gauge theory, e.g.~in the form of a non-trivial principal bundle for the gauge field. We discuss these later in this section. Among the field components $A_{0},\ldots,A_{d-1}$, only $d-2$ components correspond to physical degrees of freedom. For the purpose of quantifying complexity of the Lagrangian, we note however that all components are needed; the redundancy arising from the gauge symmetry therefore does not reduce the format $\cF$ of the theory. 

\vspace{-1mm}
\subsubsection*{Derivative interactions}
The appearance of field derivatives in the Lagrangian of equation \eqref{eq:U1} is worth discussing more generally. For the scalar theories, we assumed that all fields were canonically normalized and ignored the kinetic term. This was a strong assumption, and in general the theory could be a sigma model with a non-trivial field space metric. For the gauge theories, we are forced to rely on derivative terms by gauge invariance. Moreover, in effective field theories, one often encounters higher-derivative interaction terms in the Lagrangian. \\

Since we are focusing on the complexity of the description of the theory, the natural resolution to this issue would be to view the derivatives as auxiliary variables of the Lagrangian. That is, instead of viewing $\cL$ as a function of $\Phi$, we include derivatives as variables and view it as a function of $(\Phi,\pd_\mu\Phi,\pd_\mu\pd_\nu\Phi,\ldots)$. For the scalar theories, this allows us to treat the field space metric in the same way as the potential, namely by assigning a format and degree to each component function. In the context of abelian gauge theory this would amount to treating the components field strength $F_{\mu\nu}$ as the variables of the Lagrangian; the QFT described by \eqref{eq:U1} would then have complexity $(\cF,\cD)=\big(\tfrac{1}{2}d(d-1),2 \big)$. Taking this viewpoint naturally leads to a dependence of the format $\cF$ on the spacetime dimension $d$. We will discuss the approach of auxiliary variables for derivative interactions in more detail in chapter~\ref{ch:EFTcomplexity}. \\

As a generalization of the previous example, consider a $d$-dimensional $\text{SU}(N)$ gauge theory, whose Lagrangian is given by
\begin{equation}\label{eq:SUN}
    \cL_{\text{SU}(N)} = -\frac{1}{4g^2} \tr (F_{\mu\nu} F^ {\mu\nu}) \, ,
\end{equation}
Counting the number of real field variables, one finds that the complexity of this theory is $(\cF,\cD)=\big(\tfrac{1}{2}(N^2-1)d(d-1) , 2 \big) $. Even though the complexity of the theory grows with $N$, it is known that simpler dual descriptions emerge in large $N$ limits. This phenomenon, where the complexity grows with $N$ but reduces in the strict $N\to\infty$ limit, hints towards a realization of emergence of simplicity captured by the complexity of a QFT. 

\subsubsection*{Fermionic theories} 
Although our focus lies mainly on bosonic theories, let us nonetheless briefly comment on how these ideas may be implemented for fermions. Since format and degree are based on real-valued variables, the natural way to proceed is to look at the real and imaginary parts of the components of a fermionic field $\psi$. In particular, we view these variables as real variables multiplying a basis of Grassmann numbers. A consequence of the anti-commuting nature of fermions is that there are only finitely many non-vanishing interactions. For instance, a theory of $N$ Dirac fermions $\psi_1,\ldots,\psi_N$ is described by the general Lagrangian
\begin{equation}
    \cL(\psi_1,\ldots,\psi_N) = - \sum_{j=1}^N \overline{\psi}_j i\slashed{\pd} \psi_j - \sum_{k=1}^{D/2}\sum_{j_1,\ldots,j_k'} \lambda_{j_1\cdots j_k'}(\overline\psi_{j_1}\Gamma_{1}\psi_{j'_1}) \cdots (\overline\psi_{j_k}\Gamma_k \psi_{j'_k}) \, .
\end{equation}
Here the $\Gamma_i$ parametrize elements of the corresponding Clifford algebra.
The highest-power interaction $D$ depends on the spacetime dimension $d$ and the number of fermions. In terms of format and degree, this means that $\cD$ is bounded by a simple universal function of $\cF$. The complexity of this theory is easier to quantify due to the necessary polynomial nature of a fermionic Lagrangian.

\vspace{-1.5mm}

\subsubsection*{Topology and complexity}
So far in our discussion we have mostly dealt with theories with no non-trivial topological features. In general, these features are an important aspect of QFTs, in the form of e.g.~spacetime topology, principal bundles for gauge theories, and target spaces in non-linear sigma models. Because the concept of format and degree which we use in our notion of QFT complexity is ultimately geometric in nature, we believe that they can be generalized to QFTs exhibiting these non-trivial topological features. This may be done by subdividing spaces into trivial local patches and using logical operations to add the local complexities. Some of these topological and geometric aspects will be addressed in chapter~\ref{ch:EFTcomplexity}.

\vspace{-1.5mm}

\subsubsection*{Symmetries} 
Let us briefly comment on the role of symmetries in the complexity of the action of QFTs. At an intuitive level, the presence of symmetries may lead to the simplification of a theory, but the way in which this is seen from in the format and degree is subtle. Let us illustrate this by two examples. The QFT defined by the Lagrangian $\cL(\phi ) = -\frac{1}{2}\pd_\mu \phi \,\pd^\mu \phi  - \lambda \sin (2\pi k \phi) $
initially has infinite complexity if the field can range freely over the real line $\bbR$, because the potential then has infinitely many distinct zeros. However, it also has a global $\bbZ$ symmetry, and after taking a quotient by $\bbZ$ by identifying $\phi\sim \phi+1/k$ we obtain a simple theory with finite complexity. On the other hand, if the theory has a global symmetry group $G$ which acts in a more non-trivial way, then the theory obtained by quotienting by $G$ may have a more complicated geometry and therefore a Lagrangian of higher complexity. This illustrates that the interaction between symmetries and complexity in sharp o-minimality can be delicate.

\vspace{-1.5mm}
\subsubsection*{Vacuum structure}\label{sec:vac}
Having discussed how to use format and degree to measure the complexity of a the Lagrangian, let us highlight an application. One of the essential properties of format and degree is that they encode bounds on geometric and topological features.\\

This is made precise by the quantitative tameness theorems of chapter~\ref{ch:complexity}. For instance, given a Pfaffian function $f$ with complexity $(\cF,\cD)$, the number of connected components of the solution set to the equation $f=0$ is bounded by \cite{Fewnomials,GabVor04}
\begin{equation} \label{zero-bounds}
    b_0(\{x \,|\, f(x)=0\})   \leq 2^{\cF^2} \cF^\cF \cD ^\cF \,. 
\end{equation}
Bounds of this type, which grow polynomially in $\cD$, hold generally in a sharply o-minimal structure.
When applying this principle to QFTs, we can use this to analyze the vacuum structure of the theory. If the potential $V(\Phi)$ has complexity $(\cF,\cD)$, we can then estimate the number of vacua by considering the number of connected components of the set of critical points
\begin{equation}
    \Big\{\Phi \, \Big|\, \frac{\pd V}{\pd \Phi}=0\Big\}\,.
\end{equation}
While there is a polynomial growth in $\cD$, these estimates typically grow rapidly with $\cF$, and for simple potentials it is much more effective to count the number of vacua by hand. However, for QFTs with extremely complicated potentials, such as the effective theories arising from string compactifications, these countings are numerically very challenging. In such cases, the bounds in terms of format and degree provide simple and universal estimates for the vacuum structure of the theory. An important lesson to draw from these bounds is that the significant computational cost of increasing the format $\cF$ suggests that the minimal representation of the QFT should have the smallest possible format.

\subsubsection*{Renormalizability}
The discussion above shows how the complexity of a theory can be characterized by the format and degree of the Lagrangian, but it does not fully incorporate the quantum nature of the theory, and in particular it does not consider its renormalization, which would require the inclusion of additional terms that would contribute to the complexity. In spacetime dimension $d>2$, renormalizability restricts the Lagrangian to be polynomial of degree $\cD\leq 2d/(d-2)$. A non-renormalizable Lagrangian requires the addition of infinitely many counterterms, signaling that the theory in fact has infinite complexity.  Therefore, we expect  that non-renormalizable theories are generically incompatible with o-minimality at the quantum level. This issue can be circumvented by including an energy cutoff $\Lambda$ beyond which the theory is no longer valid, i.e.~to consider an \textit{effective field theory}. The effective perspective comes with new technical challenges, which will be addressed in chapter~\ref{ch:EFTcomplexity}.

\section{Complexity of observables}\label{sec:amp}
In the previous section we explored how to assign a complexity to the microscopic description of a QFT, focusing on the information contained in the field content and the Lagrangian. We now change our perspective, and view a QFT through its observables, focusing mostly on scattering amplitudes and correlation functions. The tameness and complexity of QFT observables have been analyzed in specific instances in the previous chapters, and now we take a more global point of view. There are four main questions on the complexity of observables which we consider:
\begin{enumerate}
    \item[(i)] What is the complexity of an individual QFT observable? 
    \item[(ii)] What is the complexity of a QFT, based on its spectrum of observables?
    \item[(iii)] What is the connection between the complexity of the Lagrangian and the complexity of observables?
    \item[(iv)] How is the complexity of a theory affected by renormalization group flow?
\end{enumerate}
Our aim is not to conclusively answer these questions, but to provide some initial ideas and observations. 

\subsection{Amplitudes and correlation functions}

\subsubsection*{General aspects of observables in QFT}
Consider a general QFT observable $\cO(\kappa,\lambda)$, which we view as a function of kinematic variables (e.g.~positions or momenta) denoted by $\kappa$, and of the parameters of the theory, collectively denoted by $\lambda$. The complexity of this observable can then be defined by its format and degree if it is sufficiently tame and definable in a sharply o-minimal structure. 
Before we proceed, let us list the different cases which we should consider. First, we should distinguish whether the observable is obtained explicitly from the quantization of an underlying microscopic theory, or whether it is viewed as an entity on its own. In the first case, we should further discern between perturbative expansions of observables, and exact observables which include all non-perturbative effects. In the second case, where  no microscopic description is available, we will analyse the complexity of observables from a bootstrap perspective. We study these case by  case, with support from the findings of the previous chapters.

\subsubsection*{Perturbative amplitudes and Feynman integrals}
As reviewed in chapter~\ref{ch:observables}, perturbative amplitudes with a fixed number of loops were shown to be definable in the o-minimal structure $\bbR_{\rm an,exp}$ in \cite{Douglas:2022ynw}. While $\bbR_{\rm an,exp}$ is not sharply o-minimal, the proof of \cite{Douglas:2022ynw} shows that perturbative amplitudes can be expressed in terms of finitely many period functions. These functions are definable in the structure $\bbR_{\rm LN,Pf}$ defined by means of log-Noetherian functions, which is conjectured to be sharply o-minimal \cite{binyamini2024lognoetherianfunctions}. 
If this conjecture holds, a systematic calculation of the complexity of perturbative amplitudes in terms of format and degree would be available. 
The complexity of perturbative amplitudes is expected to grow in at least two ways. Firstly, there is a clear dependence on the number of external particles, since each external particle introduces additional kinematic variables on which the amplitude depends, thereby increasing the format. Secondly, at higher orders in perturbation theory one must sum over a larger set of more complicated diagrams, leading to more complicated Feynman integrals and therefore functions with higher $\cF$ and $\cD$.

\subsubsection*{Exact observables and emergence of simplicity}
As we have seen in chapter~\ref{ch:observables}, perturbation theory provides, with the exception of rare cases, only asymptotic expansions to exact non-perturbative observables. While the terms in the perturbative expansion reveal much of the underlying QFT, ultimately one should consider the complexity of exact observables. Fundamentally, the exact observables arise from summing over infinitely many physical processes, and therefore there is initially no reason to expect that these observables can be described with a finite amount of information. However, upon taking infinite limits, remarkable reductions of complexity may appear, and the results of chapter~\ref{ch:observables} show that exact observables may still be tame functions. As an example, recall the asymptotic expansion of the partition function of $\phi^4$-theory on a point, 
\begin{equation}
    Z(g) \sim \sum_{k=0}^\infty  \frac{(-1)^k}{8 ^k \,k!} \Gamma(2k+\tfrac{1}{2}) g^{-k-1/2} \,.
\end{equation}
While the complexity grows order by order, the series can be resummed into the tame function
\begin{equation}
    Z(g) = \sqrt{2} \, e^g  K_{1/4}(g) \, ,
\end{equation}
which is a Pfaffian function with finite complexity, demonstrating an emergence of simplicity. 
Section~\ref{sec:0dQFTborel} also showed that the weak coupling limits of resummed functions can be formulated using classes of quasi-analytic functions.
Presently, the o-minimal structures generated by quasi-analytic functions do not support a format-degree filtration, which means that we have to resort to domains which avoid the non-analytic weak coupling limits. \\

Starting from a sufficiently simple QFT, e.g.~one whose Lagrangian has a finite complexity, it is a natural expectation that functions arising as physical observables have a limited information content, despite the growth of complexity order-by-order in perturbation theory. This leads us to conjecture that emergence of simplicity is a general phenomenon of these QFTs, or more precisely, that for a QFT with finite complexity, the exact observables can be represented with finite complexity. 

\subsubsection*{Exact observables -- bootstrapping complexity} 
A substantial part of the space of QFTs consists of theories for which no microscopic description exists, and in this case one can only rely on general physical principles such as unitarity, locality, and symmetries to constrain observables. 
It is then natural to ask whether the complexity of observables can be constrained by these principles, analogous to bootstrap approaches to QFT.\\

As an example, consider two- and three-point functions of primary fields in a conformal field theory (CFT). Conformal symmetry is sufficient to constrain these to take the exact general form 
\begin{align}
    \expval{\cO_1(x_1)\cO_2(x_2)} &= \frac{c_{12}\,\delta_{\Delta_1,\Delta_2}}{|x_1-x_2|^{\Delta_1+\Delta_2}} \,, \\
    \expval{\cO_1(x_1)\cO_2(x_2)\cO_3(x_3)} &= \frac{c_{123}}{|x_1\!-\!x_2|^{\Delta_1+\Delta_2-\Delta_3}|x_2\!-\!x_3|^{\Delta_2+\Delta_3-\Delta_1}|x_1\!-\!x_3|^{\Delta_1+\Delta_3-\Delta_2}}\,,
\end{align}
where $\Delta_i$ is the conformal weight of the field operator $\cO(x_i)$. These functions admit a simple Pfaffian chain representation, so they have a finite complexity, based on conformal symmetry. \\

A powerful feature of complexity in tame geometry is that it can be determined without having a closed-form expression. Instead, it is sufficient to find an implicit representation of the function, such as a Pfaffian chain. These differential equations can often be found by relying on underlying properties of the theory, thereby avoiding an explicit computation. The analysis of chapter~\ref{ch:CCC} provides an example of this, in the form of an algorithm underlying tree-level cosmological correlators.\\

Instead of using bootstrap methods to infer the complexity of an observable, one may also utilize complexity as a bootstrap principle. For example, we could impose a complexity bound on $n$-point correlation functions, i.e.~a maximal format and degree $\cF_n$ and $\cD_n$ depending on $n$, and study the space of QFTs satisfying these bounds without reference to a miscroscopic description. It would be interesting to explore this in future research, for instance by connecting it to Conjecture 5 of \cite{Douglas:2023fcg} on the space of CFTs satisfying bounds on the number of degrees of freedom.

\subsubsection*{Observables and complexity classes} 
So far we have considered QFT observables as individual entities and analyzed their complexity as a function of kinematic variables and theory parameters. However, our initial goal, as formulated in question (ii), was to quantify the complexity of a QFT. In principle a QFT has observables of arbitrarily high complexity, for instance scattering amplitudes with many external particles or correlation functions with many operator insertions. It is then perhaps not reasonable to expect that the complete spectrum of observables in a QFT can be described with a finite amount of information. Instead, one approach is to quantify the growth of the complexity of observables as a function of some external label $N$, such as the number of external particles or field insertions. The dependence on $N$ could then be classified by certain complexity classes, similar to computational complexity. This growth class could then be an indicator of the complexity of a QFT. \\

As an example, consider again the $\phi^4$-theory on a point, whose observables are the correlation functions
\begin{equation}\label{eq:phi4corr}
    I_N(g) = \int \dd \phi \, \phi^N \, e^{-g\phi^2 -\frac{g}{8}\phi^4}  \,.
\end{equation}
In section~\ref{sec:0dQFTdiff} it was shown that $I_N$ has complexity $(\cF,\cD)=(4,8+\lceil N/4\rceil)$, growing stepwise linearly with $N$. Following the discussion above, one would then assign to this QFT the complexity class $(\cF,\cD)\sim \big(O(1),O(N)\big)$. Note that the growth of the format and degree have to be quantified separately. In this example, only $\cD$ grows, and since computational complexity grows polynomially in $\cD$, this indicates that there are algebraic relations present among the observables which reduce their information content.

\subsubsection*{Total information of observables} 
Another approach to measuring the complexity of a QFT through its observables is to exploit algebraic relations among observables, as encountered in the example of the previous paragraph, to reduce the \textit{total} information content to a finite amount. With this in mind, we reformulate our question: is there a finite amount of data, from which all observables can be generated through a simple prescription? Here it is important to specify what we precisely mean by simple prescription; one could take an extreme viewpoint and consider a Lagrangian of finite complexity to be the finite set of initial data, and the quantization of the theory to be the prescription for obtaining the observables. Instead, we will require that the simple prescription consists of elementary logical steps, e.g.~an algebraic recursion relation. \\

In some simple settings, this question can be answered affirmatively. For example, consider a $d$-dimensional scalar QFT with polynomial interactions up to degree $D$, described by the Lagrangian of equation \eqref{eq:Lpoly} depending on $\lambda_1,\ldots,\lambda_D$, which is regularized by putting it on a finite lattice $\Gamma$. Enumerating the points on the lattice by $j=1,\ldots,|\Gamma|$, the correlation functions of this theory take the general form 
\begin{equation}
    I_{N_{1}\cdots N_{|\Gamma|}}(\lambda_1,\ldots,\lambda_D) = \int \dd\phi_1\cdots \dd\phi_{|\Gamma|} \, \phi_1^{N_1} \cdots \phi_{|\Gamma|}^{N_{|\Gamma|}} \, e^{-S[\phi]}\,.
\end{equation}
As discussed in section~\ref{sec:0dQFTdiff}, it was argued in \cite{Weinzierl:2020nhw} using twisted cohomology that all correlation functions in this theory can be expressed as a linear combination of a finite set of basis integrals of size $(D-1)^{|\Gamma|}$, where $|\Gamma|$ is the number of points on the lattice. The coefficients of this linear combination may be found algorithmically by a reduction to master integrals, as explained in \cite{Weinzierl:2020nhw}. 

\subsection{From Lagrangian to observables}\label{sec:actiontoamp}
A central question in our analysis of the complexity of QFTs concerns the relation between the complexity of the Lagrangian of the theory and the complexity of the observables, which can be quantitatively addressed using sharp o-minimality. We begin by analyzing the complexity of Feynman graphs in scalar theories.

\subsubsection*{Complexity of perturbation theory}
A natural expectation is that theories with a Lagrangian of higher complexity will have more complicated Feynman rules, and hence a more complicated perturbation theory. For example, consider a scalar QFT with an algebraic potential of complexity $(\cF,\cD)$. Every monomial in the potential yields a Feynman rule, and hence there are at most $(\cF+\cD)!/(\cF!\cD!)$ vertices in the theory. For potentials with only a few interactions, this estimate is sharpened by using the fewnomial representation.\\

The complexity of the perturbative expansion can be further probed by counting the number of diagrams $N_{\ell,n}$ for a fixed number of loops $\ell$ and external particles $n$. The growth of $N_{\ell,n}$ in $\ell$ and $n$ will depend on the format and degree $(\cF,\cD)$ of the Lagrangian. For algebraic Lagrangians, it is clear that increasing $\cF$ and $\cD$ yields a significantly faster growth of $N_{\ell,n}$. \\

Suppose now that we have a QFT with an analytic potential of finite complexity $(\cF,\cD)$. The potential then generically has infinitely many interaction terms when expanded as a power series, and perturbatively we therefore have infinitely many interactions. At any finite order in perturbation theory, the additional structure in the potential is hidden, only becoming fully visible non-perturbatively. However, there are situations in which a non-perturbative emergence of simplicity may furnish a uniform simplicity in the perturbative expansion. Recall the example from section~\ref{sec:emergentsimplicity} of the truncated exponential power series
\begin{equation}
    V_N(\phi) =  \sum_{k=0}^N \frac{1}{k!}\phi^k \,,
\end{equation}
with $\lim_{N\to\infty}V_N(\phi)=e^\phi$. As a degree $N$ polynomial this function has complexity $(1,N)$, but using the Pfaffian chain 
\begin{align}
    \zeta_1(\phi)& = \frac{1}{\phi}\,,  &&  \frac{\pd\zeta_1}{\pd \phi}=- \zeta_1^2 \,, \\
    \zeta_2(\phi)&=\phi^N\,,   && \frac{\pd\zeta_2}{\pd \phi}= N\zeta_1 \zeta_2\,, \nonumber\\ 
    \zeta_3(\phi)&=V_N(\phi) \,, && \frac{\pd\zeta_3 }{\pd \phi} = \zeta_3 -\frac{1}{N!}\zeta_2 \,, \nonumber
\end{align}
the complexity is reduced to $(4,5)$ independent of $N$. This phenomenon appears when the Pfaffian chain of the limiting function has a sufficiently low degree, so that the fewnomial representation can be used to define a Pfaffian chain for the truncated power series. \\

For analytic Lagrangians whose complexity is captured by a differential equation, the connection between the complexity of the Lagrangian and the complexity of the perturbative observables is subtle. A compelling idea is that these differential equations encode algebraic relations among the numerical coefficients appearing in the Feynman rules, which ultimately translate to algebraic relations among the terms contributing to the perturbative expansion of observables.

\subsubsection*{From Lagrangian to exact observables}
In the spirit of the emergence of simplicity discussed above, we expect the connection between the complexity of actions and exact observables to be clearer than the perturbative case. Remarkably, it is not always true that Lagrangians of high complexity yield observables of high complexity. This is illustrated in the discussion of \cite{Arkani-Hamed:2008owk}, where it was pointed out that, from the perspective of amplitudes, the simplest QFTs are those with a high amount of supersymmetry, such as $\cN=4$ super Yang-Mills and $\cN=8$ supergravity. The enormous symmetry groups of these theories result in miraculous simplifications in the amplitudes. On the other hand, the Lagrangians for these theories have a high complexity due to the complicated structure of supersymmetry, requiring a large spectrum of fields for the closure of the supersymmetry algebra.


\subsection{Renormalization group flow} A fundamental question is whether a lowering of the cutoff energy scale leads to a loss of information. Rephrased in the present setting, this question concerns the fate of QFT complexity along RG flow. Results such as the $c$-theorem \cite{Zamolodchikov:1986gt} and the $a$-theorem \cite{Komargodski:2011vj} indicate that RG flow is irreversible and therefore lead to a loss of information as measured by the central charge $c$ or anomaly coefficient $a$. In this section we study what happens to format and degree $(\cF,\cD)$, as defining another measure of information of a QFT, along RG flow.

\subsubsection*{Integrating out heavy fields} 
To start our discussion, consider a theory with several fields, one of which is considered a heavy field in the sense that its mass $m$ is much greater than the cut-off scale $\Lambda$. RG flow is then often implemented by integrating out the heavy field, thereby obtaining an effective Lagrangian for the remaining fields. This procedure reduces the number of degrees of freedom explicitly, at the expense of having more complicated interactions among the remaining fields. It will therefore be interesting to analyze what happens to $(\cF,\cD)$ when performing an integration over heavy fields. \\

In practice, one usually proceeds perturbatively and includes only finite loop corrections from integrating out the heavy field. In this approach, the complexity of the effective Lagrangian will depend on how many loops $\ell$ are included. In particular, with our previous discussion in mind, we see that the complexity of the effective Lagrangian may grow arbitrarily high by increasing $\ell$. 
To fully appreciate how complexity changes by integrating out a field, we therefore have to continue non-perturbatively and integrate out the field exactly. This clearly makes a general analysis a challenging task, and one can either proceed by considering the equations governing exact RG flow \cite{Rosten:2010vm}, or by explicitly analyzing tractable examples. \\

Here we will only study a simple model to get an intuition for what happens to complexity under integrating out a field. Consider a 0d QFT on a point with two scalars $\phi_1$ and $\phi_2$, described by the action 
\begin{equation}
    \cL(\phi,\chi) = \frac{1}{2} m^2 \phi^2 + \frac{1}{2} M^2 \chi^2 + \frac{\lambda}{4!} (\phi^2+ \chi^2)^2\,.
\end{equation}
The complexity of this theory is given by $(\cF,\cD)= (2,4)$. We assume that $m \ll M$ and now integrate out the heavy field $\chi$. This should lead to an effective Lagrangian $\cL_{\rm eff}(\phi)$ which preserves the partition function, that is, 
\begin{equation}
   \int \dd \phi \, e^{-\cL_{\rm eff}(\phi)} =  \int \dd\phi\, \dd\chi \, e^{-\cL(\phi,\chi)} \,.
\end{equation}
Since the path integral in this theory is exactly solvable, as we have seen in chapter~\ref{ch:observables}, we can evaluate the exact effective Lagrangian and find
\begin{align}
    \cL_{\rm eff}(\phi) &= \frac{1}{2} m^2 \phi^2 + \frac{\lambda}{4!}\phi^4 -\log\left[\int \dd\chi \, e^ {-\frac{1}{2} (M^2+ \lambda\phi^2/6 )\chi^2 + \frac{\lambda}{4!} \chi^2} \right]  \\
    & = 
     \frac{1}{2} m^2 \phi^2 + \frac{\lambda}{4!}\phi^ 4 -\log \left[  \sqrt{\frac{3}{\lambda}  \mu^2(\phi)    }  \exp(\frac{3\mu^4(\phi) }{4\lambda}) K_{1/4}\left( \frac{3\mu^4(\phi)}{4\lambda} \right)      \right] \,, \nonumber
\end{align}
where $\mu^2(\phi) = M^2 +\tfrac{1}{6}\lambda\phi^2$ is the effective squared mass of the $\chi$ field. The functions appearing in the effective Lagrangian can be written in terms of a Pfaffian chain, and the format and degree of the resulting theory is significantly higher than the original theory. In particular, although there are fewer variables contributing to the format, the amount of auxiliary functions needed to generate the effective dynamics for $\phi$ still leads to a higher value of $\cF$. \\

Although this is a simple example, we expect that this is a general phenomenon: integrating out a field exactly will generate complicated functions for the effective Lagrangian. This generically leads to an increase in degree, but the change in format is more subtle: there is a trade-off between the complexity loss of the integrated out fields and interactions of the UV Lagrangian, and the complexity gain due to the necessary functions needed for the IR Lagrangian. This situation may become particularly interesting upon integrating out a $N$ fields, or even an infinite tower of fields, since this may lead to an emergence of simplicity. Emergence of this type is known, for example, from the Schwinger one-loop computations of \cite{Gopakumar:1998ii,Gopakumar:1998jq} and is consistent with the emergence proposal put forward in \cite{Grimm:2018ohb,Heidenreich:2018kpg,Palti:2019pca}.

\subsubsection*{Exact RG flow equations} 
Ultimately, we are interested in the complexity of Lagrangians in $d\geq 1$. In this case, exact calculations are rare, and more powerful techniques are required. In the following we will discuss an approach using exact RG techniques. The resulting RG flows are notoriously difficult to solve, but here sharp o-minimality once again offers the advantage that the complexity of an object can already be extracted from an implicit description such as a differential equation, and an exact closed-form expression is not needed. \\

Consider a $d$-dimensional QFT describing a single scalar field $\phi$ with a cutoff energy scale $\Lambda_0$, with a Euclidean action
 \begin{equation}
     S_{\Lambda_0}[\phi] = \int \dd^ d x  \left(\frac{1}{2}(\pd_\mu \phi_k \, \pd^ \mu \phi_k + V_{\Lambda_0}(\phi) \right) \,,
 \end{equation}
 where $V_{\Lambda_0}(\phi)$ is the potential of the theory at the scale $\Lambda_0$. Suppose that we lower the energy scale to $\Lambda<\Lambda_0$ by integrating out all modes with energy between $\Lambda$ and $\Lambda_0$. Formally, the resulting effective potential $V_{\Lambda}(\phi)$ is obtained by performing the path integral over modes within this energy range. Denoting the lower-energy modes by $\phi$ and the modes with energy between $\Lambda$ and $\Lambda_0$ by $\chi$, the Wilsonian effective action becomes
  \begin{equation}\label{eq:wilsonpathint}
     S_{\Lambda}[\phi] = -\log \left[ \int D \chi \, e^ {-S_{\Lambda_0}[\phi+\chi]}   \right]\,,
 \end{equation}
 from which the effective potential $V_{\Lambda}(\phi)$ can be extracted. \\

To analyze the complexity of the effective potential, we continue in the local potential approximation (LPA), in which higher-derivative interactions generated by the path integral are projected out. This path integral can almost never be performed analytically, but upon lowering the scale infinitesimally the one-loop contribution becomes exact, and the effective potential obeys the non-perturbative Wegner-Houghton equation \cite{Wegner:1972ih}
\begin{equation}\label{eq:WegnerHoughton}
    \Lambda\frac{\pd}{ \pd \Lambda} V_{\Lambda}(\phi) = - \Lambda^d A_d \log\left(\Lambda^ 2 + \frac{\pd^2}{\pd\phi^2}V_{\Lambda}(\phi) \right) \,.
\end{equation}
Here $A_d =  \big((4\pi)^{d/2}\Gamma(d/2) \big)^ {-1}$ is an angular integration factor (see e.g.~\cite{Aoki:1996fn,Bonanno:1999ik} for a discussion).
The fact that an infinite tower of quantum corrections to the effective potential can be captured by a single differential equation suggests the compelling idea that there may be a finite-complexity description of the effective theory which is definable in an o-minimal structure. More precisely, we expect that a tame initial potential $V_{\Lambda_0}(\phi)$ flows under this equation to a tame effective potential $V_{\Lambda}(\phi)$ for $\Lambda<\Lambda_0$. The form of equation \eqref{eq:WegnerHoughton} suggests that the corresponding o-minimal structure could be the structure $\bbR_{\rm LN,Pf}$ obtained by taking the Pfaffian extension of the structure of log-Noetherian functions \cite{binyamini2024lognoetherianfunctions}.
This idea of conservation of tameness is further supported by the observation that RG flow behaves qualitatively similar to heat diffusion. Precise details on partial differential equations of this type and o-minimality are presently not known, and it is likely that a non-trivial mathematical extension is required to describe the tameness of exact RG flows via this method.

\section{Summary and discussion}
In this chapter we have shown that in order to quantify the complexity of a quantum field theory, whether one starts from the descriptive functions or the observables, one is naturally led to study the mathematical problem of assigning a well-defined complexity to functions and domains. The notion of complexity in sharp o-minimality defined by format and degree $(\cF,\cD)$ provides a solution to this problem and thereby defines a potential measure of complexity for quantum field theories. When implemented for Lagrangians, it reflects the number of degrees of freedom as well as the complexity of the interactions, and can yield a finite-complexity description even in the presence of infinitely many couplings. On the complexity of observables, we made several elementary observations regarding the growth of perturbative complexity, the emergence of simplicity in non-perturbative limits, and the relation with the complexity of the Lagrangian. \\

Through part II of this thesis we encountered several manifestations of tameness and finiteness of complexity in quantum field theories. However, these findings relied on assumptions on the structure of the theory, and settings which evade these assumptions challenge the idea that tameness may be a universal physical principle in quantum field theory. For example, if we forgo the assumption of renormalizability and consider general effective field theories, then we are forced to reckon with Lagrangians of infinitely many couplings. These Lagrangians are generally not definable in an o-minimal structure, unless an additional layer of structure is present. The space of quantum field theories therefore contains regions with theories of infinite complexity. While tameness of physical configurations and observables may arise as a consequence of the tameness of descriptive functions defining a theory, nothing seems to enforce that these descriptive functions are tame or have a finite complexity.

\subsubsection*{Towards gravity}
This situation may be resolved by one crucial physical aspect which we have not yet incorporated. Our universe includes gravity, so if we wish to focus on quantum field theories potentially describing our universe, we must include a dynamical gravitational field. This can only be done at an effective field theory level, describing physics below a finite energy cutoff. In this setting, we find a miraculous selection principle: the consistency with quantum gravity in the ultraviolet imposes powerful finiteness constraints, and appears to only admit effective field theories with finite complexity. These ideas will be uncovered in the next part of the thesis.

\setpartpreamble[u][\textwidth]{
\vspace*{1cm}
\hrulefill 
\vspace*{0.5cm}

The third and final part of this thesis is devoted to the emergence of tameness and finiteness of complexity in quantum gravity. We begin with the introductory chapter \ref{ch:string} which reviews aspects of string theory, the landscape, and the swampland. Chapter \ref{ch:volumes} is concerned with geometric aspects of moduli spaces in effective theories of quantum gravity, and proposes a novel characterization of their tame geometry in terms of tame isometric embeddings. Finally, chapter \ref{ch:EFTcomplexity} develops a notion of complexity for effective field theories, extending the discussion of chapter~\ref{ch:complexityQFT}, and culminates in the conjecture that this complexity is finite for theories consistent with quantum gravity.

\vspace*{0.5cm}
\hrulefill }

\part{Quantum gravity}\label{part4}

\chapter{Quantum gravity, string theory, and the swampland}\label{ch:string}
\setlength{\parindent}{0pt}
The aim of quantum gravity is to provide a single consistent framework into which all forces of nature are unified. 
The crux of this task is the unification of quantum field theory and general relativity. While there are strong candidates for such a theory, such as string theory, the full formulation of quantum gravity remains elusive. Nonetheless, significant progress can be made by studying these candidates or their manifestations in low-energy limits. 
At sufficiently low energy scales, any candidate theory of quantum gravity is expected to admit an effective field theory description.
At this effective level, we can therefore ask the following question:
\[
\textit{which quantum field theories are consistent with quantum gravity?}
\]
By consistent, we mean that the theory admits an ultraviolet completion to quantum gravity. The set of these effective field theories, which we call the \textit{quantum gravity landscape}, is the main concern of the following three chapters. While it is known that there is an enormous amount of effective theories in this landscape, it was realized that they are surrounded by a yet much larger \textit{swampland} of effective field theories which cannot be completed to quantum gravity \cite{Vafa:2005ui}. The consistency with quantum gravity imposes strong constraints on the permissible Lagrangians, matter spectra, symmetries, and field spaces, among which finiteness is a unifying theme. The formalization of these constraints provides an answer to the question above, and this endeavor is known as the swampland program. As we will see in this chapter, tame geometry emerges naturally in this setting, as formalized by the \textit{tameness conjecture} proposed in \cite{Grimm:2021vpn}. \\

This chapter serves as an introduction to part III, reviewing aspects of string theory, the quantum gravity landscape, and the swampland program, which are relevant for this thesis. The intention is not to provide a detailed review, but to summarize some of the main ideas in order to set the stage for chapter~\ref{ch:volumes} and chapter~\ref{ch:EFTcomplexity}.

\section{String theory and compactification}
\subsection{Bosonic strings}
We begin with a condensed introduction to string theory, with a focus toward the idea that low-energy limits of string theory produce an abundance of effective field theories consistent with quantum gravity.
We refer to \cite{BLT} for a detailed review. 
Quantum field theory rests on the assumption that the fundamental constituents of nature are points in spacetime, and the starting point of string theory is to go beyond this assumption. Therefore, string theory is essentially a theory of extended objects, among which the most basic ones are one-dimensional: the \textit{strings}.
Whereas the classical motion of a point particle can be thought of as a line in spacetime, the motion of a string forms a surface. In this way, the evolution of a string is described by the embedding $\Sigma\to M_d$ of a surface $\Sigma$, called the worldsheet, into a $d$-dimensional spacetime $M_d$. There is a natural action that can be associated to such a map, namely the area of $\Sigma$ as measured by the spacetime metric. From this point of view we obtain a field theory on the worldsheet $\Sigma$ whose degrees of freedom are the spacetime coordinates corresponding to the map $\Sigma \to M_d$. Since the spacetime coordinates are bosonic in nature, the theory formulated in this way is known as \textit{bosonic string theory}. A truly remarkable feature of bosonic string theory is that, as a quantum theory, it is only consistent for a single value of the spacetime dimension, namely $d=26$. \\

The natural way to proceed is to study the equation of motion for the string. The solutions to these equations constitute a spectrum of vibrational modes, which quantum-mechanically are interpreted as states of the string. The various states of the strings may effectively be described as fields on the spacetime $M_d$. These fields have a variety of masses and spins, and among them there is the \textit{graviton}, which is a massless spin-2 field that describes the propagation of gravity. This is a miraculous result, since it implies that a quantum theory of one-dimensional extended objects automatically incorporates gravity. This is one of the features that makes string theory so promising as a candidate theory of quantum gravity. 
There are however a number of severe problems with the bosonic string that make it unsuitable as a complete theory of nature. Firstly, its spectrum contains a tachyon, i.e.~a state with negative mass squared, which signals an instability in the theory. Secondly, there are no fermions, and a theory must have fermionic degrees of freedom for realistic matter to exist. It turns out that these problems can be simultaneously resolved by including fermions in the worldsheet field theory in a \textit{supersymmetric} way. 

\subsection{Superstrings}
Supersymmetry is a special type of symmetry which, unlike spacetime symmetries and conventional gauge symmetries, are generated by fermionic charges. The symmetry generated by these \textit{supercharges} interchanges bosonic and fermionic degrees of freedom, and is incredibly constrained by the consistency with the symmetries and spin structures of the underlying spacetime. The resulting symmetry is called supersymmetry. 
The amount of supersymmetry in a theory is measured by a number $\cN$, which counts the number of spinors whose components form the supercharges. 
Imposing a maximum spin of 2 on the degrees of freedom bounds the number of supercharges by 32, which corresponds to $\cN=8$ in four dimensions, and $\cN=2$ in ten dimensions. \\

The shortcomings of the bosonic string can be cured by including fermions and demanding that the worldsheet theory is supersymmetric. The resulting theory describes \textit{superstrings}. There are several consistent ways to implement the fermions in a supersymmetric way, depending on a handful of discrete choices one can make. However, in each case, the resulting superstring theory is only consistent in if the spacetime dimension is exactly $d=10$.

\begin{subbox}{Superstring theories}
There are five superstring theories in ten dimensions. These are known as
\[
\text{Type I}; \quad 
\text{Type IIA}; \quad 
\text{Type IIB}; \quad 
\text{Heterotic } \text{E}_8\times \text{E}_8; \quad 
\text{Heterotic }\text{SO}(32).
\]
\end{subbox}
For our purposes it is not important how exactly the various superstring theories are constructed, and we refer to \cite{BLT} for a detailed explanation. These five theories are not independent, but are related by a network of \textit{dualities} which hint towards the idea that they ultimately describe the same physics, unified by a mysterious hypothetical theory called \textit{M-theory}. \\

Upon quantization, each of the five superstrings comes with its own spectrum, and the states in the spectrum can effectively be described through fields on spacetime. The resulting supersymmetric field theories contain a gravitational field, in addition to various gauge fields and fermionic fields, and are known as \textit{supergravity} theories. Crucially, the superstring theories are free of tachyons.

\subsection{Compactification}
The apparent conflict in the number of dimensions observed in nature and required by superstring theory can be resolved by the idea of \textit{compactification}. In this construction, the spacetime $M_d$ is assumed to factor as
\begin{equation}
    M_d = M_4 \times X_D \,,
\end{equation}
where $M_4$ is a four-dimensional Lorentzian manifold representing the dimensions we observe, and $X_D$ is a compact manifold called the internal space, whose size is sufficiently small to escape detection at the length scales we are currently able to probe. Though $X_D$ is supposedly undetectable, the geometry of $X_D$ has enormous implications for the effective theory that is perceived on $M_4$. The ten-dimensional theory is said to be \textit{compactified on} $X_D$. For phenomenological purposes, we are mostly interested in compactifications where $M_4$ is a maximally symmetric space, i.e.~Minkowski space, de Sitter space (dS) or anti-de Sitter space (AdS). \\

\vspace{-5mm}

\subsubsection*{Circle compactification}\label{sec:circlecomp}
To illustrate the idea of compactification, it is instructive to consider an example which already highlights many of the general aspects. 
Consider a free massless scalar field $\phi$ defined on a five-dimensional manifold $M_5 = M_{4}\times S^1$ with metric 
\begin{equation}\label{eq:KKmetric}
g_{\hat \mu\hat\nu}\,\dd x^{\hat\mu} \, \dd x^{\hat\nu}  = \eta_{\mu\nu} \,\dd x^\mu \,\dd x^\nu  + R^2 \dd\theta^2 \,,
\end{equation}
where $R$ is the radius of the circle, and $x^{\hat\mu}=(x^\mu,\theta)$ with $\theta\sim\theta+2\pi$. The compactness of the circle factor of $M_5$ implies that any field $\phi(x,\theta)$ admits a discrete mode expansion
\begin{equation}\label{eq:KKexp}
    \phi(x,\theta) = \sum_{n\in\bbZ}\phi_n(x) \, e^{i n \theta} \,.
\end{equation}
The equation of motion for $\phi$ and the resulting equation for $\phi_n$ takes the form
\begin{equation}\label{eq:KKeom}
    \pd_\mu \pd^\mu \phi + \frac{1}{R^2}\pd_\theta^2 \phi=0 \,, \qquad
\Big(\pd_\mu \pd^\mu -\frac{n^2}{R^2}\Big)\phi_n(x) =0 \,.    
\end{equation}
In this context, the expansion in equation \eqref{eq:KKexp} is called the Kaluza-Klein (KK) expansion and the four-dimensional fields $\phi_n$ are called the KK modes. Equation \eqref{eq:KKeom} shows that each KK mode $\phi_n$ obeys the equation of motion of a massive scalar on $M_{4}$ with mass $n/R$. This shows that the physics in four dimensions is determined by the geometry of the compactification, in this case the radius $R$. \\

This theory can be coupled to gravity by promoting $g_{\hat\mu\hat\nu}$ to a dynamical field and including an Einstein-Hilbert term in the action. Since the radius $R$ appears as a component of the metric, it becomes dynamical as well, which in particular imlpies that $R$ becomes a field whose value may vary on the four-dimensional spacetime $M_{4}$.
The four-dimensional equation of motion for $R$ can be obtained by expressing the five-dimensional Ricci scalar in terms of the Ricci scalar on $M_4$, and one finds that $R$ is a massless free scalar field taking values in the space $\bbR^+$. This is an example of a \textit{modulus}: a parameter encoding the internal geometry of the compactification which appears as a free massless scalar field in the lower-dimensional theory. \\

Here we have considered the compactification of a field theory,
but ultimately we are interested in compactifications of string theory, which yield new physical phenomena. In addition to KK modes describing the momenta of the string in the internal space, there is a tower of states which describes the winding of the string around the circle. The mass of these states is proportional to the winding number, and scales positively with the radius $R$. 

\subsubsection*{Calabi-Yau compactification}
In the case of a one-dimensional compactification, the geometry is nearly uniquely determined to be a circle. For the case at hand, where six dimensions have to be compactified, there is an enormous range of possibilities for the geometry of the internal space $X$. Upon demanding that the lower-dimensional theory preserves an amount of supersymmetry in four dimensions, the permissible internal geometries are strongly constrained. In particular, in order to obtain $\cN=1$ supersymmetry, the theory must be compactified on a \textit{Calabi-Yau manifold}. These manifolds are defined as compact K\"ahler manifolds with a vanishing canonical line bundle, and their link to supersymmetry lies in their holonomy groups. Whereas the geometry of a circle is described by a single modulus, Calabi-Yau manifolds typically admit many continuous deformations. For each Calabi-Yau manifold with a fixed topology, the deformation parameters are then assembled into a \textit{moduli space} whose dimensionality is determined by the structure of the cohomology groups. The lower-dimensional effective Lagrangian contains the moduli as massless fields, and their interactions are controlled by the geometry of the Calabi-Yau manifold and its moduli space \cite{BLT}.

\subsubsection*{The string landscape}
The choices of internal space, as well as additional structures on these spaces such as branes and fluxes, result in an enormous range of consistent backgrounds of superstring theory. These backgrounds are called string vacua and constitute what has become known as the \textit{string landscape}. Each string vacuum yields a low-energy effective field theory in four dimensions, whose structure depends on the details of the compactification procedure. Whereas there are only a few consistent formulations of superstring theory in ten dimensions, the four-dimensional situation is vastly different and comes with a huge variety of quantum field theories consistent with quantum gravity.

\section{Quantum gravity and the swampland}
The magnitude of the quantum gravity landscape may suggest that essentially any gravitational effective field theory admits a UV-completion to quantum gravity through a string compactification. However, it was realized that these effective theories are exceptional rather than generic, and that the set of effective field theories inconsistent with quantum gravity, called the \textit{swampland}, is significantly larger \cite{Vafa:2005ui}. From this observation the swampland program emerged, which aims to characterize which physical principles distinguish the landscape from the swampland. 
The methodology of the swampland program essentially consists of three aspects: 
\begin{itemize}
    \item \textbf{Conjectures.} Identifying patterns in effective field theories consistent with quantum gravity. These patterns may be promoted to conjectural universal principles, known as swampland conjectures. 
    \item \textbf{Evidence.} Finding evidence for the conjectures and establishing connections between them. The evidence comes from various complementary perspectives, including for instance top-down string theory constructions and bottom-up black hole arguments.
    \item \textbf{Consequences.} Deriving phenomenological implications from the conjectures. With sufficient evidence, swampland conjectures may provide predictions for cosmology and particle physics beyond the standard model. 
\end{itemize}
To give an impression of the swampland program and a primer for the following chapters, we now briefly present some examples of swampland conjectures, and refer to \cite{Palti:2019pca,vanBeest:2021lhn,Agmon:2022thq} for detailed reviews. 

\subsection{Swampland conjectures}
\subsubsection*{Global symmetries in quantum gravity}
The first swampland conjecture that we discuss concerns global symmetries in quantum gravity, and is the oldest and most established of the conjectures \cite{Banks:1988yz,BanksSeiberg}.

\begin{subbox}{No-global-symmetries conjecture}
There are no global symmetries in a theory of quantum gravity; any symmetry must be either broken or gauged. 
\end{subbox}

Since symmetries play a central role in physics, this is a deep statement about the nature of quantum gravity and its low-energy effective field theories. The strongest evidence for this conjecture comes from arguments based on semi-classical properties of black holes (see e.g.~\cite{Palti:2019pca}). In addition, the conjecture is supported by arguments from string theory and holography, and this combination of bottom-up and top-down evidence gives the conjecture a central position in the swampland program. At the level of effective theories, the conjecture can be understood in terms of certain geometric interpolations between effective theories, captured by the cobordism conjecture \cite{McNamara:2019rup}.

\subsubsection*{Geometry of field spaces and the distance conjecture}
The distance conjecture is a statement about the asymptotic regions of field spaces in effective theories consistent with quantum gravity \cite{Ooguri:2006in}.

\begin{subbox}{Distance conjecture}
Consider an effective field theory consistent with quantum gravity with a moduli space $\cM$ parametrized by vacuum expectation values of scalar fields $\phi$. Then for any point $\phi_0\in\cM$, there exist points $\phi$ at arbitrarily large distance $d(\phi,\phi_0)$ from $\phi_0$. In addition, when the moduli approach these infinite distance limits, there is an infinite tower of states with mass scale $m_{\rm t}(\phi)$ scaling as
\begin{equation}
    m_{\rm t}(\phi) \sim m_{\rm t}(\phi_0) \,e^{-\alpha d(\phi,\phi_0)}
\end{equation}
for a certain positive constant $\alpha>0$.
\end{subbox}

The distance conjecture is thus a statement about the geometry of field spaces in quantum gravity, whose metric is determined by the kinetic term for the moduli appearing in the effective Lagrangian. Its physical significance is that it implies that effective field theories consistent with quantum gravity can only access a finite region of moduli space, since the appearance of infinite towers of light states signals the breakdown of the effective field theory description. These infinite towers can already be explicitly seen in the simplest string theory examples, such as the circle compactification discussed in the previous section. In this setting there are two infinite distance limits corresponding to the radius of the circle going to infinity or zero, and the associated towers of states are the Kaluza-Klein modes and the winding modes of the string. The distance conjecture is understood in great detail in the setting of Calabi-Yau compactifications, where Hodge theory provides powerful mathematical tools for understanding the geometry of moduli spaces \cite{Grimm:2018ohb}. For an overview of the evidence for the conjecture and a discussion of its phenomenological implications we refer to \cite{Palti:2019pca}.

\subsubsection*{Finiteness in quantum gravity}
It was already observed in the inception of the swampland program that \textit{finiteness} is an essential principle in determining which effective theories are consistent with quantum gravity \cite{Vafa:2005ui}. The idea of finiteness threads through many of the swampland conjectures and manifests itself in various distinct but interrelated ways. Below, we summarize the main finiteness conjectures on effective theories of quantum gravity.

\begin{subbox}{Finiteness conjectures}
\begin{itemize}
    \item There are finitely many distinct effective field theories consistent with quantum gravity \cite{Vafa:2005ui,Acharya:2006zw,Hamada:2021yxy}.
    \item The matter spectrum of effective field theories consistent with quantum gravity is finite and constrained by bounds on the number of fields and ranks of gauge groups \cite{Vafa:2005ui,Kumar:2010ru,Lee:2019skh,Kim:2019vuc,Katz:2020ewz,Tarazi:2021duw,Martucci:2022krl,Kim:2024eoa,Birkar:2025rcg}. 
    \item All physical amplitudes in quantum gravity are finite \cite{Hamada:2021yxy}.
    \item The accessible region in the field space of effective field theories consistent with quantum gravity has a finite volume \cite{Vafa:2005ui,Ooguri:2006in,Delgado:2024skw}.
\end{itemize}
\end{subbox}

For the moment we leave a precise explanation of these statements aside; this list serves mostly to illustrate that finiteness plays a key role in formulating quantum gravity constraints. In the following two chapters, we will encounter these conjectures in more detail.

\subsection{The tameness conjecture}
From the principle of finiteness in the swampland program, the geometric nature of effective theories in the landscape, and the appearance of o-minimality in quantum field theory, the natural question emerges whether there is a connection between tame geometry and quantum gravity. This connection turns out to be  remarkably strong. Originating from the observation that large classes of string vacua can be described with tame geometry, the following conjecture was proposed \cite{Grimm:2021vpn}.

\begin{subbox}{Tameness conjecture}
For any effective field theory consistent with quantum gravity, the scalar field space, the coupling functions, and parameter space are definable in an o-minimal structure. 
\end{subbox}

This conjecture asserts that o-minimality, as a generalized notion of finiteness, provides a mathematically rigorous unifying perspective on finiteness in the quantum gravity landscape. 
In contrast with the situation in chapter~\ref{ch:complexityQFT}, where we noted that the space of quantum field theories generally admits theories with non-tame Lagrangians, the conjecture claims that quantum gravity only permits tame Lagrangians in low-energy limits, and that the corresponding space of effective theories itself is tame as well. In the remainder of this section we briefly review the established sources of evidence for this conjecture, and in the two upcoming chapters we will discuss new evidence and connections to other swampland conjectures.

\subsubsection*{Finiteness of flux vacua}
The strongest evidence for the tameness conjecture comes from counting vacua in string compactifications. It was shown that certain classes of vacua arising in flux compactifications of Type IIB string theory\footnote{The precise result is formulated for \textit{F-theory}, a 12-dimensional geometrization of Type IIB string theory, which must be compactified on a Calabi-Yau fourfold in order to obtain a four-dimensional effective theory.} are parametrized by a set which is definable in the o-minimal structure $\bbR_{\rm an,exp}$ \cite{Grimm:2021vpn,Bakker:2023xkt}. This set of vacua therefore has finitely many connected components, which shows that there are no infinite discrete collections of vacua in this region of the quantum gravity landscape. This remarkable result provides a first instance of the use of o-minimality to address a long-standing finiteness conjecture in physics. The proof  relies on the fact that this set of string vacua can be represented as a counting problem of certain cohomology classes, which can be shown to be definable in $\bbR_{\rm an,exp}$ using the tameness of the period map in Hodge theory.
We refer to \cite{Grimm:2021vpn} for the physical details of this argument, and to \cite{Bakker:2023xkt} for the mathematical proof.

\subsubsection*{Tameness of effective Lagrangians}
A second major source of evidence for the tameness conjecture comes from the effective Lagrangians arising from compactifications of Type IIB string theory. The coupling functions in the corresponding Lagrangians, such as the field space metric, the gauge couplings, and the scalar potential, have a geometric origin which can be represented in terms of the periods of the underlying Calabi-Yau manifold. The tameness of the period map in Hodge theory then again implies that these functions are tame, demonstrating the tameness of coupling functions in a large class of effective theories consistent with quantum gravity.

\subsubsection*{Symmetries and tameness}
The absence of global symmetries in quantum gravity can be linked to tame geometry in the following way. The supergravity theories arising from string compactifications are usually accompanied by a discrete duality group encoding a symmetry in the Lagrangian and the spectrum of the theory. According to the no-global-symmetries conjecture, these symmetries must be broken or gauged, and in these theories this is accomplished by performing a quotient by the duality group, identifying configurations and states related by a symmetry operation.
By reducing to these duality quotients, periodicity is eliminated from field spaces and coupling functions. Since periodicity on unbounded domains is incompatible with o-minimality, it thus appears that the absence of global symmetries in quantum gravity protects tameness by enforcing duality quotients.\\

This concludes our brief review of finiteness and tameness in quantum gravity. In the next two chapters, we will dive deeper into certain aspects of the tameness conjecture, extending the geometric constraints on field spaces in chapter~\ref{ch:volumes} and implementing the complexity of sharply o-minimal structures in chapter~\ref{ch:EFTcomplexity}. \\

\chapter{Tame geometry of moduli spaces}
\label{ch:volumes}
\setlength{\parindent}{0pt}

In the previous chapter we have discussed how quantum gravity poses strong constraints on effective theories, and how tameness appears to emerge naturally among these constraints. The focus of this chapter is to analyze the tameness of moduli spaces of effective theories in the landscape in more detail. The chapter is split into two distinct parts. Section \ref{sec:highsugra} contains a detailed analysis of moduli spaces of supergravity theories with more than eight supercharges, and presents new results on the tameness of the topology and Riemannian geometry of these spaces in relation with quantum gravity. In section \ref{sec:embeddings}, we propose a new principle for all moduli spaces in effective theories of quantum gravity, which essentially states that these moduli spaces must admit a \textit{tame isometric embedding} into Euclidean space. We discuss how this proposal relates to recent ideas on the emergence of dualities and the complexity of moduli spaces.

\section{Moduli spaces in higher supergravity}
\label{sec:highsugra}
The goal of the following section will be to analyze the geometry of the moduli space in theories with more than eight supercharges. These theories have half-maximal or maximal supersymmetry, and we will collectively refer to them as \textit{higher supergravity} theories. Since these theories have no scalar potential, the moduli space coincides with the scalar field space, and we will use both terms interchangeably.

\subsection{Supersymmetry and symmetric spaces}
Our first step in understanding the scalar field space $\cM$ in higher supergravity theories will be to consider its local geometry, characterized by the universal cover $\widetilde \cM$ of $\cM$. 
It can be shown that the presence of more than eight supercharges in the theory requires $\widetilde\cM$ to be a symmetric space \cite{Ferrara:2008de,Cecotti1,FreedmanVanProeyen}. In general, symmetric spaces take the form $\widetilde \cM = G/K$, where $G$ is the connected component of the isometry group of $ \cM$, and $K$ is the maximal compact subgroup of $G$. \\

The scalar fields of a higher supergravity theory thus parametrize a quotient space $G/K$, which already reveals a great amount of structure of the theory. The group $G$ plays the role of a global symmetry group of the theory which is realized classically. By this we mean that the classical equations of motion of the theory are invariant under an action of $G$, but that this invariance is broken at the quantum level. The fields appearing in the supergravity Lagrangian are organized in representations of $G$.  Higher supergravity does not only require $\widetilde\cM$ to be a symmetric space $G/K$, but it also fixes the group $G$ almost uniquely upon specifying the spacetime dimension $d$ and amount of supersymmetry $\cN$. This is illustrated in Table \ref{tab:sugra} for the spacetime dimensions $d=4$, $6$, and $10$. The only non-uniqueness comes from an integer $k$, which typically counts the number of additional vector multiplets coupled to the theory. As a familiar example, the bottom entry of the table corresponds to Type IIB supergravity in ten dimensions. In this theory there are two real scalars, namely the axion and the dilaton, which are combined into the axio-dilaton $\tau$ taking values in the field space $\bbH = \text{SL}(2,\bbR)/\text{SO}(2)$. \\

\begin{table}[h]
\label{tab:sugra}
\begin{center}
\begin{tabular}{@{}  *5l  @{}}    
$d$  & $\mathcal{N}$ & $G$ & $K$    \\ \hline
4    &  3 &  $\text{SU}(3,k)$ & $\text{SU}(3)\times \text{U}(k)$\\ 
4    &  4 &  $\text{SL}(2,\bbR) \times \text{SO}(6,k)$ & $\text{SO}(2)\times \text{SO}(6)\times \text{SO}(k)$ \\ 
4    &  5 &  $\text{SU}(5,1)$  & $\text{U}(5)$ \\ 
4    &  6 &  $\text{SO}^*(12)$  & $\text{U}(6)$  \\ 
4    &  8 &  $\text{E}_{7(7)}$  & $\text{SU}(8)$  \\ 
6    &  (2,2) & $\bbR\times \text{SO}(4,k)$ & $\text{SO}(4)\times \text{SO}(k)  $ \\ 
6    &  (4,0) & $\text{SO}(5,k)$ & $\text{SO}(5)\times \text{SO}(k)$  \\ 
6    &  (4,2) & $\text{SO}(5,1)$ & $\text{SO}(5)$  \\ 
6    &  (6,0) & $\text{SU}^*(6)$ & $\text{Sp}(6) $ \\ 
6    &  (4,4) & $\text{SO}(5,5)$ & $\text{SO}(5)\times \text{\text{SO}}(5) $  \\ 
6    &  (6,2) & $\text{F}_{4(4)}$ & $\text{Sp}(6) \times \text{SU}(2) $  \\ 
6    &  (8,0) &  $\text{E}_{6(6)}$ & $\text{Sp}(8)$ \\ 
10    &  (1,0) &  $\bbR$ & -\\ 
10    &  (1,1) & $\bbR$  & - \\ 
10    &  (2,0) & $\text{SL}(2,\bbR)$ & $ \text{SO}(2)$  \\ 
\end{tabular}
 \caption{\label{tab:sugra}Classical global symmetry groups $G$ for higher supergravity theories and their compact subgroup $K$ for various choices of dimension $d$ and amount of supersymmetry $\cN$.}
 \end{center}
 \end{table}

The most general space that preserves the local field space geometry determined by $\widetilde\cM$ is a quotient of the form 
\begin{equation}
    \cM  = \Gamma \backslash \widetilde\cM \,,
\end{equation}
where $\Gamma$ is a discrete subgroup of the isometry group of $\cM$ \cite{Cecotti1}. Since $\widetilde\cM$ is a symmetric space isomorphic to $G/K$, it follows that the field space in a higher supergravity is a double quotient space of the form $\cM = \Gamma\backslash G/K$. Spaces of this form have various interesting appearances in mathematics and physics, for instance as the target space of period maps in Hodge theory \cite{Schmid}. \\

The quotient by the discrete subgroup $\Gamma$ has two interpretations. One of them is the geometric interpretation given above, in which the unspecified global geometry of $\cM$ allows for a quotient by a discrete subgroup. The second interpretation is physical and originates from quantizing the theory. At the quantum level, only symmetry transformations which preserve the Dirac quantization condition are allowed. The largest group preserving this condition is $G_\bbZ$, and this breaks the classical global symmetry group $G$ to a discrete subgroup $\cU\subseteq G_\bbZ$, which is often called the U-duality subgroup \cite{Trigiante}. In the setting of string theory, these U-dualities are expected to be exact symmetries of the quantized theory \cite{Uduality}. Taking a quotient $G/K\to \Gamma \backslash G/K$ by a subgroup $\Gamma\subseteq \cU$ amounts to gauging a subgroup of U-dualities. This notion of gauging should be understood as identifying scalar field values which lie in the same $\Gamma$-orbit.

\subsection{Symmetries, arithmetic quotients, and the swampland}
The precise nature of the group $\Gamma$ will reveal another link between tameness and the swampland program. Supersymmetry alone does not constrain $\Gamma$, other than that it should be a discrete subgroup of $G$. However, if we assume that the supergravity theory under consideration is consistent with quantum gravity, it is constrained in the following way. \\

First, we will need the completeness hypothesis, which states that an effective field theory in the landscape must have states of all possible charges consistent with Dirac quantization \cite{BanksSeiberg}. In \cite{Cecotti1} it was argued that this implies that the U-duality group $\cU$ is equal to the full group $G_\bbZ$, which is the largest group consistent with Dirac quantization. This allows a complete set of charges to be reached by U-duality transformations. In the setting of higher supergravities, the U-duality group $\cU$, which we now take to be equal to $G_\bbZ$, is the global symmetry group of the theory. In fact, it is important to note that the U-duality group is an exact symmetry of the quantum theory which is believed to extend to a UV-completion in string theory. Recall that the no-global-symmetries conjecture asserts that any apparent global symmetry must be broken or gauged \cite{BanksSeiberg}. A weaker variant of this conjecture claims that global symmetries may be present, but that the global symmetry group must be finite. As noted in \cite{Cecotti1}, consistency with the weaker no-global-symmetries conjecture now implies that almost all of $G_\bbZ$ must be gauged. Here the notion of `almost all' is made precise by requiring that the gauged subgroup $\Gamma\subseteq G_\bbZ$ must be such that the remaining global symmetry group, given by the quotient $G_\bbZ/\Gamma$, is finite. This means that we require that $\Gamma$ is a subgroup of finite index in $G_\bbZ$, which implies that  $\Gamma$ is an \textit{arithmetic} subgroup of $G$. \\

In this case, the double quotient $\Gamma \backslash G/K$ is a type of space known as an arithmetic quotient. These spaces play a central role in various areas of mathematics, and it was recently proven in \cite{BKT} that these spaces are definable in $\bbR_{\rm alg}$ and hence tame. The challenge in establishing the tameness of these spaces lies in the fact that $\Gamma$ is an infinite discrete object. The challenge of establishing tameness then lies in finding a definable fundamental set for the action of $\Gamma$. A general procedure for constructing $\bbR_\text{alg}$-definable fundamental sets for these spaces was found in \cite{BKT}, thereby proving that arithmetic quotients have a tame geometry. In conclusion, we have argued for the following statement.

\begin{subbox}{Tameness of moduli spaces in higher supergravity}
Assuming the completeness hypothesis and the absence of global symmetries in quantum gravity, the scalar field space in a supergravity theory with more than eight supercharges is an arithmetic quotient $\Gamma\backslash G/K$, which is definable in $\bbR_{\rm alg}$ and therefore has a tame geometry.
\end{subbox}

\subsection{Field space metric and gauge couplings}

In the previous subsection we have argued for the tameness of the field space of higher supergravity theories arising from quantum gravity. We now extend this analysis to also include the functions on the field space which appear in the effective Lagrangian of the theory, in particular the field space metric and the gauge coupling functions. The tameness of the metric will be especially interesting for analyzing the tame geometry of moduli spaces as Riemannian manifolds, and will be central to section \ref{sec:embeddings}.

\subsubsection*{Tameness of the invariant metric}
The kinetic term for the scalar field sector of an effective field theory takes the form
\begin{equation}
    \cL = -\frac{1}{2}\cG_{ij}(\phi) \, \pd_\mu \phi^i \,\pd^\mu \phi^j \,,
\end{equation}
where $\phi^i$ are the scalars parametrizing the field space $\cM$, and the coupling function $\cG_{ij}$ is a Riemannian metric on $\cM$. As discussed earlier in this section, the field space is of the form $\cM=\Gamma\backslash G/K$ in a higher supergravity theory. The tameness conjecture now claims that $\cG_{ij}$ is definable in an o-minimal structure. In the present context, we are thus led to consider the tameness of the metric on an arithmetic quotient. 
Here the tameness of a Riemannian metric is defined through its component functions, as explained in subsection~\ref{sec:tameDG}.
In proving that the metric $\cG_{ij}$ is tame, we provide two perspectives. We begin by giving a physical argument for why the metric is tame, making use of standard formulas in supergravity. We then proceed with a more general mathematical argument, which shows that, in fact, any invariant tensor on $\Gamma\backslash G/K$ is tame. \\

Before we begin, let us slightly refine the setting in which we work. The field space $\Gamma\backslash G/K$ is not necessarily smooth, since the discrete group $\Gamma$ may act non-freely on the homogeneous space $G/K$. However, the observation that $\Gamma\subseteq G_\bbZ$ and that $K$ is compact implies that the stabilizers of this action are finite, so that the singularities of $\cM$ are of orbifold type, admitting a finite smooth covering. In \cite{CecottiDomestic}, a more detailed discussion of the singularities of the field space is given, where it is proposed that one can work with a finite smooth cover of $\cM$. We will instead work around the singularities in a simple way by focusing on the non-singular part\footnote{It is worth emphasizing that the non-singular part $\cM_\text{s}$ is still a tame manifold, since the set of singularities which is removed is definable. Likewise, passing to a finite smooth covering also preserves tameness, since the quotient is finite.} $\cM_\text{s} \subseteq \cM\setminus \cM_\text{sing}$. This will ensure that the metric is well-defined. \\

The metric $\cG_{ij}$ on the arithmetic quotient $\Gamma\backslash G/K$ originates from the invariant metric on the group $G$. Let us therefore start by discussing the metric on $G$ itself, which we denote by $\widetilde\cG_{ij}$. Recall that $G$ is a matrix Lie group, which implies in particular that $G$ is definable in the o-minimal structure $\bbR_\text{alg}$. Consider an $\bbR_\text{alg}$-definable coordinate patch $U\subseteq G$. The definable coordinates $\phi^i$ parametrize a matrix $\cV(\phi)\in G$. In terms of the matrix $\cV(\phi)$, the metric is given by \cite{Cecotti1}
\begin{equation}\label{eq:metricG}
\widetilde \cG= \widetilde \cG_{ij} \,\dd\phi^i \,\dd\phi^j  =  -\lambda \, \text{tr}[(\cV^{-1}\pd_i \cV)(\cV^{-1}\pd_j\cV)] \, \dd\phi^i \, \dd\phi^j \,,
\end{equation}
where the indices $i,j=1,\ldots, k$ refer to the scalar fields (e.g.~$\pd_i \cV \equiv \pd \cV(\phi)/\pd \phi^i$). Here $\lambda$ is an overall normalization constant, which is fixed when the theory is obtained from a string theory compactification. This metric is invariant under global $G$-transformations of the form $\cV(\phi)\mapsto \Lambda \cV(\phi)$. Equation (\ref{eq:metricG}) shows that the invariant metric on $G$ is $\bbR_\text{alg}$-definable, since matrix inversion, differentiation, and taking the trace are operations that preserve definability. \\

The next step is to consider the metric on the quotient space $G/K$, which we will also denote by $\widetilde \cG_{ij}$. It is convenient to model objects in this space as objects on $G$ for which the group $K$ acts as a gauge symmetry. This is not a gauge symmetry that is associated to any propagating degrees of freedom, but only a procedure to remove the redundant dependence on $K$. The Lie algebra $\mathfrak{g}$ of $G$ decomposes as $\fg=\fp\oplus \fk$, where $\fk$ is the Lie algebra of the maximal compact subgroup $K$ and $\fp$ is a complementary Lie algebra, orthogonal to $\fk$ with respect to the Killing form on $\fg$. The form $\cV^{-1}\pd_i \cV$ on $G$ is $\fg$-valued, and to obtain a metric for $G/K$ we project out the $\fk$-component via a linear projection map $(\cdots)_\fp:\fg \to \fp$. In terms of a matrix parametrization $\cV(\phi)$, the metric on $G/K$ is then given by
\begin{equation}\label{eq:metricGK}
\widetilde \cG= \widetilde \cG_{ij} \,\dd\phi^i \,\dd\phi^j  =  -\lambda \, \text{tr}[(\cV^{-1}\pd_i \cV)_\fp(\cV^{-1}\pd_j\cV)_\fp] \, \dd\phi^i \, \dd\phi^j. 
\end{equation}
From this we conclude that the metric on the homogeneous space $G/K$ is definable as well, since linear projections are always definable by the axioms of o-minimal structures. \\

We note that it is possible to formulate the metric on $G/K$ in an equivalent way that avoids the need to project down to the Lie algebra $\fp$ \cite{SamtlebenLec}. The idea behind this formulation is to work with manifestly $K$-invariant objects. If $\eta$ is a $K$-invariant matrix, i.e. $\eta = k\,\eta\, k^\text{T}$ for any $k\in K$, then the matrix $\cW(\phi)=\cV(\phi) \,\eta\, \cV(\phi)^\text{T}$ is invariant under local $K$-transformations $\cV \mapsto \cV k(\phi)$. A calculation shows that, expressed using the matrix $\cW(\phi)$, the metric in (\ref{eq:metricG}) becomes
\begin{equation}\label{eq:metricGK1}
    \widetilde \cG = \frac{\lambda}{4}\, \text{tr}[(\pd_i \cW) (\pd_j \cW^{-1})] \, \dd\phi^i\, \dd\phi^j.
\end{equation}
Let us illustrate this for the familiar example of $G/K=\text{SL}(2,\bbR)/\text{SO}(2)$. We can parametrize an arbitrary element in $\text{SL}(2,\bbR)$ by using the decomposition
\begin{equation}
    \cV(t,s,\theta) = \begin{pmatrix}
t & 0 \\
0 & t^{-1} 
\end{pmatrix}
\begin{pmatrix}
1 & s \\
0 & 1 
\end{pmatrix}
\begin{pmatrix}
\cos(\theta) & \sin(\theta) \\
-\sin(\theta) & \cos(\theta) 
\end{pmatrix}\,,
\end{equation}
with $t\neq 0$. The maximal compact subgroup $K=\text{SO}(2)$ preserves the standard Euclidean metric, so the matrix $\cW=\cV \, \cV^\text{T}$ takes the form
\begin{equation}
    \cW(t,s) = \begin{pmatrix}
(1+s^2)t^2 & s \\
s & t^{-2} \\
\end{pmatrix}\,.
\end{equation}
Introducing the complex coordinate $\tau=st^2+i\,t^2\in \bbH$, this matrix becomes 
\begin{equation}
    \cW(\tau) = \frac{1}{\text{Im}\,\tau }\begin{pmatrix}
|\tau|^2 & \text{Re}\,\tau \\
\text{Re}\,\tau  & 1
\end{pmatrix}\,,
\end{equation}
which is precisely the parametrization for the $\text{SL}(2,\bbR)/\text{SO}(2)$ quotient often encountered in the supergravity literature. Using equation (\ref{eq:metricGK1}), the resulting metric is the familiar hyperbolic metric for the upper half plane,
\begin{equation}
\widetilde \cG = \frac{1}{(\text{Im}\,\tau)^2} \dd\tau \, \dd\bar{\tau} \,.
\end{equation}

The final step is to take the quotient by the discrete group $\Gamma$, and consider the metric $\cG$ on $\Gamma\backslash G/K$. In view of the discussion above, we assume that $\Gamma$ is an arithmetic subgroup of $G$ so that the field space $\cM=\Gamma\backslash G/K$ is tame. Since the quotient is discrete, the metric $\cG$ is locally identical to the metric $\widetilde \cG$ on $G/K$. This already indicates that the metric is tame, but proving this requires a more mathematical perspective which we discuss below.

\subsubsection*{Tameness of invariant tensor fields}
Let us now provide a more geometric point of view, from which the tameness of this invariant metric can be understood rigorously. Recall that any Lie group comes with a set of left-invariant vector fields, which are defined as follows. Every element $g\in G$ defines a diffeomorphism $L_g:G\to G$ by left translation by $g$. The left-invariant vector fields on $G$ are characterized by
\begin{equation}
v(gh)= (L_g)_{*h}v(h) \,,
\end{equation}
and are in one-to-one correspondence with Lie algebra elements $X\in \fg$ via
\begin{equation}
v_X(g)=(L_g)_{*e} X \,.
\end{equation}
Here $f_*$ denotes the pushforward along a smooth map $f$. We will now show that every left-invariant vector field is definable in $\bbR_\text{alg}$. To do so, fix an element $X\in \fg$. Since $G$ is a matrix Lie group, the group multiplication map $G\times G\to G$ is $\bbR_\text{alg}$-definable, and by differentiating with respect to the second factor along the vector $X$, we find that the left-invariant vector fields $v_X$ are definable, since differentiation preserves differentiability. \\

If we pick a basis $X_1,\ldots,X_m$ of the Lie algebra $\fg$, we thus obtain a definable global frame $v_{X_1},\ldots,v_{X_m}$ of the tangent bundle $TG$. It follows that there is an $\bbR_\text{alg}$-definable global trivialization
\begin{equation}
TG \to G\times \fg \,.
\end{equation}
By repeating this construction for higher rank tensors, any tensor bundle on $G$ can be globally trivialized by a definable map. From the point of view of this trivialization, a left-invariant tensor field $\cT:G\to T^{p,q}G$ is just constant, i.e.~a map
\begin{align}
G &\to G\times \fg^{ p} \otimes (\fg^*)^{ q} \\
g & \mapsto (g,\cT_0)
\end{align}
for some fixed element $\cT_0\in \fg^{ p} \otimes (\fg^*)^{ q}$. We have thus shown that any left-invariant tensor field on $G$ is $\bbR_\text{alg}$-definable. This holds in particular for the case of interest, namely the invariant metric $\widetilde \cG$. \\

We now turn to invariant tensors on the quotient $G/K$. Geometrically, these tensor fields on $G/K$ are sections of the tensor bundle $T^{p,q} (G/K)$, so the first step is to understand the structure of the tangent bundle $T(G/K)$. The tangent bundle on $G/K$ can be interpreted as an associated vector bundle to the principal $K$-bundle $G\to G/K$ and the adjoint representation\footnote{More precisely, the representation is induced by the restriction of the adjoint representation of $G$ on $\fg$ to $K$.} of $K$ on $\fg/\fk$ \cite{ParGeom}. In other words, we have the identification
\begin{equation}
T(G/K) \cong G\times_K \fg/\fk = (G\times \fg/\fk)/K\,.
\end{equation}
Here $K$ acts on $G$ via the group multiplication and on the quotient $\fg/\fk$ via the adjoint representation. This idea extends to any tensor bundle over $G/K$, and for each $p,q\geq 0$ we have the isomorphism
\begin{equation}
T^{p,q}(G/K) \cong G\times_K \big((\fg/\fk)^{ p} \otimes (\fg/\fk)^{* q}\big )\,,
\end{equation}
where the representation of $K$ on $(\fg/\fk)^{ p} \otimes (\fg/\fk)^{* q}$ is formed by taking tensor powers of the adjoint representation on $\fg/\fk$. \\

Since $G$ acts on $G/K$ by diffeomorphisms, there is again a notion of left-invariant tensor fields. We claim that  any such left-invariant tensor field on $G/K$ is definable in $\bbR_\text{alg}$. In order to prove this, let us fix the type $(p,q)$. Recall that a map is definable if its graph is definable. In the case of sections of a bundle, the graph may be interpreted as a submanifold of the bundle. From this perspective, we have to show that the image of a left-invariant $(p,q)$-tensor field is a definable subset of $T^{p,q}(G/K)$. Fix an element $\cT_0 \in (\fg/\fk)^{ p} \otimes (\fg/\fk)^{* q}$ which is invariant under the action of $K$, and consider the definable subset 
\begin{equation}
    G \times \{\cT_0\} \subseteq G \times \big((\fg/\fk)^{ p} \otimes (\fg/\fk)^{* q}\big).
\end{equation}
The tensor bundle $T^{p,q} (G/K)$ is obtained by taking the quotient by $K$, and since $K$ is a compact group acting algebraically, the projection map 
\begin{equation}
    \pi: G \times \big((\fg/\fk)^{ p} \otimes (\fg/\fk)^{* q}\big) \to \Big(G \times \big((\fg/\fk)^{ p} \otimes (\fg/\fk)^{* q}\big) \Big) /K
\end{equation}
is $\bbR_\text{alg}$-definable. The image of the slice $G\times\{\cT_0\}$ under $\pi$ is a submanifold of $T^{p,q}(G/K)$ diffeomorphic to $G/K$; it coincides with the image of the section 
\begin{align*}
    \sigma_{\cT_0} :G/K & \to  \Big(G \times \big((\fg/\fk)^{ p} \otimes (\fg/\fk)^{* q}\big) \Big) /K \\
    gK  & \mapsto (g,\cT_0)K \,.
\end{align*}
By construction, this section defines a left-invariant tensor field. By a geometric version of the Frobenius reciprocity theorem \cite{ParGeom}, there is a one-to-one correspondence between invariant sections of $T^{p,q}(G/K)$ and $K$-invariant elements in $(\fg/\fk)^{ p} \otimes (\fg/\fk)^{* q}$. In other words, all left-invariant $(p,q)$-tensor fields on $G/K$ are of the form constructed above. Since the image of this section coincides with the definable set $\pi( G \times\{\cT_0\})$, we conclude that $\sigma_{\cT_0}$ is tame. \\

As the final step in the argument, we show that the tameness of invariant tensors descends to the arithmetic quotients $\Gamma\backslash G/K$. Let $\widetilde \cT$ be an invariant $(p,q)$-tensor field on $G/K$. This defines a tensor field $\cT$ on the smooth part of $\Gamma\backslash G/K$, which is locally identical to $\widetilde\cT$. This means that, around any smooth point of $\Gamma\backslash G/K$, we can select a definable open neighbourhood $U$ which is, through the projection map $\pi:G/K\to \Gamma\backslash G/K$, identified with an open set $V\subseteq G/K$ via a definable isomorphism $f:U\to V$. The tensor field $\cT$ on $U$ is then simply given by the pullback $\cT=f^*\widetilde \cT$, and since the pullback of a definable metric along a definable map is definable we find that $\cT$ is definable on the subset $U$. \\

This local argument is however not enough to prove definability, since definability is a global property. In order to prove definability of $\cT$, we require a \textit{finite} definable open cover of $\cM=\Gamma\backslash G/K$ on which definability of $\cT$ is checked. Such a cover is obtained as follows. Since $\cM$ is definable, it can be covered by finitely many definable coordinate charts $U_i\subseteq \cM$. As a consequence of the cell decomposition theorem which we discussed in subsection \ref{sec:celldec}, we may assume that the $U_i$ are simply connected. \\

Let $F\subseteq G/K$ be a fundamental set\footnote{Following \cite{BKT}, these fundamental sets may be obtained as a finite union of Siegel sets.} for $\Gamma\backslash G/K$, and denote the projection map by $\pi: F \to \Gamma\backslash G/K$. We can then cover $F$ by the preimages $\pi^{-1}(U_i)$. We now subdivide each into $\pi^{-1}(U_i)$ connected components $V_{i,j}\subseteq G/K$, of which there are finitely many since $\pi^{-1}(U_j)$ is definable. We thus have 
\begin{equation}
\pi^{-1}(U_i) = \bigcup_j V_{i,j}.
\end{equation}
Since any connected component of a definable set is definable, each $V_{i,j}$ is definable. \\

We claim that the projection map $\pi$ is injective on the opens $V_{i,j}$, or in other words, that $V_{i,j}$ contains at most one point in every $\Gamma$-orbit. One possible way to argue this is as follows. If $\pi$ would not be injective on $V_{i,j}$, then there would be points $z$ and $\gamma z$ which are both contained in $V_{i,j}$, with $\gamma\in \Gamma$ some non-trivial group element satisfying $z\neq \gamma z$. From covering space theory it then follows that a path connecting $z$ to $\gamma z$ would project down to a path in $U_i$ with non-trivial homotopy class, which contradicts the assumption that $U_i$ is simply connected.\footnote{Note that the covering space theory applies here to the non-singular part of the arithmetic quotient, which corresponds to the removal of the fixed points of $\Gamma$.} \\

We therefore find that the restriction of $\pi$ to $V_{i,j}$ gives a definable isomorphism $f_{i,j}$ from an open set $U_{i,j} =\pi(V_{i,j})\subseteq U_i$ to $V_{i,j}$. We can now use these isomorphisms to pull back the tensor field $\widetilde \cT$ down to the arithmetic quotient. Indeed, the pullback tensor $\cT=(f_{i,j})^*\widetilde \cT$ is definable, and since the cover $U_{i,j}$ is finite this completes the proof that any invariant tensor $\cT$ on $\cM$ is definable. Finally, specializing to the case of $(0,2)$-tensor fields, we conclude that the metric $\cG$ on $\Gamma\backslash G/K$, after removing the orbifold singularities, is definable in the o-minimal structure $\bbR_\text{alg}$.

\subsubsection*{Tameness of gauge couplings}
The next sector of the higher supergravity which we consider is the abelian gauge field sector. We will focus on $p$-form gauge fields $A^a$ in a $d=2(p+1)$-dimensional spacetime, with $a=1,\ldots,n$ and $n$ denoting the rank of the gauge group. This setting is prominent in higher supergravity theories, including for instance $1$-form gauge fields in $d=4$, and it has an additional layer of structure due to the presence of electric-magnetic duality. The kinetic Lagrangian for these fields is given by
\begin{equation}
    \cL = -\half f_{ab} F^a \wedge * F^b + \half \tilde f_{ab}F^a \wedge F^b.
\end{equation}
Here $F^a = dA^a $ are the field strengths, and the $f_{ab}$ and $\widetilde f_{ab}$ are the gauge coupling functions depending on the scalar fields. A remarkable feature of higher supergravity is that the gauge coupling functions are uniquely determined. This fact will allow us to show that the gauge couplings are always tame in such theories. \\

Before we proceed, it is important to note that there are duality transformations acting on this sector. More precisely, there is a duality group $G^\text{d}$ of linear transformations which interchange the field strengths $F^a$ and their duals $G_a=*(\pd \cL/\pd F^a)$. Depending on whether $p$ is even or odd, this duality group is given by 
\begin{equation}
G^\text{d}=
\begin{cases}
\text{SL}(2n,\bbZ) &\quad \text{if }p \text{ is odd}, \\
\text{SO}(n,n) &\quad \text{if }p \text{ is even}. 
\end{cases}
\end{equation}

As a consequence, the functions $f_{ab}$ and $\widetilde f_{ab}$ are only well-defined when a duality frame is selected. Let us now briefly review the general construction of the gauge couplings. As argued in \cite{Cecotti1}, the gauge couplings are determined by a matrix
\begin{equation}
    \cE =  
    \begin{pmatrix}
    \cA & \cB \\
    \cC & \cD
    \end{pmatrix} 
    \in G^\text{d}\,,
\end{equation}
which is related to the functions $f_{ab}$ and $\widetilde f_{ab}$ through the matrix 
\begin{equation}
\cN =
\begin{cases}
i(\cB - i\cD)(\cA - i\cC)^{-1} &\quad \text{if }p \text{ is odd}, \\
(\cB - \cD)(\cA - \cC)^{-1} &\quad \text{if }p \text{ is even}.
\end{cases}
\end{equation}
The gauge couplings are then obtained by the relations
\begin{equation}
\begin{cases}
 f_{ab}= \text{Im} \, \cN_{ab} , \,\,   \tilde f_{ab}= -\text{Re} \, \cN_{ab} &\quad \text{if }p \text{ is odd}, \\
 f_{ab} = \cN_{(ab)}, \,\, \tilde f_{ab} = -\cN_{[ab]}  &\quad \text{if }p \text{ is even},
\end{cases}
\end{equation}
where the round and square brackets denote the symmetrization and antisymmetrization of indices. The matrix $\cE$ determines the gauge coupling functions $f_{ab}$ and $\widetilde f_{ab}$, but different matrices $\cE$ may give rise to the same functions $f_{ab}$ and $\widetilde f_{ab}$. It can be shown that transformations corresponding to the maximal compact subgroup 
\begin{equation}
K^\text{d} = 
\begin{cases}
\text{U}(n) &\quad \text{if }p \text{ is odd}, \\
\text{O}(n)\times\text{O}(n) &\quad \text{if }p \text{ is even}, 
\end{cases}
\end{equation}
leave the gauge couplings invariant. Therefore, the quotient space $G^\text{d}/K^\text{d}$ may be regarded as the space of possible gauge couplings. \\

Generally, the gauge couplings are functions of the scalar fields in the theory, and consequently the gauge couplings may be regarded as maps from the field space $\cM$ into the symmetric space $G^\text{d}/K^\text{d}$. In higher supergravity, this map is induced from a representation $\rho:G\to G^\text{d}$, and this representation is (up to isomorphism) uniquely determined by supersymmetry \cite{Cecotti1}. The physical interpretation of this map is that the classical global symmetry group $G$ has an action on the
gauge fields, and that this action can only be given in terms of duality transformations in $G^\text{d}$, which is
the largest possible symmetry group of the gauge kinetic terms. \\

However, in making this precise, there is a geometric obstacle. The choice of duality frame required to formulate the gauge couplings experiences monodromy. In other words, if the field space $\cM$ is not simply connected and the scalars traverse a non-trivial loop in $\cM$, then the duality frame is rotated by a $G^\text{d}$-transformation. 
The universal cover $\widetilde\cM= G/K$ of the field space is simply connected, so when viewed as a function of $\widetilde\cM$, the gauge couplings may be seen as a map 
\begin{equation}
    \tilde \mu : G/K \to G^\text{d} / K^\text{d} \,.
\end{equation}
This map was called the \textit{lifted} gauge coupling map in \cite{CecottiDomestic}. However, when we consider the true field space $\cM = \Gamma \backslash G/K$, monodromy is present and it is no longer possible to formulate a well-defined map into the space of couplings $G^\text{d}/K^\text{d}$. \\

To obtain a well-defined gauge coupling map, note that the monodromy transformations of the duality frame define a representation $\pi_1(\cM) \to G^\text{d}$. The image of this representation is the monodromy group, and it is contained in the arithmetic group $G^\text{d}_\bbZ$. We can therefore construct a well-defined gauge coupling by taking a quotient by $G^\text{d}_\bbZ$. In this way, we obtain a map 
\begin{equation}
    \mu : \cM \to G^\text{d}_\bbZ\backslash G/K\,,
\end{equation}
which we will call the \textit{intrinsic} gauge coupling map as in \cite{CecottiDomestic}. This map is related to $\tilde \mu$ through the diagram
\begin{equation}
\begin{tikzcd}
G/K \arrow[r,"\tilde\mu"]\arrow[d] & G^\text{d} /K^\text{d} \arrow[d]\\
 \Gamma \backslash G/K \arrow[r,"\mu"]& G^\text{d}_\bbZ\backslash G^\text{d} /K^\text{d}\,,
\end{tikzcd}
\end{equation}
where the vertical arrows are the projections onto the quotient spaces. 
The intrinsic gauge coupling map is a map between arithmetic quotients, originating from a map $\rho:G\to G^\text{d}$ satisfying $\rho(\Gamma)\subseteq G^\text{d}_\bbZ$. Maps of this form are called \textit{morphisms of arithmetic quotients}, and it was proven in \cite{BKT} that these morphisms are definable in $\bbR_{\rm alg}$. We therefore conclude the following.

\begin{subbox}{Tameness of gauge couplings in higher supergravity}
    The intrinsic gauge coupling map $\mu:\cM \to G^\text{d}_\bbZ\backslash G^\text{d}/K^\text{d}$ in higher supergravity theories can be geometrically interpreted as morphisms of arithmetic quotients. Consequently, the gauge coupling $\mu$, as a function of the scalar fields, is definable in $\bbR_{\rm alg}$.
\end{subbox}

Note that the gauge coupling map determines an equivalence class of the
matrix $\cE$, from which the gauge coupling functions $f_{ab}$ and $\widetilde f_{ab}$ can be extracted in an algebraic manner
which preserves tameness. We thus conclude that, regardless of whether
we choose to work with the field space $\cM$ or its universal cover $\widetilde \cM$, the gauge coupling functions in higher supergravity theories are tame, providing further evidence for the tameness conjecture.

\subsubsection*{Example: four-dimensional \texorpdfstring{$\cN=4$}{} supergravity}
We now proceed with an example of a higher supergravity theory to make the abstract constructions discussed above explicit, namely the $\cN=4$ supergravity in four dimensions. From a string theory perspective, this theory arises as the low-energy description of Type IIB compactified on $\text{K}3\times \mathbb{T}^2$ or a $\mathbb{T}^6/\bbZ_2$ orientifold. The universal cover of the scalar field space of this theory is the symmetric space 
\begin{equation}
\widetilde \cM =G/K = \text{SL}(2,\bbR)/\text{SO}(2)\times \text{SO}(6,k)/(\text{SO}(6)\times \text{SO}(k))\,.
\end{equation}
The first factor $\text{SL}(2,\bbR)/\text{SO}(2) $ is the upper half-plane $\bbH$, and the second factor depends on an integer $k$ counting the number of additional vector multiplets.

We denote an element of $\widetilde \cM$ as a pair
\begin{equation}
\cV = \left(
\begin{pmatrix}
\alpha & \beta \\
\gamma & \delta
\end{pmatrix}
,\,g
\right) \,,
\end{equation}
with $\alpha \delta - \beta\gamma =1$. The matrix $g \in \text{SO}(6,k)$ preserves the bilinear pairing $\eta = \text{diag}(-1,\ldots,-1,1,\ldots,1)$ of signature $(6,k)$, i.e.~we have $g^\text{T}\eta g = \eta$. Following the literature and the procedure explained earlier in this section, we formulate the field space metric in terms of the matrix $\cV\cV^\text{T}$. The metric on the factor $\text{SL}(2,\bbR)$ was already considered in an earlier example, and is commonly expressed in terms of $\tau$ as
\begin{equation}
\begin{pmatrix}
\alpha & \beta \\
\gamma & \delta
\end{pmatrix}
\begin{pmatrix}
\alpha & \beta \\
\gamma & \delta
\end{pmatrix}^\text{T}
=
\frac{1}{\Im\tau}
\begin{pmatrix}
|\tau|^2 & \Re\tau \\
\Re\tau & 1
\end{pmatrix} \,.
\end{equation}
The metric can then be expressed in terms of the matrix $N=gg^\text{T}$, and the resulting scalar kinetic terms take the form \cite{SchonWeidner}
\begin{equation}
-\frac{1}{4(\Im\tau)^2} \pd_\mu\tau\,\pd^\mu\bar\tau +  \frac{1}{16}\tr(\pd_\mu N_{ab}\, \pd^\mu N^{ab}) \,,
\end{equation}
Here we have restored the indices on $N$, and the upper indices indicate that an inverse is taken. From the algebraic structure of this expression it is evident that the scalar field space metric defines a tame function in these coordinates. The starting point of the construction of the gauge couplings is a representation $G \to G^\text{d}$. In this case, following \cite{SugraAuria}, the required symplectic representation of $G$ is given by 
\begin{align}
    \rho:\text{SL}(2,\bbR)\times \text{SO}(6,k) & \to \text{Sp}(2(6+n),\bbR) \\
   \left(
\begin{pmatrix}
\alpha & \beta \\
\gamma & \delta
\end{pmatrix}
,g
\right) 
& \mapsto 
\begin{pmatrix}
\alpha g & \beta \eta g^{-\text{T}}\\
\gamma \eta g & \delta g^{-\text{T}}
\end{pmatrix} \,.\nonumber
\end{align}
The matrix on the right defines the matrix $\cE$. 
Using this matrix to derive the kinetic terms for the gauge fields in terms of $\cV\cV^{\text{T}}$, one finds \cite{SamtlebenLec}
\begin{equation}
    -\frac{1}{4} \Im\tau\, N_{ab}\, F^a\wedge * F^b +\frac{1}{4}\Re\tau \, \eta_{ab} \,F^a\wedge  F^b\,.
\end{equation}
In the notation used above, we thus identify the gauge couplings in this theory as  
\begin{equation}
f_{ab} = \frac{1}{2}\Im \tau \, N_{ab}\,, \quad \widetilde f_{ab} = \frac{1}{2}\Re \tau \,\eta_{ab}\,,
\end{equation}
which shows the explicit dependence of the gauge coupling functions on the scalar fields. The tameness of these couplings follows from the algebraic group structure, as argued on general grounds above. \\

\section{Tame isometric embeddings of moduli spaces}\label{sec:embeddings}
The following section is concerned with the connection between tameness and volumetric properties of moduli spaces in the quantum gravity landscape. As reviewed in the previous chapter, it has been noted in the very inception of the swampland program that consistency with quantum gravity imposes finiteness constraints on volumes in moduli spaces \cite{Vafa:2005ui}. This was made precise more recently in \cite{Delgado:2024skw}, where a particular volume growth condition termed \textit{compactifiability} was argued to hold in the moduli space of any quantum-gravitational effective theory. In this section we demonstrate how this condition emerges from the tame geometry of moduli spaces. Remarkably, establishing this connection forces us to go beyond the mere assumption that the moduli space is a tame Riemannian manifold, and further demand that the moduli space must admit a \textit{tame isometric embedding} into Euclidean space. We subsequently introduce tame embeddability as a novel geometric tameness criterion, and conjecture in subsection \ref{sec:embconj} that this criterion holds in all moduli spaces arising from quantum gravity.

\subsection{Volumes, dualities, and compactifiability}
\label{sec:volumecompacty}
The discussion of quantum gravity constraints on volumetric properties of moduli spaces originates in \cite{Vafa:2005ui}, where it was observed that many moduli spaces in string theory have a finite volume as a consequence of dualities. The quintessential example of this phenomenon is the moduli space of Type IIB supergravity in ten dimensions, given by $\cM_{\rm IIB}=\mathbb{H}/\text{SL}(2,\bbZ)$ parametrized by the axio-dilaton $\tau$. The hyperbolic plane is a space of infinite volume, but the quotient by the duality group $\text{SL}(2,\bbZ)$ renders the volume of $\cM_{\rm IIB}$ finite. More involved examples arise from the higher supergravity theories studied in the previous section, where the arithmeticity of the duality group $\Gamma$ is precisely equivalent to the condition that the moduli space $\cM =\Gamma\backslash G/K$ has finite volume \cite{Cecotti1}. On the other hand, the moduli space of Type IIA supergravity in 10 dimensions $\cM_{\rm IIA}=\bbR$ immediately demonstrates that finiteness of volume is not true in general. \\

In order to make universal statements, we have to take into account that from the point of view of the effective theory, only a limited part of the moduli space is accessible. In asymptotic regions of moduli space, the EFT is invalidated by the infinite towers of light states which appear according to the distance conjecture. In particular, the moduli-dependent tower mass scale $m_{\rm t}(\phi)$ becomes exponentially light in the geodesic distance $D$, and the EFT description breaks down when the cutoff $\Lambda$ violates the condition $\Lambda<m_{\rm t}(\phi)$.\footnote{See \cite{Burgess:2023pnk} for a discussion of why the EFT description cannot incorporate a finite part of the emerging light tower.} For a given moduli space $\cM$, we should therefore consider the volume of the \textit{truncated moduli space}  
\begin{equation}
    \cM_{\Lambda} = \{\phi \in \cM \,|\,\Lambda < m_{\rm t}(\phi) \}
\end{equation}
in which the EFT is valid \cite{Delgado:2024skw,Baykara:2025nnc}. It was argued in \cite{Hamada:2021yxy} using the finiteness of amplitudes in quantum gravity  that the space $\cM_{\Lambda}$ must have finite volume. This analysis was generalized in \cite{Delgado:2024skw}, where it was argued that this volume is not only finite, but also does not grow too fast; in particular, it was claimed that for any $\eps>0$, the volume of the truncated moduli space is bounded as
\begin{equation}
    \text{Vol}(\cM_{\Lambda}) \ll |\log \Lambda|^{n+\eps}
\end{equation}
where $n=\dim \cM_{\Lambda}$. Using the distance conjecture to relate the cutoff $\Lambda$ to the geodesic distance as $D\sim |\log\Lambda|$, one can rephrase this bound in terms of a geodesic ball
\begin{equation}
    \cM_D = \{ \phi\in \cM \,|\,d(\phi,\phi_0) \leq D  \}
\end{equation}
for any basepoint $\phi_0\in \cM$, which then led to the following conjecture \cite{Delgado:2024skw}.

\begin{subbox}{Compactifiability conjecture}
Let $\cM$ be an $n$-dimensional moduli space of an effective field theory consistent with quantum gravity. Then $\cM$ is \textit{compactifiable}, meaning that any geodesic ball $\cM_{D}\subseteq \cM$ of radius $D$ satisfies
\begin{equation}\label{eq: compacty condition}
    \text{Vol}(\cM_{D}) \ll D^{n+\eps}
\end{equation}
for any $\eps>0$ and $D\gg 0$.    
\end{subbox}

In other words, the compactifiability conjecture asserts that the volume of a ball in moduli space grows no faster than the volume of a ball in Euclidean space. The bottom-up argument for this conjecture presented in \cite{Delgado:2024skw} proceeds by compactifying all spatial dimensions of the EFT, resulting in a 1-dimensional theory which includes a sigma model with target space $\cM$. Under the assumption of supersymmetry it is then shown that $\cM$ must be compactifiable in order for the compactified theory to have a finite ground state degeneracy, which is essential for the finiteness of quantum gravity amplitudes \cite{Hamada:2021yxy}. \\

The main motivation for the proposal of the compactifiability conjecture comes from its relation to dualities. Moduli spaces in quantum gravity typically contain regions of negative curvature, in which the volume of geodesic balls would a priori grow faster than in Euclidean space. The compactifiability criterion is then only satisfied if there is a physical mechanism which forbids excessively large asymptotic regions of negative curvature, and this is accomplished by quotienting the moduli space by a sufficiently large duality group. It can therefore be interpreted as a prediction of the emergence of dualities in theories of quantum gravity \cite{Delgado:2024skw}.

\subsection{Tameness and volume growth}
Let us now discuss how o-minimality controls the volumetric properties of tame sets. Consider an $n$-dimensional set $X\subseteq \bbR^N$ contained inside a ball $B^N(r)$ of radius $r$. How large can the volume of $X$ be? At first one might expect the volume of $X$ to be bounded by a function of the radius $r$, but if $X$ is of lower dimension than the ambient space (i.e.~$n<N$), then its $n$-dimensional volume is actually unbounded despite being contained in a finite $N$-dimensional volume. For example, if $N=2$ and $n=1$, then $X$ could be a spiral or a densely oscillating curve, as shown in figure \ref{fig:CurveVolumes} below. \\

\begin{figure}[h]
    \centering
    \includegraphics[width=1\linewidth]{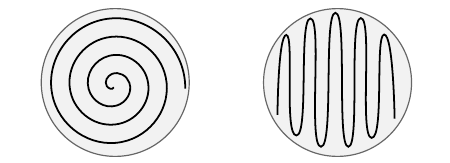}
    \caption{Two examples of curves inside a disk of radius $r$ whose volume can grow to infinity. The spiral can wind infinitely many times (left) and the frequency of the oscillation can go to infinity (right), leading to infinite length inside a bounded region.}
    \label{fig:CurveVolumes}
\end{figure}

This is precisely where tameness enters. The examples above had an unbounded volume because the set $X$ could wrap infinitely many times inside the ball $B^N(r)$, and this is exactly the sort of geometric behavior which is forbidden by o-minimality. For a tame set $X\subseteq \bbR^N$ and for any $0\leq l\leq N$ there exists an integer $b_{0,l}(X)$ such that for any $l$-dimensional affine plane $P\subseteq\bbR^N$, the number of connected components of $X\cap P$ is bounded by $b_{0,l}(X)$ \cite{yomdin2004tame}. These integers are called the \textit{Gabrielov numbers} \cite{gabrielov1968projections}, and provide a useful probe for the number of wrappings and foldings of a tame set. The fact that there exist such uniform upper bounds on the number of components of plane intersections is called the global Gabrielov property \cite{yomdin2004tame}, and this property of tame sets will ensure that their volumes are proportional to the expected Euclidean scaling. This is made precise as follows \cite{yomdin2004tame}.

\begin{subbox}{Volumes of tame sets}
    Let $X\subset \bbR^N$ be a tame set of dimension $n$. Then for any $N$-dimensional ball $B^N(r)\subseteq\bbR^N$ of radius $r$, we have
        \begin{equation}
            {\rm Vol}_n(X\cap B^N(r))\leq  c(N,n)\, b_{0,N-n}(X)\, r^n\,,
            \label{eq: Yomdin bound}
        \end{equation}
        where $b_{0,N-n}(X)$ is the $(N-n)$-th Gabrielov number of $X$, and $c(N,n)$ is a normalization constant given by
        \begin{equation}\label{eq:c-constant}
         c(N,n)={\rm Vol}_n(B^n(1)) \frac{\Gamma\left(\frac{1}{2}\right)\Gamma\left(\frac{N+1}{2}\right)}{\Gamma\left(\frac{n+1}{2}\right)\Gamma\left(\frac{N-n+1}{2}\right)}\,.
        \end{equation}
        Here ${\rm Vol}_n(B^n(1))= \pi^{n/2}/\Gamma(n/2+1)$ the volume of the $n$-dimensional ball of unit radius.
\end{subbox}

The bound \eqref{eq: Yomdin bound} is very reminiscent of the compactifiability condition \eqref{eq: compacty condition}, but there is a key difference that clouds the connection: while the discussion above concerns a subset of Euclidean space with a flat metric, the compactifiability condition is focused on the intrinsic properties of the moduli space and its curved metric. This problem can be addressed by considering an isometric embedding of the moduli space into a higher-dimensional Euclidean space, in which equation \eqref{eq: Yomdin bound} can be applied. Here we recall that an isometric embedding between two Riemannian manifolds $(\cM,g)$ and $(\cN ,h)$ is a differentiable map $f:\cM \to \cN$ which is a diffeomorphism onto its image, and satisfies $g=f^*h$. In this way, we can think of $\cM$ as a subspace of $\cN$ which inherits the ambient metric of $\cN$, and in particular the notion of distance is preserved. Such an embedding into Euclidean space is guaranteed to exist by the Nash embedding theorem \cite{nash1956imbedding}, but, as we will see in the following subsections, the tame nature of the manifold might not be preserved under the embedding.

\subsection{Tameness and compactifiability in the hyperbolic plane}
The key example that we will use to illustrate the deep connection between tameness and compactifiability is the hyperbolic plane $\bbH$, which makes a central appearance in the moduli space of Type IIB supergravity. It is a Riemannian manifold consisting of the points $\tau =x+iy\in \bbC$ with $y>0$ and metric given by
\begin{equation}\label{eq: metric H}
    \dd s^2= \frac{1}{(\Im \tau)^2}\dd \tau \,\dd \bar \tau =\frac{1}{y^2} (\dd x^2+\dd y^2)\,. 
\end{equation}
For a generic point $\tau_0=x_0+iy_0$, the associated truncated moduli space $\mathcal{M}_D(\tau_0)$ of points within distance $D$ from $\tau_0$ is given by
\begin{equation}
 \mathcal{M}_D(\tau_0)=
\big\{x+iy\in \mathbb{H} \, \big| \, (x-x_0)^2+(y-y_0{\rm cosh}(D))^2={\rm sinh}^2(D)y_0^2 \big\}\,.
\end{equation}
Note that this set is a Euclidean disk in $\mathbb{H}$ of radius $y_0 \sinh(D)$ and centered at $x_0+i y_0 \cosh(D)$. Using the hyperbolic metric \eqref{eq: metric H}, we then calculate
\begin{equation}
    \text{Vol}(\mathcal{M}_D) = \int_{\cM_D} \dd x \,\dd y \,\frac{1}{y^2} =2\pi(\cosh(D)-1) \,.
    \label{eq: vol MD growth}
\end{equation}
Note that the dependence of the volume on the origin of the ball disappears because the manifold has constant curvature. The asymptotic volume growth for large geodesic distance $D$ is exponential instead of quadratic, and we therefore conclude that the hyperbolic plane $\bbH$ is not compactifiable, as already observed in \cite{Delgado:2024skw}. \\

However, when one considers the moduli space obtained by quotienting the hyperbolic plane with the action of the standard duality group $\text{SL}(2,\bbZ)$, the volume becomes finite and thus the compactifiability condition \eqref{eq: compacty condition} is trivially satisfied. This stark contrast highlights the important interplay between the geometry of moduli spaces and the duality groups of effective theories compatible with quantum gravity \cite{Delgado:2024skw}, which we already alluded to in \ref{sec:volumecompacty}.\\

As we have seen in section \ref{sec:highsugra}, the upper half-plane $\bbH$ and its quotient $\bbH/\text{SL}(2,\bbZ)$ and their metrics are definable in the simplest o-minimal structure $\bbR_{\rm alg}$, meaning that they constitute a tame Riemannian manifold. It thus appears that, already in this elementary example, tameness is unable to distinguish between the two cases. However, here we must remember that the most powerful volume growth result at our disposal, equation \eqref{eq: Yomdin bound}, concerns subspaces of Euclidean space. This bound could therefore be used to constrain the volume growth of geodesic balls in moduli space, provided that we embed the moduli space under consideration isometrically into Euclidean space. As we will see, it is in this step that tameness is strong enough to recover the compactifiablity condition \eqref{eq: compacty condition} and even extend it beyond the asymptotic regime.

\subsubsection*{Embeddings of the hyperbolic plane -- the fundamental domain}\label{sec:hyperb}

The Nash embedding theorem \cite{nash1956imbedding} ensures the existence of an isometric embedding of the full hyperbolic plane into a Euclidean space $\bbR^n$ for large enough $n$. An explicit realization of such an embedding was found by Blanu{\v{s}}a in \cite{blanuvsa1955einbettung}, requiring $n=6$ and involving complicated non-elementary functions. Crucially, these functions contain pieces which are periodic on an unbounded domain and therefore not definable in any o-minimal structure. \\

The situation changes drastically when one restricts the moduli space to the fundamental domain of the duality group ${\rm SL}(2,\mathbb{Z})$, defined by 

\begin{align}
\label{eq: fundamental domain}
        F_{{\rm SL}(2,\mathbb{Z})}=&\{x+i y\in \bbC\ | \ -1/2\leq x \leq 0 \,, \,  x^2+y^2\geq 1\} \ \cup \\
        & \{x+i y\in \bbC\ | \ 0< x < 1/2 \,, \, x^2+y^2>1\}\,. \nonumber
\end{align}

For this fundamental domain, there exists a tame isometric embedding into $\bbR^3$ given by the section of a pseudosphere \cite{bonahon2009low}. Explicitly, the embedding can be obtained by considering the tame functions
\begin{align}
    X_0 &= \text{arccosh}\left(\frac{2y}{\sqrt{3}}\right) - \text{tanh}\left(\text{arccosh}\left(\frac{2y}{\sqrt{3}}\right)\right) \,, \\
    X_1 & = \frac{\sqrt{3}}{2y}\cos(\frac{2x}{\sqrt{3}}) \,,\\
    X_2 & = \frac{\sqrt{3}}{2y}\sin(\frac{2x}{\sqrt{3}})\,, 
\end{align}
defined on the rectangular domain $-c\pi<x<c\pi$,  $y>c$.
Using this isometric embedding into $\bbR^3$, we can apply equation \eqref{eq: Yomdin bound} to its embedded image $\widehat F_{\text{SL}(2,\bbZ)}\subseteq \bbR^3$ to bound the scaling of the volume. In this simple example, we evaluate that $c(3,2)=2\pi$ and that the Gabrielov number $b_{0,1}(2)$ , since a line can intersect the corresponding section of a pseudosphere at most two times. Therefore, the theorem predicts a polynomial bound on the scaling of the volume 
\begin{equation}
    \text{Vol}(\widehat F_{\text{SL}(2,\bbZ)} \cap B^3(r))\leq  4\pi r^2\,,
    \label{eq: F SL2Z volume bound}
\end{equation}
consistent with the expectation from the compactifiability conjecture. 

\subsubsection*{Embeddings of the hyperbolic plane -- the duality quotient}
It is important to note that the fundamental domain is a subset of $\mathbb{H}$ that does not implement the identification between points of the boundary given by the action of ${\rm SL}(2,\mathbb{Z})$. One may then ask if a tame isometric embedding of the actual moduli space $\mathbb{H}/{\rm SL}(2,\mathbb{Z})$ also exists. Though the explicit construction of such an embedding is much more complicated, there are several important results that indicate an affirmative answer. \\ 

Firstly, we recall that the quotient  $\mathbb{H}/{\rm SL}(2,\mathbb{Z})$ is topologically a punctured Riemann sphere $\mathbb{P}^1\backslash \{\infty\}$. The map from $F_{{\rm SL}(2,\mathbb{Z})}$ to $\mathbb{P}^1\backslash \{\infty\}$ that identifies the points on the boundary of the fundamental domain through the ${\rm SL}(2,\mathbb{Z})$-action is given by the Hauptmodul of the duality group. This map consists of an algebraic combination of powers of the $j$-invariant function. The $j$-function is tame when restricted to $F_{{\rm SL}(2,\mathbb{Z})}$ \cite{peterzil2004uniform}, where it is in fact injective. Consequently, the map from $\mathcal{F}_{{\rm SL}(2,\mathbb{Z})}$ to $\mathbb{H}/{\rm SL}(2,\mathbb{Z})$ represented by a punctured sphere is also tame. \\

Secondly, the hyperbolic metric of the upper half-plane is mapped to the Weil-Petersson metric of the modular curve $\mathbb{H}/{\rm SL}(2,\mathbb{Z})$ under the action of the Hauptmodul. One then conventionally removes two additional singular points, namely the elliptic points $\tau= e^{2\pi i/3}$ and $\tau= i$, and works on $\mathbb{P}^1\backslash \{0,1,\infty\}$. The metric on this space is derived from a K\"ahler potential that can be expanded as a function of the complex structure modulus in terms of the two independent periods of the modular curve
\begin{equation}
    K_{\rm cs}=-\log\left(i (\Pi^0(z)\overline{\Pi}_0(\bar{z})-\overline{\Pi}^0(\bar{z})\Pi_0(z))\right)\,.
\end{equation}
The tameness of the period map in Hodge theory \cite{BKT} implies that the Weil-Petersson metric obtained from $g_{z\bar{z}}=\partial_z\partial_{\bar{z}} K_{\rm cs}$ is tame as well. \\

The discussion above exemplifies the role of the duality group and the importance of quotienting the moduli space by its action. The identification of points in the same orbit ensures that the $j$-function and the period map are restricted to a domain where they are tame. In the example of the hyperbolic plane, this requirement results in a moduli space of finite volume.

Having established the tame nature of $\mathbb{H}/{\rm SL}(2,\mathbb{Z})$ and its metric, the only potential obstruction to the existence of a tame isometric embedding into Euclidean space are the three special limit points of the fundamental domain $F_{{\rm SL}(2,\mathbb{Z})}$. One of these limits, namely $\tau\rightarrow i\infty$, gives rise to the cusp of the punctured sphere, while the other two  correspond to elliptic points. Among these limit points, the most challenging limit is the cusp, which is the only one that describes a point at infinite distance. The tame embeddability of this geometric feature into Euclidean space can be accomplished by means of a pseudosphere, as shown earlier in this section. Consequently, we expect $\mathbb{H}/{\rm SL}(2,\mathbb{Z})$ to admit an isometric tame embedding and therefore obey similar bounds to \eqref{eq: F SL2Z volume bound}.\\

What is then the difference between the spaces $\mathbb{H}$ and $\mathbb{H}/{\rm SL}(2,\mathbb{Z})$ with regards to the isometric embedding? The complete hyperbolic plane appears to be simply too large to  be embedded in Euclidean space without folding it infinitely many times. In this sense, the negative curvature, which controls the folding, presents an obstruction to the tame embedding. The length of any bounded horizontal segment $x\in (-c,c)$ diverges when it approaches $y=0$, and such growth cannot be accounted for by any tame embedding.  This untameable behavior generally appears when approaching regions at infinite distance in moduli space. In \cite{fontenele2014complexity}, the obstruction is more rigorously formalized and extended to a certain class of simply connected Riemannian manifolds with negative curvature. The moduli space $\mathbb{H}/{\rm SL}(2,\mathbb{Z})$, after removing the elliptic points on which the metric is not defined, is not simply connected, so it evades the premise of the theorem. Moreover, it only has one limit point at infinite distance, and therefore the issue of line segments with diverging length does not arise. In general, we expect any duality group acting on $\mathbb{H}$ with a fundamental domain with a finite number of infinite distance points to be tamely embeddable. This includes Fuchsian groups of the first kind, such as the congruence subgroups of ${\rm SL}(2,\mathbb{Z})$ \cite{VoightQuaternion, Delgado:2024skw}.\\

In the following subsection we will show that this picture can be generalized significantly to provide a bound on the scaling of the volume of the original manifold with respect to the geodesic distance. Additionally, we will explore how the coefficient in front of the polynomial growth of the volume can be related to the complexity of the manifold and its embedding. 

\subsection{Conjecture on tame isometric embeddings}\label{sec:embconj}
The case of the hyperbolic plane illustrates the subtle but crucial distinction between a tame manifold and a manifold with a tame isometric embedding into Euclidean space. In this subsection we will show that this distinction holds in general, and hence present a conjecture on tame isometric embeddings of moduli spaces. 

\subsubsection*{Volume growth in embedded moduli spaces}

Consider an $n$-dimensional candidate moduli space $\cM$ (either the complete space, its fundamental domain under some duality group, or the corresponding duality quotient) with metric $g$. Suppose that we have an isometric embedding $f:\cM\to \bbR^N$ of the moduli space into an $N$-dimensional Euclidean space, and denote by $\widehat\cM =f(\cM)\subseteq \bbR^N$ the embedded moduli space. For a basepoint $\phi_0\in\cM$ and $D>0$ we then examine the truncated moduli space $\cM_D = \{ \phi\in\cM \,|\,d(\phi,\phi_0)\leq D \}$ and its embedded image $\widehat \cM_D$. Since distances along $\cM$ are preserved by the isometric embedding, $\widehat\cM_D$ constitutes a submanifold of $\widehat\cM$ consisting of points within a geodesic distance $D$. The Euclidean geometry of the ambient space then implies that we have the inclusion
\begin{equation}
  \widehat\cM_D \subseteq \widehat \cM \cap B^N(r) \quad \text{for} \quad D\leq r\,,
  \label{eq: Md subset condition}
\end{equation}
where the $N$-dimensional ball $B^N(r)$ is centered at $f(\phi_0)$.
Since the isometric embedding preserves the volumes, $ \text{Vol}\big(\widehat \cM_D\big)= \text{Vol}(\cM_D)$, and it follows that for radii $r\geq D$,
\begin{equation}\label{eq:VolDvsBemb}
   \text{Vol}(\cM_D) \leq \text{Vol}\big(\widehat \cM \cap B^N(r)\big) \,.
\end{equation}
Furthermore, the bound of equation \eqref{eq: Yomdin bound} tells us that if the embedding $f$ is tame, so that the embedded moduli space $\widehat\cM$ is a tame set, it satisfies 
\begin{equation}
   \text{Vol}\big(\widehat \cM \cap B^N(r \big)\leq C\, r^n\,,
        \label{eq: tameness vol scaling}
\end{equation}
with $n=\text{dim}(\cM)$ and $C$ a global coefficient depending on the precise nature of the embedding, on which we will comment later in this subsection. Choosing a ball of radius $r=D$ we can simply combine \eqref{eq: Md subset condition}  and \eqref{eq: tameness vol scaling} to conclude that

\begin{equation}
   \text{Vol}(\cM_D) \leq  C\, D^n 
    \label{eq: THE BOUND}
\end{equation}
which shows that if the moduli space $\cM$ is a tame Riemannian manifold admitting a tame embedding into Euclidean space, the compactifiability conjecture holds.

\subsubsection*{Statement and interpretation of the conjecture}
Motivated by the envisioned tameness of moduli spaces \cite{Grimm:2021vpn}, and the connection between tameness, compactifiability, and finiteness in quantum gravity demonstrated in this section, we propose the following conjecture.

\begin{subbox}{Tame embedding conjecture}
Let $\cM$ be an $n$-dimensional moduli space of an effective field theory consistent with quantum gravity. Then $\cM$ admits a tame isometric embedding into Euclidean space. 
\end{subbox}

This conjecture refines the idea of the original tameness conjecture \cite{Grimm:2021vpn}, where the tameness of field spaces and coupling functions was considered. As we have seen in the discussion above, the tameness of the moduli space, merely viewed as a manifold with a metric, is not strong enough to recover the volume scaling properties expected from an effective theory of quantum gravity. Let us now discuss several aspects of the conjecture in more detail. \\

First, we emphasize that the bound provided by \eqref{eq: THE BOUND}, despite depending on the global properties of the embedding, is a local constraint that holds for any point $\phi_0\in \cM$ and any value of the geodesic distance $D$. In this sense, it provides a stronger statement about the volume growth than the original compactifiability condition \eqref{eq: compacty condition}, which only holds in the asymptotic limit $D\rightarrow \infty$. \\

From a geometric perspective, the converse of equation \eqref{eq: THE BOUND} can be used to rule out the existence of a tame isometric embedding. This happens in the case of the hyperbolic plane $\mathbb{H}$, where the volume grows exponentially in the asymptotic limit, which is incompatible with the existence of a tame isometric embedding. This formalizes the observation of the previous subsection regarding the non-tame nature of Blanu{\v{s}}a's embedding and extends it from a single example to a general result applying to any geometry. Note that the existence of a tame isometric embedding requires that the bound \eqref{eq: THE BOUND} is satisfied globally. This means that to prove that such an embedding does not exist, it is sufficient to find a single asymptotic direction that violates it. Going back to our example, both the hyperbolic plane and its quotient display the same volume growth locally. One must reach points at infinite distance to observe that the growth rate of the former diverges exponentially while the latter asymptotes to a constant. \\
  
It is important to remark that moduli spaces in string theory typically have singular points, as already happens in Type IIB supergravity for the elliptic points $\tau=i$ and $\tau=e^{2\pi i/3}$ in $\bbH/\text{SL}(2,\bbZ)$. In the statement of the conjecture it is implicit that a tame isometric embedding $\cM\to \bbR^N$ is a tame map which is isometric on the non-singular parts on which a Riemannian metric is defined, whereas on the singular loci it is merely assumed to be distance-preserving. For the connection with the compactifiability conjecture it is valid to disregard the singular loci, since they are of measure zero and therefore do not contribute to the volume. \\

Finally, we note that any tame embedding for which equation \eqref{eq:VolDvsBemb} holds already implies the compactifiability condition. Therefore, the requirement that the embedding is isometric can be weakened slightly. However, this weakening is rather subtle. An embedding which contracts distances implies equation \eqref{eq: Md subset condition} but not that $\text{Vol}(\cM_D)\leq \text{Vol}(\widehat\cM_D)$, whereas an embedding which expands distances implies that $\text{Vol}(\cM_D)\leq \text{Vol}(\widehat\cM_D)$ but not that \eqref{eq: Md subset condition} holds. The embedding must be such that the rescaling of distances compensates the wrapping of the embedded manifold in the Euclidean target space, in such a way that equation \eqref{eq:VolDvsBemb} is satisfied. An isometric embedding achieves this balance, which makes it the natural notion to consider.

\subsubsection*{Scaling coefficient and complexity of the embedding}
We now turn to the coefficient $C$ appearing as a proportionality factor in equation \eqref{eq: THE BOUND}. Recall that this constant is comprised of the factor $c(N,n)$ given in equation \eqref{eq:c-constant} depending on the dimension of the moduli space $n$ and the ambient Euclidean dimension $N$, and the Gabrielov numbers $b_{0,N-n}(\widehat \cM \cap B^N(D) )$ bounding the number of connected components of codimension-$n$ hyperplanes intersecting $\widehat\cM \cap B^N(D)$. If the embedding is not only tame, but definable in a sharply o-minimal structure, then it is possible to go beyond qualitative finiteness statements and set explicit bounds on the coefficients of the volume scaling. In particular, we can bound the Gabrielov numbers in terms of the complexity of the embedding. \\

One of the key properties of sharply o-minimal structures is that there exists a universal function $\text{poly}_{\cF}(\cD)$ that bounds the number of connected components of any definable set $X$ in $\Omega_{\cF,\cD}$. Suppose that the embedded moduli space $\widehat\cM$ has complexity $(\cF,\cD)$. Meanwhile, any $(N-n)$-dimensional plane $P\subseteq \bbR^N$ satisfies $P\in\Omega_{N,n}$, since it is defined by $n$ equations of degree $1$ in $N$ variables. It then follows that $\widehat\cM\cap P \in \Omega_{\cF,\cD+n}$, and hence
\begin{equation}
    b_{0,N-n}(\widehat\cM)\leq {\rm poly}_{\cF}(\cD+n)\,
\end{equation}
for any codimension-$n$ plane $P$. We can therefore refine equation \eqref{eq: THE BOUND} for the moduli spaces admitting a tame embedding in a sharply o-minimal structure by
\begin{equation}
    C \leq c(N,n) \, \text{poly}_{\cF}(\cD+n)\,.
    \label{eq: complexity bound}
\end{equation}
Note that this bounding polynomial $\text{poly}_{\cF}(\cD+n)$ is universal for sets in $\bbR^N$ for fixed $N$, within a given sharply o-minimal structure. Therefore, once the ambient dimension $N$ has been established, the dependence on the particular choice of moduli space is present only through its dimensionality $n$ and the complexity $(\cF,\cD)$ of the tame isometric embedding. 

\subsection{Outlook on the conjecture}\label{sec:emboutlook}
In this section we have observed that the existence of a tame isometric embedding of the moduli space into Euclidean is a sufficient condition to recover the compactifiability condition \eqref{eq: compacty condition}. Extending the tameness conjecture of \cite{Grimm:2021vpn} and the conjecture that compactifiability holds for all quantum gravity moduli spaces \cite{Delgado:2024skw}, we proposed the tame embedding conjecture. Reformulating the compactifiability condition in terms of tameness offers several advantages. First, it ties this condition into a general framework and thereby unifies it with other finiteness conjectures in quantum gravity. Second, it provides a sharp local characterization of the volume growth that goes beyond asymptotic statements. Despite its local validity, universal bounds on the coefficients of the scaling are formulated in terms of global geometric properties of the moduli space and its embedding. When extending the tameness principle to sharp o-minimality, the coefficients of the volume growth can be recast as functions of the complexity of the moduli space. We thus can establish a quantitative connection to complexity and finiteness of information through the study of the coefficient present in the polynomial growth. Extending this point, a notion of complexity for quantum field theory as discussed in chapter~\ref{ch:complexityQFT}
could be used to connect the tame embedding conjecture with other aspects of finiteness in quantum gravity. This will be the starting point of the next chapter.

\subsubsection*{Tameness and emergence of dualities}

We emphasize that our findings also demonstrate a direct link between tame embeddability and the existence of dualities. The example of the hyperbolic plane highlighted that the moduli space, without taking the duality quotient, is too large to be tamely isometrically embedded into Euclidean space. In fact, in this example both T- and S-transformations are required to render the quotiented moduli space small enough to admit a tame embedding, while T- and S-transformations individually are not sufficient. In general, recall from section \ref{sec:hyperb} that the main result of \cite{fontenele2014complexity} essentially states that simply connected negatively curved manifolds cannot admit a tame embedding. Discrete duality quotients break simply connectedness,\footnote{To be precise, it breaks simply connectedness upon removing the singular loci. Alternatively, one may consider simply connectedness through the orbifold fundamental group as remarked in \cite{Delgado:2024skw}.} so that the assumption of the theorem is evaded. In fact, as noted in the context of marked moduli spaces in \cite{Raman:2024fcv},  the breaking of simply connectedness is always a consequence of the existence of dualities, which further establishes the connection between tameness and dualities. Furthermore, the tameness of the isometric embedding is reminiscent of the tameness of the period map on the moduli space, for which the quotient by a sufficiently large duality group is an essential part of the proof \cite{BKT}. Moreover, the proper consideration of duality quotients is also essential in the finiteness proof of \cite{Bakker:2023xkt}. In the future, it would be interesting to consider more generally what additional properties of the duality groups could be inferred from tame embeddability. In reference \cite{Delgado:2024skw}, the duality groups of algebraically compactifiable moduli spaces are proved to be semisimple. Algebraic compactifiability is stronger than standard compactifiability and seems closely related to tame isometric Euclidean embeddings. The latter has the added advantage that it holds beyond the cases where the moduli space is a complex manifold, implying that tame geometry could further extend these results.

\subsubsection*{Taming the Nash embedding theorem}

From a purely mathematical perspective, our results highlight the distinction between tame Riemannian manifolds and Riemannian manifolds admitting a tame isometric embedding. The connection with compactifiability also suggests that the latter is the more relevant notion for physical applications. Stating the precise mathematical conditions that ensure that a tame Riemannian manifold admits a tame isometric embedding into Euclidean space would require the formulation of a tame version of the Nash embedding theorem. 

\begin{subbox}{Tame embedding problem}
Under which geometric assumptions does a tame Riemannian manifold $(X,g)$ admit a tame isometric embedding into Euclidean space $\bbR^N$? 
\end{subbox}

We pose the establishment of such as theorem as an open mathematical problem which requires a deep synergy between Riemannian geometry and o-minimality.

The existing proof of the Nash embedding theorem fundamentally relies on operations which are completely incompatible with o-minimality: the isometric embedding is constructed through a limiting procedure in a perturbative scheme with oscillations of successively smaller length scales \cite{nash1956imbedding}. This indicates that entirely new methods may be needed for the proof of such a theorem. \\

An intermediate result would be to establish the theorem for compact tame manifolds. In this setting, we can already show that the theorem would hold for the subclass of compact analytic manifolds. In this setting, tameness can be inferred from the o-minimal structure of restricted analytic functions $\bbR_{\rm an}$ \cite{van1994real}. To establish the statement of tame embeddability, we can then use the analytic Nash embedding theorem~\cite{greene1971analytic}, which guarantees the existence of an analytic isometric embedding for any Riemannian manifold with an analytic metric. The definability in $\bbR_{\rm an}$ then implies the tameness of the isometric embedding.\\

Returning to moduli spaces, it is clear that compactness would be too strong of an assumption. In fact, it was conjectured that moduli spaces in quantum gravity should be non-compact in the original formulation of the distance conjecture \cite{Ooguri:2006in}. Nevertheless, from the relation between compactness and tameness of the embedding we infer that infinite distance limits are the main source of potential conflict. Given the general result of \cite{fontenele2014complexity}, spaces with negative curvature and infinite distance boundaries are especially problematic when considering tame embeddings. Consequently, the conjecture suggests a deep connection between the curvature of moduli spaces \cite{Marchesano:2023thx, Marchesano:2024tod, Castellano:2024gwi}, the distance conjecture \cite{Ooguri:2006in}, and the finiteness of complexity and information. A characterization of the properties that infinite distance limits must satisfy in order to verify the tame embedding conjecture could further enhance the understanding of all these topics. With section \ref{sec:highsugra} in mind, a viable mathematical strategy for the tame embedding conjecture would be to consider the most controlled moduli spaces in the quantum gravity landscape: the arithmetic quotients of higher supergravity.

\chapter{Complexity of effective field theories in the quantum gravity landscape}
\label{ch:EFTcomplexity}
\setlength{\parindent}{0pt}


Finiteness is a central theme in the quantum gravity constraints on low-energy effective field theories. It manifests itself in various ways, and constrains the size of the quantum gravity landscape \cite{Vafa:2005ui,Acharya:2006zw}, the matter spectrum \cite{Kumar:2010ru,Lee:2019skh,Kim:2019vuc,Katz:2020ewz,Tarazi:2021duw,Martucci:2022krl,Kim:2024eoa,Birkar:2025rcg}, geometric features of moduli spaces \cite{Ooguri:2006in,McNamara:2019rup}, and amplitudes \cite{Hamada:2021yxy}. In the previous two chapters we have seen how tameness provides a unifying perspective on finiteness in quantum gravity. Formalized by the tameness conjecture, o-minimality limits the permissible Lagrangians and geometries of effective theories consistent with quantum gravity, and has deep connections to several other swampland conjectures. \\

The aim of this chapter is to argue that the compatibility with quantum gravity enforces a bound on the complexity in effective field theories. In order to do this, we continue the ideas of chapter~\ref{ch:complexityQFT} and use sharp o-minimality to develop a measure of complexity for effective field theories. This is a significant conceptual step, since the data defining a Wilsonian effective field theory contains infinite towers of higher-dimensional operators and couplings, and appears a priori to have an infinite complexity. Using the ideas from part II of this thesis, we show based on several exactly solvable examples that in special cases the data defining the effective theory can be reorganized into an equivalent description of finite complexity. Subsequently, we use insights from the swampland program to argue that these special cases are realized in effective theories compatible with quantum gravity, and propose a \textit{finite complexity conjecture} which captures this idea. We first formulate a local version of the conjecture, which essentially states that each effective field theory in the quantum gravity landscape admits a description of finite complexity. Combining this with the hypothesis that there are finitely many distinct quantum-gravitational effective theories, we propose a stronger global version of the conjecture which asserts that the complexity of these theories is uniformly bounded across the quantum gravity landscape. We present evidence for the conjectures based on examples from string theory and connections to other finiteness conjectures.

\section{Complexity bounds on effective field theories}\label{sec:conjecture}
In the previous chapters of this thesis we introduced a method to characterize the complexity of mathematical objects in physical theories based on sharp o-minimality, and in chapter~\ref{ch:complexityQFT} we have implemented these ideas in the Lagrangians defining QFTs. The aim of this section is to generalize this idea to EFTs, which will require us to deal with various technical aspects such as the complexity of infinite sums, cutoffs and parameter dependence. Eventually, the goal is to develop a consistent complexity framework capable of describing EFTs in terms of sharp o-minimality and to capture finiteness properties of EFTs in the quantum gravity landscape. In order to do this, we have to address two essential conceptual points:  
\begin{itemize}
    \item The Lagrangian of an EFT in principle contains all local operators consistent with symmetries constructible from the fields and their derivatives. 
    \item EFTs frequently come in families, depending on a set of parameters which also contribute to the overall complexity of the effective theory. 
\end{itemize}
By considering these challenges we will develop a deeper understanding of how the information of an  EFT can be represented, how this information gets repackaged as the cutoff evolves and how different descriptions are needed to fully characterize a family of EFTs across its parameter space.

\subsection{Complexity and EFT Lagrangians}
Consider a general $d$-dimensional EFT Lagrangian
\begin{equation}
\cL = \sum_{n} \Lambda^{d-\Delta_n}c_n \cO_n(\Phi,\pd \Phi,\ldots) \,,    
\end{equation}
where $\Phi$ collectively denotes the field content, $\Lambda$ is the cutoff scale, and $\cO_n$ is a local dimension-$\Delta_n$ operator with (dimensionless) Wilson coefficient $c_n$. It appears at first that such a Lagrangian generically has infinite complexity, since it requires infinitely many independent Wilson coefficients to be specified. However, sharp o-minimality provides a resolution: in special cases, there may be algebraic relations among the set $\{c_n\}$ such that there are only finitely many independent coefficients. In these cases, the Lagrangian, as a function of $\Phi$, could potentially be resummed into a function which is definable in a sharply o-minimal structure with finite complexity. \\

Fully uncovering this structure requires a non-perturbative understanding of the theory under consideration. In the following we will consider a few non-perturbative examples where the exact effective Lagrangian is available, showing that it is in principle possible for these functions to have finite complexity. We first focus on the part of the Lagrangian that only depends on the fields, i.e.~the potential, and discuss higher-derivative interactions at the end of the subsection.

\subsubsection*{Zero-dimensional QFT}
As we have seen in chapter \ref{ch:observables}, in zero dimensions path integrals reduce to ordinary integrals, which enables exact calculations that can shed light on how effective actions may be described with finite complexity. We will illustrate this by considering a simple 0d QFT with two scalars, described by the Euclidean Lagrangian
\begin{equation}
    \cL(\phi,\chi) = \frac{1}{2} m^2 \phi^2 + \frac{1}{2} M^2 \chi^2 + \frac{\lambda}{4!} (\phi^2+ \chi^2)^2\,.
\end{equation}
Integrating out the $\chi$ field yields an effective Lagrangian for $\phi$, which is given by
\begin{equation}
    \cL_{\rm eff}(\phi) = \sum_{n\geq 0}  c_{2n} \phi^ {2n} = \frac{1}{2} m^2 \phi^2 + \frac{\lambda}{4!}\phi^4 -\log\left[\int \dd\chi \, e^ {-\frac{1}{2} (M^2+ \lambda\phi^2/6 )\chi^2 + \frac{\lambda}{4!} \chi^2} \right] \,,
\end{equation}
with an infinite set of Wilson coefficients $c_{2n}$. These coefficients can be obtained perturbatively by expanding the interaction term inside the integral over $\chi$, and collecting the contributions to each effective interaction in $\phi$. Performing this procedure, we find
\begin{equation}
    c_0 =- \log(Z_0)\,,  \quad  c_2 = \frac{1}{2}m^2 + a_2\,, \quad  c_4 = \frac{\lambda}{4!} + a_4 \quad\text{and} \quad c_{2n}= a_{2n} \quad \text{for  }n\geq3\,, 
\end{equation}
where for $n\geq 1$ the sequence $a_{2n}$ can be expressed as
\begin{equation}
     a_{2n} =  \sum_{\ell=1}^n  \tfrac{1 }{\ell} \big(- \tfrac{2}{Z_0}\big)^ \ell \!\! \sum_{   \substack{k_1+\cdots+k_\ell =n\\ k_i\geq 1}} \prod_{i=1}^ \ell \Big(  \sum_{k=0}^ \infty     \tfrac{ (-1)^ {k+k_i} }{ k!k_i! }   \left( \tfrac{ \lambda }{6}   \right)^{k+k_i}M^{-4k-2k_i-1} \,\Gamma\big(2k+k_i+\tfrac{1}{2}\big)  \Big) \,,
\end{equation}
and $Z_0$ is the partition function of $\chi$ evaluated at $\phi=0$, 
\begin{equation}
    Z_0 = \int \dd\chi \, e^ {-\frac{1}{2} M^2\chi^2 + \frac{\lambda}{4!} \chi^4} \,.
\end{equation}
Note that the expression for $a_{2n}$ is written in terms of asymptotic series due to the divergent nature of the perturbative expansion. \\

The expressions for the Wilson coefficients are rather complicated, and it is not immediately apparent how the resulting effective Lagrangian $\cL_{\rm eff}(\phi)$ could have a finite complexity. However, the 0d path integral can be performed exactly, yielding an analytic expression for the effective Lagrangian given by 
\begin{equation}
    \cL_{\rm eff}(\phi) = \frac{1}{2} m^2 \phi^2 + \frac{\lambda}{4!}\phi^ 4 -\log \left[  \sqrt{\frac{3}{\lambda}  \mu^2(\phi)    }  \exp(\frac{3\mu^4(\phi) }{4\lambda}) K_{1/4}\left( \frac{3\mu^4(\phi)}{4\lambda} \right)      \right] \,,
\end{equation}
where $\mu^2(\phi) = M^2 +\tfrac{1}{6}\lambda\phi^2$ is the effective squared mass of the $\chi$ field, and $K_{1/4}$ is a modified Bessel function of the second kind. It was shown in chapter \ref{ch:observables} that this function can be described in terms of a Pfaffian chain of differential equations, meaning that it is definable in the o-minimal structure $\bbR_{\rm Pf}$, and in particular that it has finite complexity. This simple example illustrates the following point: even if the effective Lagrangian has infinitely many independent Wilson coefficients, each of which takes a complicated form, it can still potentially be resummed into an object of finite complexity, provided that the UV theory is sufficiently simple and under computational control. The finiteness of complexity relies on a differential equation obeyed by the effective action, which reorganizes the information content of the infinite set of Wilson coefficients into a finite amount of information. 

 \subsubsection*{Exact renormalization group in higher-dimensional EFTs} 
 Another perspective on exact effective actions, which is valid in any spacetime dimension, is provided by exact RG equations such as the Wegner-Houghton equation discussed in chapter~\ref{ch:complexityQFT}. The current understanding of o-minimality is unable to incorporate the type of partial differential equation encountered in RG flows, but we expect that these equations are sufficiently tame to be definable in a novel o-minimal structure. 
 We leave a detailed analysis of these ideas for future work, but before we proceed, it is interesting to note that non-perturbative RG flow techniques have also been developed in the context of quantum-gravitational effective field theories \cite{Reuter:1996cp,Lauscher:2001ya,Reuter:2007rv,Saueressig:2023irs}.

\subsubsection*{Supersymmetric theories -- Seiberg-Witten example}
In EFTs with supersymmetry, the structure of the effective Lagrangian is much more constrained, and  non-perturbative statements can often be inferred with the help of non-renormalization theorems and geometric constructions. 
This happens for instance for BPS states in supersymmetric theories, whose masses robustly survive quantum corrections. In some cases, higher perturbative corrections vanish completely due to the holomorphy or symmetry constraints and one can obtain exact expressions of the Wilson coefficients of the effective theory, characterizing them completely with a finite amount of information. \\

We will illustrate this with an example, namely Seiberg-Witten theory \cite{Seiberg:1994rs} (we refer to \cite{Seiberg:1994aj,klemm:1995wp, Lerche:1996xu,Alvarez-Gaume:1996ohl} for reviews). This theory is a geometric description of  $\text{SU}(2)$ pure gauge theory with $\mathcal{N}=2$ supersymmetry in four dimensions. In the absence of matter hypermultiplets, the effective theory can be decomposed into two $\mathcal{N}=1$ chiral multiplets $W_\alpha$ and $A$, and we denote by $a$ the scalar component of $A$. The presence of $\mathcal{N}=2$ supersymmetry allows us to express the effective Lagrangian in terms of a single holomorphic function $F(A)$, the prepotential, as 
\begin{equation}
\begin{aligned}
        \cL = & \frac{1}{4\pi} \mathrm{Im} \Bigg[ 
\int \dd^2\theta \,\dd^2\bar{\theta} \,\frac{\partial F(A)}{\partial A} \bar{A} + \int \dd^2\theta \, \Big( \frac{1}{2} \sum_{\alpha} \frac{\partial^2 F(A)}{\partial A^2 } W^{\alpha} W_{\alpha} \Big) 
\Bigg]\,.
\end{aligned}
\label{eq: SW lagrangian}
\end{equation}
The quantum corrections to the classical prepotential $F=\frac{1}{2}\tau_0 a^2$ were found to be perturbatively exact at one-loop order, with non-perturbative corrections from a series of instantons \cite{Seiberg:1988ur}. The resulting full expression takes the form
    \begin{equation}
    F(a)=\frac{1}{2}\tau_0 a^2+\frac{i}{\pi}a^2\log\left(\frac{a}{\Lambda}\right) +\frac{1}{2\pi i}a^2\sum_{n=1}^\infty c_n \left(\frac{\Lambda}{a}\right)^{4n}\,,
    \label{eq: SW prepotential}
    \end{equation}
where the $c_{n}$ parametrize the instanton corrections and $\Lambda$ is a dynamically generated scale. The main realization of \cite{Seiberg:1994rs, Seiberg:1994aj} was to identify the moduli space of $\text{SU}(2)$ Yang-Mills with that of an elliptic curve, which allows for an exact calculation of the coefficients $c_n$. This description thereby yields full control over the non-perturbative effective action. It was shown in \cite{Matone:1995rx} that they can be encoded in terms of the function $G(a)=\pi i \big(F(a)-\frac{1}{2}a\partial_a F(a)\big)$, which satisfies the differential equation 
\begin{equation}
    (1-G^2)\frac{\dd^2G}{\dd a^2}+\frac{a}{4}\left(\frac{\dd G}{\dd a}\right)^3=0\,.
\end{equation}
Using these two differential equations the prepotential $F$ can be represented as a log-Noetherian function, so that the effective action in equation  \eqref{eq: SW lagrangian} is definable in an o-minimal structure and admits a description with finite complexity.\footnote{To fully describe the Lagrangian in terms of log-Noetherian functions over the whole moduli space, one needs to cover the moduli space by special regions called cells. We will study this point more in-depth in section \ref{sec:parameters}.} From this differential equation, one can recover the coefficients $c_n$ by means of a recursion relation \cite{Matone:1995rx}, which further supports the viewpoint that infinitely many a priori independent coefficients can be captured with finite information.

\subsubsection*{Geometry of higher-derivative interactions}
So far, we have focused on the potential of the EFT. However, a general EFT Lagrangian contains an infinite series of higher-derivative interactions. In order to consider derivative interactions in the framework of o-minimality, which in principle only describes functions and sets of finitely many real variables, we have to introduce an auxiliary real variable for each spacetime derivative. For example, in a theory with a scalar $\phi$ we introduce auxiliary variables $X_\mu = \pd_\mu \phi$, $X_{\mu\nu}=\pd_{\mu}\pd_{\nu}\phi$, etc.~so that a higher-derivative operator in the Lagrangian becomes a monomial in the auxiliary variables, for example
\begin{equation}
    \phi^{p} \big (\pd_\mu\phi  \,\pd^ {\mu }\phi\big)^ {q} \big(\pd_{\nu}\pd_\rho\phi \, \pd^\nu \pd^\rho \phi  \big)^{r} \longrightarrow \phi^{p} (X_\mu X^ \mu)^ {q} (X_{\nu\rho} X^ {\nu\rho} )^{r} \,.
\end{equation}
Recently, this approach was implemented for EFT Lagrangians in \cite{Craig:2023wni,Craig:2023hhp}, where the Lagrangian was formulated as a function on the jet bundle over field space, parametrized by the fields and the auxiliary derivative variables. In particular, it was argued in \cite{Craig:2023hhp} that the jet bundle EFT Lagrangian has a great amount of geometric structure which persists to any order in the derivative expansion.

\subsubsection*{Complexity of the higher-derivative expansion}
In this formulation, it is a well-defined mathematical question to consider the tameness and complexity of the EFT Lagrangian as a function of the fields and the auxiliary variables including higher-derivative interactions. The Wegner-Houghton equation, studied in chapter~\ref{ch:complexityQFT}, can be systematically modified beyond the local potential approximation to describe the non-perturbative RG evolution of higher-derivative interactions \cite{Bonanno:1999ik}, so that the discussion earlier in this section extends to higher derivatives. \\

When extending this discussion to the full EFT Lagrangian including all orders of the derivative expansion, this approach runs into a fundamental problem for tameness. The expansion would require the inclusion of infinitely many auxiliary variables, meaning that we would have to formulate the Lagrangian on the infinite jet bundle, while o-minimality is only defined in finite-dimensional settings. One way of addressing this issue would be to truncate the derivative expansion, introducing only finitely many auxiliary variables, and consider the complexity of the truncated EFT Lagrangian. Note that one can then still consider general functions of these auxiliary variables, which means that effectively some of the infinite towers of derivative interactions remain. For example, if one includes the first and second derivatives of the spacetime metric as auxiliary variables, one can still consider the complexity of general functions $f(R,R_{\mu\nu},R_{\mu\nu\rho\sigma})$ of the curvature invariants in a gravitational EFT. An alternative approach would be to formulate the Lagrangian in momentum space, in which the replacement $\pd_{\mu}\to p_\mu$ completely circumvents the introduction of auxiliary variables. We could then consider the complexity of the function $\cL(\widehat\Phi,p_\mu)$, where $\widehat\Phi$ collectively denotes the Fourier transforms of the field content. This approach would in principle allow one to quantify the complexity of the complete EFT expansion. Since we focus on the conventional position-space Lagrangians, we will not consider this approach here. In the remainder of this chapter, we focus mainly on the two-derivative sector of the EFT Lagrangian.

\subsection{Complexity, parameter spaces, and EFT domains}\label{sec:parameters}
In the previous subsection, we have focused on the complexity of the EFT Lagrangian as a function of the fields. We now extend this discussion to also include the complexity associated to the external parameters. Consider an EFT with a single scalar field $\phi$, whose two-derivative Lagrangian takes the general form 
\begin{equation}
    \cL(\phi,\pd\phi) = -\frac{1}{2}\pd_\mu \phi\, \pd^ \mu \phi + \sum_{n=0}^\infty c_n \phi^n \,.
\end{equation}
The complete parameter space of this family of EFTs is the infinite-dimensional space $\cM$ spanned by the Wilson coefficients $\{c_n\}$. As we have learned in the previous subsection, a generic point in this space determines an EFT with infinite complexity. However, on special subsets of $\cM$, the Lagrangian may be recast into a tame function of finite complexity. With this idea in mind, we can formally cover $\cM$ by subsets on which the Lagrangian has a \textit{fixed} complexity. We will call these sets \textit{EFT domains}, and formalize this concept at the end of this section.

\subsubsection*{Example: complexity and symmetries in 0d}
To illustrate this notion in a simple example, consider the Lagrangian
\begin{equation}\label{eq:SON-Lagrangian}
    \cL(\phi_1,\ldots,\phi_N) = \frac{1}{2}\sum_{j=1}^N m_j^2 \phi_j ^2  +  \frac{\lambda}{4!} \Big(\sum_{j=1}^N  \phi_j^2\Big ) ^2 \,,
\end{equation}
describing a 0d theory of $N$ scalars $\phi_1,\ldots,\phi_N$ with a quartic interaction. The parameter space $\cM$ of this family of theories is the $(N+1)$-dimensional space given by the values of $m_1^2,\ldots,m_N^2$ and $\lambda$. At a generic point in $\cM$, the complexity of the Lagrangian is $(\cF,\cD)= (N,4)$. The subset given by $\lambda=0$ then defines an $N$-dimensional EFT domain on which the complexity reduces to $(N,2)$, since the degree of the polynomial potential is lowered. Consider now the loci on which $n$ masses coincide, e.g.~$m_1^2=\cdots = m_n^2=m^2$. Within this set, the theory acquires an $\text{SO}(n)$ symmetry, and by rewriting the fields $\phi_1,\ldots,\phi_n$ in spherical coordinates, they can be fully described in terms of a radial field $\rho$. Here the Lagrangian becomes\footnote{Note that at the quantum level, there is an additional term $(N-1)\log \rho$ in the effective Lagrangian due to the Jacobian of the path integral measure $\dd \phi_1\cdots\dd\phi_N =\rho^{N-1} \dd \rho \, \dd \Omega_{N-1}$. This term can be included as a Pfaffian function, and increases the complexity slightly but independently of $N$ and $n$.} 
\begin{equation}
    \cL(\rho,\phi_{n+1},\ldots,\phi_N) = \frac{1}{2} m^2\rho^2 + \frac{1}{2}\sum_{j=n+1}^N m_j^2 \phi_j ^2   +  \frac{\lambda}{4!} \Big(\rho^2+\sum_{j=n+1}^N  \phi_j^2\Big ) ^2\,,
\end{equation}
so that this subset of the parameter space defines an EFT domain on which the Lagrangian has reduced complexity $(N-n+1,4)$. The parameter space $\cM$ can in this way be fully decomposed into EFT domains with varying complexity. In this example, the EFT covering of $\cM$ is associated to symmetries of the Lagrangian.

\subsubsection*{EFT domains and local Lagrangians} 
In the discussion above, the parameter space was taken to be the space of couplings or a subset thereof, and there was a globally valid Lagrangian over the entire parameter space. In many cases, for instance for the EFTs arising from quantum gravity, the parameter space of a family of EFTs instead consists of a set of vacua formed by vacuum expectation values (vevs) of fields, and the values of the Wilson coefficients depend on these vevs. In these cases, an EFT covering of the parameter space is further constrained by the fact that there may not be a globally valid Lagrangian. For example, consider the $\text{SU}(2)$ Seiberg-Witten theory, for which the Coulomb branch moduli space coincides with the moduli space of an elliptic curve, given by a Riemann sphere with three marked points. In this setting it is known that there is no globally valid EFT Lagrangian \cite{Seiberg:1994aj}. Instead, the moduli space must be covered by at least three regions, each located around a different singularity and endowed with its own locally valid EFT description. In this case, the underlying physical mechanism is the convergence of the instanton expansions around the various singularities. The complexity of the Lagrangians of Seiberg-Witten theory, including the EFT covering of the moduli space, is studied in great detail in \cite{Carrascal:2025vsc}.\\

\begin{figure}[h!]
    \centering
    \includegraphics[width=0.95\linewidth]{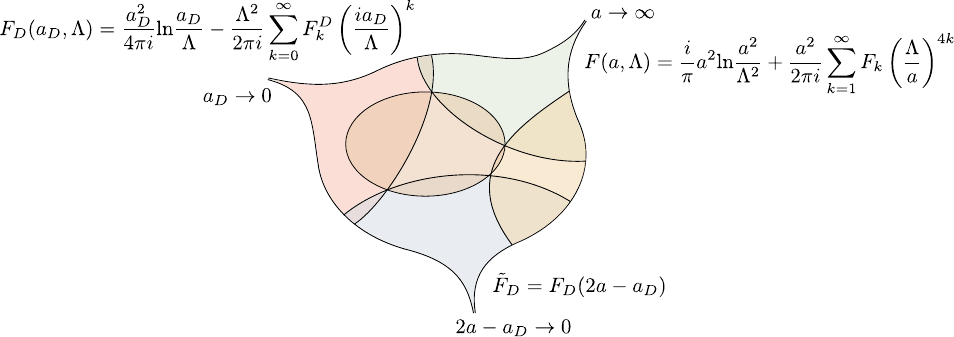}
    \caption{EFT covering of the $\mathcal{N}=2$ $\text{SU}(2)$ Seiberg-Witten moduli space with finitely many EFT domains. Around each of the three infinite distant limits there exists a different EFT description of finite complexity in terms of different combinations of variables $a$ and $a_D$, capturing the electric-magnetic duality discussed in \cite{Seiberg:1994rs}. In addition, three other discs centered around regular points are included to cover the full moduli space.}
    \label{fig: SW patches}
\end{figure}

Finally, let us note that the parameter space $\cM$ of an EFT can be a complicated manifold which is not naturally embedded as a subset of Euclidean space. In order to implement the framework of sharp o-minimality, we must therefore view $\cM$ as a tame manifold, and assign a complexity $(\cF,\cD)$ through the tame coordinate charts. Hence, in order to define the complexity of a locally valid Lagrangian in parameter space, the notion of EFT domain must be compatible with the tame coordinates on $\cM$.

\subsubsection*{Definition of EFT coverings}
The discussion of this section suggests that it is useful to introduce a notion of EFT domains and EFT covering, and we formalize this as follows. 

\begin{subbox}{EFT domains and EFT coverings}
  Given a set of EFTs parametrized by a space $\cM$, we define an \textit{EFT domain} as a subset $U\subseteq \cM$ on which the EFTs are described by a single Lagrangian $\cL_U$ with a fixed complexity $(\cF_U,\cD_U)$. 
An \textit{EFT covering} $\mathcal{U}$ is a collection of EFT domains $U_i$ which cover $\cM$.   
\end{subbox}
These concepts reflect the cell decomposition theorem, as well as Hardt's decomposition theorem for definable families upon viewing the local Lagrangians as defining the fibers of a bundle \cite{VdDries}.

We require that EFT domains and coverings have the following properties:
\begin{itemize}
    \item Each EFT domain has a well-defined dimension. Different domains within an EFT covering may have different dimensions. In contrast to the cells of a tame set, we allow the $U_i$ to overlap. 
    \item Every EFT domain $U$ is assumed to have a tame coordinate map $U\to V\subseteq \bbR^n$, and the complexity of the Lagrangian is evaluated on the set $V$.
\end{itemize}
On regions in the parameter space where there is no Lagrangian with finite complexity, we formally set $\cF$ and $\cD$ to infinity.
Intuitively, an EFT covering is simply a parametrization of Lagrangian descriptions which together cover the parameter space of a set of EFTs.

\section{Complexity bounds and quantum gravity}
\label{sec:complexity and quantum gravity}

In this section, we build on the ideas of the previous section and combine them with quantum gravity constraints on effective theories. This will lead us in section~\ref{sec:complexity_conjecture} to propose two  finite complexity conjectures on EFTs in the quantum gravity landscape. In section~\ref{sec:complexity_comp} we provide further motivation for these conjectures by discussing how finite complexity EFTs can arise in string theory and compactifications of higher-dimensional theories. Finally, in section~\ref{sec:complexity_volume_counting} we discuss the connection of the finite complexity conjectures with other swampland conjectures related to volumes in moduli spaces and counting of theories. 

\subsection{Swampland conjectures on complexity bounds} \label{sec:complexity_conjecture}

We now turn to the formulation of the two conjectures. First we focus on an individual EFT, and propose a local version of the conjecture. We then take a more global perspective, and propose a stronger version of the conjecture which states that the complexity of EFTs is uniformly bounded across the landscape.\\

We have seen that completely generic EFT Lagrangians require an infinite amount of information to be described, encoded in the infinite set of Wilson coefficients. However, in top-down cases where the UV origin of the effective theory is sufficiently well-understood, either by non-perturbative techniques, supersymmetry, or other geometric constraints, it appears that this information can be reorganized into an object of finite complexity. 

\subsubsection*{Finiteness of light states and the swampland} 
Let us make a clearer connection between the finiteness of complexity and finiteness constraints proposed in the swampland program. From the inception of the program, it was argued that quantum gravity bounds the ranks of gauge groups and massless spectra of EFTs \cite{Vafa:2005ui,Kumar:2009ae,Kumar:2010ru, Morrison:2011mb,Park:2011wv,Grimm:2012yq}; an idea that is supported by the claimed finiteness of compact Calabi-Yau threefolds \cite{reid1987moduli,yau2008survey}, which has seen much progress when restricting to elliptic fibrations \cite{Gross1993AFT, MR4939522,MR4801611,Birkar:2025gvs}. In particular, bounding the field content in the setting of six-dimensional supergravity has recently been a very active topic ~\cite{Lee:2019skh,Kim:2019vuc,Katz:2020ewz,Tarazi:2021duw,Kim:2024eoa,Birkar:2025rcg} (see also \cite{Martucci:2022krl} for a discussion on bounds in four dimensions). The finiteness of massless and light states is further supported by the general expectation that quantum gravity imposes a fundamental cutoff scale $\Lambda_{\rm QG}$ beyond which the EFT description is no longer valid. Depending on the mass hierarchies of the fundamental states of the underlying UV theory, this scale can have a different origin as has been investigated in many recent works \cite{Dvali:2007hz,Dvali:2007wp,Dvali:2008ec,Grimm:2018ohb,Castellano:2021mmx,vandeHeisteeg:2022btw,Cribiori:2022nke,Castellano:2022bvr,vandeHeisteeg:2023dlw,Blumenhagen:2023yws,Burgess:2023pnk,Bedroya:2024uva,Calderon-Infante:2025ldq,ValeixoBento:2025bmv}. 
A universal upper bound on $\Lambda_{\rm QG}$ is expected to be given by the species scale $\Lambda_{\rm sp}$, which is parametrically lowered when the number of light species increases. Consequently, EFTs coupled to gravity cease to be valid when too many light species are included. In particular, for a given EFT cutoff $\Lambda$, we must have  $\Lambda<\Lambda_{\rm QG}$, which bounds the number of fields. These two ideas together imply that the contribution to the complexity of the Lagrangian coming from the number of fields is bounded for EFTs in the quantum gravity landscape.  

\subsubsection*{Finiteness of complexity -- individual EFTs} 
Even if one assumes the finiteness of states in the EFT, the Lagrangian could still depend on infinitely many independent Wilson coefficients. Supporting our expectation to have a finite complexity representation, it is argued in  \cite{Heckman:2019bzm} that the Wilson coefficients of an EFT consistent with quantum gravity are determined by finitely many parameters. This implies that the infinitely many terms of the EFT Lagrangian are actually related to each other, hinting towards the existence of a finite complexity representation of the EFT expansion as discussed in section~\ref{sec:conjecture}.\footnote{A typical way to encode relations among the Wilson coefficients is to describe the series with a differential equation with finitely many controllable coefficients, as exemplified for Pfaffian structures throughout this thesis.} 
Taken together, these observations lead us to propose a local version of the finite complexity conjecture:

\begin{subbox}{Local finite complexity conjecture}
    Every EFT consistent with quantum gravity has a description in which the two-derivative Lagrangian is definable in a sharply o-minimal structure and has finite complexity $(\cF_{\rm EFT},\cD_{\rm EFT})$.
\end{subbox}

Let us emphasize that this conjecture is formulated at the level of an individual EFT in the quantum gravity landscape. It formalizes the idea that each of these EFTs is non-generic and admits a description of finite complexity $(\cF_{\rm EFT},\cD_{\rm EFT})$, with format and degree depending on the considered EFT.
It can be viewed as a quantitative refinement of the claim that the Lagrangian of an  EFT consistent with quantum gravity is a tame function, upgrading the corresponding part of the tameness conjecture \cite{Grimm:2021vpn} by replacing o-minimality with sharp o-minimality. The conjecture is also strictly stronger than the finiteness claim for Wilson coefficients proposed in \cite{Heckman:2019bzm}, since functions depending on only finitely many parameters may still be non-tame while appearing to carry finite information. The distinction arises because sharp o-minimality imposes finiteness constraints that are stable under all logical operations, rather than merely at the level of parametrization. \\

It is important to highlight that the conjecture is only formulated for effective field theories admitting a Lagrangian description, and that the evaluation of the complexity is obtained by restricting the Lagrangian to the two-derivative level. Rather than being motivated by physical arguments, these conditions arise from the limitations of the application of the formalism of sharp o-minimality to quantum field theories detailed in chapter \ref{ch:complexityQFT} and section \ref{sec:conjecture}. However, the picture is by no means complete or unique, and alternative approaches could allow for a more general characterization of complexity that enable an extension of the regime of validity of the conjecture. In particular, we expect that including a finite number of higher-order derivative fields by the procedure explained in the previous section would increase the complexity bounds while maintaining the core statement of the conjecture. Accounting for the complete higher-derivative expansion is much more challenging and requires developing new procedures for the evaluation of the complexity, as discussed in the previous section. Finally, extending the validity of the conjecture to non-Lagrangian theories may be feasible by  
using sharp o-minimality to formulate appropriate finite complexity bounds from a different perspective, for instance by analyzing quantum gravity constraints on their observables, geometric aspects of moduli spaces, or dualities relating them to Lagrangian theories.

\subsubsection*{Finiteness of complexity -- landscape of EFTs} 
While the local finite complexity conjecture proposes that each effective theory in the landscape admits a finite, EFT-dependent complexity $(\cF_{\rm EFT},\cD_{\rm EFT})$, it does not exclude the possibility that there exist families of consistent quantum gravitational EFTs whose complexity becomes unbounded. However, it is widely believed that the landscape of EFTs arising from string theory, and more generally from any consistent theory of quantum gravity, is finite once suitable cutoffs are imposed \cite{DouglasStringsTalk,Vafa:2005ui,Acharya:2006zw,Hamada:2021yxy}. \\

For example, it has been conjectured in \cite{Hamada:2021yxy} that there are only finitely many distinct EFTs consistent with quantum gravity that are valid up to a chosen cutoff scale $\Lambda$ if one identifies two EFTs in the same connected moduli space. Such finiteness statements are supported by the aforementioned broad body of results \cite{Kumar:2009ae,Kumar:2010ru, Morrison:2011mb,Park:2011wv,Grimm:2012yq,reid1987moduli,yau2008survey,Gross1993AFT,MR4939522,MR4801611,Birkar:2025gvs,Lee:2019skh,Kim:2019vuc,Katz:2020ewz,Tarazi:2021duw,Kim:2024eoa,Birkar:2025rcg,Martucci:2022krl} as well as by the recent general proofs of the finiteness of flux vacua in Type IIB string theory and F-theory \cite{Bakker:2023xkt,Grimm:2021vpn}. A first conceptual refinement of these finiteness results was formulated in the tameness conjecture \cite{Grimm:2021vpn,Douglas:2023fcg}, which asserts that the field spaces and parameter spaces of EFTs valid below a cutoff $\Lambda$ are definable in an o-minimal structure. While this guarantees finiteness in a qualitative sense, it does not assign quantitative bounds and therefore does not yet allow for explicit, computable measures of complexity. 
Extending the tameness conjecture of \cite{Grimm:2021vpn,Douglas:2023fcg} and the local finite complexity conjecture stated above, we arrive at the hypothesis that there is a uniform bound on the complexity of EFTs in the landscape. We formalize this in our main conjecture as follows:

\begin{subbox}{Finite complexity conjecture}
For fixed spacetime dimension $d$, consider a finite energy cutoff $\Lambda>0$ in $d$-dimensional Planck units.
\begin{itemize} 
\item[(i)] The set of $d$-dimensional EFTs consistent with quantum gravity with energy cutoff at least $\Lambda$ is definable in a sharply o-minimal structure and has finite complexity $(\cF_\Lambda,\cD_{\Lambda})$.

\item[(ii)] There exists a pair of positive integers $(\mathfrak{F}_\Lambda,\mathfrak{D}_\Lambda)$, such that any $d$-dimensional EFT consistent with quantum gravity with energy cutoff at least $\Lambda$ has a description for which the two-derivative Lagrangian is definable in a sharply o-minimal structure and has a complexity $(\cF_{\rm EFT},\cD_{\rm EFT})$ satisfying $\cF_{\rm EFT}\leq \mathfrak{F}_\Lambda$ and $\cD_{\rm EFT}\leq\mathfrak{D}_\Lambda$.
\end{itemize}
\end{subbox}

Informally speaking, the conjecture proposes that there is an information limit on effective theories of quantum gravity, implemented by a uniform complexity bound across the landscape. It integrates various perspectives on finiteness ideas that have been suggested in the swampland program and opens the possibility to formalize these through quantitative bounds.  
Let us now discuss the statement of the conjecture in more detail.

\subsubsection*{Part (i) of the conjecture}
Let us begin by discussing the first part of the conjecture. It concerns a set $\cM_{\rm QG;\Lambda}$ which is assumed to parameterize the space of EFT Lagrangians that are valid at least up to some cutoff scale $\Lambda$. In general, the space $\cM_{{\rm QG};\Lambda}$ is expected to be a very complicated object, consisting of a formal union of parameter spaces, moduli spaces, and field spaces of the EFTs. Its precise definition will require to address several important points, e.g.~when two EFTs are considered to be equivalent. Furthermore, the structure and boundaries of $\cM_{{\rm QG};\Lambda}$ will be enforced by inherent consistency of the theories and in particular by the requirement that the considered EFTs are compatible with quantum gravity. The conjecture that the full space $\cM_{\rm QG; \Lambda}$ is definable in a sharply o-minimal structure therefore constitutes a great challenge to verify in generality. However, the complexity constraints translate also to all smaller subspaces obtained, for example, by appropriate restrictions.\footnote{Note that this aligns with the idea that finiteness in sharp o-minimality is compatible with operations such as linear projections or intersections.} This fact allows for testing the conjecture in many settings or finding concrete counter-examples. We will highlight some basic examples in favor of the conjecture in section~\ref{sec:complexity_comp}. \\

We also emphasize that even though the conjecture claims that $\cM_{{\rm QG};\Lambda}$ has finite complexity, it leaves open how precisely one assigns the complexity to this set. In principle, a format and degree can only be assigned to sharply o-minimal subsets of Euclidean space. For components of $\cM_{{\rm QG};\Lambda}$ that have more structure, such as a Riemannian metric, it is tempting to introduce a notion of sharply o-minimal manifolds. A natural definition of such a manifold is obtained by 
tracing through the definition of a tame manifold, and quantifying the complexity of each step in the construction. In this way, the complexity would correspond to the sum of the complexities of the tame coordinate regions and their coordinate transition functions. 
An alternative perspective is to embed $\cM_{{\rm QG};\Lambda}$ into Euclidean space, so that it directly admits a well-defined format and degree $(\cF_{\Lambda}^{\rm emb},\cD_{\Lambda}^{\rm emb})$. At first, the additional requirement of an embedding may appear unnatural. However, as shown in chapter~\ref{ch:volumes}, if one combines the embedding with the idea that field spaces have an intrinsic notion of distance, and one takes this metric into account by demanding that the embedding is isometric, one recovers various known quantum gravity constraints. We will come back to these interpretations in 
section~\ref{sec:complexity_volume_counting}, where we will comment on how complexity is related to volume growth and counting. 

\subsubsection*{Moduli space of quantum gravity theories}
Given our notion of $\cM_{{\rm QG};\Lambda}$, we now introduce the set of all EFTs consistent with quantum gravity denoted by $\cM_{\rm QG}$ as being given by the $\Lambda\to 0$ limit of $\cM_{{\rm QG};\Lambda}$. This limiting procedure captures all EFTs consistent with quantum gravity, valid up to an arbitrarily small $\Lambda$. We can attempt to assign a complexity $(\cF_{\rm QG},\cD_{\rm QG})$ to $\cM_{\rm QG}$ by a limiting procedure, if we are able to find representations for all $\cM_{{\rm QG};\Lambda}$ that have a uniform bound $\cF_{\Lambda} \leq \cF_{\rm QG}$ and $\cD_{\Lambda}  \leq \cD_{\rm QG}$.\footnote{This aligns with the notion of limit sets in tame geometry \cite{van2005limit}.} It is then tempting to extend the part (i) of the conjecture to the whole space $\cM_{\rm QG}$ and claim that the complexity $(\cF_{\rm QG},\cD_{\rm QG})$ is finite. It turns out, however, that one then would have to refine the notion of $\cM_{\rm QG}$, since otherwise there are immediate counter-examples arising when the parameter set is infinite and discrete, such as for Type IIB string theory backgrounds like AdS$_5\times S^5$ with $N$ units of flux. 
We will discuss a possible way to deal with these cases in section~\ref{sec:complexity_volume_counting} and refrain from extending the conjecture at this point. \\

It is also instructive to compare our definition of $\cM_{\rm QG}$ with the notion of moduli spaces used in \cite{Ooguri:2006in} and subsequent follow-ups. In string theory settings one often introduces a moduli 
space $\cM$ without setting a cutoff scale, and therefore does not work within a single EFT. In our picture, we view these moduli spaces $\cM$ as components of $\cM_{\rm QG}$, which might be labeled by additional (potentially discrete) parameters. This allows us 
to extend the conjectures about viable moduli spaces $\cM$ to $\cM_{\rm QG}$. In particular, extending \cite{Ooguri:2006in} one obtains the statement that $\cM_{\rm QG}$ is parametrized by inequivalent expectation values of fields and there are no other free parameters.\footnote{Note that this is a familiar fact within string theory but, there is no a priori reason this should be the case in every theory of quantum gravity.} In section~\ref{sec:complexity_volume_counting} we
will further comment on how some of the swampland conjectures apply to $\cM_{\rm QG}$ and interplay with the tameness of $\cM_{{\rm QG};\Lambda}$. 

\subsubsection*{Part (ii) of the conjecture}
The tame geometry of $\cM_{{\rm QG};\Lambda}$ and the local finite complexity conjecture imply that the parameter space $\cM_{{\rm QG};\Lambda}$ has a \textit{finite} EFT covering, i.e.~a covering by finitely many domains in which an EFT Lagrangian can be defined. This formalizes and generalizes the conjecture on the finiteness of EFTs valid up to a cutoff $\Lambda$:

\begin{subbox}{Finiteness of EFTs}
The space $\cM_{{\rm QG};\Lambda}$ of EFTs with cutoff at least $\Lambda$ has a finite EFT covering. 
\end{subbox}

In particular, for a given $\Lambda$, there should in principle be a mininum number $\cN_\Lambda$ of EFT domains required to cover $\cM_{{\rm QG};\Lambda}$, which would depend on the complexity $(\cF_\Lambda,\cD_\Lambda)$ as well as on the complexity of the corresponding EFTs. Part (ii) of the conjecture then implies the all of these EFTs have a common upper bound $(\mathfrak{F}_\Lambda,\mathfrak{D}_\Lambda)$ on their complexity. Assuming that the explicit values for $(\mathfrak{F}_{\Lambda},\mathfrak{D}_\Lambda)$ are known, any EFT with cutoff $\Lambda$ that violates these bounds by not having any description with $\cF_{\rm EFT}<\mathfrak{F}_\Lambda$ and $\cD_{\rm EFT}<\mathfrak{D}_\Lambda$ belongs to the swampland.

\subsection{Finite complexity from string theory and compactification} \label{sec:complexity_comp}

In this section we discuss a number of examples supporting the finite complexity conjecture. 
In order to do that, we start from string theory as a candidate theory of quantum gravity and explore the landscape of its low-energy EFTs. In addition to supporting the statement of the conjecture, some of the examples will 
clarify why the assumptions of the conjecture are necessary. 

\subsubsection*{Scales in quantum gravity} 
Before turning to explicit examples, let us discuss in more detail the role of the cutoff scale. String theory generally contains infinite towers of excitations, such as string modes and Kaluza–Klein states in lower-dimensional compactifications. Since an EFT of finite complexity cannot include infinitely many individual states, one must introduce a cutoff $\Lambda$ that restricts the effective description to finitely many degrees of freedom. Validity of the EFT requires $\Lambda\le \Lambda_{\rm QG}$, where $\Lambda_{\rm QG}$ denotes the intrinsic quantum gravity scale at which the EFT breaks down. Determining the appropriate estimate for $\Lambda_{\rm QG}$ is therefore important for the finite complexity conjecture. The cutoff lies below the $d$-dimensional Planck mass $M_{\rm pl}$, and minimally has to account for the impact of light species, leading to the species scale $\Lambda_{\rm sp}<M_{\rm pl}$ \cite{Dvali:2007hz,Dvali:2007wp,Dvali:2008ec}. However, further obstructions can arise \cite{Burgess:2023pnk,Bedroya:2024uva}. In the presence of an interacting tower of states, such as a Kaluza–Klein tower, a natural choice for the cutoff is the mass of the lightest state in the tower, defining the tower scale $\Lambda_{\rm t}$. This is well-motivated near infinite-distance limits of moduli spaces, where towers of states are expected to become light \cite{Ooguri:2006in}, but $\Lambda_{\rm t}$ might fail to control quantum-gravitational effects in the bulk. To address this, the black hole scale $\Lambda_{\rm BH}$ was introduced in \cite{Bedroya:2024uva}, signaling a breakdown of the EFT description through black hole instabilities. It satisfies $\Lambda_{\rm BH}\le \Lambda_{\rm sp}$ and approaches $\Lambda_{\rm t}$ in asymptotic regions. In the examples below we focus on the asymptotic regime and adopt the tower scale $\Lambda_{\rm t}$ as a natural cutoff (avoiding partial truncations of the tower); using $\Lambda_{\rm BH}$ would mainly modify the numerical complexity estimates while leaving the qualitative conclusions unchanged.

\subsubsection*{Ten-dimensional supergravity theories} 
The five consistent supersymmetric string theories reduce to supergravity theories with $\cN=2$ (Type IIA/Type IIB) and $\cN=1$ (heterotic/Type I) supersymmetry. These theories have no free parameters other than the ten-dimensional Planck mass. Nevertheless they admit non-trivial moduli spaces $\cM_{\Lambda}$. We will now briefly discuss how these theories admit a description of finite complexity using Type IIA and Type IIB as examples.\\

The starting point to understand Type IIA is 11-dimensional supergravity, a theory with no scalars and no free parameters. Upon compactification on a circle, we obtain Type IIA supergravity in ten dimensions. The resulting theory has a single scalar, the dilaton $\phi$, that is related to the radius of the circle through $R_{11}=e^{\frac{2}{3} \phi}$ in 11-dimensional Planck units. Consequently, the moduli spaces $\cM_{\Lambda}$ are finite-length intervals parameterizing the vacuum expectation values $g_s$ of the dilaton field $e^{\phi}$. The action of the massless fields including both fermionic and bosonic components can be found for instance in \cite{Bergshoeff:2001pv} up to four fermionic terms, from which it is straightforward to verify that both the field content and the functions involved have finite complexity. Even if we account for higher fermionic interactions, these are polynomial and thus have a bounded complexity, as argued in chapter~\ref{ch:complexityQFT}. The massless spectrum of the theory is complemented by towers of massive states,  generated by string excitations (obtained from M2 branes wrapping the circle in the 11-dimensional picture) and the non-perturbative D0 states (which are the Kaluza-Klein modes of the compactification). \\

\begin{figure}[h!]
    \centering
    \includegraphics[width=0.8\linewidth]{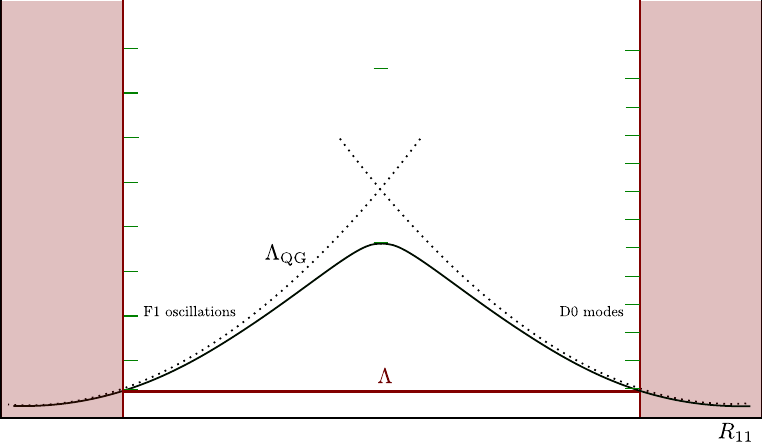}
    \caption{Depiction of the moduli space $\cM_{\Lambda}$ of Type IIA supergravity in ten dimensions. The boundaries of the region are located at the points in which $\Lambda_{\rm QG}$ (given by in this case by the lightest state of the tower) falls below $\Lambda$. The shaded areas correspond to inaccessible regions of the moduli space of the compactification that do not admit an EFT with the chosen cutoff. }
    \label{fig: diagram IIA in 10d}
\end{figure}

The masses of these excited states are given by
\begin{equation}
     M^2=\frac{k}{R_{11}^2}+\frac{2}{\alpha'}(N_L+N_R+E_0)\,,
    \label{eq: mass excited states}
\end{equation}
where $k$ and $(N_L,N_R)$ count the Kaluza-Klein momentum and the string excitations respectively. From equation \eqref{eq: mass excited states}, it is clear that the cutoff $\Lambda_{\rm t}$, set by the mass of the lightest field in the towers, changes for every point of the moduli space spanned by $e^\phi$.\footnote{For constant Planck mass, this means that $\alpha'$ changes as we move in moduli space since the $d$-dimensional string coupling, the $d$-dimensional Planck mass and the string mass are related by $1/g_{s,d}\sim \alpha'^2 M_{{\rm pl},d}^4$.} The complexity of the EFT becomes finite when considering a fixed cutoff $\Lambda$. If $\Lambda<\Lambda_{\rm t}$,
we are only left with the massless content and the associated Lagrangian can be written in terms of sharply o-minimal functions with bounded complexity, in agreement with the finite complexity conjecture. Above $\Lambda_{\rm t}$, no EFT description exists.  Therefore, for a given choice of probe cutoff $\Lambda$, there is a patch $\mathcal{M}_{\Lambda}$ in the radial moduli space bounded by the values at which the Kaluza-Klein modes (in the large radius limit) or the winding modes (in the small radius limit) become lighter than the mass scale $\Lambda_{\rm QG}$, as seen in figure \ref{fig: diagram IIA in 10d}.  It is inside this patch where there exists a well-defined Lagrangian description with finite complexity and cutoff valid at least up to $\Lambda$. \\

\begin{figure}[h!]
    \centering
    \includegraphics[width=0.7\linewidth]{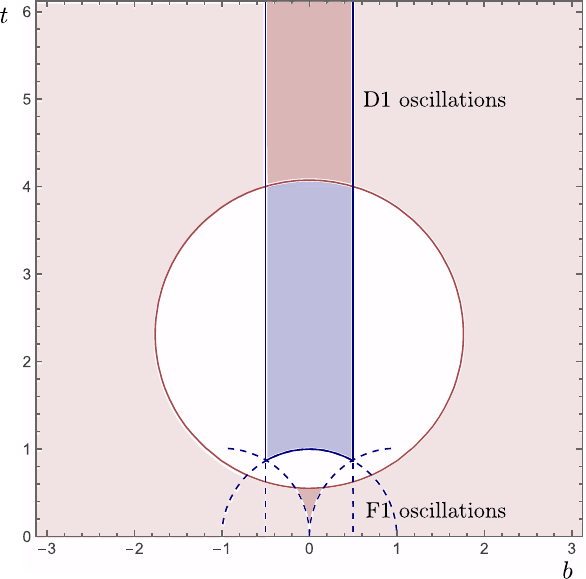}
    \caption{Depiction of the moduli space $\cM_{\Lambda}$ of Type IIB supergravity in ten dimensions as a subset of the hyperbolic plane. The solid blue lines depict the boundaries of the fundamental domain of $\mathbb{H}/\text{SL}(2,\mathbb{Z})$. The region of the fundamental domain where the EFT supergravity description is valid is highlighted with shaded blue, and the dark shaded red corresponds to the region of the fundamental domain where the masses of the tower of D1 oscillations fall below the cutoff $\Lambda$. Finally, the region outside the fundamental domain where the masses of the S-dual string tower fall below the cutoff has also been highlighted. }
    \label{fig: diagram IIB in 10d}
\end{figure}

Type IIB supergravity in ten dimensions displays many of the same general features, but the realization is considerably different, since in addition to the dilaton there is another free scalar field, namely the RR 0-form $c_0$. Together they generate the two-dimensional axio-dilaton moduli space $\cM_{\rm IIB} = \mathbb{H}/\text{SL}(2,\mathbb{Z})$, which we studied in detail in the previous chapter. Writing $\mathbb{H}$ as a group quotient this space can be expressed as the arithmetic quotient
\begin{equation} \label{eq:IIBquotient}
   \cM_{\rm IIB} = \text{SL}(2,\mathbb{Z}) \backslash \text{SL}(2,\bbR)/\text{SO}(2) \,.
\end{equation}
As it was the case in the Type IIA discussion, the complexity of the Lagrangian of the massless content is finite and towers of states become exponentially light in the infinite distance limits of moduli space. When fixing a cutoff, we can determine a region of finite geodesic distance where the EFT description is valid. After accounting for the quotient by the duality group there is a single infinite distance limit and the allowed region at a fixed cutoff takes the shape shaded in blue in figure \ref{fig: diagram IIB in 10d}. Later in this section we will return to the discussion of complexities of quotient spaces such as \eqref{eq:IIBquotient}.

\subsubsection*{M-theory on a torus}
\begin{figure}[h!]
    \centering
    \includegraphics[width=0.9
    \linewidth]{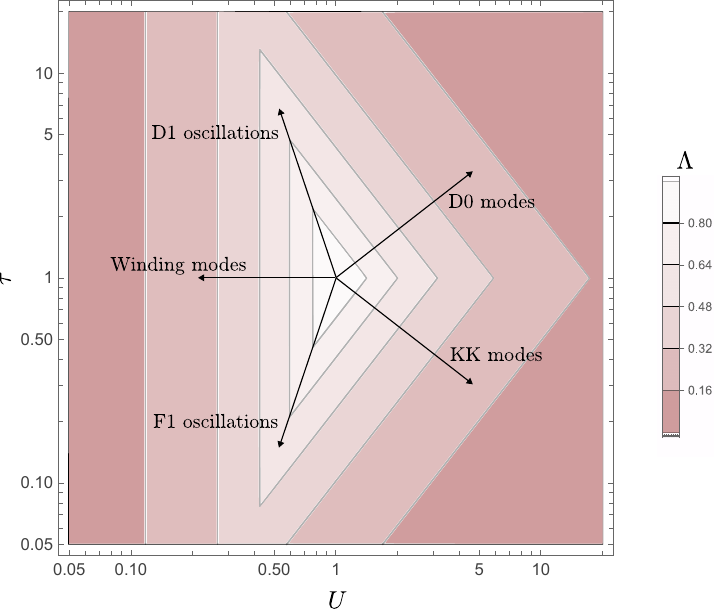}
    \caption{Shape of the moduli space $\cM_{\Lambda}$ of the nine-dimensional theory obtained from compactifying M-theory on a torus for different values of $\Lambda$ (in nine-dimensional Planck units). The moduli space has been parametrized on logarithmic axes using the volume of the torus $U=R_{10}R_{11}$ and its complex structure $\tau=R_{11}/R_{10}$. The towers whose light states set the boundary of validity of the effective theory along each direction are highlighted.}
    \label{fig: M torus moduli space}
\end{figure}

We now consider theories in lower dimensions, starting with compactifications of Type IIA and Type IIB on a circle. Both of them can be discussed together as different limits of M-theory compactified on a torus $\mathbb{T}^2$. In this case, the moduli space becomes three-dimensional, since the nine-dimensional massless spectrum now includes three scalars: the radius of the circle $R_{10}$, the lower-dimensional dilaton $\phi$, and its axionic partner (either the RR 0-form of Type IIB or the component along the compact dimension of the RR 1-form of Type IIA). The resulting moduli space is the space \cite{Schwarz:1995dk, Aspinwall:1995fw}
\begin{equation}
    \mathcal{M}_{\mathbb{T}^2}=\mathbb{R}_{+}\times \text{SL}(2,\mathbb{Z}) \backslash \text{SL}(2,\bbR)/\text{SO}(2)\, .
\end{equation}
The complexity of the Lagrangian of the massless sector increases as the theory is compactified to lower dimensions, but it can still be bounded with an appropriate description. As in the previous examples, such a description is only valid in a finite region of the moduli space, determined by the behavior of the massive towers of states along the different directions and a choice of cutoff. 
To see this more explicitly, note that now there are two non-compact directions to consider, namely the two radial directions of the toroidal compactification.\footnote{Since the axion does not play a prominent role in the asymptotic behavior of the towers, from this point onward we will assume it is fixed to a constant value, with different choices being related through modular transformations \cite{Etheredge:2022opl,Calderon-Infante:2023ler,vandeHeisteeg:2023dlw}.} Along these two directions, in addition to the three towers of the ten-dimensional theories (F1, D1 and D0 modes) one must also add the Kaluza-Klein and winding towers that become light in the limits of large and small radius $R_{10}$ respectively. Together, for a fixed cutoff, they bound a compact region $\mathcal{M}_\Lambda$ of the two-dimensional moduli space, as displayed in figure \ref{fig: M torus moduli space}. Outside of this region, the complexity of the original nine-dimensional EFT diverges since the full tower must be included, and the description ceases to be valid. \\

These simple examples illustrate how the connected components of moduli spaces arising in quantum gravity can be covered by a finite number of patches, each associated with an EFT description of finite complexity, and how this covering is rooted on the network of dualities that permeates string theory. We will discuss the connection to dualities in more detail in  section \ref{sec:complexity_volume_counting}. In more complicated examples, we expect the required number of EFT domains to grow rapidly, but that the core underlying features remain the same.

\subsubsection*{Complexity of higher supergravity} 
The EFT examples discussed so far are instances of supergravity theories with a large amount of supersymmetry. 
They can be viewed as part of the EFT landscape with maximal or half-maximal supersymmetry. As demonstrated in section \ref{sec:highsugra}, the geometry of these theories is under strong control, and their field spaces and coupling functions are definable in the o-minimal structure $\bbR_{\rm alg}$,
as a consequence of the invariance of the Lagrangian under the algebraic group $G$. Since the structure $\bbR_{\rm alg}$ is sharply o-minimal, this means that this field space has an explicitly computable finite complexity $(\cF,\cD)$ depending on $G,K,$ and $\Gamma$. In principle, this complexity may be calculated by counting the formats and degrees of the algebraic functions required to define the tame coordinate charts on $\cM$, following the construction of \cite{BKT} using Siegel sets \cite{BorelJi}. In addition, let us note that many of the higher Wilson coefficients in maximal supergravity theories coming from string corrections are understood in detail \cite{Pioline:1998mn,Green:1999pv,Green:1999pu,Green:2005ba,Green:2010kv,Green:2010wi}, and can be captured by differential equations \cite{Aoufia:2025ppe,Ibanez:2026yyk}. \\

One may go beyond this analysis and consider gauging a subgroup $G_0\subseteq G$, which turns the theory to a gauged supergravity theory and introduces a scalar potential on $\cM$. This construction can be formulated in terms of the embedding tensor, which covariantly embeds the gauged Lie algebra into the Lie algebra of the group $G$ and satisfies a number of algebraic constraints \cite{SamtlebenLec,Gallerati:2016oyo}. Given the algebraic nature of this gauging procedure, it is a natural expectation that the local gauged supergravity Lagrangian, including the generated scalar potential, is still definable in $\bbR_{\rm alg}$. We leave it as a problem for future research to check for which values of the embedding tensor the finite complexity conjecture can be established in higher supergravity, and how these values relate to the UV-embedding of the theory.

\subsubsection*{Compactifications of Type IIB string theory}
A next natural generalization is to look at lower-dimensional compactifications of string theory and test the finite complexity conjecture for the arising EFTs. Such EFTs become quickly very complicated and, in the absence of supersymmetry or other controlling symmetry groups, quantum corrections modify the complexity computation significantly. However, in Type IIB string theory and F-theory compactifications on Calabi-Yau manifolds, we expect that at least certain sectors of the EFT admit computable finite complexity descriptions. In particular, many couplings in the subsector controlled by the complex structure moduli space $\cM_{\rm cs}$
of a Calabi–Yau manifold are determined by the period integrals of the relevant holomorphic forms over an integral basis of homology cycles. These integrals are encoded by the period map
\begin{equation}
    \Phi: \cM_{\rm cs}  \rightarrow \Gamma \backslash  G/ K \, .
\end{equation}
We are therefore led to analyze the complexity of the moduli space $\cM_{\rm cs}$, the space\footnote{If the $\Gamma$ is chosen to be the monodromy group, then $\Gamma \backslash  G/ K $ may fail to be an arithmetic quotient.} 
$\Gamma \backslash  G/ K $, and the transcendental map $\Phi$. Despite these substantial challenges, recall from chapter~\ref{ch:complexity} that it was conjectured in~\cite{beyondo-min,binyamini2024lognoetherianfunctions} that the period integrals are definable in a sharply o-minimal structure, allowing one to assign to them a well-defined complexity $(\cF,\cD)$. This conjecture constitutes a quantitative refinement of the tameness of period maps proven in \cite{BKT}. It is expected that the o-minimal structure $\bbR_{\rm LN}$ plays a central role in this analysis \cite{binyamini2024lognoetherianfunctions}. 
Establishing explicit bounds on the complexity of such effective field theories is likely to be highly non-trivial, and we expect that meaningful results can initially be obtained only for relatively simple examples. Building on the recent advances of \cite{binyamini2024lognoetherianfunctions,Carrascal:2025vsc}, it appears to be feasible to derive complexity bounds for Calabi–Yau moduli spaces with a small number of moduli. Due to the intrinsic geometric nature of sharp o-minimality, we generally expect that the complexity $(\cF_{\rm EFT},\cD_{\rm EFT})$ of the lower-dimensional theory depends on the format and degree of the compactification manifold. 
Ultimately the framework should also accommodate additional sectors and quantum corrections, and a complete determination of the complexity of general string-theoretic EFTs currently seems to be out of reach. \\

\begin{figure}
    \centering
\includegraphics[width=0.65\linewidth]{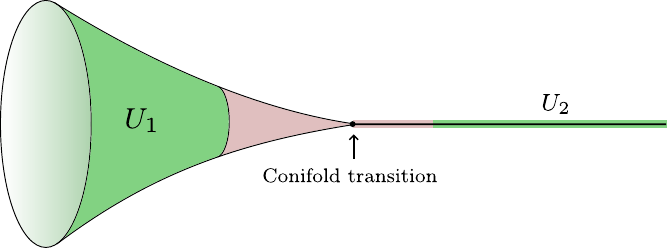}
    \caption{Depiction of the moduli space near a conifold transition between two topologically distinct compact Calabi-Yau manifolds. For a fixed cutoff, there are EFT domains $U_1$ and $U_2$ that provide four-dimensional EFT descriptions of their respective compactifications near the conifold point without reaching it.}
    \label{fig: conifold transition}
\end{figure}

Nevertheless, the complexity perspective provides a powerful lens through which to investigate quantitative properties of the string theory landscape. As noted above, the finite complexity conjecture is naturally aligned with the expected finiteness of compact Calabi-Yau manifolds, reflecting the idea that putative discrete infinities are constrained by a uniform complexity bound on EFTs. To draw a precise connection, we emphasize that this requires more than purely referring to potentially infinite discrete sets of manifolds.  Since all Calabi-Yau manifolds are believed to be connected via flop transitions and conifold transitions \cite{reid1987moduli}, there may even exist a single theory that allows one to explore all Calabi-Yau vacua simultaneously \cite{Sen:2025bmj,Sen:2025oeq}. If such an idea can be made precise and one has infinitely many topologically distinct compact Calabi-Yau manifolds, we expect to find a contradiction with the finite complexity conjecture.  Intuitively, this stems from the infinitely many geometric transitions in the field space that appear in conflict with having a finite cell decomposition. Our perspective using EFT domains is consistent with the expectation that there are indeed finitely many compact Calabi-Yau threefolds that are connected by transitions. In this case we can consider the four-dimensional EFTs arising from compactifications of Type II string theory that constitute expansions around a Minkowski vacuum. Depending on the cutoff, we encounter EFTs on distinct domains, as depicted schematically for the EFTs near a conifold transition in figure \ref{fig: conifold transition}. The moduli space $\cM_{\rm QG}= \lim_{\Lambda \rightarrow 0} \cM_{{\rm QG};\Lambda}$ is then naturally identified with the irreducible moduli space proposed by Reid \cite{reid1987moduli}, and has finite complexity. This will be contrasted with AdS compactifications below.   \\

To further illustrate how the finite complexity conjecture restricts discrete infinities, consider four-dimensional 
$\text{SU}(N)$ Yang–Mills theory with $\cN=2$ supersymmetry. While for small values of 
$N$ such theories are expected to admit embeddings into string theory \cite{Kumar:2010ru,Morrison:2011mb,Kim:2019ths,Tarazi:2021duw,Martucci:2022krl}, the finite complexity conjecture implies the existence of an upper bound on $N$. Indeed, as $N\rightarrow \infty$, not only does the number of light fields increase, but the complexity of the associated Coulomb branch EFTs grows as well \cite{Carrascal:2025vsc}. This is consistent with general expectations on bounds of ranks gauge groups in quantum gravity. 

\subsubsection*{Partial towers and the species scale}
In this chapter we take the perspective that the infinite towers of states predicted by the distance conjecture cannot partially be included in the EFT in a consistent manner. Nonetheless, it is worth briefly discussing the finite complexity conjecture in the setting of partially truncated towers of states. As the moduli of an EFT are varied, there may be fields whose mass falls below the fixed cutoff $\Lambda$, and when these are included the EFT changes qualitatively, requiring a modified Lagrangian valid on a different EFT domain. The condition that the species scale $\Lambda_{\rm sp}$ must remain above the cutoff $\Lambda$, and the general scaling relation 
\begin{equation}\label{eq:speciesscaling}
    \Lambda_{\rm s} \sim \frac{ M_{\rm pl} }{ N^{\frac{1}{d-2}} }
\end{equation}
where $N$ denotes the number of light species \cite{Dvali:2007hz,Dvali:2007wp}, together imply that the spectrum of EFTs in the quantum gravity landscape remains bounded even if partial towers are included. This in turn implies that, even in this broader perspective, the contribution to the complexity $(\cF_{\rm EFT},\cD_{\rm EFT})$ of a theory coming from the number of fields remains finite, strengthening the conjecture. Conversely, in this point of view the scaling relation \eqref{eq:speciesscaling} shows that the species scale is lowered when the format of the theory increases. This hints towards the compelling idea that the species scale is controlled by the complexity of the EFT. It would be interesting to investigate whether there is a deeper connection between the quantum gravity scale and the complexity of the spectrum and interactions of an effective theory.

\subsubsection*{AdS compactifications}
Let us now turn to compactifications of string theory and M-theory that lead to EFTs only valid around an AdS vacuum. These compactifications are formulated on backgrounds of the form AdS$_d \times X$, where $X$ is positively curved and supported by background fluxes. Famously, there are infinite series of such vacuum solutions within string theory, as demonstrated for example by the AdS$_5\times S^5$ compactifications of Type IIB string theory with $N$ units of $F_5$-flux threading the $S^5$. Denoting by $M_{\rm pl}$ the five-dimensional Planck mass, we note that the effective theory will admit an unavoidable cutoff $\Lambda_{\rm t}\sim M_{\rm pl} N^{-2/3}$, since at this scale Kaluza-Klein modes of $S^5$ will become relevant. Requiring that $\Lambda_{\rm t}>\Lambda$ for a fixed cutoff $\Lambda$, we find that only finitely many values of $N$ contribute to the EFT landscape parametrized by $\cM_{{\rm QG};\Lambda}$. Recalling our definition of $\cM_{\rm QG} = \lim_{\Lambda \rightarrow 0} \cM_{{\rm QG};\Lambda}$, we recognize that this would imply that $\cM_{\rm QG}$ contains a discrete infinite set of EFTs. Using the AdS/CFT duality it was argued that these theories are genuinely different EFTs \cite{Banks:2025nfe,Sen:2025oeq}, which sheds doubt on the tameness of the full moduli space $\cM_{\rm QG}$. This does, however, not contradict our conjecture about the individual $\cM_{{\rm QG};\Lambda}$, but indicates that in the AdS case, tameness requires additional assumptions. This is consistent with the conjectures of \cite{Douglas:2023fcg} for the tameness of the space of CFTs and the recent discussion in \cite{Baykara:2025nnc}. In fact, tameness and complexity may provide a powerful tool to analyze and \textit{count} AdS vacua, as we discuss in the next subsection. 

\subsection{Counting EFTs, volume growth, and complexity}
\label{sec:complexity_volume_counting}
In this section we discuss how the finite complexity conjecture relates to other recently stated swampland constraints, such as the volume growth in moduli spaces discussed in section~\ref{sec:embeddings} and the counting of vacua. We then suggest how these results can be used to turn the abstract bounds claimed by the conjecture into explicit numbers. Note that we will only focus on the complexity of the spaces $\cM_{{\rm QG};\Lambda}$ and $\cM_{\rm QG} = \lim_{\Lambda \rightarrow 0} \cM_{{\rm QG};\Lambda}$ and leave a study of the complexities of the EFT Lagrangians to future work.

\subsubsection*{Distance functions on \texorpdfstring{$\cM_{\rm QG}$}{}}
To apply the finite complexity conjecture we have to gain a better handle on the geometry of $\cM_{\rm QG}$ and $\cM_{{\rm QG};\Lambda}$. These spaces are rather abstractly defined and have a known structure only in specific sectors. For example, we have recalled in section~\ref{sec:complexity_comp} that $\cM_{\rm QG}$ can contain the moduli spaces of complex structure deformations of Calabi-Yau manifolds. These components are Riemannian manifolds with a well-defined physical metric, the Weil-Petersson metric \cite{Candelas:1990pi}. In contrast, $\cM_{\rm QG}$ may contain discrete points associated to a flux lattice, or disconnected components of varying dimension. This happens, for example, in the AdS compactifications discussed in the previous subsection. In these general cases there is no naturally associated notion of distance that indicates how close two disconnected pieces of $\cM_{\rm QG}$ are to each other. Candidate distance functions on such disconnected or non-Riemannian spaces have already been discussed in \cite{Douglas:2010ic} and recently in \cite{Li:2023gtt,Shiu:2023bay,Basile:2023rvm,Mohseni:2024njl,Debusschere:2024rmi,Palti:2025ydz}. While a complete picture has yet to be developed, these constructions motivate the expectation that $\cM_{\rm QG}$ admits a natural description as a metric space with a well-defined distance function.

\subsubsection*{Counting of EFTs}
Let us now assume that the finite complexity conjecture is satisfied and consider the space $\cM_{{\rm QG};\Lambda}$. We can then attempt to count the number of distinct effective theories existing on this space. While this question appears to be straightforward to pose, it is a deep and subtle challenge to decide how to perform the counting. One natural way to proceed is to consider the finite EFT covering  $\mathcal{U}=\{ U_i\}_{i=1,...,n_{\mathcal{U}}}$ as defined at the end of section~\ref{sec:conjecture}. A natural way to define the number of EFTs is the number of domains $n_{\mathcal{U}}$, which corresponds to the number of EFT Lagrangians needed to cover $\cM_{\rm QG;\Lambda}$. To get a universal result, we can then take the infimum over all coverings and define 
\begin{equation} \label{eq:number-EFTs}
n(\cM_{{\rm QG};\Lambda}) = \inf_{\cU} (n_{\cU})\,.
\end{equation}
As emphasized above, the finite complexity conjecture for $\cM_{\rm QG;\Lambda}$ ensures that $n(\cM_{{\rm QG};\Lambda})$ is finite.
While this counting takes into account the validity of the EFTs, via our definition of $\cM_{{\rm QG};\Lambda}$ and the notion of an EFT covering, it does not account for the fact that some EFTs may have larger field spaces and hence can accommodate wider ranges of couplings. 

\subsubsection*{Weighted counting of EFTs}
In this spirit, it is a long-standing challenge to suggest a physically plausible and mathematically well-defined weight factor modifying \eqref{eq:number-EFTs}. A novel suggestion was proposed in \cite{Baykara:2025nnc}, where it was argued that a natural choice of weight factor is obtained by accounting for the volume of the field space in \textit{flat and tachyonic directions}. Assuming it to be physically well-motivated, one faces various challenges to formalize such a counting measure. It turns out that the finite complexity conjecture ensures that such a weight-assignment can be made precise. \\

Firstly, while volume is a well-defined notion for a Riemannian manifold, we have to specify  how volumes are measured in a general metric space. The finite complexity conjecture ensures that the so-called Hausdorff measure, defined on a metric space by covering the space with metric balls and minimizing over all possible covers, becomes a well-defined volume measure. In fact, if the considered metric space is tame, it assigns finite volumes to bounded sets \cite{Barroero2014}. \\

Secondly, in the presence of a generic scalar potential it can be difficult to distinguish flat and tachyonic directions without a clear handle on the structure of the potential, and it generally does not lead to finitely many regions with distinct vacua. This situation can again be resolved by the finite complexity conjecture, which implies the sharp o-minimality of the potential. There are finitely many regions of the field space on which $\partial V = 0$ \cite{VdDries} and finitely many regions with Hessians of fixed rank and definite signature. Sharp o-minimality implies that these regions are tame sets and the number of such regions can be constrained by the complexity of the Lagrangian $(\cF_{\rm EFT},\cD_{\rm EFT})$, which is universally bounded by $(\mathfrak{F}_\Lambda,\mathfrak{D}_\Lambda)$.
For simplicity we will consider only flat directions in the following, leaving the more general discussion to future work. 

\subsubsection*{Counting with flat potential}
Let us now consider the space of EFTs that are valid at least up to $\Lambda$ and have a vanishing potential, whose moduli space we denote by $\cM_{\text{flat};\Lambda}$. We introduce an EFT covering $\cU = \{U_i\}_{i=1,...,n_\cU}$ for this space, subject to the requirements outlined at the end of section~\ref{sec:conjecture}. We now replace the direct count of equation \eqref{eq:number-EFTs} by 
\begin{equation}\label{eq:counting1}
   \mathfrak{m}_{\mathcal{U}}(\cM_{\Lambda}) = \sum_{i=1}^{n_{\cU}} \text{Vol}_{d_i}(U_i)\, . 
\end{equation}
This expression counts flat directions with a volume factor as suggested in \cite{Baykara:2025nnc}, and is well-defined and finite due to the finite complexity conjecture. In particular, here it is essential that each EFT domain $U_i$ has a fixed dimension $d_i$, that the covering is finite, and that we assign Vol$_{0}(U_i) = |U_{i}|$ to zero-dimensional sets. To remove the dependence on $\cU$, we again take the infimum, but this time by minimizing the weighted counting:
\begin{equation}
\label{eq:frak-n}
 \mathfrak{n}(\cM_{\text{flat};\Lambda})
 =\inf_{\mathcal{U}}\sum_i \text{Vol}_{d_i}(U_i)\ .
\end{equation} 
Since the Hausdorff measure agrees with the standard volume in Riemannian manifolds, we expect that \eqref{eq:frak-n} gives a formal generalization of the counting function introduced in \cite{Baykara:2025nnc}. In the following, we discuss how the expression \eqref{eq:frak-n} relates to complexity and comment on the expected scaling with $\Lambda$ based on \cite{Baykara:2025nnc}. 

\subsubsection*{Complexity and embeddings}
In order to evaluate equation \eqref{eq:frak-n} in terms of complexity, and to study the finite complexity conjecture itself, we face the challenge to assign a complexity to the spaces $\cM_{\text{flat};\Lambda}$ and $\cM_{\text{flat}} = \lim_{\Lambda\to 0}\cM_{\text{flat};\Lambda}$. This complexity should account for their metric structure. Since a complexity theory for metric spaces and Riemannian manifolds has not been developed so far, one way to nevertheless use the complexity theory of sharply o-minimal structures is to attempt to embed $\cM_{\text{flat};\Lambda}$ and $\cM_{\text{flat}}$ into a Euclidean space $\bbR^N$ of sufficiently high dimension while preserving the metric structure. In particular, this embedding should be isometric for all components of $\cM_{\text{flat}}$ that are Riemannian manifolds. Upon embedding $\cM_{\text{flat};\Lambda}$ and $\cM_{\text{flat}}$ into $\bbR^N$, there is a natural way to assign a complexity to these spaces if the image is definable in a sharply o-minimal structure. As explained in chapter~\ref{ch:volumes}, the tameness of the embedding precisely reproduces the compactifiability criterion proposed in \cite{Delgado:2024skw}, which states that the volume of a geodesic ball in moduli space grows asymptotically no faster than in Euclidean space. Eventually, we envision that it is possible to avoid a geometric detour through isometric embeddings and define the complexity intrinsically. Nonetheless, we will see that the available results from Euclidean tame geometry enable us to make the relation between volume and complexity  explicit. 

\subsubsection*{Volume growth and complexity} 
Let us first consider a component $\cM \subseteq \cM_{\text{flat}}$ which is a $\kappa$-dimensional Riemannian manifold with metric $g$ and therefore admits an isometric embedding $\widehat \cM\subseteq \bbR^N$ for sufficiently large $N$ by the Nash embedding theorem. We now assume that $\widehat \cM$ is definable in a sharply o-minimal structure as required by the finite complexity conjecture, which brings us back to the setting of section~\ref{sec:embeddings}. In particular, recall that we have shown that a geodesic ball $\cB^\kappa(r)\subseteq \cM$ of radius $r$ defined with respect to the metric on $g$ satisfies the inequality 
\begin{equation} \label{eq:volumegrowth}
\text{Vol}_\kappa(\cB^\kappa(r)) \leq c(N,\kappa) \, \cC(\cF,\cD)\,   r^\kappa\,.
\end{equation}
Here $\cC(\cF,\cD)$ is a coefficient depending on the complexity of the embedding, and  
\begin{equation} \label{eq:c-coeff}
    c(N,\kappa) = \text{Vol}_\kappa(B^\kappa(1))\,  \frac{\Gamma \big(\frac{1}{2} \big)\Gamma \big(\tfrac{N+1}{2} \big) }{\Gamma \big( \tfrac{ \kappa+1}{2} \big)\,\Gamma \big( \tfrac{N-\kappa-1}{2} \big)} \, , 
\end{equation}
where $\text{Vol}_\kappa(B^\kappa(1))$ is the volume of the $\kappa$-dimensional ball of radius 1. The prefactor $c(N,\kappa)$ can be bounded by a function of the complexity $(\cF,\cD)$, since the $\cC^k$ Nash embedding theorem gives an upper bound $N\leq\frac{1}{2}\kappa(\kappa + 1)(3\kappa + 11)$, and by $\kappa \leq \cF$ we have $N\leq O(\cF^3)$. 
This implies that there is always a bound of the form 
\begin{equation} \label{eq:volumegrowth3}
  \text{Vol}_\kappa(\cB^\kappa(r)) \leq \mathfrak{C}(\cF,\cD)\,   r^\kappa\,
\end{equation}
in which the prefactor no longer has an explicit dimensional dependence and is only a function of the complexity $(\cF,\cD)$. Sharp o-minimality guarantees that there exists a universal functional form of the coefficient $\mathfrak{C}(\cF,\cD)$, which may be inferred from the embedding or by analyzing the volumes of geodesic balls in $\cM$.

\subsubsection*{Complexity dependence} 
The volume bound \eqref{eq:volumegrowth3} provides a natural relation between the counting formula \eqref{eq:frak-n} and the complexity of $\cM_{\text{flat};\Lambda}$. By the general properties of sharp o-minimality, we expect that the complexity coefficients grow polynomially in $\cD$ and generically exponentially in $\cF$. 
There is, however, a remarkable way to circumvent an exponential growth in $\cF$. This can be seen by examining the dimensional coefficient $c(N,\kappa)$ given in \eqref{eq:c-coeff}, in which the volume $ B^\kappa(1)$ decays super-exponentially as $\kappa^{-\kappa/2}$ in the limit $\kappa \rightarrow \infty$. 
This fast fall-off may still be compensated by the growth of the other factors in \eqref{eq:c-coeff}, but there are simple examples for which the complexity coefficient $\mathfrak{C}(\cF,\cD)$ inherits this asymptotic behavior. For example, when $\cM$ is given as a degree-$\cD$ algebraic hypersurface in $\bbR^{\cF}$, one can show that
\begin{equation}\label{eq:suppressed_C}
  \mathfrak{C}(\cF,\cD) \sim \cF^{-\cF/2} \cD \, .   
\end{equation}
This super-exponential suppression was key in the arguments of \cite{Baykara:2025nnc} to argue for a specific cutoff dependence of $\mathfrak{n}(\cM_{\text{flat};\Lambda})$ in the limit $\Lambda \rightarrow 0$. It would be interesting to investigate whether this decay in the complexity coefficient can be realized in moduli spaces coming from quantum gravity or string theory.

\subsubsection*{Cutoff dependence and distances}
Let us close this subsection by combining the volume growth bound \eqref{eq:volumegrowth3} and the weighted counting of vacua \eqref{eq:frak-n}. Consider a finite EFT covering $\cM_{\text{flat};\Lambda}=\bigcup_i U_i$ on a moduli space of EFTs with vanishing potential. 
We denote the diameter of each $U_i$ by $2r_i$, and place each $U_i$ into a $d_i$-dimensional geodesic ball $\mathcal{B}^{d_i}_i(r_i)$, i.e.~we have $U_i \subseteq \mathcal{B}^{d_i}_i(r_i) \subseteq\cM_{\text{flat}}$ for each $i$. We then denote the complexity of the ball $\mathcal{B}^{d_i}_i(r_i)$ by $(\cF_i,\cD_i)$, which yields the upper bound 
\begin{equation}
   \text{Vol}_{d_i}(U_i)\leq  \text{Vol}(\mathcal{B}^{d_i}_i(r_{i}))  \leq \mathfrak{C}(\cF_i,\cD_i)\,   r_i^{d_i} \, . 
\end{equation}
Note that the exponential nature of the distance conjecture implies that the radii in the moduli space are bounded by $r_{\rm max} \sim \frac{1}{\alpha}|\log \Lambda|$, where $\alpha$ is an order-one constant that we will suppress in the following. Upon inserting this into \eqref{eq:frak-n}, we find 
\begin{equation} \label{eq:nM-explicit}
  \mathfrak{n}(\cM_{\text{flat};\Lambda}) \lesssim \inf_{\mathcal{U}}\sum_{i=1}^{n_\cU} \mathfrak{C}(\cF_i , \cD_i) |\log \Lambda|^{d_i} \, . 
\end{equation}
Given that $\cM_{\text{flat};\Lambda} =\bigcup_i \mathcal{B}^{d_i}_i(r_{i})$, the axioms of sharp o-minimality imply that $\cF_\Lambda \leq\max(\cF_i) $ and $\cD_\Lambda \leq \sum_i\cD_i$, where $(\cF_\Lambda,\cD_\Lambda)$ is the complexity of $\cM_{\text{flat};\Lambda}$.
Assuming that $\cM_\text{flat}$ is a moduli space of finite complexity $(\cF,\cD)$, we note that the expression \eqref{eq:nM-explicit} is a polynomial in $|\log \Lambda|$ with cutoff-independent coefficients, which reproduces the claim of \cite{Baykara:2025nnc} that $\mathfrak{n}(\cM_{\text{flat};\Lambda}) \lesssim a |\log \Lambda|^b $, where $a,b$ are complexity-dependent coefficients. \\ 

As mentioned in section~\ref{sec:complexity_comp}, the definability of $\cM_\text{flat}$ in a sharply o-minimal structure is plausible for the present case of compactifications of string theory without potential. In case that $(\cF_{\Lambda},\cD_{\Lambda})$ diverges in the limit $\Lambda \rightarrow 0$ limit, as is the case for many classes of AdS vacua as discussed in section~\ref{sec:complexity_comp}, one needs to refine and extend the argument. In such an extension the strong suppression as inequation~\eqref{eq:suppressed_C} will play a key role, since it indicates that small values of $\cF_i$ dominate the sum. \\

\pdfbookmark[-1]{Conclusions}{}

{
\renewcommand{\chapterformat}{\relax} 
\addchap{Conclusion}
}
\setlength{\parindent}{0pt}

\subsection*{Summary}

In this thesis we have studied the notions of tameness and complexity, realized by the framework of o-minimality, across various physical applications in quantum field theory and quantum gravity. We have demonstrated that many classes of physical functions are mathematically tame, and in certain cases we have been able to use sharp o-minimality to quantify the complexity of physical objects and theories. Finally, we have argued that the emergence of tameness in physics aligns with manifestations of finiteness in quantum gravity, converging towards the idea that effective theories of quantum gravity admit a description of finite complexity.

\subsubsection*{Part I -- Tame geometry}
In part I of the thesis we have reviewed the theory of tame geometry, laying the mathematical foundations of the thesis. We began in chapter~\ref{ch:tameness} by introducing o-minimal structures, and discussed the tameness theorems governing the geometry of the sets in these structures. These include foundational theorems such as the cell decomposition theorem, as well as theorems linking o-minimality to arithmetic and algebraic geometry. We then presented various examples of o-minimal structures generated by tame function classes, and explained how the framework can be generalized to geometry beyond Euclidean space. \\

Chapter~\ref{ch:complexity} centered around sharp o-minimality, which defines a quantitative version of tame geometry. We showed how tame sets in these structures admit a measure of complexity, and explained the interpretation of this notion of complexity. The main example was provided by the class of Pfaffian functions, which are tame functions whose complexity is controlled by systems of differential equations.
This class of functions, and their generalizations, form the basis of many applications to physics discussed in this thesis. \\

\subsubsection*{Part II -- Quantum field theory}
In part II of the thesis we have studied how these tame mathematical structures arise in physics, focusing on quantum field theory. Chapter~\ref{ch:classical} began with a general discussion of tameness in physics, highlighting some of the conceptual challenges arising in the implementation of o-minimality. 
This was followed by an analysis of tameness and complexity in quantum mechanics and classical field theory. Here we showed that bound state wavefunctions in one-dimensional quantum systems are tame under moderate assumptions, and discussed physical interpretations of the complexity of classical field configurations. \\

In chapter~\ref{ch:observables} we turned to non-perturbative observables in quantum field theory, and in particular to zero-dimensional theories where exact calculations are available. Relying on the differential chains introduced in the previous part, we showed that non-perturbative observables in selected models admit a description of finite complexity. We then discussed how these methods may be generalized to general lattice quantum field theories. 
The limitation of this approach is that the differential chains cannot capture the non-analytic nature of weak-coupling limits, and we addressed this problem by using the theory of Borel resummation. We demonstrated that the resummed asymptotic expansions of observables in various models are Gevrey functions, which shows that they are tame on a domain which include the non-analytic weak coupling limit. \\

In chapter~\ref{ch:CCC}, we analyzed the tameness and complexity of cosmological correlators, which are the natural observables in cosmological quantum field theories. Focusing on a conformally coupled scalar in a spacetime with a power-law scale factor, we demonstrated that tree-level wavefunction coefficients contributing to the correlators can be captured by a Pfaffian chain. We then calculated the complexity of these functions, and discussed the physical interpretation of the results. \\

Finally, in chapter~\ref{ch:complexityQFT} we explored how tame geometry may be used to define an intrinsic notion of complexity for quantum field theories. We discussed various perspectives based on the complexity of the Lagrangian and the complexity of the observables, and investigated the physical interpretations of this proposal through connections to symmetries, perturbation theory, and renormalization group flow. \\

\subsubsection*{Part III -- Quantum gravity}
In part III of the thesis we focused towards effective quantum field theories coupled to gravity. In chapter~\ref{ch:string} we reviewed aspects of quantum gravity and string theory, with emphasis on the landscape of effective theories consistent with quantum gravity in the ultraviolet. 
The universal constraints satisfied by these effective theories, formulated through the swampland program, are rooted in various manifestations of finiteness. We then reviewed the tameness conjecture, which provides a unifying view on finiteness in quantum gravity by proposing that these effective theories can be described within o-minimal structures. \\

In chapter~\ref{ch:volumes}, we studied the emergence of tameness in the geometry of moduli spaces of effective theories consistent with quantum gravity. We began with an analysis of supergravity theories with more than eight supercharges, and showed that the field spaces and Lagrangians of these theories are tame as a consequence of the geometric structure of supersymmetry. In particular, we demonstrated the semi-algebraic definability of the field space metric and the gauge coupling functions.
We then used quantum gravity constraints on the volume growth in moduli spaces to argue that all moduli spaces in effective theories of quantum gravity must admit a tame isometric embedding into Euclidean space.  
Through the connection between volumes in moduli spaces and the emergence of dualities, our results indicate a link between tame geometry and dualities in quantum gravity. \\

Finally, in chapter~\ref{ch:EFTcomplexity} we used sharp o-minimality to develop a quantitative framework for capturing the complexity of effective field theories in the quantum gravity landscape. We argued  that in special cases the infinite Wilsonian data of an effective Lagrangian can be encoded in a finite amount of information, often through implicit structures such as differential constraints, yielding a description of finite complexity. Combining insights from the previous chapters and finiteness constraints governing the landscape, we argued that these special cases are realized in quantum gravity, culminating in the proposal of the \textit{finite complexity conjecture}. We provided additional evidence for the conjecture in several examples coming from string theory and compactifications, and studied the connection with volumes in moduli spaces, dualities, and the counting of vacua.

\subsection*{Outlook}
We now conclude with a discussion of the scientific results of the thesis and provide an outlook on future research directions which have emerged.

\subsubsection*{The language of physics}
We began the thesis with the following question: \textit{what is the language of physics?} Having reached the end of the thesis, it is time to reflect on this question. In a variety of physical theories, we have shown that the physical functions appearing in the description of the theory can be defined in an o-minimal structure. In essence, this means that the physical laws described by these functions can be specified with a finite amount of information. The geometric interpretation of this idea is that the graphs of these physical functions have finitely many geometric features. On the other hand, the logical interpretation of this idea is that these functions can be composed from a finite number of simple logical symbols and formulas. In addition, the measure of complexity defined by sharp o-minimality provides a method of strengthening this interpretation, by quantifying the complexity of the laws of physics both geometrically and logically. We may therefore draw the ambitious conclusion that o-minimality has the potential for playing a central role in formulating the language of physics.  \\

However, we must carefully delineate the scope of this idea. The prospect that every quantity in physics can ultimately be defined within an o-minimal structure is not true, as already illustrated by simple counterexamples such as oscillations in infinite domains of time and space.
Sometimes, the formulation of a theory prompts us to rely on objects of infinite complexity, even if the resulting observable quantities are constrained by finiteness. The tameness of physics is therefore necessarily limited to a subset of physical theories or physical functions. In order to address to what extent \mbox{o-minimality} is the language of physics, it is then an important objective to determine these limitations and to find the boundary between finiteness and infinity in physics. This thesis shows that this boundary is at times unexpected.
For example, non-perturbative observables in quantum field theory receive contributions from infinitely many virtual physical processes, and effective Lagrangians contain an infinite series of interactions. Yet, our results demonstrate instances of emergent simplicity and provide a proof of principle that these quantitites may nonetheless admit a description of finite complexity. \\

We have argued that these effects are the strongest in the presence of gravity, and formalized our observations in the finite complexity conjecture. This conjecture proposes that the effective physical theories which are consistent with our universe, in the sense that they describe viable low-energy limits of quantum gravity, 
can be described with the language of tame geometry. This is an ambitious idea, and it is conceivable that the finite complexity conjecture is not true.
Regardless of whether the conjecture holds or not, a deeper understanding of it may teach us profound lessons on finiteness and infinity in physics, and the limitation of describing the laws of nature with a finite amount of information. 
This thesis only provides the first steps towards these ideas, and many questions remain unanswered, bringing us to the outlook, where we present various directions for future research.

\subsubsection*{Structure and complexity of observables in quantum field theory}
There are several key questions concerning the tame geometry of observables in quantum field theory. Firstly, it remains an open mathematical question for which theories the spectrum of observables can be defined within a sharply \mbox{o-minimal} structure, enabling a quantitative complexity analysis. At the perturbative level, currently the main obstruction is to establish the sharp o-minimality of periods, since perturbative amplitudes can be defined using period integrals in a large class of quantum field theories. Resolving the conjectured sharp o-minimality of periods \cite{binyamini_tameness_2023} therefore unlocks many new applications in quantum field theory. The introduction of the effectively o-minimal log-Noetherian functions constitute significant progress towards this goal \cite{binyamini2024lognoetherianfunctions}, providing a promising prospect for the future. In the case of non-perturbative observables, a key question is whether a sharply o-minimal structure hosting manifestly non-analytic functions exists. As an intermediate step, one could attempt to prove whether the structure generated by Gevrey functions or the quasi-analytic structures introduced in \cite{rolin2007quasi} are effectively o-minimal. \\

Another intriguing direction to investigate further is the connection between the complexity of the descriptive functions of a quantum field theory and the complexity of its observables. A reduction in complexity of the Lagrangian signals that there are algebraic relations among the couplings of the interactions in the theory, which implies algebraic relations in the Feynman rules. 
We therefore expect that these relations will manifest themselves in a reduction of complexity in the perturbative and non-perturbative observables. As a first step, this idea can be realistically tested in zero-dimensional models, and may ultimate be generalized to observables in higher-dimensional quantum field theories. \\

\subsubsection*{Tameness theorems in physical theories}
In this thesis we have shown that many physical quantities are tame, but we have not yet made use of the full power of tame geometry. In particular, the application of the tameness theorems, such as the cell decomposition and the Pila-Wilkie theorem, has not been carried out explicitly and remains open for exploration. We expect that these tools may be especially useful for analyzing the geometric structure of solution sets of complicated systems of equations. For example, the Pila-Wilkie theorem could be used to further probe the arithmetic structure of the set of flux vacua in string theory compactifications, extending the ideas of \cite{Grimm:2023lrf,Grimm:2024fip}.

\subsubsection*{Finiteness of complexity in quantum gravity}
The finiteness of complexity in quantum gravity is one of the central ideas proposed in this thesis, and this conjecture opens many avenues for further research. First, it is an essential task to establish more evidence for the conjecture.
From a top-down perspective, a key objective is to establish precise values of the complexity of effective field theories arising from examples in string theory.
In this context, a proof of the conjectured sharp o-minimality of periods, which control the effective Lagrangians of a huge class of low-energy limits of string theory, would unlock a plethora of possibilites for quantifying the complexity bounds in the conjecture.
The status of the conjecture would be significantly strengthened by finding more bottom-up evidence, for instance coming from black hole physics, the existence of dualities, or finiteness of quantum gravity amplitudes. \\

Another promising direction to explore further is the connection between the complexity of an effective field theory and the energy scale $\Lambda$. In particular, the complexity bound $(\cF_\Lambda,\cD_\Lambda)$ appearing in the finite complexity conjecture depends on the cutoff, and it is an important challenge to understand how these bounds scale as $\Lambda$ approaches $0$ or the quantum gravity scale $ \Lambda_{\rm QG}$.
Furthermore, it would be interesting to investigate whether there is a complexity-induced species scale mechanism which lowers the quantum gravity scale $\Lambda_{\rm QG}$ when $\cF_{\rm EFT}$ or $\cD_{\rm EFT}$ tends to infinity. \\

\vspace{1cm}

Finally, a major extension of our proposal would be to formulate finiteness of complexity without the restricting the effective field theories to have a fixed space-time dimension. This could, in principle, allow one to move beyond the cutoff-restricted quantum gravity landscape, at the price of systematically incorporating dual descriptions that may change the effective dimensionality or even require non-Lagrangian descriptions.
An illustrative example is a decompactification limit in moduli space, where an infinite tower of Kaluza-Klein modes becomes light, the lower-dimensional effective theory ceases to be valid, and the apparent complexity diverges.
Yet, the same physics can be captured by a higher-dimensional effective theory with finitely many fields, providing an effective description with finite complexity.   
Similarly, in other asymptotic regimes where weakly coupled asymptotically tensionless strings are expected to emerge \cite{Lee:2019wij}, the worldsheet theory may furnish the finite-complexity description underlying an infinite tower of string excitations.
This realization of emergence of simplicity and incorporation of dual descriptions may ultimately provide a perspective on finiteness in the complete quantum gravity landscape.

\subsubsection*{Quantum information theory and holography}
While both sharp o-minimality and quantum information theory provide a measure of complexity for quantum states, the precise connection between these measures remains unclear. The notion of quantum computational complexity originating from quantum information theory plays a significant role in holography, and is believed to be dual to the volume of a certain spatial slice in the bulk theory \cite{Susskind:2014rva,Brown:2015bva}.  Since format and degree may in principle be assigned to any sufficiently tame mathematical object, it
would be fascinating to gain a deeper understanding of the connection between complexity in tame geometry and quantum computational complexity, and to investigate whether the format and degree of a quantum state has any holographic interpretation.

\subsubsection*{The logic of physics}
As explained in the first part of this thesis, tame geometry has its origins in logic and model theory. Through the connections with o-minimality and other logical principles characterizing mathematical languages and theories, the results of this thesis may help to study  
the logical structure underlying the laws of physics. For instance, using the decidability of the structure $\bbR_{\rm alg}$ \cite{Tarski,Seidenberg}, it follows that first-order logical statements formulated in $\bbR_{\rm alg}$-definable physical theories, such as the higher supergravity theories considered in section~\ref{sec:highsugra}, are decidable. This is in contrast with recently formulated manifestations of undecidability in physics \cite{Cubitt:2015xsa,Tachikawa:2022vsh,Perales-Eceiza:2024qhd,Mir:2026lpn}.
It would be interesting to explore the connection further and to attempt to chart the logical properties of physical theories.

\subsubsection*{Complexity, approximation, and geometric entropy}
From a geometric perspective, one shortcoming of the measure of complexity in tame geometry is that it is insensitive to the size of geometric features. For instance, suppose that we have a simple geometric object to which we apply a complicated perturbation. The complexity of the resulting object will then increase significantly, regardless of how small the perturbation is. To address this issue, we believe that it is worthwhile to explore the implementation of geometric approximation schemes that disregard geometric features below a certain error threshold. In this manner, one could assign an approximate format and degree to tame sets which captures geometric features up to a desired level of accuracy. Similar ideas appear in quantum computational complexity, where target states may be produced up to a specified norm error. 
A natural starting point for such a notion is the concept of geometric $\varepsilon$-entropy developed in \cite{yomdin2004tame}.

\subsubsection*{Tame geometry and embeddings}
The discussion of chapter~\ref{ch:volumes} provides a new tameness criterion for tame manifolds based on the existence of a tame isometric embedding into Euclidean space. This criterion remarkably excludes many spaces which appear to be tame, such as the hyperbolic plane and its higher-dimensional analogs, and is strongly linked to constraints emerging from quantum gravity. From a physical perspective, it is then important to study the tame embedding conjecture in more general quantum gravity moduli spaces in order to fully uncover the links between dualities, tameness, and gravity. Furthermore, from a purely mathematical perspective we believe that the tame embedding problem, which consists of formulating the geometric assumptions for the existence of tame isometric embeddings and proving an associated tame Nash embedding theorem, is a fascinating direction to explore further.


{
\renewcommand{\chapterformat}{\relax} 
\addchap{Samenvatting}
}
\setlength{\parindent}{0pt}
In dit proefschrift hebben we de wiskundige begrippen tamheid en complexiteit bestudeerd, gedefinieerd via het principe van o-minimaliteit, en deze toegepast in het natuurkundige kader van de kwantumveldentheorie en de kwantumzwaartekracht. We hebben aangetoond dat veel klassen van fysische functies wiskundig gezien tam zijn, en in zekere gevallen hebben we door middel van de scherpe o-minimaliteit de complexiteit van natuurkundige objecten en theorie\"en in kaart gebracht. Uiteindelijk hebben we geponeerd dat de aanwezigheid van tamme meetkunde in de natuurkunde overeenkomt met de manifestaties van eindigheid in kwantumzwaartekracht, wat resulteerde in het idee
dat de effectieve veldentheorie\"en van kwantumzwaartekracht beschreven kunnen worden met eindig veel complexiteit.

\subsubsection*{Deel I -- Tamme meetkunde}
Deel I van dit proefschrift omvat een uitgebreide inleiding tot de tamme meetkunde, en legt daarmee de wiskundige basis van dit onderzoek. In hoofdstuk~\ref{ch:tameness} zijn we begonnen met het introduceren van o-minimale structuren, die de kern van deel I vormen. We hebben vervolgens de tamheidsstellingen behandeld die de meetkundige eigenschappen van tamme verzamelingen beheren. Hiertoe behoren fundamentele stellingen, zoals de celdecompositiestelling, alsmede stellingen die de o-minimaliteit verbinden met getaltheorie en algebra\"ische meetkunde. We hebben een breed scala aan voorbeelden van o-minimale structuren behandeld, en uitgelegd hoe de tamme meetkunde kan worden gegeneraliseerd naar niet-Euclidische meetkunde. \\

Het onderwerp van hoofdstuk~\ref{ch:complexity} is de zogeheten scherpe o-minimaliteit, een kwantitatieve versie van tamme meetkunde. We hebben uitgelegd hoe tamme verzamelingen in dergelijke structuren over een maat van complexiteit beschikken, en hoe deze complexiteit ge\"interpreteerd kan worden. Het belangrijkste voorbeeld bestaat uit de klasse van Pfaffiaanse functies, die de basis vormen voor veel van de natuurkundige toepassingen die zijn ontwikkeld in dit proefschrift.

\subsubsection*{Deel II -- Kwantumveldentheorie}
In deel II van dit proefschrift hebben we bestudeerd hoe deze tamme wiskundige structuren opduiken in de natuurkunde, met name in de kwantumveldentheorie. Hoofdstuk~\ref{ch:classical} begon met een algemene beschouwing over tamheid in natuurkunde, gevolgd door een analyse van o-minimaliteit in kwantummechanica en klassieke veldentheorie. Hier hebben we aangetoond dat onder milde aannames de golffuncties van gebonden toestanden in eendimensionale kwantumsystemen tamme functies zijn, en hebben we de fysische interpretatie van complexiteit van klassieke veldconfiguraties besproken. \\

In hoofdstuk~\ref{ch:observables} hebben we ons gericht tot niet-perturbatieve observabelen in kwantumveldentheorie, en in het bijzonder in nuldimensionale theorie\"en waar exacte berekeningen mogelijk zijn. Met behulp van ketens van differentiaalvergelijkingen hebben we in enkele modellen laten zien dat niet-perturbatieve observabelen een beschrijving van eindige complexiteit hebben. Vervolgens hebben we besproken hoe deze methode kan worden uitgebreid naar roosterkwantumveldentheorie\"en. Deze aanpak is beperkt tot analytische functies, en daardoor niet toepasbaar op de zwakke koppelingslimieten die niet-analytisch van aard zijn. Om dit probleem aan te pakken hebben we gebruikgemaakt van de theorie van Borelresommatie. We hebben aangetoond dat de hersomde asymptotische expansies van observabelen in verscheidene modellen tot de klasse van Gevrey-functies behoren, wat betekent dat ze tam zijn op een domein dat de niet-analytische zwakke koppelingslimiet omvat. \\

In hoofdstuk~\ref{ch:CCC} bestudeerden we de tamheid en complexiteit van kosmologische correlatiefuncties, de belangrijkste observabelen in kosmologische kwantumveldentheorie\"en. We hebben ons gefocust op  een hoekgetrouw gekoppeld scalair veld in een ruimtetijd met een machtswet-schaalfactor, en aangetoond dat golffunctiecoefficiënten die op boomniveau bijdragen aan deze correlatiefuncties kunnen worden beschreven met een Pfaffiaanse ketting. Met dit resultaat hebben we de complexiteit van deze functies berekend, en de fysische interpretatie van de uitkomst besproken. \\

Ten slotte hebben we in hoofdstuk~\ref{ch:complexityQFT} onderzocht hoe tamme meetkunde kan worden gebruikt om een intrinsiek begrip van complexiteit voor kwantumveldentheorie\"en te defini\"eren. We hebben verscheidene perspectieven verkend, gebaseerd op de complexiteit van de Lagrangiaan dan wel de complexiteit van de observabelen, en hebben de fysische interpretatie van dit voorstel bestudeerd door middel van verbanden met symmetrie\"en, storingsrekening, en renormalisatiegroepsstroming. 

\subsubsection*{Deel III -- Kwantumzwaartekracht}
Deel III van dit proefschrift betreft kwantumveldentheorie\"en die zijn gekoppeld aan zwaartekracht. Hoofdstuk~\ref{ch:string} bestaat uit een inleiding tot enkele aspecten van kwantumzwaartekracht en snaartheorie, met nadruk op het landschap van effectieve veldentheorie\"en die consistent zijn met kwantumzwaartekracht in het ultraviolet. De universele eisen waar deze theorie\"en aan voldoen, geformuleerd door het zogeheten moerasprogramma, zijn geworteld in verschillende manifestaties van eindigheid. Het tamheidsvermoeden geeft een verenigende blik op eindigheid in kwantumzwaartekracht, en stelt dat deze effectieve theorie\"en kunnen worden beschreven binnen o-minimale structuren. \\

In hoofdstuk~\ref{ch:volumes} hebben we bestudeerd hoe tamheid opduikt in de meetkunde van moduliruimten in effectieve theorie\"en die consistent zijn met kwantumzwaartekracht. We begonnen met een analyse van superzwaartekrachtstheorie\"en met meer dan acht superladingen, en hebben laten zien dat de veldenruimten en Lagrangianen van deze theorie\"en tam zijn als gevolg van supersymmetrie. Vervolgens hebben we gebruikgemaakt van beperkingen op de volumegroei binnen moduliruimten om te beargumenteren dat alle moduliruimten in effectieve veldentheorie\"en van kwantumzwaartekracht over een tamme isometrische inbedding in de Euclidische ruimte moeten beschikken. Door de relatie tussen volumes in moduliruimten en de aanwezigheid van dualiteiten tonen onze resultaten aan dat er een verband is tussen tamme meetkunde en dualiteiten in kwantumzwaartekracht. \\

Ten slotte hebben we in hoofdstuk~\ref{ch:EFTcomplexity} gebruikgemaakt van de scherpe o-minimaliteit om een kwantitatief raamwerk te ontwikkelen die de complexiteit van effectieve veldentheorie\"en van kwantumzwaartekracht beschrijft. We hebben beredeneerd dat in bijzondere gevallen de oneindige Wilsoniaanse data van een effectieve Lagrangiaan kan worden omvat met eindig veel informatie, door middel van impliciete structuren zoals differenti\"ele relaties die resulteren in een representatie van eindige complexiteit. Door dit idee samen te voegen met de inzichten uit de voorgaande hoofdstukken en de eindigheidsbeperkingen van kwantumzwaartekracht, hebben we beargumenteerd dat deze bijzondere gevallen tot stand komen in het kwantumzwaartekrachtslandschap.
Deze idee\"en zijn verwezenlijkt in het \textit{vermoeden van eindige complexiteit}. We hebben vervolgens aanvullend bewijsmateriaal voor dit vermoeden geleverd vanuit snaartheorie en compactificaties, en hebben tot slot de verbanden met volumes in moduliruimten, dualiteiten, en het tellen van vacua bestudeerd.

{
\renewcommand{\chapterformat}{\relax} 
\addchap{Acknowledgements}
}

\setlength{\parindent}{0pt}
The completion of a PhD feels like a long journey, yet like a fleeting period of time.
It feels like a solitary undertaking, yet one ultimately shaped by the people along the way. It is certain that my time as a PhD student would not have been nearly as wonderful as it was without the amazing people in my life. Here I wish to take a moment to express my heartfelt gratitude to them. \\

First and foremost, I want to sincerely thank my advisor Thomas, for giving me the opportunity to follow my dreams and for guiding me over the past years. I greatly admire your courage and vision in fully committing to a completely new line of research, and your ability to inspire when research inevitably gets stuck. There have been several times when I had doubts about what we were doing, but in our discussions you invariably renewed my motivation. You gave me tremendous freedom to explore my ideas, and I am proud of what we developed together.\\

Next, I want to thank my collaborators. David, you have been one of my closest collaborators in the past years, and we have shared many highs and lows of research, office spaces, distractions, dinners, and deep discussions about o-minimality, life, ancient civilizations, and random interests. I am grateful that you were always there to patiently listen to me when research got frustrating, and  I fondly look back on our time in Utrecht and Boston. Arno, I really enjoyed our venture into cosmological correlators: it was always fun and exciting to work together, even when our problems reduced to something as deceptively simple as counting lines and circles. Giovanni, you were the first master student I supervised, and I really enjoyed our collaboration. I admire your diligence and precision, and I am very happy that your project led to a publication.
Lorenz, it was a great pleasure to collaborate on my first publication. Your vast knowledge always impressed me, and I enjoyed our time together at the institute in the first year of my PhD. \\

I want to thank the many wonderful people at the ITP. Shradha, thank you for being my friend and paranymph, and for the countless laughs, memorable moments, valuable conversations about life, and fun activities.  
You brought a lively energy to the institute and these years would not have been the same without you. \\

Next I would like to thank my amazing office neighbours: no day at the institute went by without checking if one of you was around for a coffee.
Artim, you are the strongest physicist I know, and also one of the kindest. I really enjoyed your consistent presence at the institute, our summer break games (usually a strange variation of chess or mastermind), and our inside jokes. 
Nico, we discovered that we are essentially isomorphic and could talk endlessly about our common interests. There were a few tough moments during which you truly listened to me, and I am very grateful for that. 
Raúl, I really enjoyed your warm friendship, your humour, and your fun stories on an incredibly diverse range of topics. 
Mexx, I fondly look back on our coffee chats and (black hole) chess games; defeating you once in a classical game remains my proudest chess achievement. \\

There are many more people in the ITP who I would like to thank.
Pedro (Arthur), you joined the institute in the second half of my PhD, and I really value our friendship, shared humour, and all the Portugese words you taught me. 
Jeroen, your kindness meant a lot to me in the early stages of my PhD when I had just joined the string theory group. In addition, you introduced me to bouldering, which I have been passionate about ever since.
Lumen, I am happy to have you as a friend ever since we moved from Twente to Utrecht. Together we conquered many courses, shared two households, and even founded our own institute (of metals and strings). I want to thank Anouar, David, Dimos, Emanuele, Hossein, Iris, Joren, Lucas,  Maaike, Nuriye, Oscar, Pablo, Robin, Sandro, Tomas, and Will,
for all the fun interactions during breaks, walks in the botanical garden, and DRSTP schools. \\

I would like to thank the (past) members of the string theory group: Casey, César, Claire, Damian, Edwan, Erik, Filippo, George, Guoen, Guim, Huaxuan, Javi, Nava, Ro, Stefano, Thorsten, Tuna, and Wilke,
for the valuable discussions and enjoyable conversations during string lunch and coffee breaks, where we discussed the state of the world, languages, artificial intelligence, Dutch food traditions, and the heterotic string. Stefan, thank you additionally for being my second promotor and for all the good advice you have given me. 
Umut, thank you for teaching me quantum field theory and for the warm energy you brought to the institute. You are greatly missed. \\

I would like to thank Mariëlle, Olga, and Annette for always ensuring that the institute runs smoothly, for the cookies on Thursdays, and being there for me whenever I had a question or forgot my office key. \\

During my time as a PhD student I had the opportunity to supervise several students, which has been a very enjoyable and enriching experience. Lucia, Max, Tibor, Victor, thank you for your trust in me and for the great enthusiasm you brought to your thesis projects. In addition, Martín and Ferdy, it was great to see how you extended the scope of tameness and complexity. Lars and Luuk, thank you for the fun interactions and canonical meetings; I hope that one day somebody continues with our idea of compactifying M-theory on the Cantor set. \\

I am grateful for the many people I got to meet at conferences and schools: thank you to all of those who made these events so inspiring and enjoyable. I wish to thank Lou van den Dries; it was a great honour and pleasure to meet one of the founders of o-minimality, and I will never forget how he proudly introduced me as a fellow speaker from Utrecht at the model theory conference in Singapore. 
I would like to thank Gal Binyamini for the insightful discussions, sharing his deep understanding of sharp o-minimality. I thank Dan Freed and Cumrun Vafa for their warm hospitality during my time at Harvard. \\

\vspace{2mm} 

My PhD began with the Solvay school, and these three months formed a special time in my life. Thank you to everyone who made this experience so memorable. 
Jose, Sahaja, and Séb, thank you for the close bond we formed in this time. \\

\vspace{2mm} 

Frank, Lucas, and Sam, thank you for your friendship and the Springer Heist meetings. Our time as master students brought great joy, be it by grinding through innumerable homework sets or virtual fishing, and formed the foundation of my PhD. \\

\vspace{2mm} 

Leon and Moniek, Thyrza and Fabian, thank you for warmly welcoming me into the Kors family and immediately making me feel at home, for the many enjoyable gatherings and dinners, and for the wonderful ski trips. \\

\vspace{2mm}

Levi, I am grateful to have you as my brother, and that we understand each other better and better as we get older.
After I had finally finished this thesis, we traveled to Golfe-Juan, where we spent much of our childhood, and this nostalgic trip meant a lot to me.  I'm proud of you.  \\

\newpage 
I am incredibly grateful for my best friends Jordi, Noah, Quint, and Wietze. 
We have shared countless phases of our lives, travels around the world, years of layered jokes that nobody else would understand, and unforgettable moments.
You ensure that I always have people to laugh with and things to look forward to.
We made so many memories that I could write a book longer than this thesis; thank you for everything. \\

A special thanks goes to my parents, Ronald and Joan. As time passes, I realize ever more that I am where I am because of you. You raised me with an incredible amount of love and dedication, and it is for this reason that I was able to develop myself and uncover my potential as a person. I am forever grateful for this. While I spend most of my time in Utrecht, visiting you in Nijverdal always makes me feel like home. \\

Finally, a special thanks goes to Isolde. Meeting you has made my life more \mbox{wonderful} and colorful in a way that I can hardly express in words. Thank you for always patiently listening to me, for embracing me as I am, for the amazing \mbox{adventures} we went on, for the conversations that make anything interesting, and for the many special moments we shared. 
You are the love of my life, and I look forward to everything that lies ahead of us.

{
\renewcommand{\chapterformat}{\relax} 
\addchap{About the author}
}
\setlength{\parindent}{0pt}
Mick van Vliet was born in July 1998
and grew up in Nijverdal. He obtained his high school diploma cum laude from Reggesteyn Noetsele in 2016. In that year he started studying Applied Physics and Applied Mathematics at the University of Twente, and he graduated cum laude from both programs in 2019. His bachelor thesis on topological insulators was supervised by Geert Brocks and Matthias Schlottbom. He then moved to Utrecht to continue his studies in Theoretical Physics and Mathematical Sciences at Utrecht University. Under the supervision of Thomas Grimm and Gil Cavalcanti, he wrote his master thesis on tame geometry and the string landscape, and graduated both master programs cum laude in 2022. \\

Afterwards, he continued his research at the interface of tame geometry and physics as a PhD candidate supervised by Thomas Grimm. This thesis presents the results of the research conducted in this time, and was successfully defended on the 11th of September 2026, after which his PhD was awarded cum laude. Aside from physics and mathematics, he is passionate about bouldering, playing piano, and chess. \\

\cleardoublepage
{
\renewcommand{\chapterformat}{\relax} 

\makeatletter \renewcommand*{\bib@heading}{%
\clearpage \thispagestyle{\chapterpagestyle}%
\global\@topnum\z@ \@@makechapterhead{\bibname}%
\markboth{\bibname}{\bibname}%
} \makeatother

\addcontentsline{toc}{chapter}{Bibliography}

\bibliographystyle{JHEP}

\bibliography{references.bib}
}

\backmatter

\end{document}